\documentclass[twocolumn,aps,pra,nofootinbib,superscriptaddress,10pt,floatfix]{revtex4-2}
\setcitestyle{numbers, round, square}

\usepackage[english]{babel}
\usepackage[utf8]{inputenc}
\usepackage{amsmath}
\usepackage{amssymb}
\usepackage{amsthm}
\usepackage{braket}
\usepackage{dsfont}
\usepackage{bbold}
\usepackage{nicefrac}
\usepackage{siunitx}
\usepackage{bm}

\usepackage{array} 
\usepackage{booktabs} 
\usepackage{diagbox}
\usepackage{graphicx}
\usepackage{makecell}
\usepackage{multirow} 
\usepackage{multirow}
\usepackage{tikz}
\usetikzlibrary{patterns}

\usepackage[linesnumbered,ruled,vlined,algo2e]{algorithm2e}
\usepackage{appendix}
\usepackage{etoolbox}
\usepackage{lipsum}
\usepackage[newcommands]{ragged2e}
\usepackage[normalem]{ulem}
\usepackage{xcolor}

\usepackage[linesnumbered,ruled,vlined]{algorithm}
\usepackage{algpseudocode}

\makeatletter
\newenvironment{protocol}[1][htb]{%
    \renewcommand{\ALG@name}{Protocol}

    \begin{algorithm}[#1]%
    }{\end{algorithm}
}
\makeatother

\usepackage[font=small,labelfont=bf,justification=justified]{caption}
\usepackage{float}
\usepackage{subfigure} 
\usepackage[subfigure]{tocloft}

\usepackage[linktocpage]{hyperref} 
\hypersetup{
  colorlinks=true,
  citecolor=teal,
  urlcolor=teal,
  linkcolor=blue,
  linktocpage=true
}

\usepackage{soul}

\newcommand*\diff{\mathop{}\mathrm{d}}

\newcommand{\DE}[1]{\left\{#1\right\}}

\newtheoremstyle{myplain} 
  {}                     
  {}                     
  {\normalfont}         
  {}                     
  {\bfseries}            
  {.}                    
  { }                    
  {\thmname{#1}\thmnumber{ #2}\thmnote{ (#3)}} 

\theoremstyle{myplain}
\newtheorem{thm}{Theorem}[section]

\newtheorem{dfn}[thm]{Definition} 
\newtheorem{exam}[thm]{Example}

\newtheorem{opt}[thm]{Optimization}

\newtheorem{prop}[thm]{Proposition}

\numberwithin{equation}{section}

\newcommand{\addparttoc}[1]{%
    \addtocontents{toc}{\protect\vspace{0\baselineskip}}%
    \addcontentsline{toc}{section}{\protect\numberline{}#1}%
    \addtocontents{toc}{\protect\vspace{0\baselineskip}}%
}

\makeatletter

\renewcommand*\l@section[2]{%
  \vspace{1\baselineskip}%
  \@dottedtocline{1}{0em}{1.5em}{#1}{#2}%
}
\makeatother

\def\be{\begin{equation}}
\def\ee{\end{equation}}

\begin{document}

\title{Quantum Key Distribution with Imperfections: Recent Advances in Security Proofs}

\author{Patrick Andriolo}
\thanks{Corresponding author: \href{mailto:patrick.andriolo@tuwien.ac.at}{patrick.andriolo@tuwien.ac.at}}
\affiliation{TU Wien, Atominstitut \&  Vienna Center for Quantum Science and Technology, Stadionallee 2, 1020 Vienna, Austria}

\author{Esteban Vasquez}
\affiliation{TU Wien, Atominstitut \&  Vienna Center for Quantum Science and Technology, Stadionallee 2, 1020 Vienna, Austria}

\author{Elizabeth Agudelo}
\affiliation{TU Wien, Atominstitut \&  Vienna Center for Quantum Science and Technology, Stadionallee 2, 1020 Vienna, Austria}

\author{Max Riegler}
\affiliation{Quantum Technology Laboratories GmbH,
Clemens-Holzmeister-Str. 6/6, A-1100 Vienna, Austria}

\author{Matej Pivoluska}
\affiliation{Quantum Technology Laboratories GmbH,
Clemens-Holzmeister-Str. 6/6, A-1100 Vienna, Austria}

\author{Gláucia Murta}
\email{glaucia.murta@tuwien.ac.at}
\affiliation{TU Wien, Atominstitut \&  Vienna Center for Quantum Science and Technology, Stadionallee 2, 1020 Vienna, Austria}

\begin{abstract}


In contrast to classical public-key cryptosystems, where the security of encoded messages relies on on computational assumptions, Quantum Key Distribution (QKD) enables two distant parties to establish a shared secret key that, when combined with a one-time pad, provides information-theoretically secure encryption, provided that the QKD protocol is supported by a rigorous security proof. In the last decades, security proofs robust against a wide range of eavesdropping strategies have established the theoretical soundness of several QKD protocols. 
However, most proofs are based on idealized models of the physical systems involved in such protocols and often include assumptions that are not satisfied in practical implementations. This mismatch creates a gap between theoretical security guarantees and actual experimental realizations, making QKD protocols vulnerable to attacks. 
To ensure the security of real-world QKD systems, it is therefore essential to account for imperfections in security analyses. In this article, we present an overview of recent analytical and numerical developments in QKD security proofs, which provide a versatile approach for incorporating imperfections and re-establishing the security of quantum communication protocols under realistic conditions.
\end{abstract}

\maketitle

\begin{center}
\textbf{CONTENTS}
\end{center}
\vspace{-1cm}

{\small
\renewcommand{\contentsname}{}
\hypersetup{linkcolor=teal}
\tableofcontents
}

\section{Introduction}
\label{sec: introduction}

Quantum Key Distribution (QKD) \cite{bennett2014quantum,ekert1991quantum} is a cryptographic primitive that enables two parties, Alice and Bob, to establish a shared secret key with information-theoretic security. 
When this key is used in a one-time pad scheme~\cite{Vernan}, it provides security even against an all-powerful eavesdropper, Eve. 
This stands in contrast to classical cryptographic schemes, which rely on computational assumptions and are therefore vulnerable to retroactive attacks. 
The removal of computational assumptions in QKD, however, comes at a cost: the security proofs now depend critically on the physical properties of the underlying cryptosystem.

As QKD matures as a technology, theoretical protocols derived under idealized assumptions about states, devices, and channels start to meet practical implementations in a wide range of platforms~\cite{Diamanti2016, pirandola2020advances}. 
In practice, however, the assumptions underlying QKD security proofs are rarely satisfied exactly. While theoretical analyses typically model sources and detectors as well‑characterized and memoryless devices obeying fixed quantum descriptions, real implementations inevitably deviate from these idealized models. Such deviations include imperfect phase randomization \cite{nahar2023imperfect}, mode mismatch \cite{marcomini2025loss}, detector inefficiencies and dark counts \cite{trushechkin2022security}, basis‑dependent losses \cite{winick2018reliable}, and other practical imperfections. Although these imperfections may be small, they can invalidate the assumptions of the security proof and open side channels that are not captured by the idealized model. As a consequence, the information‑theoretic security guaranteed at the theoretical level may not hold for the implemented system.

The practical relevance of this mismatch between theory and implementation has been highlighted by a series of experimental quantum‑hacking demonstrations, which exploit device behaviors not accounted for in the security proofs (see~\cite{xu2020secure} for an extensive overview of practical quantum hacking attacks).
A prominent example is the detector control attack based on tailored bright illumination, which enables full key extraction from commercial QKD systems by manipulating avalanche photodiode detectors in regimes not described by the assumed measurement model~\cite{lydersen2010hacking}. Similarly, source vulnerabilities have been demonstrated in realistic implementations, such as the laser‑seed control attack, which exploits imperfect phase randomization in practical laser sources~\cite{Sun2015}. These works illustrate that security vulnerabilities arising due to deviations of the  physical devices from the assumptions made in the idealized protocol represent a real threat to quantum cryptosystems.

The aim of this work is to provide an overview of recent analytical and numerical techniques that allow realistic imperfections to be incorporated into QKD security proofs, thereby bridging the gap
between practical implementations and theoretical models. 
More precisely, the recent analytical developments allow the consideration of finite-sized cryptographic protocols \cite{scarani2008quantum, scarani2008security,dupuis2020entropy,dupuis2019entropy,metger2024generalised,renner2007symmetry,renner2008security} against very general adversary eavesdropping strategies, which previously could be analyzed only in asymptotic conditions. In parallel, numerical approaches appear in the context of reliably and efficiently optimizing entropic quantities relevant for the key rate estimation \cite{winick2018reliable,araujo2023quantum,lorente2025quantum}, a task that usually can only be done without numerics for very symmetric protocols. 
After presenting such tools, we present a table synthesizing security proofs for different protocols which combine these tools to incorporate imperfections in realistic device models.

The manuscript is organized as follows. After the introduction, Sec.~\ref{sec: related works} briefly lists other reviews on QKD and related topics, clarifying which aspects are already well covered in the literature and which ones we focus on here. The remainder of the manuscript is divided into two main parts. \\

\textbf{Part I} (Sections \ref{sec: keyrate}–\ref{sec: device trutability}) is used to provide a general QKD background and to set up the framework in which we later formulate tools for security proofs. In Sec.~\ref{sec: keyrate} we introduce the figure of merit of a QKD protocol. In Sec.~\ref{sec: security} we present a formal definition of security for general QKD protocols in terms of correctness, secrecy, and robustness parameters. In Sec.~\ref{sec: QKD protocols} we model generic QKD schemes as sequences of quantum and classical maps applied to a distributed state, considering measurement, announcement of results, sifting, parameter-estimation, information-reconciliation, and privacy-amplification steps. 
The map-based description is constructed with the purpose of being applied later in the numerical implementations. 
In Sec.~\ref{sec: eavesdropping strategies} we review different eavesdropping strategies (individual, collective, and coherent attacks) and show some of the typical assumptions made about the eavesdropper in modern security proofs, where we also connect Eve's power with the property of a protocol to be modeled in terms of identical and independently distributed states. 
In Sec.~\ref{sec: degrees of freedom} we discuss the use of discrete- and continuous-variable degrees of freedom in QKD with some examples. Finally, in Sec.~\ref{sec: device trutability} we present QKD scenarios according to the level of trust in the devices, covering fully characterized (device-dependent), device-independent (based on nonlocal correlations), and one-sided device-independent (based on steerable states) protocols, showing how quantum correlations in each scenario relate to QKD protocols. \\

\textbf{Part II} (Sections \ref{sec: algorithms}–\ref{sec: connecting}) presents analytical and numerical tools for incorporating imperfections into realistic security proofs for QKD. In Sec.~\ref{sec: algorithms}, we introduce several numerical methods built on top of standard tools of semidefinite programming for estimating asymptotic key rates for characterized scenarios, and explain how they relate to the map-based description of Sec.~\ref{sec: QKD protocols}. 
In Sec.~\ref{sec: finite keys} we examine analytical techniques for finite-key proofs, with particular emphasis on the application of the quantum de Finetti theorem for QKD (the postselection technique), entropic uncertainty relations, and entropy accumulation theorems. Finally, in Sec.~\ref{sec: connecting} we illustrate how the analytical and numerical tools can be combined in concrete examples to obtain realistic key-rate estimates under experimental imperfections. \\

Four appendices complement the main text: Appendix~\ref{sec: entropic quantities} summarizes the entropic quantities used throughout the text and fixes their notation. Appendix~\ref{app: secretkey} provides details on the derivation of the secret key length. Appendix~\ref{sec: basic optimizations} recalls basic notions of convex and semidefinite programming that underpin the numerical methods discussed in Part II, and \ref{sec: DI methods} details how optimizations for the asymptotic key rate are performed in the device-independent case.

\section{Related work}
\label{sec: related works}

There are many review papers covering different facets of QKD. 
For a general introduction to the subject, the reader may consult the books \cite{wolf2021quantum, vidick2023introduction}, as well as the broad overviews \cite{scarani2009security, gisin2002quantum, pirandola2020advances,Diamanti2016, zhang2025towards,xu2020secure,sena2025tutorial}
The formal definition of security of quantum key distribution and its composable treatment are discussed in \cite{renner2008security,portmann2014cryptographic, portmann2022security, ferradini2025defining}.
Self-contained security proofs for various QKD protocols can be found in \cite{tomamichel2017largely, Wiesemann2024, Mizutani2025,tupkary2025qkd,tupkary2026rigorous}. 
On the side of different trust models, device-independent protocols and their implementation are broadly discussed in \cite{ghoreishi2025future, arslan2025device, primaatmaja2023security, buhrman2009non, murta2019towards}, and measurement-device-independent variants are reviewed in \cite{xu2014measurement}. 
Continuous-variable QKD protocols are overviewed in \cite{usenko2025continuous,juvencio2025digital,wang2025advances,
laudenbach2018gaussianMod,diamanti2015distributing,anka2025introductoryreviewtheorycontinuousvariable}, and recent progress in multipartite QKD (often referred to as conference key agreement) is detailed in \cite{murta2020quantum}. 
On the experimental side, recent developments are summarized in \cite{Diamanti2016,pirandola2020advances,lo2014secure,chou2023satellite,xu2020secure}. 
The numerical approaches for evaluating key rates are built on top of optimization techniques, which are discussed in detail in  \cite{zhou2022numerical,tavakoli2024semidefinite,skrzypczyk2023semidefinite, boyd2004convex}. 
Finally, some specific protocols receive special attention due to their practical relevance, such as decoy-state QKD which are discussed in details in~\cite{tupkary2025qkd,wang2005review}. 

In contrast to the above works, our focus is on reviewing recent numerical and analytical techniques for QKD security analysis, with an emphasis on how they can be combined in practice. We explain the underlying methods, highlight their respective strengths and limitations, and detail works that integrate them to obtain robust security proofs for realistic protocols, including finite-size effects and experimental imperfections under general attacks.

\begin{center}
\vspace{5mm}
\textbf{PART I: BACKGROUND}
\addparttoc{Part I: Background}
\vspace{5mm}
\end{center}

In this first part, we detail the steps of a general quantum key distribution protocol, introduce the fundamental concepts underlying different QKD protocols, and define the figure of merit -- the secret key rate -- that quantifies a protocol's efficiency.
We will also address how different characteristics of secret communication scenarios (e.g., the power of eavesdroppers and the trust level of authenticated parties over their laboratories) modify the evaluation of key rates; provide a rigorous definition of security for general protocols; and distinguish how the use of different degrees of freedom affects the key rates. 
The definition of different entropic quantities used in this section can be found in Appendix \ref{sec: entropic quantities}.

\section{Key rates} 
\label{sec: keyrate}

The objective of a QKD protocol with $n$ rounds is to enable two authenticated parties, Alice and Bob, whose systems are denoted by $A$ and $B$, to obtain identical and private strings, $K_A$ and $K_B$, of length $\ell$ bits. 

The performance of a QKD protocol is commonly quantified by the \emph{secret key rate},  defined as
\begin{align}\label{eq:keyrate}
    r\equiv \frac{\ell}{n}.
\end{align}
The parameter $r$ in Eq.~\eqref{eq:keyrate} is measured in bits/round and indicates the average number of secret bits generated per protocol iteration. 
Since quantum states are generally a limited and non-free resource, it is often useful to consider a more practical figure of merit that accounts for the generation rate of states $\Gamma$ (rounds/second) when comparing protocols implemented in different platforms. 
In this case, the key rate in bits per second is given by 
\begin{align}
    r=\Gamma \frac{\ell}{n} \text { bits} / \mathrm{s}.
\end{align}
In order to benchmark the potential of a QKD protocol, one often considers the \emph{asymptotic key rate}
\begin{align}
\label{eq: asymptotic key rate}
    r_\infty \equiv \lim_{n\rightarrow\infty} \frac{\ell (n)}{n},
\end{align}
which is typically much simpler to compute and quantifies the asymptotic behavior of the protocol.
In this limit, convoluted entropic quantities reduce to single-round expressions, making the security analysis considerably simpler than in the finite-size scenario. 

In Sec.~\ref{sec: finite keys} we will see that the more realistic treatment of finite-sized keys can be approached with different techniques. These tools can also be combined with numerical techniques consisting of efficient and reliable optimization methods (Sec.~\ref{sec: algorithms}) in order to benchmark the performance of a wide range of QKD protocols.

\section{Formal definition of security}
\label{sec: security}

The concept of ``security'' of a certain protocol against an adversary, an eavesdropper which is located and can act outside the laboratories of Alice and Bob (the so-called public domain), is often informally referred but can be rigorously defined. 
The purpose of this section is to provide the definition for $\varepsilon$-security of a QKD protocol, which gives us a quantitative measure for how close a protocol is from an idealized description of the task~\cite{wolf2021quantum,ferradini2025defining}.

The two main properties of interest in a QKD protocol are  \textit{correctness} and  \textit{secrecy}. 
Correctness quantifies the similarity of the keys $K_A$ and $K_B$ that the authenticated parties have at the end of the protocol. 
Ideally, both parties should hold identical bit strings at the end of a QKD protocol. 
However, due to possible failures during its execution, Alice and Bob may, with a small probability, end up with distinct keys. 
This motivates the following definition of $\varepsilon_\text{c}$-correctness.

\noindent
\dfn{\textbf{($\boldsymbol{\varepsilon}_\text{c}$-correctness)}} {We denote by $\Omega$ the event that a certain QKD protocol does not abort. 
The protocol is said to be $\varepsilon_\text{c}$-correct, if the probability that the keys possessed by Alice and Bob---$K_A$ and $K_B$--- differ is smaller than $\varepsilon_\text{c}$, i.e.
\begin{align}\label{eq: correcntess}
    p((K_A \neq K_B) \wedge \Omega) \leq \varepsilon_\text{c}.
\end{align}} 
Secrecy, on the other hand, accounts for the knowledge the eavesdropper, Eve, may have about the final key. 
This can be quantified by the trace distance between the ideal state---when a potential eavesdropper is completely uncorrelated with the final key---and the final state that characterizes the systems at the end of a QKD protocol.

\noindent
\dfn{\textbf{($\boldsymbol{\varepsilon}_\text{s}$-secrecy)}} {Let $p(\Omega)$ denote the probability that a QKD protocol does not abort and $\varepsilon_\text{s}$ be a real parameter in $ [0,1]$. A protocol is said to be $\varepsilon_{\text{s}}$-secret\footnote{In older definitions of security, instead of considering the uncorrelation between quantum states, this property was defined in terms of distances between the probability distributions associated with $K_A$ and $E$. This definition has been proven not to be \textit{composable} (i.e. protocols which individually are secure, when composed, result in a global unsecure protocol) \cite{ferradini2025defining}.} if
\begin{align}\label{eq: secrecy def}
p(\Omega) \cdot\left\|\rho_{K_A E \vert \Omega}-2^{-\ell}\mathbb{1}_{K_A} \otimes \rho_{E \vert \Omega}\right\|_{\mathrm{Tr}} \leq \varepsilon_\text{s},
\end{align}
where $\Vert \cdot \Vert_\text{Tr}$ denotes the trace distance, $\mathbb{1}_{K_A}= \sum_k\ket{k}\bra{k}_{K_A}$ is the identity operator in the space of strings $K_A \in \{0,1\}^{\ell}$.} \\

The definition of secrecy provided above quantifies the distance between the state $\rho_{K_A E \vert \Omega}$---in which an attack has been performed, so that Eve's system is generally correlated with Alice's string---and the ideal state where the string $K_A$ is completely uncorrelated with the eavesdropper, who has no information about the key---yielding the separability across the partition $K_A: E$, represented by the uncorrelated state $\mathbb{1}_{K_A} \otimes \rho_{E \vert \Omega}$.

The conditioning $E \vert \Omega$ in Eq. (\ref{eq: secrecy def}) is important for the emphasis that we only ask for secrecy when the protocol actually outputs a key\footnote{
In fact, both conditions, Eq.~\eqref{eq: correcntess} and Eq.~\eqref{eq: secrecy def}, arise from a single requirement: that the final state of the protocol be close to an ideal state, $\|\rho_{K_AK_BE}-\sigma^{\rm ideal}_{K_AK_BE}\|_\text{Tr}$.
The two conditions, \eqref{eq: secrecy def} and~\eqref{eq: correcntess}, are obtained by decomposing this trace-distance expression into separate contributions and by taking into account that, for the event where the protocol aborts, the corresponding terms vanish. We refer the reader to~\cite{ferradini2025defining} for a recent discussion of security definitions.}. 
In other words, we compare the real state of Eve conditioned on the protocol not aborting with an ideal state in which her system is completely uncorrelated with Alice’s key. 
Outside $\Omega$, no key is generated, and therefore there is no secrecy condition to be satisfied\footnote{Alternatively, one can think that secrecy is trivially satisfied when a key of size zero is generated.}. 

As we will see later, the parameters of correctness and secrecy are determined by classical processing steps of a QKD protocol (see Protocol \ref{alg:QKD steps}). Furthermore, correctness and secrecy compose the security of a QKD protocol.

\noindent
\thm{\label{thm: security} 
If a protocol is $\varepsilon_{\text{c}}$-correct and $\varepsilon_{\mathrm{s}}$-secret, then it is $\varepsilon$-secure with $\varepsilon=\varepsilon_{\text{c}}+\varepsilon_{\text{s}}.$
}\\

\noindent
\textbf{Proof.} For a proof of Theorem \ref{thm: security} the reader can check \cite{ferradini2025defining, wolf2021quantum, renner2008security}. \\

In addition to security, another requirement for a useful QKD protocol is \textit{robustness} (also denoted \textit{completeness}). 
A given protocol may be provably secure, however, if all practical implementations lead to a high probability of aborting, the protocol becomes of limited usefulness in practice, as it rarely produces a usable secret key. 
This aspect is quantified through the notion of protocol robustness, defined as follows.

\noindent
\dfn{\textbf{($\boldsymbol{\varepsilon_\text{r}}$-robustness).}} {A QKD protocol is $\varepsilon_\text{r}$-robust if there exist a physical implementation in which, in the absence of an eavesdropper, the probability that the protocol aborts is at most $\varepsilon_\text{r}$.} \\

\section{Quantum key distribution}
\label{sec: QKD protocols}

 In this section, we detail the steps of a standard QKD protocol, see Protocol~\ref{alg:QKD steps}. To be precise, we restrict ourselves to the class of protocol in which the secret key is generated by measurements of individual distributed quantum systems, followed by classical post-processing of the data\footnote{There are other types of protocols that differ substantially from this blueprint, e.g., by distilling entanglement and then generating a key, or the protocols where the measurement basis itself encode the key bits \cite{abushgra2022variations,sun2024enhancing,du2024advantage}.}. The individual building blocks of the protocol will be described next.
 The specific subsections with the detailed description are indicated in the protocol outline.\\

\begin{protocol}[H]
\caption{Phases of QKD}\label{alg:QKD steps}
\begin{algorithmic}[1]
\State \textbf{Distribution of states:} A source is used to send states to Alice and Bob (Sec. \ref{sec: Distribution of states});
\State \textbf{Measurements:} Alice and Bob randomly select bases to measure their share of the distributed state (Sec. \ref{sec: measurements});  
\State \textbf{Sifting:} Measurement basis is announced and the parties discard rounds in which different measurement bases were used (Sec. \ref{sec: sifting});  
\State \textbf{Parameter estimation:} Alice and Bob estimate the amount of errors in their strings (Sec. \ref{sec: parameter estimation});  
\State \textbf{Information reconciliation:} Classical error correction is performed to ensure that Alice and Bob have identical keys (Sec. \ref{sec: information reconciliation}); 
\State \textbf{Privacy amplification:} The partially secret key is distilled to a smaller, more secret string (Sec. \ref{sec: privacy amplification}).
\end{algorithmic}
\end{protocol} 
\vspace{0.25cm}

The ``quantum phase" of Protocol~\ref{alg:QKD steps} consists of the first two steps.
The remaining steps involve only classical post-processing of the string of bits generated by the measurement outcomes in the quantum phase.
These classical steps are however crucial to ensure security of a QKD protocol.

Earlier developments of the modern security-proof framework employed in this work~\cite{renner2008security,tomamichel2017largely} established the relationship between the security parameters (defined in Sec.~\ref{sec: security}), the total number of rounds $n$ in the distribution, the measurement phases of the protocol, and the length $\ell$  of the secret key that can be distilled from a QKD protocol of the form described in Protocol~\ref{alg:QKD steps}.

As a general statement, we can say that a QKD protocol of the form of Protocol~\ref{alg:QKD steps} establishes an  $\varepsilon_\text{IR}$-correct and an ($\varepsilon_\text{PA}+2\epsilon$)-secret key of size
\begin{align}
\label{eq: key general formula}
        \ell=H_{\min }^\epsilon\left(A_1^n \vert E\right)_\rho- \vert \text{leak}_\text{IR} \vert - \mathcal{O}\left(\log \varepsilon_\text{PA}^{-1}\right).
\end{align}
Here the parameter $\varepsilon_\text{IR}$ is determined by the specific information reconciliation protocol used in step 5, and $\vert \text{leak}_\text{IR}\vert$ is the respective information leaked by Alice in this step, in order to correct the keys. 
The term $\mathcal{O}\left(\log \varepsilon_\text{PA}^{-1}\right)$ is a penalty associated with the privacy amplification step (step 6). 
More details on the derivation of the Eq.~\eqref{eq: key general formula} and the relation to the security parameters can be found in Appendix~\ref{app: secretkey}.

Moreover, in the asymptotic limit $n\rightarrow \infty$ of infinitely many  independent and identically distributed (iid) repetitions of the state distribution and measurement, we can derive 
a simple expression for the \emph{asymptotic key rate} in terms of conditional entropies of a single round. The so called Devetak-Winter formula \cite{devetak2005distillation} relates the information of Alice's key possessed by Eve and Bob:
\begin{align}
    r_\infty = H(A\vert E) - H(A \vert B).
\end{align}

To construct the quantum phase of Protocol~\ref{alg:QKD steps}, i.e. distribution and measurement, we will follow the approach introduced in \cite{winick2018reliable}, where it is shown that QKD protocols based on \textit{fully characterized} scenarios can be synthesized in terms of very general maps acting over the distributed states. Our description of the quantum phase will be performed in terms of individual rounds - assuming so-called independent and identically distributed (iid) states -, while more general (non-iid) rounds will be explained in Sec.~\ref{sec: eavesdropping strategies}. We also start our discussion about general QKD protocols assuming characterized devices for each party (and other device trustability levels will be detailed through Sec.~\ref{sec: device trutability}). The maps describing the quantum phase of QKD, which will be constructed along Sections \ref{sec: Distribution of states} and \ref{sec: measurements}, will afterwards be used for numerical optimization algorithms for asymptotic key rate calculations (Sec.~\ref{sec: algorithms}).

\subsection{Distribution of states}
\label{sec: Distribution of states}

QKD protocols of the form described in Protocol~\ref{alg:QKD steps} can be broadly classified into two categories according to the method used to prepare and distribute quantum states:  prepare-and-measure (PM) schemes and entanglement-based (EB) schemes. 
In PM scenarios, one of the authenticated parties, Alice, prepares a (typically idealized) single-party quantum state and sends it through an insecure quantum channel to Bob. 
In contrast, in EB scenarios, a source---possibly located in the public domain and thus potentially controlled by the eavesdropper---distributes entangled states to both Alice and Bob.
The signals that reach Alice and Bob in practice differ from the idealized states. 
In QKD it is considered that deviations from an idealized pure state are caused by the disturbances introduced by the eavesdropper in her interaction with the insecure quantum channel (PM scheme) -- see Fig.~\ref{fig:PM protocol}, or the preparation of the quantum state (EB scheme) -- see Fig.~\ref{fig:EBprotocol}.

\subsubsection{Prepare-and-measure (PM) protocols}
\begin{itemize}
    \item[(PM)] Alice prepares a quantum state randomly chosen from the set $\{\ket{\phi_{a,x}} \}$ according to probability distribution $\{p(a, x) \}$. 
    The information of the produced state ($a$ and $x$) is classically stored, and the prepared states are sent to Bob through the quantum channel. 
    This is repeated $n$ times. 
\end{itemize}

\begin{center}
\begin{figure}[t]
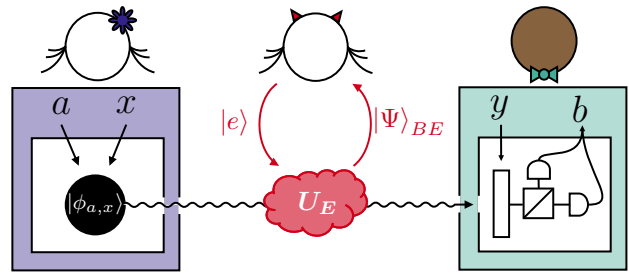


  
\tikzset {_rrp6a3dx5/.code = {\pgfsetadditionalshadetransform{ \pgftransformshift{\pgfpoint{0 bp } { 0 bp }  }  \pgftransformrotate{-90 }  \pgftransformscale{2 }  }}}
\pgfdeclarehorizontalshading{_72akbtq1l}{150bp}{rgb(0bp)=(1,1,1);
rgb(37.5bp)=(1,1,1);
rgb(50bp)=(1,1,1);
rgb(62.5bp)=(1,1,1);
rgb(100bp)=(1,1,1)}
\tikzset{every picture/.style={line width=0.75pt}} 


    
    \caption{Distribution of states in a prepare-and-measure  QKD protocol. In a PM scenario, Eve intercepts the signal $\ket{\phi_{a,x}}$ emitted by the trusted source in Alice's laboratory. The malicious party can couple an auxiliary state $\ket{e}$ to it, and through an unitary operation $U_{E}$ it acquires information carried in $\ket{\Psi}_{BE}$.}
    \label{fig:PM protocol}
\end{figure}
\end{center}

Therefore, in a generic PM protocol, Alice encodes the bit $a$ and the preparation basis $x$ in a state $\ket{\phi_{a,x}}$  with probability $p(a, x)$. The description of the state sent through the channel for a fixed preparation basis $x$, $\rho_{x}^\text{PM}$, is then given by 
\begin{align}\label{eq: state 1 pm}
    \rho_{x}^\text{PM}
    =\sum_{a} p(a\vert x) \ket{a, x}\bra{a, x}_A \otimes \ket{\phi_{a, x}} \bra{\phi_{a, x}}_B .
\end{align}
The state $ \ket{\phi_{a, x}}$ is what leaves Alice's laboratory, but it may be disturbed by interactions with the eavesdropper in the insecure quantum channel. Eve may apply an arbitrary unitary operation $U_E$ that coherently couples the system sent by Alice to an auxiliary system under her control ($\ket{\Psi}_{BE}=U_E(\ket{\phi_{a,x}}\otimes \ket{e})$), see Fig.~\ref{fig:PM protocol}.

As an example, we can consider the PM BB84 protocol~\cite{bennett2014quantum}:
\noindent
\exam{\textbf{(Distribution of states in PM-BB84)}} {In a PM description for the BB84 protocol, $x=0$ labels quantum states prepared in $Z$ basis (the parameter $a=0,1$ determines the states $\ket{\phi_{0,0}}=\ket{0}$ and $\ket{\phi_{1,0}}=\ket{1}$, respectively), while $x=1$ corresponds to encoding in $X$ basis (with $\ket{\phi_{0,1}}=\ket{+}$ and $\ket{\phi_{1,1}}=\ket{-}$).} \\

\subsubsection{Entanglement-based (EB) protocols}
\begin{itemize}
        \item[(EB)] The entanglement-based realization consists of a source---potentially in control of Eve---distributing an entangled state to Alice and Bob.
        This is repeated $n$ times.
\end{itemize}

\exam{\textbf{(Distribution of states in EB-BB84)}} {In an EB description of BB84, in the ideal honest case, a source distributes a bipartite maximally entangled state (a Bell pair) to Alice and Bob, for instance, the state
\begin{align}\label{eq: bell pair}
    \ket{\Phi^+}= \frac{1}{\sqrt{2}}(\ket{00}+\ket{11}).
\end{align}}

\begin{center}
\begin{figure}[t]
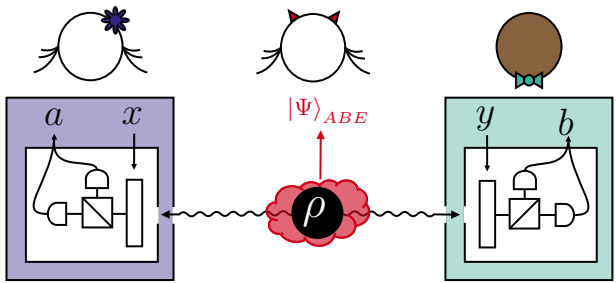


  
\tikzset {_t7b1fjjz9/.code = {\pgfsetadditionalshadetransform{ \pgftransformshift{\pgfpoint{0 bp } { 0 bp }  }  \pgftransformrotate{-90 }  \pgftransformscale{2 }  }}}
\pgfdeclarehorizontalshading{_zj3qso3u2}{150bp}{rgb(0bp)=(1,1,1);
rgb(37.5bp)=(1,1,1);
rgb(50bp)=(1,1,1);
rgb(62.5bp)=(1,1,1);
rgb(100bp)=(1,1,1)}

  
\tikzset {_lv9wban71/.code = {\pgfsetadditionalshadetransform{ \pgftransformshift{\pgfpoint{0 bp } { 0 bp }  }  \pgftransformrotate{-90 }  \pgftransformscale{2 }  }}}
\pgfdeclarehorizontalshading{_ekj74iyrh}{150bp}{rgb(0bp)=(1,1,1);
rgb(37.5bp)=(1,1,1);
rgb(50bp)=(1,1,1);
rgb(62.5bp)=(1,1,1);
rgb(100bp)=(1,1,1)}
\tikzset{every picture/.style={line width=0.75pt}} 


  
    \caption{Distribution of states in an EB-QKD protocol. 
    Eve controls both the source and the quantum channel and is therefore modeled as holding the purification $\ket{\Psi}_{ABE}$ of the quantum state $\rho$, the resulting distributed states to Alice and Bob.}
    \label{fig:EBprotocol}
\end{figure}
\end{center}

As in the PM scheme, the signal received by Alice and Bob in the EB description may be a mixed state $\rho$. The resulting state is mixed due to interactions with the environment, which in the worst case may correspond to actions of Eve in the source and the quantum channel (Fig.~\ref{fig:EBprotocol}).

In the following, we will denote by $M_{a_x \vert x}$ ($N_{b_y \vert y}$) the positive operator-valued measure (POVM) used by Alice (Bob) when an observable indexed by $x$ ($y$) leads to an outcome $a_x$ ($b_y$)\footnote{Note that we keep the measurement index as a subscript in the outcome label. This will be useful when presenting the numerical security proof techniques in Sec. \ref{sec: algorithms}, since we will introduce maps that need to make explicit which input $x$ leads to a certain output $a_x$, demanding the notation $M_{a_x\vert x}$ instead of the more usual $M_{a\vert x}$.}. Let us remember that a POVM is the most general mathematical description of a quantum measurement, represented by a set of operators which associate a matrix to the pair $(a_x,x)$ satisfying
\begin{align}
    M_{a_x\vert  x} \geq 0  \quad \forall \quad  a_x,x; \quad \sum_{a_x} M_{a_x\vert x} = \mathbb{1} \quad \forall \quad x.
\end{align}

The result of measuring a quantum system and recording the classical outcome is described by a classical-quantum state  (cq-state) (cf. Appendix \ref{sec: entropic quantities} for a comprehensive description of cq-states). 
In order to describe such a situation, a measurement described by a set of POVM elements $\left\{M_{a_x\vert x}\right\}$ paired with the classical register that records the respective inputs and outputs characterizes a \textit{measure-and-prepare channel}, defined by:
\begin{align}\label{eq: MP map}
\Lambda_\text{MP}(\rho) &\equiv\sum_a\ket{a,x}\bra{a, x} 
\,\operatorname{Tr} (M_{a_x\vert x} \rho M_{a_x\vert x}^{\dagger}) \\
&=\sum_a p(a\vert x)\ket{a, x}\bra{a, x} 
\end{align}
where 
$p(a\vert x)=\text{Tr}(M_{a_x\vert x} \rho M_{a_x\vert x}^{\dagger})$. 

In contrast with Eq. (\ref{eq: state 1 pm}), after the source $S$ distributes the state $\rho$ and one of the parties (let's say, Alice) performs a measurement labelled by $x$, the state of Alice and Bob is described by
\begin{align}\label{eq: map rho 1 eb}
    \rho_x^\text{EB} =  \sum_{a}p(a|x)\ket{a,x}\bra{a,x}_A \otimes \rho_B^{a|x},
\end{align}
where $\rho_B^{a|x}=\operatorname{Tr}_A (\ket{a,x}\bra{a,x} \otimes \mathbb{1}_B \cdot \rho )$.

The most general way to model the action of Eve in an EB protocol is to give her access to the purification of the state $\rho$ that is distributed to Alice and Bob, see Fig.~\ref{fig:EBprotocol}. 
Therefore, the distribution of a single round can be described by a \textit{pure} tripartite state $\ket{\Psi}_{ABE}$, where $E$ denotes the auxiliary system held by the eavesdropper. In this situation, the mixed state shared among the two authenticated parties is $\rho = \operatorname{Tr}_E(\ket{\Psi}\bra{\Psi}_{ABE})$. 
The numerical approaches described in Sec.~\ref{sec: algorithms} allow the constraints imposed by the observed statistics and the collective key-rate expression (Eq.~(\ref{eq: devetak-winter})) to be written entirely in terms of the bipartite density matrix $\rho$ and the classical registers generated during the protocol (Eq.~(\ref{optimization alpha})).

\subsubsection{PM-EB equivalence}

In general, PM protocols can be converted into EB schemes~\cite{bennett1992quantum}. 
This is especially useful for the security analysis because the description of Eve in an EB scheme is simply given by the purification of the distributed state, rather than by modelling the arbitrary unitary operations she could apply in a PM scenario (see Figs.~\ref{fig:PM protocol} and~\ref{fig:EBprotocol}).
Some PM protocols (e.g. the BB84~\cite{bennett2014quantum} and the six-state protocol~\cite{bruss1998optimal}) have well-known equivalent EB schemes, where the preparation of states by Alice is equivalent to Alice measuring a maximally entangled state. 
A general condition for such type of equivalence is given by the following proposition.

\prop{\textbf{(Condition for EB protocol implementation)}} {\label{thm: PM to EB}Consider a PM protocol in which Alice encodes the value $a$ by preparing the state $\left|\phi_{a, x}\right\rangle$  with probability $p(a \vert x)$, where $x$ is the basis used in the preparation. 
The protocol admits an EB implementation where the honest source distributes the following entangled state 
\begin{align}\label{eq:PM-EB1}
        \ket{\psi}=\sum_a \sqrt{p(a|0)}\ket{a}_A\otimes \ket{\phi_{a,0}}_B,
    \end{align}
if it holds that:
\begin{align}\label{eq:PM-EB}
\sum_a p(a \vert x)\ket{\phi_{a, x}}\bra{\phi_{a, x}}=\sum_a p\left(a \vert x^{\prime}\right)\ket{\phi_{a, x^{\prime}}} \bra{\phi_{a, x^{\prime}}} \forall x, x^{\prime}.
\end{align}
\label{prop: equivalence between PM and EB protocols} \\
\textbf{Proof.} A proof of Proposition \ref{thm: PM to EB} can be found in \cite{wolf2021quantum}. \\

Some protocols may not satisfy the condition given by Eq.~\eqref{eq:PM-EB}; however, an EB description should still be possible since entanglement is a necessary condition for the existance of a secure key~\cite{curty2004entanglement}. Indeed, we can always define a (perhaps more artificial) EB scheme associated to any PM protocol by using a  source-replacement scheme that considers a physical system in higher dimension. In general the state describing the distributed system in a PM protocol when Alice prepares states in the basis labelled by $x$, Eq.~\eqref{eq: state 1 pm}, can be described by Alice and Bob sharing the entangled state 
\begin{align}\label{eq: expanded state sourcerep}
    \ket{\psi} = \sum_{a,x} \sqrt{p(a, x)}\ket{a,x}_{A} \otimes \ket{\phi_{a,x}}_{B},
\end{align}
and Alice measuring part of the subsystem $A$ in the basis $\{\ket{x}\bra{x}\}$. 
Here $A$ is Alice's internal system---which cannot be accessed by the eavesdropper and contains the registers of the classical data used to select the emitted state---while the signal that leaves her laboratory and will be sent to Bob lives in the system $B$. 
This source-replacement scheme for a general PM protocol will be used in the numerical techniques described in Sec. \ref{sec: algorithms}.

In practice, PM protocols are often easier to implement, as they do not require the generation and distribution of entanglement, which can significantly limit performance due to low generation rates. 
Consequently, in long-distance scenarios (e.g., earth-satellite communication \cite{chou2023satellite, yin2017satellite, liao2017satellite}), PM protocols are generally a more feasible option.
We note, however, that despite the PM–EB equivalence discussed above, entanglement-based protocols can offer stronger security guarantees. 
In particular, the EB description equivalent to a PM protocol (see Eq.~\eqref{eq:PM-EB1} and Eq.~\eqref{eq: expanded state sourcerep}) relies on a precise characterization of the state-preparation device, which is assumed to generate a single-particle state $\ket{\phi_{a,x}}$ to be sent through the quantum channel. 
In photonic implementations, this assumption translates into the requirement of an ideal single-photon source---an approximation that is only imperfectly realized in practice and that opens the door to potential security loopholes. 
To address such imperfections, the decoy-state method, discussed in Sec.~\ref{sec:decoy}, was introduced in~\cite{lo2005decoy}. 
More generally, security threats arising from source imperfections can be mitigated by fully EB implementations, which relax assumptions on state preparation and thereby strengthen security.

From now on, unless otherwise stated, we will focus on EB protocols. In remaining of this section, we work explicitly with the (mixed) bipartite state distributed to Alice and Bob in an EB protocol, and omit the register $E$ of the eavesdropper, with all channels implicitly understood acting as the identity on $E$. 

\subsection{Measurements}
\label{sec: measurements}

After the state distribution, Alice and Bob measure their share of the distributed quantum state with their respective POVMs, $\DE{M_{a_x\vert x}}$ and $\DE{N_{b_y\vert y}}$. The experimental data produced in the measurement phase consist of the joint probabilities 
\begin{align}\label{eq: probabilities EB}
p(ab\vert xy) = \text{Tr} \left( M_{a_x\vert x} \otimes N_{b_y\vert y} \, 
\rho\right).
\end{align}

In order to model the measurement process in a coherent way (i.e. keeping a pure state representation of the total state of Alice, Bob and Eve), we  can enlarge the Hilbert space of the protocol, which until now consisted of $\mathcal{H}_A \otimes \mathcal{H}_B\otimes \mathcal{H}_E$, to include registers that store information about the measurement bases and outcomes. 
The new parts of the total Hilbert space represent classical registers\footnote{A register here should be understood as a unit that temporarily stores information to be used for processing. 
In the context of QKD, a quantum state is often referred to as a quantum register.}, in a way that at the end of the protocol the information publicly available to an eavesdropper is composed of $\tilde{A}\tilde{B}$, where $\tilde{A}$ ($\tilde{B}$) labels the basis measured by Alice (Bob). 
The specific outcome of each measurement is stored in the register $\overline{A}$ ($\overline{B}$), containing values that remain unknown to Eve. The Kraus operators \cite{nielsen2010quantum} -- objects representing measurements, announcement of inputs, and registration of outputs -- acting on the enlarged Hilbert space are given by
\begin{subequations}
\begin{align}\label{eq: kraus operators}
K_x^A &=\sum_{a_x} \sqrt{M_{a_x \vert x}} \otimes|x\rangle_{\widetilde{A}} \otimes\left|a_x\right\rangle_{\overline{A}}, \\
K_y^B &=\sum_{b_y} \sqrt{N_{b_y \vert y}} \otimes|y\rangle_{\widetilde{B}} \otimes\left|b_y\right\rangle_{\overline{B}}.
\end{align}
\end{subequations}
With this modelling, the post-measurement state is synthesized by the map
\begin{align}\label{eq: map A}
\begin{aligned}
\mathcal{M}(\rho)  & =\sum_{x,y}(K_x^A \otimes K_y^B) \rho(K_x^A \otimes K_y^B)^{\dagger}\in \mathcal{H}_{A\tilde{A}\overline{A}B\tilde{B}\overline{B}}.
\end{aligned}
\end{align}

\exam{\textbf{(BB84 Kraus operators)}} {As Alice and Bob measure the same observables $X$ and $Z$, their POVMs are projectors onto the computational and Hadamard bases. 
In the projective case, eaadd{for Alice,} $ \sqrt{M_{a_x \vert x}} =  M_{a_x \vert x}$ \ and similarly for Bob.
In the Kraus-operator description Eq.~\eqref{eq: kraus operators}), measurement, announcement, and registration of Alice’s outcome in the basis $x=0,1$ are explicitly given by
\begin{subequations}
\begin{align}\small
  K^A_0
  &= \ket{0}\bra{0} \otimes \ket{0}_{\tilde A} \otimes \ket{0_0}_{\overline{A}}
   + \ket{1}\bra{1} \otimes \ket{0}_{\tilde A} \otimes \ket{1_0}_{\overline{A}}, \label{eq:BB84_KA0} \\
  K^A_1
  &= \ket{+}\bra{+} \otimes \ket{1}_{\tilde A} \otimes \ket{0_1}_{\overline{A}}
   + \ket{-}\bra{-} \otimes \ket{1}_{\tilde A} \otimes \ket{1_1}_{\overline{A}}, \label{eq:BB84_KA1}
\end{align}
\end{subequations}
and similarly for Bob.
The first factor in each term acts on the physical qubit, the middle factor stores the publicly announced basis, and the last factor stores the private measurement outcome.}

The efficiency of the protocol can be improved, without affecting its security, by choosing the measurement bases with a biased distribution (i.e. different measurement bases are chosen with different probabilities). 
In practice, this means that one basis is selected with higher probability and is used in the majority of rounds. 
The less-likely basis is primarily used for parameter estimation (Sec. \ref{sec: parameter estimation}), while the rounds measured in the preferred basis are kept for key generation. 
For this reason, these are commonly referred to as the \emph{test basis} and the \emph{key-generation basis}, respectively.

\subsection{Sifting}
\label{sec: sifting}

Alice and Bob must now announce their measurement choices in order to be able to sift the rounds in which they chose different measurement settings. 
This consists of making the registers $\tilde{A}$ and $\tilde{B}$ publicly available, and therefore accessible to the eavesdropper Eve.
The sifting phase involves both parties discarding rounds according to a predefined strategy. 

For instance, in BB84, the discarded rounds are those  in which the parties measured different bases. In order to formalize this idea, we consider a set $\widehat{AB}$ consisting of all measurement choices that are kept. 
If we want to take the state with announcement information $\mathcal{M}(\rho)$ and select the kept announcements, then we define a projector $\Pi$ acting on the expanded Hilbert space $\mathcal{H}_{A\tilde{A}\overline{A}B\tilde{B}\overline{B}}$ selecting only these results. 
The projector is explicitly given by
\begin{align}\label{eq: map Pi}
    \Pi=\sum_{(x, y) \in \widehat{AB}} \ket{x}\bra{x}_{\widetilde{A}} \otimes \ket{y} \bra{y}_{\widetilde{B}}.
\end{align}
\noindent
The action of the sifting operation, defined above, in an arbitrary state $\sigma$ is
\begin{align}\label{eq: sifting map}
    \sigma \mapsto \frac{\Pi \sigma \Pi}{p_{\text {pass }}} \in \mathcal{H}_{A\tilde{A}\overline{A}B\tilde{B}\overline{B}}.
\end{align}
Where $p_\text{pass}$ refers to the probability of $\sigma$ passing the sifting phase and is given by
\begin{align}\label{eq: sifting probability}
    p_\text{pass} = \operatorname{Tr} (\Pi \sigma \Pi).
\end{align}

In order to register the information obtained so far in the quantum phase, we define a function mapping the triple composed by Alice's basis $x$; her announcement $a_x$; and Bob's measured basis $y$ into the alphabet $\mathcal{X}$ of the protocol, i.e.
\begin{align}
    g: \widetilde{A}\times\bar{A}\times \widetilde{B} &\rightarrow \mathcal{X}, \\
    (x, a_x, y) &\mapsto \{ 0,\dots, \vert \mathcal{X}\vert -1 \}.
\end{align}
This key map is stored in a (quantum) register $R$, whose configuration is incorporated in the density matrix of the system with the action of an isometry $V$ defined as
\begin{align}\label{eq: map V}
    V=\sum_{x, a_x, y}\ket{g\left(x, a_x, y\right)}_R \otimes\ket{x}\bra{x}_{\widetilde{A}} \otimes \ket{ a_x}\bra{a_x}_{\bar{A}} \otimes 
    \ket{y} \bra{y}_{\widetilde{B}}.
\end{align}
Under the action of the isometry above, an arbitrary state $\sigma$ is mapped into
\begin{align}
\label{eq: rho4}
    \sigma \mapsto V \sigma V^{\dagger} \in \mathcal{H}_{R A \widetilde{A} \bar{A} B \widetilde{B} \bar{B}}.
\end{align}

\exam{\textbf{(BB84 sifting and key map)}} For the standard EB BB84 protocol, Alice and Bob keep only those rounds in which they measure in the same basis. Hence, the set of kept announcements is$\widehat{AB} = \{(0,0),(1,1)\}$, and the sifting projector in Eq.~\eqref{eq: map Pi} becomes
\begin{align}
  \Pi_{\mathrm{BB84}}
  &= \ket{0}\bra{0} \otimes \ket{0}\bra{0}
   + \ket{1}\bra{1}\otimes \ket{1}\bra{1}.
\end{align}
The key alphabet is $\mathcal{X}=\{0,1\}$, and the key map is chosen such that Alice’s raw key bit is simply her measurement outcome (whenever the bases match).
Explicitly,
\begin{align}
  g(0,0_0,0) &= 0, & g(0,1_0,0) &= 1, \nonumber\\
  g(1,0_1,1) &= 0, & g(1,1_1,1) &= 1,
\end{align}
Furthermore, The corresponding isometry $V$ storing the raw key in register $R$ (Eq.~\eqref{eq: map V}) takes the form 
\begin{widetext}
\begin{align}
\begin{aligned}
  V_{\mathrm{BB84}}
  &= \ket{0}_R \otimes
     \Big(
        \ket{0}\bra{0}_{\tilde{A}} \otimes \ket{0_0}\bra{0_0}_{\overline{A}} \otimes \ket{0}\bra{0}_{\tilde{B}}
      + \ket{1}\bra{1}_{\tilde{A}} \otimes \ket{0_1}\bra{0_1}_{\overline{A}} \otimes \ket{1}\bra{1}_{\tilde{B}}
     \Big) \\
  &\quad
   + \ket{1}_R \otimes
     \Big(
        \ket{0}\bra{0}_{\tilde{A}} \otimes \ket{1_0}\bra{1_0}_{\overline{A}} \otimes \ket{0}\bra{0}_{\tilde{B}}
      + \ket{1}\bra{1}_{\tilde{A}} \otimes \ket{1_1}\bra{1_1}_{\overline{A}} \otimes \ket{1}\bra{1}_{\tilde{B}}
     \Big).
\end{aligned}
\end{align}
\end{widetext}
After the combined action of the measurement map $\mathcal{M}$, the sifting projector $\Pi_{\mathrm{BB84}}$ and the isometry $V_{\mathrm{BB84}}$, the key information is stored in the classical register $R$. 

The action of the maps defined so far to describe the measurements ($\mathcal{M}$, Eq. (\ref{eq: map A})), sifting\footnote{More recent numerical security proofs eschew $\Pi$, which we maintain here for completeness. In later papers, the rounds that are not kept for the key are mapped to a special symbol $\perp$. This is because one needs to worry about the privacy amplification claim if the raw key length can vary.} ($\Pi$, Eq. (\ref{eq: map Pi})) and key storage ($V$, Eq. (\ref{eq: map V})) can be conveniently composed into a single CPTP map $\mathcal{G}$ defined as
\begin{align}\label{eq: map G}
    \mathcal{G}(\sigma) \equiv V \Pi \mathcal{M}(\sigma) \Pi V^{\dagger}.
\end{align}
At this point, the key-mapped information lives in the quantum register $R$. 
In order to consider the entire classical bit string in subsequent phases (parameter estimation, information reconciliation, and privacy amplification), we need to model the transition from the quantum register to a classical storage $Z^{R}$. 
This can be done through the creation of a cq-state, in which the classical part $Z^{R}$, is correlated with the quantum part $A \widetilde{A} \bar{A} B \widetilde{B} \bar{B}$. 

Following \cite{winick2018reliable}, we introduce an operation that removes all coherences in the classical register while leaving the rest of the subsystems untouched. 
The classical register obtained from $R$ is modelled through the action of a ``pinching channel'' \cite{tomamichel2015quantum}
\begin{align}\label{eq: map Z}
    &\mathcal{Z}: \mathcal{H}_{R A \widetilde{A} \bar{A} B \widetilde{B} \bar{B}} \rightarrow \mathcal{H}_{Z^R A \widetilde{A} \bar{A} B \widetilde{B} \bar{B}} \\
    &\sigma \mapsto \sum_j (| j \rangle \langle j | _ { R } \otimes \mathbb { 1 } ) \sigma \left(|j\rangle\left\langle j\right|_R \otimes \mathbb{1}\right)
\end{align}
which selects only the key-mapped entries and organizes them in a cq-density matrix.

As mentioned before, the maps $\mathcal{G}$ and $\mathcal{Z}$ (constructed initially in \cite{winick2018reliable}) are of particular importance, as they will allow us to write the calculation of key rates as tractable optimization problems in terms of relative entropies (Sec. \ref{sec: algorithms}). 
At the end of the sifting phase, the state shared among the parties is
\begin{align}
    \mathcal{Z}(\mathcal{G}(\rho)) \in \mathcal{H}_{Z^{R}A\tilde{A}\overline{A}B\tilde{B}\overline{B}}
\end{align}

\subsection{Parameter estimation}
\label{sec: parameter estimation}

So far, we have been describing the state of a  single-round. 
The global space $\mathcal{H}_1^{n}$ of all $n$ rounds of the protocol is
\begin{align}
    \mathcal{H}_1^{n} = \bigotimes_{i=1}^{n} \mathcal{H}_{Z^{R}_iA_i\tilde{A}_i\overline{A}_iB_i\tilde{B}_i\overline{B}_i} .
\end{align}
For the purpose of classical postprocessing, from now on we will use the notation $A_1^{n} \equiv (A_1, \dots, A_n)$ and $B_1^{n} \equiv (B_1, \dots, B_n)$ for the strings possessed by Alice and Bob, respectively.
In order to determine the quality of the data $A_{1}^{n}$ and $B_{1}^{n}$ contained in the state after sifting, $\mathcal{Z}(\mathcal{G}(\rho))$, the \emph{parameter estimation} step is performed. 
Two general ways to perform parameter estimation of finite-size QKD protocols have been considered in recent security proofs: \textit{fixed-} \cite{wolf2021quantum} and \textit{variable-length} \cite{tupkary2024security,ben2005universal,hayashi2012concise} protocols. 

\subsubsection{Fixed-length protocols}

In fixed-length finite-size QKD protocols, Alice and Bob decide in advance what will be the length of the final key, and parameter estimation is used to estimate if a certain test is passed or not. 
Operationally, they look at the sampled data and either accept or abort if the observed statistics are not ``good enough'' (i.e. the number of errors is consistent with an eavesdropper attack leaking a lot of information). 
In this setting, parameter estimation is naturally phrased as a hypothesis test (or threshold check): from a random sample of disclosed symbols, they use concentration bounds, e.g. Chernoff-Hoeffding bounds \cite{mannalath2025sharp, hoeffding1963probability} --- for instance the one provided by Serfling's inequality --- to upper bound the relevant global error or other, more fine-grained parameters and compare them to an abort threshold. 

\exam{\textbf{(Serfling's inequality \cite{wolf2021quantum})} \label{exam: serfling}} {Given a set of binary random variables $\{x_i\}_{i=1,\dots,n}$ with arithmetic average
\begin{align}
    \overline{x}=\frac{1}{n} \sum_{i=1}^n x_i,
\end{align}
a sample (without replacements) $\{x^{\prime}_j\}_{j=1}^m \subseteq \{x_i\}_{i=1}^n$ of size  $m< n$ has its average given by
\begin{align}
    \overline{x}^{\prime}=\frac{1}{m} \sum_{j=1}^n x^{\prime}_j.
\end{align}
Considering the real parameter $\beta \in [0,1]$, 
$\overline{x}$ and $\overline{x}^{\prime}$ satisfy
\begin{align}\label{eq: serfling bound}
    p(\overline{x}^{\prime} \geq \overline{x} +\beta) \leq \text{exp}\left(-\frac{2 \beta^2 nm}{n-m+1}\right).
\end{align}
} \\
In a QKD protocol, $\overline{x}$ corresponds to the (unknown) QBER over all sifted rounds, while $\overline{x}'$ is the observed QBER on the $m$ test rounds used in parameter estimation to infer the global QBER's.

\subsubsection{Variable-length protocols}

By contrast, variable-length protocols treat the observed data as input to a key-length function: rather than a binary ``pass/abort'' decision with a fixed output size.
They choose the final key length adaptively as a function of the measured statistics, with length $0$ included as a legitimate outcome for the case where the estimated parameters show that no secret key can be obtained. 
From the parameter-estimation perspective, this means the finite-statistics machinery is used to build a confidence region for different possible underlying parameters, but instead of only certifying pass (or abort), it directly determines how much secret key can be safely extracted. 

The parameters estimated during this step of a QKD protocol serve two distinct purposes:
(i) some are used to determine the amount of information that must be exchanged to successfully correct the raw key during the information reconciliation step, and
(ii) others are used to bound the information that may be available to Eve.
It is important to distinguish between these two types of estimated parameters. 
The second type is security-critical, whereas the first mainly concerns the engineering of an efficient information reconciliation scheme. 
If the latter is inaccurately estimated, the primary consequence is reduced efficiency or a failed reconciliation, which is detected during the error-correction verification step, as will be  discussed in the next section. An incorrect estimation of Eve’s information, however, directly leads to a secrecy failure.

\subsection{Information reconciliation}
\label{sec: information reconciliation}

Information reconciliation (IR) is the terminology in QKD for the use of a (classical) error-correction method that turns the strings $K_A$ and $K_B$ into identical strings, which is why it is also referred to as the error-correcting phase. 
Here, we focus on the particular case considered in most QKD protocols: one-way information reconciliation.

By now, the classically registered information in $\overline{A_{1}^{n}}$ and $\overline{B_{1}^{n}}$ is partially secret and partially correlated. Therefore, it is necessary to ensure that both parties possess identical strings through error correction. 
In order to perform such a task, Alice and Bob will communicate information about their data to correct their strings. 
If the process is performed in such a way that one of the strings remains fixed, say Alice's, and only Bob changes his string according to the communicated information, then this is called a ``one way" process\footnote{In a two-way QKD error correction process, both Alice and Bob change their strings and one looses the asymptotic bound on the minimal leak being $H(\overline{A}|\overline{B})$. 
The reader interested in two-way error correction may check \cite{tupkary2023using,gottesman2003proof}.}.

A key tool for ensuring the success of error correction --- and, as we will see, of privacy amplification as well --- is the use of \textit{hash functions}. 
This kind of function maps input strings of length $n$ to output strings of length  $\ell\leq n$.
A particular characteristic of the used hash functions must be the property of \textit{2-universality}.

\dfn{\textbf{(2-universal hash functions)}} {A family of hash functions $\mathcal{F}=\{f: \left.\{0,1\}^n \rightarrow\{0,1\}^{\ell}\right\}$ is called 2-universal if for every two strings $X_1^{n}, Y^{ n}_{1} \in\{0,1\}^n$ with $X_1^{n} \neq Y^{n}_{1}$ we have
\begin{align}
p\left(f(X_1^{n})=f(Y^{n}_{1})\right) \leq 2^{-\ell} ,
\end{align}
where $f$ is \textit{chosen uniformly at random in} $\mathcal{F}$.} \\

The property of 2-universality essentially guarantees a good distribution of outputs. 
The existence of the 2-universal family of hash functions is secured for all $\ell < n$ \cite{carter1977universal,wegman1981new}, and the uniformly random choice of hash functions in $\mathcal{F}$ ensures that the information reconciliation does not depend on computational assumptions regarding the one-wayness of a specific $f\in \mathcal{F}$.

The information publicly communicated during the information reconciliation step ($\text{leak}_\text{IR}$), used by Bob to evaluate whether his guess of Alice's string is correct, is typically composed by \textit{syndromes} (computed from Alice's string), and we denote them as $\text{synd}(\overline{A_{1}^{n}})$. 
An ideal information reconciliation protocol, i.e. one that leaks the minimal information, is described as Protocol~\ref{alg:info-recon}.

\begin{protocol}[H]
\caption{Information reconciliation}\label{alg:info-recon}
\begin{algorithmic}[1]
\Require $\text{synd}(\overline{A_{1}^{n}})$, $\overline{B_{1}^{n}}$, $\epsilon_{\text{IR}} \in \mathbb{R}$.
\Ensure Proceed or abort protocol
\State Alice sends $\text{synd}(\overline{A_{1}^{n}})$ to Bob.\\
Using $\overline{B^{n}_{1}}$ and $\text{synd}(\overline{A_{1}^{n}})$, Bob computes a guess $\overline{B^{n}_{1}}^{\prime}$ for Alice's string.\\
Alice choses a function $f_{\rm IR}$ from a two-universal family of hash funcations and communicates $f_{\rm IR}$ and $f_{\rm IR}(\overline{A_{1}^{n}})$ such that $\log \left( \epsilon_{\text{IR}}^{-1} \right) = \vert f_\text{IR}(\overline{A_{1}^{n}}) \vert $ and sends it to Bob. \\
Bob checks if $f_{\rm IR}(\overline{B_1^{n}}^{\prime}) = f_{\rm IR}(\overline{A_1^{n}})$ \\
\eIf{$f_{\rm IR}(\overline{B_1^{n}}^{\prime}) = f_{\rm IR}(\overline{A_1^{n}})$}{
    \Return \text{proceed protocol}
}{
    \Return \text{abort protocol}
    
}
\end{algorithmic}
\end{protocol}

The highlight of Protocol~\ref{alg:info-recon} is in step 2: the eavesdropper does gain some information through the reconciliation process, but this information only allows Alice  and Bob  to recover a common key because their strings are already highly correlated.
It was shown in \cite{renner2005simple,brassard1993secret} that the minimum leakage $\text{leak}_\text{IR}$ of a one-way information reconciliation protocol is bounded by 
\begin{align}
\label{eq: minimum leak 2}
\begin{aligned}
    \vert \text{leak}_\text{IR} \vert  \leq\, &H_{\max }^{\nicefrac{\epsilon_{\text{IR}}^{\prime}}{2}}(\overline{A_1^n} \vert \overline{B_1^n})\,+ \\
& \log \left(\frac{8}{\epsilon_{\mathrm{IR}}^{\prime}{ }^2}+\frac{2}{2-\epsilon_{\mathrm{IR}}^{\prime}}\right)+\log \left(\epsilon_{\mathrm{IR}}^{-1}\right),
\end{aligned}
\end{align}
where $H_\text{max}^{\varepsilon}$ is the smooth max-entropy (defined in Eq. (\ref{eq: def max entropy}), Appendix \ref{sec: entropic quantities}).  

In practice, the public information exchanged during information reconciliation can be generated by different families of error-correction procedures. From the viewpoint of the security proof, what matters is that the total communicated information is properly included in $|\text{leak}_{\mathrm{IR}}|$, and that the residual disagreement probability is bounded by $\epsilon_{\mathrm{IR}}$. A general figure of merit that can be used to benchmark different error correction codes is the \textit{error correction efficiency}, defined as \cite{mueller2025performance}
\begin{align}
    f \equiv \frac{\vert \text{leak}_\text{IR} \vert}{n H(A \vert B)}.
\end{align}
The numerator quantifies the practical amount of information to perform the information reconciliation phase. The denominator, on the other hand, corresponds to the minimum amount of information which ideally would be leaked and is a consequence of the so-called Slepian-Wolf bound, whose derivation can be found in \cite{slepian2003noiseless} and corresponds to the asymptotic limit of Eq. (\ref{eq: minimum leak 2}) (see Section \ref{sec: collective attacks} for more details about the limit of asymptotic regime). This makes $f$ to be generally lower bounded by unit, and any practical error correction code has values of $f>1$. Below we provide two examples of common error correction codes for information reconciliation \cite{mueller2025performance}.

\exam{(\textbf{LDPC codes} \cite{martinez2012blind}) A standard one-way information reconciliation protocol is obtained with low-density parity-check (LDPC) codes. In this case, Alice computes the syndrome $\mathrm{synd}(\overline{A_1^{N}}^{\prime}) = H \overline{A_1^{m}}^{\prime}$, where $H$ is a sparse parity-check matrix of dimensions $N \times m$ (where $N\geq n$ is the number of columns and $m\geq n$ is the number of rows) and $\overline{A_1^{m}}^{\prime}$ is an extended string - an expanded raw key - whose expanded bits are filled randomly or optimized beforehand, and sends it to Bob through the classical channel. Using $B_1^n$ and $\mathrm{synd}(\overline{A_1^{N}}^{\prime})$, Bob computes the guess $\overline{B}_1^{\prime n}$ for Alice's string. LDPC codes can be tailored for different values of \textit{coding rates}, denoted by $R$ and given by
\begin{align}
    R \equiv 1- \frac{m}{N}.
\end{align}
The error correction efficiency of an LDPC code with a certain coding rate generally depends on $N$, $m$ and the QBER of the protocol \cite{mueller2025performance}
\begin{align}
    f = \frac{m}{N h(\text{QBER})}.
\end{align}
}}

\exam{(\textbf{Cascade} \cite{brassard1993secret}) In this case, the inputs for information reconciliation are Alice's and Bob's strings and the estimated QBER's. Instead of sending a single syndrome $\mathrm{synd}(\overline{A_1^{n}})$, Alice and Bob iteratively reveal parities of blocks of their strings (with length defined according to the QBER), and Bob updates his current guess $\overline{B}_1^{\prime n}$ by locating errors through binary searches. Despite being a iterative protocol with information exchanged by both Alice and Bob, Cascade is still considered a one-way information reconciliation protocol. Indeed in the Cascade, the string of Alice remains fixed, and only Bob corrects his sequence. In practice, Cascade is efficient because it can achieve leakage close to the ideal limit, at the cost of requiring many rounds of classical communication. To the date, the best achievable error correction efficiency for a (modified version of) Cascade protocol is $f = 1.025$ \cite{pacher2015information}.

\subsection{Privacy amplification}
\label{sec: privacy amplification}

The final step of a general QKD protocol consists of converting the (ideally) identical strings possessed by Alice and Bob --- partially known by Eve --- into smaller strings which are completely unknown by the malicious party \cite{bennett1988privacy}. 
The use of 2-universal hash functions by the authenticated parties, similar to the ones used in the information reconciliation step, allows an exponential reduction in the knowledge that Eve has about the key. 
This fact is ensured by the \emph{quantum leftover hashing lemma} (which is discussed in more detail in Appendix \ref{app: secretkey}). 
The final keys shared by Alice and Bob are given, respectively, by
\begin{align}
  K_{A} = f_\text{PA}(\overline{A_1^{n}})\;,\; K_{B} = f_\text{PA}(\overline{B_1^{n}}').
\end{align}

\begin{protocol}[H]
\caption{Privacy amplification}\label{alg:PA}
\begin{algorithmic}[1]
\Require $\overline{A_{1}^{n}}$, $\overline{B_1^{n}}^{\prime}$, $\ell$.
\Ensure $K_A$, $K_B$
\State 
Alice choses a function $f_\text{PA}$ from a two-universal family of hash funcations and communicates  $f_{\rm PA}$ to Bob.
\State Alice sets $K_A= f_\text{PA}(\overline{A_{1}^{n}})$ and Bob sets $K_B= f_\text{PA}(\overline{B_1^{n}}^{\prime})$ such that $\vert f_\text{PA}(\overline{A_{1}^{n}}) \vert = \ell$.
\end{algorithmic}
\end{protocol}

The parameters in information reconciliation and privacy amplification define the penalty terms associated to the final distilled key of length $\ell$ according to Eq. \eqref{eq: key general formula}. 
We provide more details regarding the derivation of the secret key in Appendix \ref{app: secretkey}.

\subsection{Assumptions in QKD}
\label{sec: final remarks}

Several tacit assumptions underlie the different steps of a QKD protocol described in the previous sections. 
Ensuring that these assumptions hold in an actual implementation is crucial for guaranteeing the security of the generated key.
In this section, we list the assumptions inherent in a QKD protocol. 
We note that Assumptions 2–5 are common to any cryptosystem, including classical ones. 
The strength of QKD lies in the fact that, with only these minimal assumptions, one can, in principle, achieve information-theoretic security, in contrast to the classical case. 
Assumption 6 is specific to QKD, and it is precisely here that potential security vulnerabilities may arise. 
In section~\ref{sec: device trutability}, we discuss how this last assumption can be relaxed in realistic scenarios involving imperfect devices.

\begin{enumerate}
    \item \textbf{Correctness and completeness of Quantum Theory:} Security proofs for QKD strongly rely on the principles of quantum theory (such as linearity and hermiticity) to be not only correct, but also complete. 
    An incomplete theory could allow an eavesdropper to exploit unknown behaviours of quantum systems to breach the security;
    
    \item \textbf{Authenticated classical channel:} Alice and Bob have access to a public but authenticated\footnote{The possibility of authentication itself is an underlying assumption in QKD \cite{tupkary2026authentication}.} classical channel. 
    This ensures that all classical messages exchanged during the protocol can be publicly read but cannot be modified or forged by Eve.

    \item \textbf{Isolated labs}: Alice and Bob must guarantee that no information is leaked from their laboratories, except for what is specified in the protocol. 
    The possibility of unintended information leak can allow an eavesdropper to broadcast (in real time, or \textit{a posteriori}) information that is thought by Alice and Bob to be secure;

    \item \textbf{Random number generators}: Alice and Bob should have access to sources of randomness to implement their measurements (which could be even generated with quantum resources \cite{ma2016quantum}). 
    Otherwise, an eavesdropper can exploit sequences that look random, but are deterministic from her perspective.

    \item \textbf{Trusted classical post-processing}: The classical post-processing of the raw key -- parameter estimation, error correction, and privacy amplification --  is assumed to be trusted.    

    \item \textbf{Trusted devices}: Devices of Alice and Bob are capable of implementing correctly the devices (or state preparation) specified in the protocol. 
    This assumption is dropped
    in device-independent scenarios, as we will describe in Sec. \ref{sec: device trutability}.
\end{enumerate}

\section{Eavesdropping strategies}
\label{sec: eavesdropping strategies}

An adversary interested in eavesdropping on the secret communication established between Alice and Bob can act according to different possible strategies \cite{ekert1991quantum}. A general attack\footnote{By ``attack" we refer to operations that the eavestropper can perform striclty in the public domain - the authenticated clasical channel or the insecure quantum channel -, while being unable to disable, damage or incapacitate the devices of the authenticated parties.  Strategies in which Eve can make these devices not to work according to their expected behaviour characterize the so called \textit{quantum hacking}. Examples of these attacks can be found in \cite{lydersen2010hacking, jain2014trojan, gerhardt2011experimentally, jain2014risk, bugge2014laser}} can be described in an EB perspective considering that Eve holds a purification of the distributed state among the honest parties (as depicted in Fig. \ref{fig:EBprotocol}), whose mixedness is considered to be a consequence of the correlation of the eavesdropper itself with the signal. This means that any deviation from the idealized state is introduced by disturbances of the malicious party, whose information is then collected by the eavesdropper by performing a POVM $\{Z_e\}$, with a particular outcome $e$ and with corresponding probability $p(e) = \text{Tr} (Z_e \rho_E)$. If the eavesdropper is assumed to have memory, it can also perform transformations in its marginal states before the measurement.

In what follows we characterize the possible strategies of Eve according to (i) the amount of emitted states that she can interact simultaneously with and (ii) her capacity to manipulate or store the information acquired by them in one or many rounds). These distinctions account for the so-called \textit{individual}, \textit{collective} and \textit{coherent} attacks, which are also synthesized in Table \ref{tab:attacks_comparison_compact}

\subsection{Individual attacks}
\label{sec: individual attacks}

These are the weakest attacks that can be performed, and are characterized by an Eve that always has the same purifications, and as a consequence the distribution of states between the authenticated parties is identical in every round, corresponding to the iid $n$-copy $\rho_{AB}^{\otimes n}$. Therefore, she can only interact \textit{individually} with each distributed signal and measuring her side information and recording its outcome
\begin{align}\label{eq: individual attack 1}
        \rho_E^{\otimes n}=\text{Tr}_{AB}( \ket{\Psi}\bra{\Psi}_{ABE}^{\otimes n}).
\end{align}
In individual attacks, the eavesdropper has no access to a quantum memory \cite{rosset2018resource}, so operations over her side channel are restricted by local operations (measurements or local unitaries) $\mathcal{E}^{(i)} \in \mathcal{H}_{i = 1,\dots, n}$ for every round indexed by $i$:
\begin{align}\label{eq: individual attack 2}
    \rho_E^{(i)} \mapsto \mathcal{E}_i( \rho^{(i)} ) \quad \forall  \quad i = 1,\dots , n.
\end{align}
Security against these types of attacks was first established in \cite{slutsky1998security}. 

\subsection{Collective attacks}
\label{sec: collective attacks}

In this situation, Eve is still restricted to work round-by-round (so that distributed states are also iid), but now is capable of doing global measurements or operations in her quantum side information. This provides Eve with information in the form of a state similar to the one encountered in individual attacks, Eq. (\ref{eq: individual attack 1}), but allowed to perform global operations, for example arbitrary global unitaries or measurements $\mathcal{E} \in \mathcal{H}^{\otimes n}$:}
\begin{align}
   \rho_E^{\otimes n} \mapsto \mathcal{E}( \rho_E^{\otimes n}).
\end{align}

Importantly, the iid assumption of collective attacks allows us to use the \textit{quantum asymptotic equipartition property} (AEP) to derive a simple expression for the key rate \cite{tomamichel2009fully, cover1999elements}. 
The AEP essentially allows us to treat both the smooth min- and max-entropies as homogeneous with respect to the number of rounds. 
In other words, if an iid distribution $\rho_{AE}^{\otimes n}$ is employed in the protocol, then the smooth entropies are given by the product of the number of rounds and the single-round conditional entropy $H(A \vert E)$. 
This property is then useful to, for instance, derive a simple formula to compute the asymptotic key rate of a QKD protocol (Eq. (\ref{eq: asymptotic key rate})) in terms of entropic quantities --- this is known as the Devetak-Winter formula 
\cite{devetak2005distillation}:
\begin{align}
\label{eq: devetak-winter}
    r_{\infty}=H(A \vert E)_\rho-H(A \vert B)_\rho .
\end{align}

\thm{\textbf{(Quantum asymptotic equipartition property, Theorem 1 of \cite{tomamichel2009fully})} \label{thm: AEP}} {For an iid distribution of $n \geq \frac{8}{5} \log(2 \epsilon^{-2})$ rounds, the following inequality 
holds
\begin{align}
H_{\min }^\epsilon\left(A_1^n \vert E_1^n\right)_{\rho_{A E}^{\otimes n}} \geq n H(A \vert E)_{\rho_{A E}}-\sqrt{n} \delta\left(\epsilon, \eta_{A E}\right), \\
H_{\max }^\epsilon\left(A_1^n \vert E_1^n\right)_{\rho_{A E}^{\otimes n}} \leq n H(A \vert E)_{\rho_{A E}}+\sqrt{n} \delta\left(\epsilon, \eta_{A E}\right),
\end{align}
where $\delta\left(\epsilon, \eta_{A E}\right)=4\left(\log \eta_{A E}\right) \sqrt{\log (2\epsilon^{-2})}$ and $\eta_{A E}=\sqrt{2^{-H_{\min }(A \vert E)_\rho}}+\sqrt{2^{H_{\max }(A \vert E)_\rho}}+1$.} \\

\noindent
\textbf{Proof.} For a proof of the AEP we refer the reader to \cite{tomamichel2009fully}. \\

A particular consequence of the AEP is the limit
\begin{align}
& \lim_{n\rightarrow \infty}H_{\min }^\epsilon\left(A_1^n \vert E_1^n\right)_{\rho_{A E}^{\otimes n}} = H(A \vert E)_{\rho_{A E}},
\end{align}
which ensures that the key rate (Eq. (\ref{eq: key general formula})) of a protocol with infinite rounds can be simplified to the Devetak-Winter formula (Eq.\eqref{eq: devetak-winter}) in terms of the conditional entropy.

In contrast with the assumption of an infinite number of rounds, the AEP starts to give us a hint of what happens in the finite-round regime (aspect which we will explore in more detail in Sec. \ref{sec: finite keys}). 
We can see that for a relatively small number of rounds ($<10^6$), terms of order $\mathcal{O}(\sqrt{n})$ become relevant, so that it is possible that the key vanishes in this regime under noisy parameters. 
Therefore, it is necessary to perform a finite number of rounds in order to allow the distillation of a positive key. 
This increases the difficulty of implementing QKD protocols in practice, as it is often the case that entanglement sources have comparatively low generation rates to support a useful application.

\subsection{Coherent attacks}
\label{sec: coherent attacks}

This is the most general type of attack, in which the states distributed to Alice and Bob can exhibit arbitrary correlations across the different rounds, i.e., the state of $n$ rounds is non-iid $\rho_{A_1^n B_1^n} \neq \rho_{A B}^{\otimes n}$. With this, the global state in the eavesdropper side is
\begin{align}
        \rho_{E_1^{n}}=\text{Tr}_{AB}( \ket{\Psi}\bra{\Psi}_{A_1^nB_1^nE_1^n} ).
\end{align}
As in the collective attack case, Eve can perform an arbitrary global operations $\mathcal{E}$ on her purification of the global state of $n$-rounds,
\begin{align}
   \rho_{E_1^n} \mapsto \mathcal{E}( \rho_{E_1^n}).
\end{align}

\begin{table}
\centering
\caption{Eavesdropping attacks organized according to the marginal state of Eve and its capacity to operate in the side channel.}
\label{tab:attacks_comparison_compact}
\renewcommand{\arraystretch}{3} 
\setlength{\tabcolsep}{3.5pt} 

\begin{tabular}{c c c} 
\toprule
\textbf{Attack} &
\makecell{\textbf{Side state $\rho_{E_1^n}$}} &
\makecell{\textbf{Side} \\  \textbf{operations}} \\
\midrule 

\makecell{\textbf{Individual}}
&
\makecell{ $\bigotimes\limits_{i=1}^n \rho_E^{(i)}$ \vspace{0.1cm}\\ (iid)}
&
\makecell{$\mathcal{E}_i (\rho_E^{(i)})$ \vspace{0.1cm}\\ (local)}
\\[12pt]

\makecell{\textbf{Collective}}
&
\makecell{$\bigotimes\limits_{i=1}^n \rho_E^{(i)}$ \vspace{0.1cm}\\ (iid)}
&
\makecell{$\mathcal{E}(\rho_E^{\otimes n})$ \vspace{0.1cm}\\ (global)}
\\[12pt]

\makecell{\textbf{Coherent}}
&
\makecell{$\mathrm{Tr}_{A_1^nB_1^n}(\ket{\Psi}\bra{\Psi}_{A_1^nB_1^nE_1^n})$ \vspace{0.1cm}\\ (non-iid)}
&
\makecell{$\mathcal{E}(\rho_{E_1^n})$ \vspace{0.1cm}\\ (global)}
\\

\bottomrule
\end{tabular}
\end{table}

From a theoretical perspective, individual, collective and coherent attacks, are the different classes of attacks that Eve can perform.  
To transition from theory to practice, the next section shows different physical degrees of freedom that can serve to encode key bits in a QKD protocol.

\section{Discrete and continuous degrees of freedom}
\label{sec: degrees of freedom}

Real implementations of QKD can be built on two distinct encoding paradigms: discrete-variables (DV) and continuous-variables (CV) \cite{usenko2025continuous}. 
While many QKD protocols use light as the encoding resource, the fundamental distinction lies in which degree of freedom of light is used to encode the secret key.

DV-QKD protocols, such as BB84, typically use the polarization of single photons. 
The detection events are inherently discrete, registering the presence or absence of a photon in a specific state (e.g., a “click" in a detector). 
In contrast, CV-QKD \cite{usenko2025continuous, anka2025introductoryreviewtheorycontinuousvariable} encodes information onto the continuous quadratures of the electromagnetic field (amplitude and phase) of a light wave, using {e.g.} coherent or squeezed states. 
The measurement is performed using {phase-sensitive methods, as} homodyne or heterodyne detection {(also called eight-port homodyne detection)} outcome. 

The use of CV or DV protocols leads to a technological divergence: DV-QKD traditionally relies on single-photon detectors, which can be challenging to operate, while CV-QKD can, in principle, be implemented with standard telecommunications-grade coherent detectors, offering potential cost and integration advantages.
However, CV protocols face their own challenges, such as the need for more complex error correction and a higher  channel transmission {rate} to achieve comparable performance at long distances. 
In what follows, we provide a few examples.

\subsection{Discrete variables}
\label{sec: discrete variables}

\exam{\textbf{Polarization encoding \cite{grunenfelder2018polarization}}}  
This is a canonical and intuitive method for implementing {DV-}QKD protocols like BB84. 
{In its idealized form, information is encoded into perfect single photons through their polarization states, exploiting the quantum-mechanical property that measurements in an incompatible basis disturb the state.} 
The protocol uses two conjugate bases, typically the rectilinear basis (horizontal $\ket{H}$ and vertical $\ket{V}$ polarization) and the diagonal basis (diagonal $\ket{+}$ and anti-diagonal $\ket{-}$ polarization). 
In PM schemes, Alice randomly prepares and sends photons in one of these four states, and Bob measures each incoming photon in a randomly chosen basis. 
Alternatively, in EB schemes, both Alice and Bob perform polarization measurements in the $X$ or $Z$ basis chosen randomly.

\exam{\textbf{Phase encoding} \cite{pathak2023phase}.} {is a primary method for {DV-}QKD that is particularly well-suited for long-distance fiber-optic channels. 
In its fundamental form, information is encoded into the relative phase of light waves. 
A prominent protocol is the Differential Phase Shift (DPS) QKD scheme \cite{inoue25difphase}, where Alice sends a train of coherent pulses and encodes the secret key by applying a random phase shift of $0$ or $\pi$ to each pulse relative to the previous one.
Bob decodes the information using a one-bit-delay Mach-Zehnder interferometer, which causes adjacent pulses to interfere. 
The interference result (constructive or destructive) reveals the phase difference and thus the encoded bit, directing the photon to one of two detectors. 
A key advantage of phase encoding over polarization encoding is its inherent robustness against polarization drift in optical fibers, a major source of instability. Security proofs against coherent attacks for this sort of protocol can be found in \cite{mizutani2023finite,sandfuchs2025security}.}

\exam{\textbf{Time-bin encoding \cite{boaron2018timebin}}} {is a widely used {DV-}QKD technique that is inherently robust for fiber-optic transmission and free space {links} \cite{Tang2023, Cocchi2025}. 
In its fundamental form, information is encoded into the temporal location of a single photon. 
A qubit is defined by two-time slots: an ``early" ($E$) bin and a ``late" ($L$) bin. 
Alice prepares a single photon in a superposition state or in a definite-time-bin state and sends it to Bob. 
For the BB84 protocol, the $Z$ basis states are the classical time bins: $\ket{E}$ for a bit $0$ and $\ket{L}$ for a bit $1$. 
The $X$ basis states are superpositions, such as $\ket{+} = (\ket{E} + \ket{L})/\sqrt{2}$, created using an interferometer to generate coherence between the two time bins. 
Bob measures in the $Z$ basis by simply detecting the photon's arrival time, or in the $X$ basis using his own interferometer to measure the phase relationship between the bins. 
The key advantage of time-bin encoding is its resilience to polarization fluctuations in optical fibers, as it relies on stable temporal distinctions.}

\subsection{Continuous variables}
\label{sec: cv}

\exam{\textbf{Gaussian-modulated coherent state encoding \cite{grosshans2003gaussianMod}}} In this PM scheme, the sender (Alice) encodes information in the continuous quadratures of a light field. 
Specifically, she prepares coherent states whose complex amplitudes are randomly sampled according to a Gaussian distribution. 
The receiver (Bob) then uses homodyne detection to measure one of the two quadratures at random. 
This method is significant because it achieves security using only standard laser sources and conventional photodetectors, without the need for single-photon detectors or nonclassical light sources such as {those of} squeezed states. 
The Gaussian nature of the modulation is crucial, as it makes Eve's optimal attack a Gaussian iid strategy,
forming the basis for the security analysis.

\exam{\textbf{Discrete Modulation for CV-QKD \cite{ghorai2019discreteMod}}} Discrete Modulation (DM) for PM CV-QKD simplifies the requirement of Gaussian modulation encoding by having the sender (Alice) choose from a small, finite set of coherent states. 
A common and practical implementation is the four-state protocol, also known as quadrature phase-shift keying, where Alice first randomly prepares one of the states $\{\ket{\alpha}, \ket{i\alpha}, \ket{-\alpha}, \ket{-i\alpha}\}$. 
For quadrature measurements, Bob uses homodyne/heterodyne detection to make quadrature measurements \cite{Guo2025}. 
The key advantage of this approach is its experimental simplicity: the modulators required to switch between these few states are less demanding than those needed for the precise, continuous Gaussian modulation required before. 
Furthermore, the subsequent classical error-correction step is more efficient. 

The key challenge for DM CV-QKD is proving security, as Eve’s optimal attacks are harder to model than in Gaussian-modulated protocols. However, recent advances provide evidence that DM protocols can achieve key rates and distances comparable to those of their Gaussian-modulated counterparts \cite{Pan2022, sayat2024}, making them highly attractive and practical candidates for real-world deployment.

\section{QKD with imperfect devices}
\label{sec: device trutability}
In section~\ref{sec: final remarks}, we discussed the different assumptions present in the security proof of a QKD protocol. In particular, standard security proofs rely on the physical properties of the underlying system and therefore depend on a precise description of the devices used in the protocol. However, achieving such precise characterization in practice is challenging, and deviations from the assumed device behavior can create potential vulnerabilities in the protocol.
To maintain the strong, information-theoretic security offered by QKD, there are two main approaches: (i) refine the system description by explicitly accounting for known device imperfections in the security proofs; or (ii) relax assumptions about device characterization, treating the devices in a more black-box manner.

\begin{center}
\begin{figure}[h]
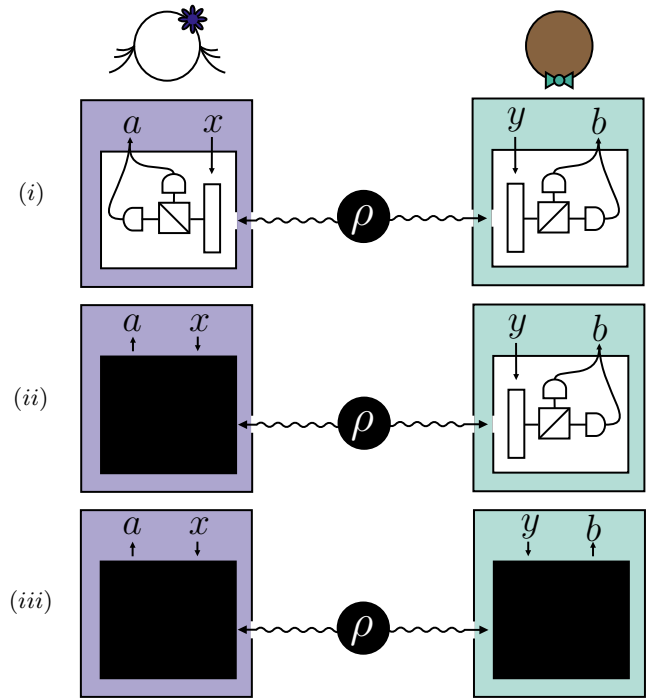

  
\tikzset {_wnu84f5t5/.code = {\pgfsetadditionalshadetransform{ \pgftransformshift{\pgfpoint{0 bp } { 0 bp }  }  \pgftransformrotate{-90 }  \pgftransformscale{2 }  }}}
\pgfdeclarehorizontalshading{_8h7qrw44r}{150bp}{rgb(0bp)=(1,1,1);
rgb(37.5bp)=(1,1,1);
rgb(50bp)=(1,1,1);
rgb(62.5bp)=(1,1,1);
rgb(100bp)=(1,1,1)}

  
\tikzset {_grqjtsvut/.code = {\pgfsetadditionalshadetransform{ \pgftransformshift{\pgfpoint{0 bp } { 0 bp }  }  \pgftransformrotate{-90 }  \pgftransformscale{2 }  }}}
\pgfdeclarehorizontalshading{_lhwr0eoc5}{150bp}{rgb(0bp)=(1,1,1);
rgb(37.5bp)=(1,1,1);
rgb(50bp)=(1,1,1);
rgb(62.5bp)=(1,1,1);
rgb(100bp)=(1,1,1)}

  
\tikzset {_f4b6jck53/.code = {\pgfsetadditionalshadetransform{ \pgftransformshift{\pgfpoint{0 bp } { 0 bp }  }  \pgftransformrotate{-90 }  \pgftransformscale{2 }  }}}
\pgfdeclarehorizontalshading{_pfitp9f23}{150bp}{rgb(0bp)=(1,1,1);
rgb(37.5bp)=(1,1,1);
rgb(50bp)=(1,1,1);
rgb(62.5bp)=(1,1,1);
rgb(100bp)=(1,1,1)}

  
\tikzset {_p7uppw1ny/.code = {\pgfsetadditionalshadetransform{ \pgftransformshift{\pgfpoint{0 bp } { 0 bp }  }  \pgftransformrotate{-90 }  \pgftransformscale{2 }  }}}
\pgfdeclarehorizontalshading{_f5ik2vt4j}{150bp}{rgb(0bp)=(1,1,1);
rgb(37.5bp)=(1,1,1);
rgb(50bp)=(1,1,1);
rgb(62.5bp)=(1,1,1);
rgb(100bp)=(1,1,1)}

  
\tikzset {_2nbl3c28j/.code = {\pgfsetadditionalshadetransform{ \pgftransformshift{\pgfpoint{0 bp } { 0 bp }  }  \pgftransformrotate{-90 }  \pgftransformscale{2 }  }}}
\pgfdeclarehorizontalshading{_42zkxo596}{150bp}{rgb(0bp)=(1,1,1);
rgb(37.5bp)=(1,1,1);
rgb(50bp)=(1,1,1);
rgb(62.5bp)=(1,1,1);
rgb(100bp)=(1,1,1)}

  
\tikzset {_sr4pmsg2q/.code = {\pgfsetadditionalshadetransform{ \pgftransformshift{\pgfpoint{0 bp } { 0 bp }  }  \pgftransformrotate{-90 }  \pgftransformscale{2 }  }}}
\pgfdeclarehorizontalshading{_yd95cwkwv}{150bp}{rgb(0bp)=(1,1,1);
rgb(37.5bp)=(1,1,1);
rgb(50bp)=(1,1,1);
rgb(62.5bp)=(1,1,1);
rgb(100bp)=(1,1,1)}
\tikzset{every picture/.style={line width=0.75pt}} 


    \caption{Characterization of devices in cryptographic scenarios, according to the knowledge of agents about the mechanism of their devices. In $(i)$ the devices of Alice and Bob are assumed to be fully characterized; $(ii)$ relies on the assumption that one of the laboratories cannot be characterized and the devices may be untrusted, being therefore treated as a black-box, and encoded in a steering scenario. 
    The situation $(iii)$ depicts the protocol in which both laboratories are untrusted, matching a Bell scenario.}
    \label{fig:Devices}
\end{figure}
\end{center}

In Sec.~\ref{sec: device dependent protocols}, we provide examples of how specific mathematical models of device imperfections can be incorporated into security proofs, such as accounting for multiple-photon emissions during state preparation or dark counts in the detectors. 
In Sec.~\ref{sec: DI protocols}, we discuss a more drastic approach in which the devices’ trustworthiness is completely relaxed, effectively treating them as black boxes (see Fig.~\ref{fig:Devices}).
In this scenario, the only information available is the devices’ inputs, outputs, and observed statistics, and we detail how such descriptions are formalized. 
Finally, in Sec.~\ref{sec: one-sided device independent protocols}, we examine the intermediate case in which the devices of only one party are treated as untrusted, while the other party’s devices remain characterized.

\subsection{Imperfections in fully characterized protocols}
\label{sec: device dependent protocols}

The situation when the devices of different agents can be assumed perfectly trusted is usually referred to as characterized or device-dependent scenario, as it is possible to make assumptions about the internal mechanisms, degrees of freedom, states and measurements (and their dimensions) together with possible imperfections regarding each of these aspects. 
The majority of works in QKD is concentrated in this class, as it allows one to explore different combinations of components tailored for very specific practical implementation (e.g., ground-satellite communication \cite{li2014space, liao2017satellite, liao2018satellite, yin2017satellite} and decoy-state communication \cite{hu2021decoy, meyer2011implement}).

In what follows, we provide a few examples of how imperfections can be modeled for device-dependent protocols, following conventions and notation introduced in \cite{winick2018reliable, araujo2023quantum, lorente2025quantum}.

\subsubsection{No-click events}
\label{sec: no clicks}

A no-click event occurs when a photon is emitted but not detected. 
In this case, the absence of detection can be associated with a valid measurement, i.e., there is a POVM element that represents such an event. To model this, the Hilbert space dimension may be expanded to include the detected one-photon subspace. 
The inclusion of this sector can be encoded with a direct sum composition of the click (the detection event), labeled by $\boldsymbol{0}$, with the usual measurement operators $M_{a_x \vert x}$. In this way, ``successful" measurements correspond to
\begin{align}\label{eq: click}
    M_{a_x \vert x} \mapsto M_{a_x \vert x} \oplus \boldsymbol{0},
\end{align}
so that the the dimension of the measurement operators increases by 1 unit. 
For example, qubit BB84 measurements would live in a qutrit space, as described above. 
The POVM element associated with the no-click event is then determined by the completeness of the measurement operators and is represented by the outcome $a_x+1$.
Its matrix representation is given by
\begin{align}\label{eq: no click}
    M_{a_x+1 \vert x} = \mathbb{1} - \sum_{a_x} M_{a_x \vert x}.
\end{align}
An example of a numerical security proof that uses this model can be found in \cite{winick2018reliable}.

\subsubsection{Efficiency mismatch}

Efficiency mismatch refers to the situation in which the probability of obtaining a click depends not only on the measurement basis, but also on the specific detector outcome $a_x$, i.e. which click within that basis was registered (e.g. the clicks for outcomes 0 and 1 have different efficiencies). 
In a device-dependent model, this is incorporated by multiplying measurement operators conditioned on the detection event, described according to the enlarged Hilbert space (Eq.~\eqref{eq: click}, by its efficiency $\eta_{a_x}$, leading to a subnormalized operator
\begin{align}
    M_{a_x \vert x} \oplus \boldsymbol{0} \mapsto \eta_{a_x}\cdot M_{a_x \vert x} \oplus \boldsymbol{0}.
\end{align}
As a consequence, conditioned on a detection event, the effective measurement statistics become biased, and this bias can be efficiently considered within the numerical key-rate optimization methods discussed in Sec. \ref{sec: algorithms}, which otherwise would be hard to solve analytically for general protocols.

\exam{\textbf{(BB84 measurements with efficiency mismatch \cite{winick2018reliable}})} {As an example, we consider the measurements of the standard BB84 protocol. 
Denoting as $p_z$ the probability of measuring in basis $Z$, the effect of efficiency mismatch in different detectors is encoded in the modified POVM's
\begin{align}\label{eq: eff-mismatch-povm}
    M_{0_0 \vert 0} &= \eta_{0_0}|0\rangle\langle 0| \oplus \boldsymbol{0}, \\
    M_{1_0\vert 0} &= \eta_{1_0}|1\rangle\langle 1| \oplus \boldsymbol{0}, \\
    M_{0_1 \vert 1} &= \eta_{0_1}\,|+\rangle\langle +| \oplus \boldsymbol{0}, \\
    M_{1_1 \vert 1} &= \eta_{1_1}\,|-\rangle\langle -| \oplus \boldsymbol{0},
\end{align}
where each $\eta$ quantifies the mismatch between the efficiencies of the detectors associated with its respective outcome index.
The remaining POVM element (related to the no-click event) is then fixed by completeness, as discussed in Eq. (\ref{eq: no click}), so that the mismatch is treated here purely as an imbalance between the click operators rather than as an additional new outcome.} \\
Further imperfections typically incorporated in POVM elements, such as efficiency and dark count rate, are discussed in \cite{van2017photodetector}. We proceed with a description of imperfections of the source.

\subsubsection{Partial state characterization}

In a PM protocol, it is assumed that the signal that leaves the source is a well-characterized state, often the pure eigenstates of some operator (e.g. Pauli eigenstates in the BB84 and six-state protocol), as described in Eq.~\eqref{eq: state 1 pm}. 
In reality, the source itself distributes imperfect states, a condition that has been addressed with different techniques \cite{gottesman2004security, tamaki2013loss, pereira2019quantum, pereira2020quantum}. A possible approach to partially characterize the distributed state is to consider its overleap with the idealized pure signal that should leave the source \cite{mizutani2019quantum,mizutani2020quantum,koashi2003secure}. This approach has been used in recent security proofs for PM scenarios \cite{pereira2025optimal, curras2025numerical}. This source imperfection has been addressed through the inclusion of the inequality
\begin{align}\label{eq: partial state characterization mixed state}
    \bra{\psi} \rho \ket{\psi} \geq 1- \epsilon ,
\end{align}
where $\ket{\psi}$ is the (pure) state that ideally leaves the source, and $\rho$ is the effective shared state, with $\epsilon$ being a sort of efficiency parameter. Such imperfections can be easily included in optimization algorithms, thus being a useful tool for numerical security proofs based on the methods described in Sec. \ref{sec: algorithms}. 

\subsubsection{Decoy state protocols}\label{sec:decoy}

The assumption that the source generating single-qubit states in the BB84 protocol is very strong and rarely met in practice, since single-photon sources aren't a fully developed technology. 
Current single-photon sources typically rely on probabilistic processes, resulting in low rates, which in turn make idealized, single-photon protocols, far from practical implementations.

Instead of relying on ideal single-photon sources, a far more realistic and practical option is the use of weak coherent light pulses. 
In protocols such as the BB84, information is  encoded in the polarization degree of freedom of these weak pulses. 
However, this approach comes with an important drawback: there is a non-negligible probability that the source emits multiple photons in the same quantum state.
While replacing demanding single-photon sources with practical ones addresses a key experimental challenge; it simultaneously allows new eavesdropping strategies. 
If a pulse of many photons is used, an eavesdropper can intercept some of them and acquire information without leaving tracks that Alice and Bob can detect through their QBERs, a strategy known as \textit{photon number splitting} attack \cite{huttner1995quantum, scarani2004quantum}. 
The density operator $\rho_\mu$ that describes a weak coherent pulse of intensity $\mu \ll 1$ with $n$ photons is given by
\begin{align}\label{eq: weak coherent pulses}
    \begin{aligned}
\rho_\mu & =\frac{1}{2 \pi} \int\limits_0^{2 \pi} \diff  \theta\left|\sqrt{\mu} e^{i \theta}\right\rangle\left\langle\sqrt{\mu} e^{i \theta}\right|  \approx \sum_{n=0}^{\infty} e^{-\mu} \frac{\mu^n}{n!}|n\rangle\langle n| .
\end{aligned}
\end{align}
The infinite sum approximating\footnote{If the phase is randomized, then Eq. (\ref{eq: weak coherent pulses}) becomes an equality.} the coherent state can be seen as a convex mixture of Fock states following a Poissonian distribution 
\begin{align}
    p(n) = e^{-\mu}\frac{\mu^n}{n!},
\end{align}
evidencing that the use of weak coherent pulses allows  non-zero probability events  involving multiple photons. To avoid photon-number-splitting attacks, it was proposed \cite{gottesman2004security} that the security of this protocol can be guaranteed only for rounds with single-photon events. 
Taking into account that the multiple-round events are fully insecure, leads to the following asymptotic key rate
\begin{align}\label{eq: devetak winter decoy}
    r_{\infty}=\Gamma_Z^{(1)}(1-h(q_X^{(1)}))-\Gamma_Z h(Q_Z),
\end{align}
where $\Gamma_Z$ is Bob's gain (i.e. the probability of a detection to happen), $\Gamma_Z^{(1)}$ is the probability of a single-photon to be sent and detected, $Q_Z$ is the QBER in the $Z$ basis, and $q_X^{(1)}$ is the QBER in the $X$ basis for single-photon events.

While $\Gamma_Z$ and $Q_Z$ can be empirically estimated, $\Gamma_Z^{(1)}$ and $q_X^{(1)}$ are not directly observed. 
The decoy-state technique then involves sending weak coherent pulses with different intensities to probe the quantum channel and bound the unobserved parameters. 
This allows one to obtain a reliable lower bound for the asymptotic key rate of Eq. (\ref{eq: devetak winter decoy}). 

The crucial point is that the decoy states differ from the key-generation states only in their intensities (while other properties remain the same). 
In this way, Eve cannot distinguish between signals used as decoy states and key generation states. 
As a consequence, the yields and QBERs are independent of the chosen intensities.

With the use of the decoy state method, one can bound the unobserved parameters, related to the single photon events, and ensure significant key rates for the BB84 based on weak coherent pulses\footnote{Note that with the use of decoy states we are encoding a discrete variable in a CV system. This requires a description of measurements in this sort of setting, which is considerably more difficult than the case of POVM's in DV.  A common approach to treat infinite-dimensional measurements in the context of decoy-states is the so-called \textit{squashing model}. 
The interested reader is invited to check \cite{li2020application} for extra details.}. 
In fact, the use of two intensities already allows performances close to ideal \cite{ma2005practical,wang2005beating}. 
More details regarding decoy states can be found in \cite{grasselli2021quantum}, and a review on existing security proofs can be found in \cite{tupkary2025qkd}.

\subsubsection{Confidence intervals of feasible states}\label{sec: confidence regions}

In both PM and EB QKD protocols, the ideal behavior of the system is specified in terms of probability distributions satisfying Born’s rule, as in Eq.~(\ref{eq: probabilities EB}). In practice, however, experimental data are obtained in the form of \emph{empirical frequencies}, which fluctuate around the underlying probabilities due to finite statistics. As a consequence, these observed frequencies cannot, in general, be expected to satisfy the ideal constraints exactly, and may even yield statistics represented by spurious or unphysical objects (e.g. density matrices with negative eigenvalues).

A standard way to address this issue is to adopt a statistical description based on \emph{confidence regions}. Given a state and set of measurements leading to a theoretical behaviour $\boldsymbol{p}$ derived from Born's rule (Eq. (\ref{eq: probabilities EB})), and a real amount of data given by empirical relative frequencies encoded in $\boldsymbol{f}$, the use of confidence regions consists of, rather than attempting to find a single quantum state compatible with the probability distribution from finite data, one considers a subset of states that is compatible with the empirical observed statistics. Security is then established against the worst-case element within this set.

Several approaches to construct such confidence regions have been developed in the QKD literature. The simplest way to make use of these confidence regions is with concentration inequalities - such as Hoeffding, Chernoff, or Serfling bounds - which we exemplified in Eq. (\ref{eq: serfling bound}) with Serfling's inequality. These bounds can be tailored for different sorts of data (i.e. to derive confidence intervals for selected parameters such as standard error rates or Bell parameters \cite{renner2008security, tomamichel2012tight}), and can vary depending on whether the protocol is assumed to be iid or not. In general, concentration bounds have the following structure: A parameter $\theta=\theta(\boldsymbol{p}) \in \mathbb{R}$ (e.g. the quantum bit error rate) is estimated from its empirical value $\theta^{\prime}=\theta(\boldsymbol{f})$, characterizing a concentration bound defined by the probability of the real and empirical value to be sufficiently close up to a probability $1-\epsilon$
\begin{equation}\label{eq: concentration bound}
p(\vert \theta-\theta^{\prime}\vert \le \delta(\epsilon,N)) \geq 1-\epsilon ,
\end{equation}
where $\delta(\epsilon,N)$ is a function depending on the chosen inequality with monotonically decreasing behaviour in the number of samples considered. 

Concentration bounds of the the form of Eq. (\ref{eq: concentration bound}) allow one to define the confidence region $\mathcal{C}_\text{CB}$ as
\begin{align}\label{eq: CB confidence region}
\mathcal{C}_\text{CB} =\{\,\boldsymbol{p} : |\theta(\boldsymbol{p})-\theta^{\prime}| \le \delta(\epsilon,N) \}.
\end{align}

An example of security proof which used this techniques can be found in \cite{winick2018reliable}. Another interesting case in shown in \cite{mannalath2025sharp}, where an extension of Serfling's inequality for finite sampling is derived which performs, and appears useful in regimes typical for QKD. These examples are used in security analyzes based on the entropic uncertainty relation~\cite{tomamichel2012tight,tomamichel2017largely}, an analytical technique for finite-size key rates which we detail in Sec. \ref{sec: EUR}.

More recent works have explored confidence regions defined with the relative entropy (or Kullback--Leibler divergence) constraints involving the distributions $\mathbf{f}$ and $\mathbf{p}$, defined by
\begin{align}\label{eq: kullbackleibler}
    D_{\mathrm{KL}}(\boldsymbol{f}\Vert\boldsymbol{p}) =\sum_i f_i \log\frac{f_i}{p_i},
\end{align}
which often yield tighter bounds in the finite-size regime. In this approach, the confidence region is defined with the full probability distributions as
\begin{equation}
\mathcal{C}_{\text{KL}}
=\{\,\boldsymbol{p} \;:\; D_{\mathrm{KL}}(\boldsymbol{f}\Vert\boldsymbol{p})
\le \delta(\epsilon,N) \},
\end{equation}
in contrast with Eq. (\ref{eq: CB confidence region}), where the probabilities lead to the estimated parameters $\theta$. Some works that make use of this sort of confidence region are \cite{bunandar2020numerical,sano2010secure,sasaki2015key}.

In the context of numerical security proofs, a particularly convenient class of confidence regions consists of those that can be expressed as convex-linear constraints, allowing them to be included as constraints in optimization algorithms for key rates (detailed in Sec. \ref{sec: algorithms}). This sort of approach can be found in ~\cite{araujo2023quantum, lorente2025quantum}, where the confidence region is based on requiring that the deviation $\boldsymbol{p}-\boldsymbol{f}$ lies within an ellipsoid characterized by a covariance matrix\footnote{The covariance matrix is defined as the matrix of second central moments of a set of observables. It characterizes the spread of each observable and the correlations between different observables by containing information of the first and second moments of the measured states. In this way, it provides a complete description of Gaussian states, as they're determined by first and second moments, and encodes their noise properties and correlations} $\Sigma$~\cite{anka2025introductoryreviewtheorycontinuousvariable} and a parameter $\chi^2$ that defined the radius of the ellipsoid. This leads to the following confidence region
\begin{align}\label{eq: confidence region SDP}
\begin{aligned}
\mathcal{C}_{\text{SDP}}=\{\,\boldsymbol{p}\;:\;
&\bigl\langle \boldsymbol{p}-\boldsymbol{f},\, 
\Sigma^{-1}(\boldsymbol{p}-\boldsymbol{f}) \bigr\rangle
\le \chi^2 \}.
\end{aligned}
\end{align}
Such confidence regions are particularly well suited for numerical key-rate evaluations based on semidefinite programming, as they allow one to directly optimize security quantities over all quantum states compatible with the observed data.

\subsection{Device-independent protocols}
\label{sec: DI protocols}

In this section, we adopt a different approach to dealing with imperfections and present a method  tailored to the most adversarial scenario. Rather than fine-tuning detailed models of the devices, we fully relax any assumptions about their internal workings.

Device-independent (DI) protocols \cite{brunner2014bell, zapatero2023advances, ghoreishi2025future} are the ones in which no assumption is made about internal characteristics of any of the laboratories (e.g., discrete or continuous degrees of freedom, dimensions, specific components, etc.). This sort of setup matches the description supplied by \textit{Bell non-locality} scenarios \cite{einstein1935can, bell1964einstein, fine1982hidden, brunner2014bell, scarani2019bell}, so that nonlocality  is incorporated as a resource for the security of the generated key \cite{acin2006bell, acin2007device, masanes2011secure, gonzales2021device, schwonnek2021device, huber2013weak}.

The test for this type of correlation is performed during the parameter estimation step, in which certain rounds of the protocol are used to estimate the value of a Bell test.  
Moreover, Alice and Bob have one pair of inputs for which they expect strong correlations. 
They use a subset of the rounds corresponding to this input pair to estimate the QBER, and the remaining rounds are used for key generation.

The data obtained in the Bell test is used to ensure that the correlations shared among Alice and Bob are nonlocal, i.e., they're capable of violating Bell inequalities, which are defined as linear combinations of the measured joint statistics $p(ab\vert xy)$,
\begin{align}
    \mathcal{B} \equiv \sum_{abxy} c_{a b x y}\, p(ab\vert xy).
\end{align}
A Bell inequality generally satisfies the chain of inequalities
\begin{align}\label{eq: bell inequality}
    \mathcal{B} \leq \beta_\mathcal{L}^{\mathcal{B}} \leq \beta_\mathcal{Q}^{\mathcal{B}},
\end{align}
where $\beta_\mathcal{L}^{\mathcal{B}}$ is an upper bound achievable with in classical theory. 
This threshold can be violated up to a value $\beta_\mathcal{Q}^\mathcal{B}$ allowed by quantum theory. 
Specific conditions must be met in order to surpass the local bound $\beta_\mathcal{L}^{\mathcal{B}}$, such as the use of entangled states \cite{werner1989quantum} (particularly implying that DI protocols are always formulated in the EB description) and incompatible measurements within the laboratory of each party \cite{fine1982hidden}. 
Correlations that can be classically explained, i.e. that satisfy the local bound $\mathcal{B} \leq \beta_\mathcal{L}^{\mathcal{B}}$, are of the form
\begin{align}\label{eq: lhv}
    p(ab\vert xy) = \sum_{\lambda}p(\lambda) p(a\vert x\lambda) p(b\vert y\lambda)\ ,
  \ \forall \ a,b,x,y.
\end{align}
In the equation above, $\lambda$ represent \textit{local hidden variables} (LHV), latent unknown factors in the experiment that, if known, could explain the correlations observed between Alice and Bob without quantum theory.

In a QKD perspective, if a classical model exists for a certain set of data produced by Alice and Bob, then there is a description of the experiment in terms of LHV which could be controlled by the eavesdropper, that can determine the outcomes of Alice and Bob with certainty. Violating a Bell inequality implies that no such classical model exists, and therefore, intuitively, we can infer that an eavesdropper cannot have full information about the outcomes in a Bell inequality test. This is the basis for DI quantum cryptography.

These sorts of inequalities can be thought geometrically as hyperplanes in the space of joint probabilities $p(ab\vert xy)$, which characterize the so-called local polytope\footnote{A polytope is a convex set whose interior points can be written as a convex mixture of a finite number of extremal elements.} and are witnesses of nonlocal statistics --- the probabilities that cannot be classically explained according to Eq. (\ref{eq: lhv}). The simplest Bell inequality, introdued in \cite{clauser1969proposed} by Clauser, Horne Shimony and Holt (CHSH), is associated to the scenario in which both Alice and Bob perform two dichotomic measurements and its given by
\begin{align}
\begin{aligned}
    \mathcal{B}_\text{CHSH}&=\left\langle A_0 B_0\right\rangle+\left\langle A_0 B_1\right\rangle+\left\langle A_1 B_0\right\rangle-\left\langle A_1 B_1\right\rangle \\
    &\hspace{0.5cm}\leq 2 \leq 2\sqrt{2},
\end{aligned}
\end{align}
where $\beta_\mathcal{L}=2$ and $\beta_\mathcal{Q}=2\sqrt{2}$ and the correlators are defined by
\begin{align}
    \left\langle A_x B_y\right\rangle=p(a=b \vert x y)-p(a \neq b \vert x y).
\end{align}
In what follows, we show how the violation of this inequality can be used to parametrize the key rate in a DIQKD protocol.

\exam{\textbf{(DIQKD with CHSH inequality \cite{pironio2009device,masanes2011secure})}} {In a protocol based on the CHSH inequality, Alice has inputs $x\in\{0,1\}$ while Bob can choose between $y\in\{0,1,2\}$. In parameter estimation, the rounds with $y\in\{0,1\}$ are used to estimate the CHSH value, while the pair of inputs $(x,y)=(0,2)$ is used for key generation and the estimative of the QBER $Q$. For any state and measurements compatible with the expected value of the Bell inequality $\mathcal{B}_\text{CHSH}$, the asymptotic key rate is given by
\begin{align}\label{eq:diqkd_rate_chsh_example}
    r_\infty \geq
    1 - h\!\left(
    \frac12+\frac12\sqrt{\left(\frac{\mathcal{B_\text{CHSH}}}{2}\right)^2-1}
    \right) - h(Q),
\end{align}
where $h(\bullet)$ denotes the binary entropy.} \\

The incorporation of Bell nonlocality into realistic QKD protocols is very challenging \cite{nadlinger2022experimental, liu2022toward, zhang2022device, murta2019towards}, as the certification of violation of such inequalities in a detection-loophole-free setting is itself a remarkably difficult task. A consequence of this is that the efficiency of sources and detectors must be considerably better than those  typically found in characterized protocols. In order to make DIQKD closer to real implementations from a theoretical perspective, new strategies have been explored to enhance the robustness of these protocols. 
One promising direction is the use of routed Bell tests to make these protocols more resilient to loss of distributed states and allow for implementations along larger distances \cite{lobo2023certifying, schwonnek2021device, chaturvedi2024extending, le2025device, kossmann2025routed, tan2024entropy}.

\subsection{One-sided device-independent protocols}
\label{sec: one-sided device independent protocols}

In a one-sided device independent protocol (1SDI) we have the intermediate situation containing both characterized and uncharacterized devices. 
In this case, we consider that one party possesses a black box, while the other can control a characterized machine, being therefore compatible with a description in terms of \emph{quantum steering} \cite{uola2020quantum, cavalcanti2016quantum, vsupic2020self}, an intermediate form of quantum correlation lying between entanglement and nonlocality. 
``Intermediate'' here should be understood in the sense that Bell nonlocality always implies steering, and steering always implies entanglement, while the converses fail in general, as there exist steerable states that do not violate any Bell inequality, and entangled states that are unsteerable \cite{wiseman2007steering}.

In a 1SDI protocol, we will consider Alice as the untrusted party: her device receives a classical input $x$ and outputs $a$, while Bob’s device is fully characterized (trusted measurements on his side). 
This setting can be seen as Alice’s black box prepares a collection of states with a given probability: $p(a|x)$. 
From Bob’s point of view, the received systems can be described by (sub-normalized) conditional states $\{\sigma_{a|x}\}_{a,x}$, an object known as an \textit{assemblage}. 
In terms of the unknown POVMs $\{M_{a_x|x}\}_{a,x}$ on Alice’s side, we can write an assemblage as
\begin{equation}
  \sigma_{a|x}
  = \operatorname{Tr}_A\left[(M_{a_x|x} \otimes \mathbb{1}_B)\,\rho_{AB}\right] ,
  \ \forall \ a,x.
  \tag{8.4}
\end{equation}

Analogously to the notion of classicality being synthesized in LHV models for Bell scenarios, in steering scenarios the assemblages admit a classical description if no genuine ``steering'' can happen from Alice to Bob. 
This is described by a local-hidden-state (LHS) model: we have again a hidden latent variable $\lambda$ with probability $p(\lambda)$ defining a pre-existing quantum state $\rho_\lambda$ to Bob, and instructs Alice’s black box to output $a$ upon input $x$ with probability $p(a|x\lambda)$. 
In this case, Bob’s assemblage takes the form
\begin{equation}\label{eq: LHS model}
  \sigma_{a|x}^{\mathrm{LHS}}
  = \sum_{\lambda} p(\lambda)\,p(a|x\lambda) \rho_\lambda \ ,
  \ \forall\ a,x.
\end{equation}
Whenever an assemblage $\{\sigma_{a|x}\}$ can be decomposed as in Eq. (\ref{eq: LHS model}), the correlations between Alice’s outputs and Bob’s conditional states can be classically explained.

In this sort of scenario, in contrast with the device-dependent case, it is necessary to ensure that the correlations encoded in joint probabilities $p(ab\vert xy)$ are compatible with the existence of steering. 
This is usually certified with the violation of \emph{steering inequalities} \cite{mukherjee2021role}.

Analogously to Bell inequalities in the fully DI setting, steering inequalities are linear functionals that separate LHS assemblages from genuinely steerable ones. 
Given a family of Hermitian operators $\{F_{a|x}\}_{a,x}$ acting on Bob’s system, we define
\begin{align}
  \mathcal{S} \equiv \sum_{a,x} \operatorname{Tr}\left(F_{a|x}\,\sigma_{a|x}\right)
  \leq \beta_{\mathrm{LHS}}^{\mathcal{S}} \leq \beta_\mathcal{Q}^{\mathcal{S}},
\end{align}
where $\beta_{\mathrm{LHS}}^{\mathcal{S}} $ is the maximal value of $\mathcal{S}$ over all LHS assemblages of the form in Eq. (\ref{eq: LHS model}). 
Violation of this bound by the observed assemblage implies that no LHS model exists, i.e., the data demonstrate steering. 
In the context of 1SDI-QKD, part of the data is used to test such steering inequalities, ensuring that the Alice–Bob correlations cannot be simulated by an eavesdropper who only distributes pre-determined states to the trusted party while controlling the untrusted device \cite{branciard2012one, masini2024one, xin2020one, tomamichel2013one, gehring2015implementation}.

\begin{center}
\textbf{PART II: TOOLS FOR SECURITY PROOFS}
\addparttoc{Part II: Tools for security proofs}
\end{center}

In the second part of this review, we provide a brief introduction to different information-theoretic analytical and numerical tools that allow the inclusion of imperfections and more realistic assumptions in the security proofs of QKD protocols. 
The numerical tools consist of recent algorithms that extend the standard theory of semi-definite programming (SDP), allowing the efficient calculation of key rates in the asymptotic regime, considering different imperfections in the device-dependent protocols. 
The numerical methods are more than mere key estimating algorithms, as they are equipped with theorems that prove their optimality and therefore are capable of provide a reliable \textit{numerical security proof} for the key rate obtained for a certain protocol model and the imperfections associated with it. On the other hand, the analytical tools are constituted by theorems that allow one to treat the distribution of keys of finite length which are secure against collective attacks. Combining the analytical and numerical tools, one can construct strong security proofs for realistic QKD protocols.

\section{Asymptotic key rate optimization algorithms}
\label{sec: algorithms}

In this section, we detail how the asymptotic key rate $r_\infty$ of QKD protocols can be efficiently bounded using different algorithmic strategies, enabling the numerical security analysis of various protocols. These algorithms mainly differ in how they implement the conditional entropy minimization in the Devetak–Winter formula
\begin{align}\label{eq: devetak winter minimization}
    r_{\infty} \geq \min_{\rho}(H(A \vert E)_\rho)-H(A \vert B)_\rho ,
\end{align}
where the minimization represents the worst-case scenario compatible with the protocol. The state over which the optimization is performed corresponds to the single-round (iid) reduction. A key rate valid against collective attacks obtained through the minimization of Eq. (\ref{eq: devetak winter minimization}) can be converted to a key rate against coherent attacks with analytical techniques explained of Sec. \ref{sec: finite keys}, which in turn will depend on the number $n$ of rounds.

Notice that $H(A \vert B)$ depends on the amount of leak required for each specific protocol. 
This term does not need to be included in the minimization, as analytical upper bounds can usually be computed from the estimated parameters. For simple protocols, it is possible to obtain tight, and in some cases fully analytical, lower bounds on $H(A \vert E)$. 
However, as the protocol increases in complexity --- with more measurement settings, higher dimensions, and the inclusion of imperfections --- deriving such bounds analytically quickly becomes infeasible.

The minimization of $H(A \vert E)$ is a difficult optimization task due to the complexity of finding global optimal points for the conditional von Neumann entropy.
As it is a non-linear function, the usual tools of SDPs (for which we provide an introduction in Appendix \ref{sec: basic optimizations}) must then be extended to tackle these QKD rates, and such algorithms must also be compatible with the assumptions present in each protocol. 
In other words, a numerical method tailored for device-independent scenarios is not applicable for a fully characterized protocol (see Sec. \ref{sec: device trutability} for a description of device-independence), as they represent completely different optimization problems. In particular, fully characterized protocols demand the optimization to happen over a state with characterized dimension\footnote{This contrasts with DI scenarios, where the numerical techniques are agnostic regarding the dimensions of states and measurements.}, being a $d \times d$ matrix (so that CV protocols demand either a dimension cutoff, or a squashing procedure to treat it similarly to a finite-dimensional system \cite{li2020application}. More precisely squashing is used in DV protocols where measurements are implemented with threshold detectors. The problem there is that the actual POVMs of threshold detectors are sensitive to the photon number and hence infinite dimensional. Squashing is a solution to this infinite dimensionality \cite{li2020application}. In contrast, the CV is dealt with differently -- the cutoff is performed at the level of SDPs, i.e. the SDP is formulated in the original infinite dimensional space and another, finite SDP is found. Solutions to the finite dimensional SDP are shown to lower bound the solution to the infinite dimensional SDP (This method is referred to as "dimension-reduction
method" \cite{upadhyaya2021dimension}).

Following the notation in \cite{winick2018reliable}, we denote by $\mathbf{S}$ the set of \textit{feasible} density operators of the initial distribution stage of the protocol, i.e. the set of normalized, hermitian an positive semi-definite (PSD) matrices satisfying the probabilistic constraints of the cryptographic scenario, reflecting the information that can be obtained through measurements in the protocol (as shown in Eq.~(\ref{eq: probabilities EB})):
\begin{align}\label{Eq: measurement equality constraints}
\mathbf{S}=\left\{\rho \in \mathbf{H}_{+} \vert \operatorname{Tr}\left(\Gamma_i \rho\right)=\gamma_i, \forall i\right\}.
\end{align}
In the definition above, $\{\Gamma_i\}_i$ is a set of measurements, e.g. $ M_{a_x\vert x} \otimes N_{b_y\vert y}$ for an EB protocol, with $i$ being an index enumerating the measurement events $a_xb_y\vert xy$ (outcomes $a_x$ and $b_y$ are obtained in the measurement of $x$ and $y$, respectively), or measurements giving marginal statistics in Alice's laboratory in the case of PM protocols. 
We include $\mathbb{1}$ in the set $\{\Gamma_i\}_i$ as a convention, to be able to write the normalization of $\rho$ as a similar constraint equation.

Furthermore, instead of performing the optimization over the conditional entropy $H(A\vert E)$, most numerical techniques make use of the \textit{relative entropy}, which has interesting properties that can be exploited to construct these reliable numerical security analyses (i.e. to work with a tractable objective function). The relative entropy, often referred to as Umegaki divergence, is defined as follows.

\dfn{\textbf{(Quantum relative entropy)}\label{dfn: relative entropy}}
{Let $\rho$ and $\sigma$ be density operators on the same Hilbert space such
that $\operatorname{supp}(\rho) \subseteq \operatorname{supp}(\sigma)$\footnote{If $\operatorname{supp}(\rho) \nsubseteq \operatorname{supp}(\sigma)$  we set
$D(\rho \Vert \sigma) = +\infty$.}, where $\operatorname{supp}(\bullet)$ denotes the support of operator $\bullet$. The quantum relative entropy\footnote{For mixed states described by ensembles following distributions $f_i$ and $p_i$, the relative entropy reduces to the Kullback-Leibler divergence defined in Eq. (\ref{eq: kullbackleibler}).} of $\rho$ with respect to $\sigma$ is defined as
\begin{align}\label{eq: relative entropy}
    D(\rho \Vert \sigma)
    \equiv \text{Tr}\left[\rho \bigl(\log \rho - \log \sigma\bigr)\right].
\end{align}
} 

The methods that will be discussed in Sec. \ref{sec: Winick} and \ref{sec: Lorente}, proposed respectively by Winick et al. \cite{winick2018reliable} and Lorente et al. \cite{lorente2025quantum}, use of the following theorem, which allows the removal of Eve's side information from the relative entropy due to the action of the pinching channel $\mathcal{Z}$ (Eq. (\ref{eq: map Z})), so that the optimization does not need to take into account the measurement operators of this party.

\thm{\textbf{(\textbf{Removal of Eve's system with pinching map, adapted from Theorem 1 of \cite{coles2011information}})}}{\label{thm: removing eve}
Let $\rho$ denote the bipartite state shared by Alice and Bob, and let $\mathcal{G}$ be the completely positive map defined in Eq. (\ref{eq: map G}) that represents the quantum-classical processing up to (and including) the sifting step and the isometry that coherently copies the raw key to the register $R$. Let $\mathcal{Z}$ be the pinching channel on $R$ in the raw-key basis (Eq. (\ref{eq: map Z})), tensored with the identity on all other registers. Then, for the classical raw-key register $Z^R$, the equality
\begin{align}\label{eq: equality conditional entropy relative entropy}
    p_{\mathrm{pass}}\cdot H(Z^R \vert E \widetilde{A} \widetilde{B}) =  D(\mathcal{G}(\rho)\,\|\, \mathcal{Z}(\mathcal{G}(\rho)))
\end{align}
holds, with $p_\text{pass}$ as defined in Eq. (\ref{eq: sifting probability}).} \\

\noindent
\textbf{Proof.} A proof of Theorem \ref{thm: removing eve} can be found in \cite{coles2012unification}. \\

\noindent
On the other hand, the numerical method shown in Sec. \ref{sec: Araujo}, proposed by Araújo et al. in \cite{araujo2023quantum}, relies on the identity
\begin{align}\label{eq: relative entropy conditional entropy}
    H(A \vert E)
    &= - D\bigl(\rho_{AE} \Vert \mathbb{1}_A \otimes \rho_E\bigr),
\end{align}
where the measurement operators of Eve are kept.

The method proposed in \cite{winick2018reliable} and discussed in Sec.\ref{sec: Winick}, uses a property that allows for the removal of the eavesdropper side information (see Theorem \ref{thm: removing eve}), avoiding the necessity of optimizing over measurements of this party.
Another case: the method proposed in \cite{lorente2025quantum} and discussed in Sec. \ref{sec: Araujo} explores the possibility of globally bounding the Umegaki divergence in terms of a specific series (see Theorem \ref{thm: gauss radau quantum}), an idea that was firstly introduced for DI methods (which we discuss in detail in Appendix \ref{sec: appendix gauss radau DI}). The last numerical method that we discuss for device-dependent protocols, developed in \cite{lorente2025quantum} and discussed in Sec. \ref{sec: Lorente} consists in defining a conic space based on the structure of \ref{eq: relative entropy}, where the optimization can be efficiently performed. From now on, we denote with $\alpha$ the optimal solution of the minimization problem
\begin{align}\label{optimization alpha}
   \alpha \equiv &\min_{\rho \in \mathbf{S}} H(A \vert E)_\rho . 
\end{align}

\subsection{Reliable key rates via quadratic optimization and duality}
\label{sec: Winick}

Winick \textit{et al.} show in \cite{winick2018reliable} a set of theorems that ensure a reliable lower bound (which in certain cases may be tight) for the optimization problem posed in Eq. (\ref{optimization alpha}). Here, the word reliable refers to the possibility of finding \textit{provable} lower bounds, contrasting with heuristic methods in which solutions can be found, but in general cannot be certified as lower bounds for the key-rate problem (for instance, see-saw algorithms).

For the method proposed in \cite{winick2018reliable}, it is important to write the relative entropy defining the objective function without the eavesdropper's information. This is allowed by the result of Theorem \ref{thm: removing eve}, proven in \cite{coles2012unification}, which provides an identity involving the conditional von Neumann entropy considering the total (public and private) information available to the eavesdropper $E\tilde{A}\tilde{B}$ in terms of the quantum relative entropy taking into account only the information in the shared state among the authenticated parties. The possibility of describing a device-dependent QKD protocol in terms of $\mathcal{G}$ and $\mathcal{Z}$ (as detailed in Section~\ref{sec: QKD protocols}) is essential to work with $f(\rho) \equiv D(\mathcal{G}(\rho)\Vert \mathcal{Z}(\mathcal{G}(\rho))$ as the objective function instead of the conditional entropy.

The numerical technique of this method consists of using Theorem \ref{thm: removing eve} to remove the eavesdropper's measurements from the function to be optimized, having therefore a convex objective function. The convexity then allows the use of duality and linearization arguments to be applied in the optimization, according to Theorem \ref{thm: theorem 1 winick}. Furthermore, this result can be strengthened with the inclusion of perturbations in the objective function which map the optimal solutions to the interior of the feasible set $\mathbf{S}$, as detailed in Theorem \ref{thm: perturbation winick}. it is interesting to guarantee interior point solutions instead of boundary optimal variables, as the latter case often produces the spurious \texttt{infeasible} outcomes due to numerical precision errors. Techniques that ensure such interior point optimality are referred to as \textit{facial reductions}, which we will discuss more along this section.

To construct this numerical application, one needs to define a gradient for operator-valued functions, holding in particular for the objective function given in Eq. (\ref{optimization alpha}). A standard definition for this is the element-wise derivative written in basis $\{\ket{j}\}$. 
\begin{align}
\nabla f(\rho)\equiv\sum_{j, k} d_{j k}|j\rangle\langle k|,
\end{align}
with $d_{jk}\equiv\left.\partial_{\sigma_{j k}} f(\sigma) \right|_{\sigma=\rho}$ and $\sigma_{j k}\equiv\langle j| \sigma|k\rangle$.

The method of Winick et al. \cite{winick2018reliable} works in two steps: In the first, an initial state $\rho_0 \in \mathbf{S}$ passes through a quadratic minimization (an adaptation of the so-called Frank-Wolfe's algorithm \cite{frank1956algorithm} for density matrices) in order to approach another point $\rho$, over which we apply theorems involving linearization and duality techniques in order to find a certifiable lower bound for the key rate corresponding to the global optimum $\rho^{*}$. This means that the quadratic optimization itself is useful to reach a near-optimal point, but the reliability of the method is guaranteed by the theorems \ref{thm: theorem 1 winick} and \ref{thm: perturbation winick}. The following pseudocode shows how the Frank-Wolfe algorithm works \cite{frank1956algorithm,winick2018reliable}. \\

\begin{protocol}[H]
\caption{Frank-Wolfe for density matrices}\label{alg:step1-minimization}
\begin{algorithmic}[1]
\State 
Let $\epsilon > 0$, $\rho_0 \in \mathcal{S}$ and set $i = 0$.
\State Compute $\Delta \rho \equiv  \arg\min_{\Delta \rho} \operatorname{Tr}\left[(\Delta \rho)^T \nabla f(\rho_i)\right]$ subject to $\Delta \rho + \rho_i \in \mathcal{S}$ 
\State \textbf{If} $\operatorname{Tr}\left[(\Delta \rho)^T \nabla f(\rho_i)\right] < \epsilon$ then \textbf{STOP}.
\State \textbf{Else} Find $\lambda \in (0,1)$ that minimizes $f(\rho_i + \lambda \Delta \rho)$.
\State Set $\rho_{i+1} = \rho_i + \lambda \Delta \rho$, $i \leftarrow i + 1$ and go to step 2.
\end{algorithmic}
\end{protocol}

Given a near-optimal point, the second part of the method consists of making use of duality (see Appendix \ref{sec: basic optimizations}) of the problem of Eq. (\ref{optimization alpha}) to find a provable lower bound for $D(\mathcal{G}(\rho^*)\,\|\, \mathcal{Z}(\mathcal{G}(\rho^{*})))$. This bound is certified by the following theorem. \\

\noindent
\thm{(Theorem 1 of \cite{winick2018reliable}) \label{thm: theorem 1 winick}} {Given any $\rho \in \mathbf{S}$, if $\nabla f(\rho)$ exists, then $\alpha \geqslant \beta(\rho)$, with $\alpha$ defined in Eq. (\ref{optimization alpha}) and
\begin{align}\label{eq: winick lower bound 1}
\beta(\sigma) \equiv f(\sigma)-\operatorname{Tr}\left(\sigma^T \nabla f(\sigma)\right)+\max _{\vec{y} \in \mathbf{S}^*(\sigma)} \vec{\gamma} \cdot \vec{y}.
\end{align}
The optimization to be solved then is the maximization in the right-hand side, where $\mathbf{S}^*$ denotes the dual space (see Appendix \ref{sec: basic optimizations} for an explanation of duality), defined as 
\begin{align}
\mathbf{S}^*(\sigma) & \equiv\left\{\vec{y} \in \mathbb{R}^n \vert \sum_i y_i \Gamma_i^T \leqslant \nabla f(\sigma)\right\} .
\end{align}} 

\noindent
\textbf{Proof.} For the proof of Theorem~\ref{thm: theorem 1 winick} we refer the reader to~\cite{winick2018reliable}.

\begin{center}
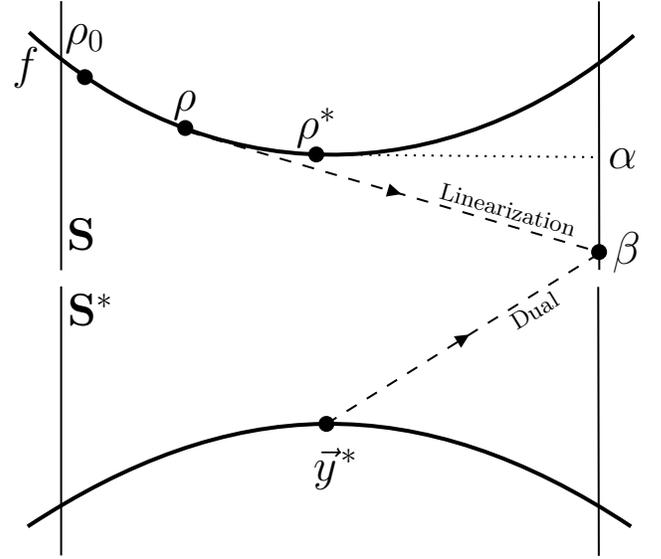
\begin{figure}
  
\tikzset {_68avrqgtv/.code = {\pgfsetadditionalshadetransform{ \pgftransformshift{\pgfpoint{0 bp } { 0 bp }  }  \pgftransformrotate{-270 }  \pgftransformscale{2 }  }}}
\pgfdeclarehorizontalshading{_cl6e9qro2}{150bp}{rgb(0bp)=(0.29,0.56,0.89);
rgb(37.5bp)=(0.29,0.56,0.89);
rgb(62.5bp)=(0.29,0.56,0.89);
rgb(100bp)=(0.29,0.56,0.89)}
\tikzset{_jcvm64xl8/.code = {\pgfsetadditionalshadetransform{\pgftransformshift{\pgfpoint{0 bp } { 0 bp }  }  \pgftransformrotate{-270 }  \pgftransformscale{2 } }}}
\pgfdeclarehorizontalshading{_hy9c2alb8} {150bp} {color(0bp)=(transparent!100);
color(37.5bp)=(transparent!100);
color(62.5bp)=(transparent!0);
color(100bp)=(transparent!0) } 
\pgfdeclarefading{_eot014lz0}{\tikz \fill[shading=_hy9c2alb8,_jcvm64xl8] (0,0) rectangle (50bp,50bp); } 

  
\tikzset {_ak8gj7nxw/.code = {\pgfsetadditionalshadetransform{ \pgftransformshift{\pgfpoint{0 bp } { 0 bp }  }  \pgftransformrotate{-270 }  \pgftransformscale{2 }  }}}
\pgfdeclarehorizontalshading{_5nhxfwy4j}{150bp}{rgb(0bp)=(0.39,0.15,0.96);
rgb(37.5bp)=(0.39,0.15,0.96);
rgb(62.5bp)=(0.72,0.91,0.53);
rgb(100bp)=(0.72,0.91,0.53)}
\tikzset{_7w7vgbums/.code = {\pgfsetadditionalshadetransform{\pgftransformshift{\pgfpoint{0 bp } { 0 bp }  }  \pgftransformrotate{-270 }  \pgftransformscale{2 } }}}
\pgfdeclarehorizontalshading{_7z57ss0x5} {150bp} {color(0bp)=(transparent!0);
color(37.5bp)=(transparent!0);
color(62.5bp)=(transparent!100);
color(100bp)=(transparent!100) } 
\pgfdeclarefading{_0gia58gmq}{\tikz \fill[shading=_7z57ss0x5,_7w7vgbums] (0,0) rectangle (50bp,50bp); } 
\tikzset{every picture/.style={line width=0.75pt}} 

\begin{tikzpicture}[x=0.75pt,y=0.75pt,yscale=-1,xscale=1,scale=0.8]

\draw  [draw opacity=0][shading=_cl6e9qro2,_68avrqgtv,path fading= _eot014lz0 ,fading transform={xshift=2}] (169.5,2) -- (509.88,2) -- (509.88,174) -- (169.5,174) -- cycle ;
\draw    (170.75,2) -- (170.75,172) ;
\draw    (509.75,2.25) -- (510,171.67) ;
\draw  [fill={rgb, 255:red, 0; green, 0; blue, 0 }  ,fill opacity=1 ] (327.2,98.77) .. controls (327.2,96.34) and (329.17,94.37) .. (331.6,94.37) .. controls (334.03,94.37) and (336,96.34) .. (336,98.77) .. controls (336,101.2) and (334.03,103.17) .. (331.6,103.17) .. controls (329.17,103.17) and (327.2,101.2) .. (327.2,98.77) -- cycle ;
\draw  [dash pattern={on 4.5pt off 4.5pt}]  (251,84.4) -- (509.87,160.33) ;
\draw [shift={(385.23,123.77)}, rotate = 196.35] [fill={rgb, 255:red, 0; green, 0; blue, 0 }  ][line width=0.08]  [draw opacity=0] (8.93,-4.29) -- (0,0) -- (8.93,4.29) -- cycle    ;
\draw  [draw opacity=0][shading=_5nhxfwy4j,_ak8gj7nxw,path fading= _0gia58gmq ,fading transform={xshift=2}] (169.75,179.75) -- (509.7,179.75) -- (509.7,352) -- (169.75,352) -- cycle ;
\draw    (170.75,182) -- (170.75,351.75) ;
\draw    (509.33,182.33) -- (509.7,352) ;
\draw  [fill={rgb, 255:red, 0; green, 0; blue, 0 }  ,fill opacity=1 ] (505.47,160.33) .. controls (505.47,157.9) and (507.44,155.93) .. (509.87,155.93) .. controls (512.3,155.93) and (514.27,157.9) .. (514.27,160.33) .. controls (514.27,162.76) and (512.3,164.73) .. (509.87,164.73) .. controls (507.44,164.73) and (505.47,162.76) .. (505.47,160.33) -- cycle ;
\draw [line width=1.5]    (150.5,21.67) .. controls (261.67,122.67) and (413,126) .. (531.83,23.67) ;
\draw [line width=1.5]    (149.67,333.5) .. controls (276.33,250.17) and (395.33,244.17) .. (530,333.5) ;
\draw  [fill={rgb, 255:red, 0; green, 0; blue, 0 }  ,fill opacity=1 ] (244.6,82.1) .. controls (244.6,79.67) and (246.57,77.7) .. (249,77.7) .. controls (251.43,77.7) and (253.4,79.67) .. (253.4,82.1) .. controls (253.4,84.53) and (251.43,86.5) .. (249,86.5) .. controls (246.57,86.5) and (244.6,84.53) .. (244.6,82.1) -- cycle ;
\draw  [fill={rgb, 255:red, 0; green, 0; blue, 0 }  ,fill opacity=1 ] (181.3,50) .. controls (181.3,47.57) and (183.27,45.6) .. (185.7,45.6) .. controls (188.13,45.6) and (190.1,47.57) .. (190.1,50) .. controls (190.1,52.43) and (188.13,54.4) .. (185.7,54.4) .. controls (183.27,54.4) and (181.3,52.43) .. (181.3,50) -- cycle ;
\draw  [fill={rgb, 255:red, 0; green, 0; blue, 0 }  ,fill opacity=1 ] (333.6,268.77) .. controls (333.6,266.34) and (335.57,264.37) .. (338,264.37) .. controls (340.43,264.37) and (342.4,266.34) .. (342.4,268.77) .. controls (342.4,271.2) and (340.43,273.17) .. (338,273.17) .. controls (335.57,273.17) and (333.6,271.2) .. (333.6,268.77) -- cycle ;
\draw  [color={rgb, 255:red, 255; green, 255; blue, 255 }  ,draw opacity=1 ][fill={rgb, 255:red, 255; green, 255; blue, 255 }  ,fill opacity=1 ] (170,172.5) -- (511.25,172.5) -- (511.25,181) -- (170,181) -- cycle ;
\draw  [dash pattern={on 4.5pt off 4.5pt}]  (338,268.77) -- (509.87,161.33) ;
\draw [shift={(428.17,212.4)}, rotate = 147.99] [fill={rgb, 255:red, 0; green, 0; blue, 0 }  ][line width=0.08]  [draw opacity=0] (8.93,-4.29) -- (0,0) -- (8.93,4.29) -- cycle    ;
\draw  [dash pattern={on 0.84pt off 2.51pt}]  (331.6,98.77) -- (509.75,100.5) ;

\draw (172.53,137.33) node [anchor=north west][inner sep=0.75pt]  [font=\LARGE]  {$\mathbf{S}$};
\draw (138.17,29.07) node [anchor=north west][inner sep=0.75pt]  [font=\LARGE]  {$f$};
\draw (317.17,67.13) node [anchor=north west][inner sep=0.75pt]  [font=\LARGE]  {$\rho ^{*}$};
\draw (410.11,114.07) node [anchor=north west][inner sep=0.75pt]  [rotate=-15.86] [align=left] {Linearization};
\draw (173,184.73) node [anchor=north west][inner sep=0.75pt]  [font=\LARGE]  {$\mathbf{S^{*}}$};
\draw (293.07,7.75) node [anchor=north west][inner sep=0.75pt]  [color={rgb, 255:red, 255; green, 255; blue, 255 }  ,opacity=1 ] [align=left] {Primal space};
\draw (298.57,333.42) node [anchor=north west][inner sep=0.75pt]  [color={rgb, 255:red, 255; green, 255; blue, 255 }  ,opacity=1 ] [align=left] {Dual space};
\draw (450.07,200.53) node [anchor=north west][inner sep=0.75pt]  [rotate=-328.53] [align=left] {Dual};
\draw (516.5,145.9) node [anchor=north west][inner sep=0.75pt]  [font=\LARGE]  {$\beta $};
\draw (171.3,14.7) node [anchor=north west][inner sep=0.75pt]  [font=\LARGE]  {$\rho _{0}$};
\draw (513.9,93.15) node [anchor=north west][inner sep=0.75pt]  [font=\LARGE]  {$\alpha $};
\draw (328,281.8) node [anchor=north west][inner sep=0.75pt]  [font=\LARGE]  {$\vec{y}^{\hspace{0.05cm}*}$};
\draw (239.6,56.6) node [anchor=north west][inner sep=0.75pt]  [font=\LARGE]  {$\rho $};

\end{tikzpicture}
    \caption{Schematic representation of the lower bound given in Eq. (\ref{eq: winick lower bound 1}). In the first step, the quadratic Frank-Wolfe algorithm is used to go from an initial guess $\rho_0$ to a near-optimal variable $\rho$. The provable lower bound for the objective function evaluated in $\rho^{*}$ is achieved by considering contributions from a linearization term in the near-optimal $\rho$ together with a term appearing from duality of SDP's. The reliability of the method is guaranteed by the possibility to certify that the linearization and duality terms are always lower bounds for the primal objective function.}
    \label{fig:FrankWolf}
\end{figure}
\end{center}

The property of weak duality - a constructive property which always holds for convex optimization problems \cite{skrzypczyk2023semidefinite,boyd2004convex} - tells us that in a primal minimization problem the dual optimal value is always lower or equal than the primal optimum (as illustrated in Fig. \ref{fig:FrankWolf}) -- or equivalently, the dual minimum is always greater or equal than the primal maximum in a maximization primal program --, so that the reliability of Eq. (\ref{eq: winick lower bound 1}) is mathematically sound, and the accuracy of the lower bound $\beta$ is defined by the linearization term. 

The reader might notice that a condition for Theorem \ref{thm: theorem 1 winick} to hold is the existence of the derivative of $f(\rho)$. A situation in which such derivative is ill-defined is when $\mathcal{G}(\rho)$ isn't full rank. In order to ensure this condition for the existence of the derivative, one can introduce in $\rho$ a small amount of white noise (represented by the normalized full rank operator $\mathbb{1}$) through the convex combination
\begin{align}\label{eq: perturbed map}
\mathcal{G}_\epsilon(\rho)\equiv\left(\mathcal{D}_\epsilon \circ \mathcal{G}\right)(\rho), \, \mathcal{D}_\epsilon(\rho)\equiv (1-\epsilon) \rho+\epsilon \frac{\mathbb{1}}{d'},
\end{align}
with $\epsilon \in [0,1]$. This allows the definition of the perturbed objective function
\begin{align}
    f_\epsilon(\rho)\equiv D\left(\mathcal{G}_\epsilon(\rho) \| \mathcal{Z}\left(\mathcal{G}_\epsilon(\rho)\right)\right),
\end{align}
so that the gradient
\begin{align}
    \left[\nabla f_\epsilon(\rho)\right]^T=\mathcal{G}_\epsilon^{\dagger}\left(\log \mathcal{G}_\epsilon(\rho)\right)-\mathcal{G}_\epsilon^{\dagger}\left(\log \mathcal{Z}\left(\mathcal{G}_\epsilon(\rho)\right)\right)
\end{align}
exists for every $\rho \geq 0$. The consequence of this perturbation to the right side of  Eq. (\ref{optimization alpha}) is the corresponding modification of the lower bound $\beta$, previously defined in Eq. (\ref{eq: winick lower bound 1}), and the introduction of a correction term $\zeta_\varepsilon$ to ensure reliability. This mapping is depicted in Fig. \ref{fig: interior sets} and is contrasted with the facial reduction approach used in other methods.

\thm{(Theorem 2 of \cite{winick2018reliable}) \label{thm: perturbation winick}} {Let $\rho \in \mathbf{S}$, where $\rho$ is $d \times d$ and $\mathcal{G}(\rho)$ is $d^{\prime} \times d^{\prime}$. Let $\epsilon \in \mathbb{R}$. If $0<\epsilon \leqslant 1 /\left[e\left(d^{\prime}-1\right)\right]$, where $e$ is the base of the natural logarithm, then
\begin{align}
    \alpha \geqslant \beta_\epsilon(\rho)-\zeta_\epsilon
\end{align}
where
\begin{subequations}
\begin{align}
\beta_\epsilon(\sigma) & \equiv f_\epsilon(\sigma)-\operatorname{Tr}\left(\sigma^T \nabla f_\epsilon(\sigma)\right)+\max _{\vec{y} \in \mathbf{S}_\epsilon^*(\sigma)} \vec{\gamma} \cdot \vec{y}, \\
\mathbf{S}_\epsilon^*(\sigma) &\equiv \left\{\vec{y} \in \mathbb{R}^n \vert \sum_i y_i \Gamma_i^T \leqslant \nabla f_\epsilon(\sigma)\right\}, \\
\zeta_\epsilon & \equiv 2 \epsilon\left(d^{\prime}-1\right) \log \frac{d^{\prime}}{\epsilon\left(d^{\prime}-1\right)} .
\end{align}
\end{subequations}} 

\noindent
\textbf{Proof.} The proof of Theorem~\ref{thm: perturbation winick} can be found in~\cite{winick2018reliable}. \\

The strength of Theorem \ref{thm: perturbation winick} resides in the fact that it can provide a lower bound for the relative entropy, even when the matrix gradient $\nabla f$ does not exist, due to the use of Eq. (\ref{eq: perturbed map}), whose action is illustrated in Fig. \ref{fig: interior sets}. Winick et al. \cite{winick2018reliable} further generalize the previous results by allowing the inclusion of measurement data constraints based on inequalities (in contrast with the equalities satisfied in Eq. (\ref{Eq: measurement equality constraints})), which are bounded by a numerical imprecision error $\epsilon'$:
\begin{align}
    &\vert \text{Tr}\left(\Gamma  \rho\right)-\gamma_i\vert  \leq \epsilon^{\prime} & \forall i.
\end{align}

\begin{center}
\begin{figure}
\begin{centering}

  
\tikzset {_8dasvjbwl/.code = {\pgfsetadditionalshadetransform{ \pgftransformshift{\pgfpoint{0 bp } { 0 bp }  }  \pgftransformrotate{0 }  \pgftransformscale{2 }  }}}
\pgfdeclarehorizontalshading{_0waat9jbi}{150bp}{rgb(0bp)=(0.18,0.18,0.84);
rgb(37.5bp)=(0.18,0.18,0.84);
rgb(45.357142857142854bp)=(0.18,0.18,0.84);
rgb(50.714285714285715bp)=(0.18,0.18,0.84);
rgb(53.39285714285714bp)=(0.18,0.18,0.84);
rgb(57.03543526785714bp)=(0.18,0.18,0.84);
rgb(62.5bp)=(0.18,0.18,0.84);
rgb(100bp)=(0.18,0.18,0.84)}
\tikzset{_zr8r9hnhk/.code = {\pgfsetadditionalshadetransform{\pgftransformshift{\pgfpoint{0 bp } { 0 bp }  }  \pgftransformrotate{0 }  \pgftransformscale{2 } }}}
\pgfdeclarehorizontalshading{_6tsv3t9o0} {150bp} {color(0bp)=(transparent!80);
color(37.5bp)=(transparent!80);
color(45.357142857142854bp)=(transparent!70);
color(50.714285714285715bp)=(transparent!60);
color(53.39285714285714bp)=(transparent!51);
color(57.03543526785714bp)=(transparent!40);
color(62.5bp)=(transparent!0);
color(100bp)=(transparent!0) } 
\pgfdeclarefading{_mgeols1vj}{\tikz \fill[shading=_6tsv3t9o0,_zr8r9hnhk] (0,0) rectangle (50bp,50bp); } 

  
\tikzset {_51qegrd2k/.code = {\pgfsetadditionalshadetransform{ \pgftransformshift{\pgfpoint{0 bp } { 0 bp }  }  \pgftransformrotate{0 }  \pgftransformscale{2 }  }}}
\pgfdeclarehorizontalshading{_xyv8tkkmj}{150bp}{rgb(0bp)=(0.18,0.18,0.84);
rgb(37.5bp)=(0.18,0.18,0.84);
rgb(45.357142857142854bp)=(0.18,0.18,0.84);
rgb(50.714285714285715bp)=(0.18,0.18,0.84);
rgb(53.39285714285714bp)=(0.18,0.18,0.84);
rgb(57.03543526785714bp)=(0.18,0.18,0.84);
rgb(62.5bp)=(0.18,0.18,0.84);
rgb(100bp)=(0.18,0.18,0.84)}
\tikzset{_6vlh3wglr/.code = {\pgfsetadditionalshadetransform{\pgftransformshift{\pgfpoint{0 bp } { 0 bp }  }  \pgftransformrotate{0 }  \pgftransformscale{2 } }}}
\pgfdeclarehorizontalshading{_a9g5ydk9e} {150bp} {color(0bp)=(transparent!80);
color(37.5bp)=(transparent!80);
color(45.357142857142854bp)=(transparent!70);
color(50.714285714285715bp)=(transparent!60);
color(53.39285714285714bp)=(transparent!51);
color(57.03543526785714bp)=(transparent!40);
color(62.5bp)=(transparent!0);
color(100bp)=(transparent!0) } 
\pgfdeclarefading{_jhpqyh4nn}{\tikz \fill[shading=_a9g5ydk9e,_6vlh3wglr] (0,0) rectangle (50bp,50bp); } 
\tikzset{every picture/.style={line width=0.75pt}} 

\begin{tikzpicture}[x=0.75pt,y=0.75pt,yscale=-1,xscale=1,scale=1.05]

\draw  [dash pattern={on 4.5pt off 4.5pt}]  (307.83,32.75) -- (308.33,217) ;
\path  [shading=_0waat9jbi,_8dasvjbwl,path fading= _mgeols1vj ,fading transform={xshift=2}] (308.18,127.41) -- (258.73,195.12) -- (179.04,169.02) -- (179.25,85.17) -- (259.06,59.45) -- cycle ; 
 \draw   (308.18,127.41) -- (258.73,195.12) -- (179.04,169.02) -- (179.25,85.17) -- (259.06,59.45) -- cycle ; 

\draw  [fill={rgb, 255:red, 208; green, 2; blue, 27 }  ,fill opacity=1 ] (304.5,127.35) .. controls (304.5,125.21) and (306.23,123.48) .. (308.37,123.48) .. controls (310.51,123.48) and (312.25,125.21) .. (312.25,127.35) .. controls (312.25,129.49) and (310.51,131.23) .. (308.37,131.23) .. controls (306.23,131.23) and (304.5,129.49) .. (304.5,127.35) -- cycle ;
\draw    (178.41,27.12) -- (303.62,27.1) ;
\draw [shift={(306.62,27.1)}, rotate = 179.99] [fill={rgb, 255:red, 0; green, 0; blue, 0 }  ][line width=0.08]  [draw opacity=0] (5.36,-2.57) -- (0,0) -- (5.36,2.57) -- cycle    ;
\draw [color={rgb, 255:red, 255; green, 255; blue, 255 }  ,draw opacity=1 ]   (302.58,127) -- (295.58,127) ;
\draw [shift={(292.58,127)}, rotate = 360] [fill={rgb, 255:red, 255; green, 255; blue, 255 }  ,fill opacity=1 ][line width=0.08]  [draw opacity=0] (3.57,-1.72) -- (0,0) -- (3.57,1.72) -- cycle    ;
\draw [color={rgb, 255:red, 255; green, 255; blue, 255 }  ,draw opacity=1 ]   (282.83,158.5) -- (275.91,152.89) ;
\draw [shift={(273.58,151)}, rotate = 39.04] [fill={rgb, 255:red, 255; green, 255; blue, 255 }  ,fill opacity=1 ][line width=0.08]  [draw opacity=0] (3.57,-1.72) -- (0,0) -- (3.57,1.72) -- cycle    ;
\draw [color={rgb, 255:red, 255; green, 255; blue, 255 }  ,draw opacity=1 ]   (219.33,179.5) -- (222.52,171.06) ;
\draw [shift={(223.58,168.25)}, rotate = 110.7] [fill={rgb, 255:red, 255; green, 255; blue, 255 }  ,fill opacity=1 ][line width=0.08]  [draw opacity=0] (3.57,-1.72) -- (0,0) -- (3.57,1.72) -- cycle    ;
\draw [color={rgb, 255:red, 255; green, 255; blue, 255 }  ,draw opacity=1 ]   (219.83,75) -- (222.85,83.67) ;
\draw [shift={(223.83,86.5)}, rotate = 250.82] [fill={rgb, 255:red, 255; green, 255; blue, 255 }  ,fill opacity=1 ][line width=0.08]  [draw opacity=0] (3.57,-1.72) -- (0,0) -- (3.57,1.72) -- cycle    ;
\draw [color={rgb, 255:red, 255; green, 255; blue, 255 }  ,draw opacity=1 ]   (282.83,96.5) -- (275.26,101.99) ;
\draw [shift={(272.83,103.75)}, rotate = 324.06] [fill={rgb, 255:red, 255; green, 255; blue, 255 }  ,fill opacity=1 ][line width=0.08]  [draw opacity=0] (3.57,-1.72) -- (0,0) -- (3.57,1.72) -- cycle    ;
\draw [color={rgb, 255:red, 255; green, 255; blue, 255 }  ,draw opacity=1 ]   (181.33,127) -- (191.33,127) ;
\draw [shift={(194.33,127)}, rotate = 180] [fill={rgb, 255:red, 255; green, 255; blue, 255 }  ,fill opacity=1 ][line width=0.08]  [draw opacity=0] (3.57,-1.72) -- (0,0) -- (3.57,1.72) -- cycle    ;
\draw  [color={rgb, 255:red, 255; green, 255; blue, 255 }  ,draw opacity=1 ][dash pattern={on 4.5pt off 4.5pt}] (286.12,127.35) -- (251.96,174.13) -- (196.92,156.09) -- (197.06,98.18) -- (252.19,80.41) -- cycle ;
\draw    (206.54,117.77) -- (227.58,129.75) ;
\draw [shift={(219.32,125.05)}, rotate = 209.65] [fill={rgb, 255:red, 0; green, 0; blue, 0 }  ][line width=0.08]  [draw opacity=0] (5.36,-2.57) -- (0,0) -- (5.36,2.57) -- cycle    ;
\draw    (227.58,129.75) -- (245.83,132) ;
\draw [shift={(239.29,131.19)}, rotate = 187.03] [fill={rgb, 255:red, 0; green, 0; blue, 0 }  ][line width=0.08]  [draw opacity=0] (5.36,-2.57) -- (0,0) -- (5.36,2.57) -- cycle    ;
\draw    (245.83,132) -- (264.33,131.75) ;
\draw [shift={(257.68,131.84)}, rotate = 179.23] [fill={rgb, 255:red, 0; green, 0; blue, 0 }  ][line width=0.08]  [draw opacity=0] (5.36,-2.57) -- (0,0) -- (5.36,2.57) -- cycle    ;
\draw    (264.33,131.75) -- (286.12,127.35) ;
\draw [shift={(277.78,129.04)}, rotate = 168.59] [fill={rgb, 255:red, 0; green, 0; blue, 0 }  ][line width=0.08]  [draw opacity=0] (5.36,-2.57) -- (0,0) -- (5.36,2.57) -- cycle    ;
\draw  [fill={rgb, 255:red, 0; green, 0; blue, 0 }  ,fill opacity=1 ] (204.33,117.77) .. controls (204.33,116.55) and (205.32,115.56) .. (206.54,115.56) .. controls (207.76,115.56) and (208.75,116.55) .. (208.75,117.77) .. controls (208.75,118.99) and (207.76,119.98) .. (206.54,119.98) .. controls (205.32,119.98) and (204.33,118.99) .. (204.33,117.77) -- cycle ;
\draw  [dash pattern={on 4.5pt off 4.5pt}]  (467.83,32.55) -- (468.33,216.8) ;
\path  [shading=_xyv8tkkmj,_51qegrd2k,path fading= _jhpqyh4nn ,fading transform={xshift=2}] (468.18,127.21) -- (418.73,194.92) -- (339.04,168.82) -- (339.25,84.97) -- (419.06,59.25) -- cycle ; 
 \draw   (468.18,127.21) -- (418.73,194.92) -- (339.04,168.82) -- (339.25,84.97) -- (419.06,59.25) -- cycle ; 

\draw  [fill={rgb, 255:red, 208; green, 2; blue, 27 }  ,fill opacity=1 ] (464.5,127.15) .. controls (464.5,125.01) and (466.23,123.28) .. (468.37,123.28) .. controls (470.51,123.28) and (472.25,125.01) .. (472.25,127.15) .. controls (472.25,129.29) and (470.51,131.03) .. (468.37,131.03) .. controls (466.23,131.03) and (464.5,129.29) .. (464.5,127.15) -- cycle ;
\draw    (338.41,26.92) -- (463.62,26.9) ;
\draw [shift={(466.62,26.9)}, rotate = 179.99] [fill={rgb, 255:red, 0; green, 0; blue, 0 }  ][line width=0.08]  [draw opacity=0] (5.36,-2.57) -- (0,0) -- (5.36,2.57) -- cycle    ;
\draw    (364.04,118.07) -- (387.58,129.55) ;
\draw [shift={(378.15,124.95)}, rotate = 205.99] [fill={rgb, 255:red, 0; green, 0; blue, 0 }  ][line width=0.08]  [draw opacity=0] (5.36,-2.57) -- (0,0) -- (5.36,2.57) -- cycle    ;
\draw    (387.58,129.55) -- (405.83,131.8) ;
\draw [shift={(399.29,130.99)}, rotate = 187.03] [fill={rgb, 255:red, 0; green, 0; blue, 0 }  ][line width=0.08]  [draw opacity=0] (5.36,-2.57) -- (0,0) -- (5.36,2.57) -- cycle    ;
\draw    (405.83,131.8) -- (424.33,131.55) ;
\draw [shift={(417.68,131.64)}, rotate = 179.23] [fill={rgb, 255:red, 0; green, 0; blue, 0 }  ][line width=0.08]  [draw opacity=0] (5.36,-2.57) -- (0,0) -- (5.36,2.57) -- cycle    ;
\draw    (424.33,131.55) -- (452.01,127.26) ;
\draw [shift={(440.74,129.01)}, rotate = 171.2] [fill={rgb, 255:red, 0; green, 0; blue, 0 }  ][line width=0.08]  [draw opacity=0] (5.36,-2.57) -- (0,0) -- (5.36,2.57) -- cycle    ;
\draw  [fill={rgb, 255:red, 0; green, 0; blue, 0 }  ,fill opacity=1 ] (361.83,118.07) .. controls (361.83,116.85) and (362.82,115.86) .. (364.04,115.86) .. controls (365.26,115.86) and (366.25,116.85) .. (366.25,118.07) .. controls (366.25,119.29) and (365.26,120.28) .. (364.04,120.28) .. controls (362.82,120.28) and (361.83,119.29) .. (361.83,118.07) -- cycle ;
\draw  [color={rgb, 255:red, 255; green, 255; blue, 255 }  ,draw opacity=1 ][dash pattern={on 4.5pt off 4.5pt}] (345.5,90.25) .. controls (349.25,84.75) and (408.75,66.5) .. (416,68.75) .. controls (423.25,71) and (459,117.25) .. (458.75,126.25) .. controls (458.5,135.25) and (421.5,185.5) .. (415.75,187.25) .. controls (410,189) and (348.75,168.5) .. (345.5,164.25) .. controls (342.25,160) and (341.75,95.75) .. (345.5,90.25) -- cycle ;
\draw  [fill={rgb, 255:red, 208; green, 2; blue, 27 }  ,fill opacity=1 ] (282.25,127.35) .. controls (282.25,125.21) and (283.98,123.48) .. (286.12,123.48) .. controls (288.26,123.48) and (290,125.21) .. (290,127.35) .. controls (290,129.49) and (288.26,131.23) .. (286.12,131.23) .. controls (283.98,131.23) and (282.25,129.49) .. (282.25,127.35) -- cycle ;
\draw  [fill={rgb, 255:red, 208; green, 2; blue, 27 }  ,fill opacity=1 ] (448.14,127.26) .. controls (448.14,125.12) and (449.87,123.39) .. (452.01,123.39) .. controls (454.15,123.39) and (455.89,125.12) .. (455.89,127.26) .. controls (455.89,129.4) and (454.15,131.14) .. (452.01,131.14) .. controls (449.87,131.14) and (448.14,129.4) .. (448.14,127.26) -- cycle ;

\draw (173,10) node [anchor=north west][inner sep=0.75pt]   [align=left] {{ direction of optimization}};
\draw (223.08,229.25) node [anchor=north west][inner sep=0.75pt]   [align=left] {(a)};
\draw (201.33,101.9) node [anchor=north west][inner sep=0.75pt]  [font=\normalsize]  {$\rho _{0}$};
\draw (313.33,117.4) node [anchor=north west][inner sep=0.75pt]  [font=\normalsize]  {$\rho ^{*}$};
\draw (332.58,10) node [anchor=north west][inner sep=0.75pt]   [align=left] {{ direction of optimization}};
\draw (383.08,229.05) node [anchor=north west][inner sep=0.75pt]   [align=left] {(b)};
\draw (358.83,99.45) node [anchor=north west][inner sep=0.75pt]  [font=\normalsize]  {$\rho _{0}$};
\draw (473.58,116.9) node [anchor=north west][inner sep=0.75pt]  [font=\normalsize]  {$\rho ^{*}$};

\end{tikzpicture}   \caption{Illustration of two methods that guarantee interior point solutions for the optimization of Eq. (\ref{optimization alpha}): (a) shows the perturbation method used by Winick et al. \cite{winick2018reliable}, while (b) exhibits the use of a barrier method used by Lorente et al. \cite{lorente2025quantum}, which we detail in Sec. \ref{sec: Lorente}. The purple pentagon represent the feasible set of the optimization, with the gradient increasing toward the opaque region, which represents the direction of the objective function at each step. The red dot represents the optimal variable ($\rho^{*}$), while the black dot represents the initial point of the optimization ($\rho_{0}$). The barrier function (wite dashed line) is used to avoid the convergence of the objective function to the boundary of the (reduced) feasible set. In the illustrations above, gap between the barrier and the facets of the feasible set is out of scale to evidence the action of the facial reduction methods. In practice, the reduced sets aren't restrictive enough to prevent a solution from being too far from a boundary solution.}
\label{fig: interior sets}
\end{centering}
\end{figure}
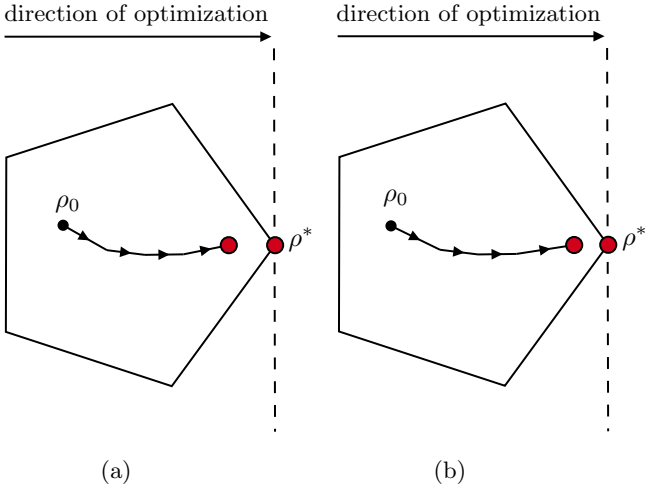
\end{center}

In a further work, the method is generalized to a Gauss-Newton interior point algorithm, which increases the efficiency of calculations and avoids the need for the aforementioned perturbations \cite{hu2022robust}. In the next section, we shift to a different method to evaluate key rates, which was initially introduced for key rates of DI settings \cite{brown2024device} (which we detail in \ref{sec: appendix gauss radau DI}) and then adapted for fully-characterized protocols \cite{araujo2023quantum}.

\subsection{Lower bounding key rates with Gauss-Radau expansions}
\label{sec: Araujo}

Gauss-Radau quadratures are a particular sort of series expansion which are interesting in the context of QKD and randomness expansion due to the fact that, by construction, they can be used as global bounds for the relative entropy in every point of its domain, providing a method which reliably computed lower bounds for the key rate. In this way, we are sure to be computing a worst-case scenario for any number of terms considered in such expansion.

This technique was originally developed to bound the conditional von Neumann entropy in DI scenarios \cite{brown2024device} as an alternative for the use of $H_\text{min}$ as a lower bound \cite{masanes2011secure}, and afterwards was adapted for fully characterized protocols in \cite{araujo2023quantum}. Here we focus in Gauss-Radau expansion for the device-dependent case, and explain the DI methods in Appendix \ref{sec: DI methods}.

We will start looking how a Gauss-Radau expansion is done for a single-variable real-valued function, and then we show it is generalization for density matrices.

\thm{\textbf{(Theorem 3.8 of \cite{brown2024device}, Gauss--Radau quadratures).}\label{thm: gauss radau}}{
For any integer $m \geq 1$, there exist nodes $t_1,\ldots,t_{m-1} \in (0,1)$ and positive weights 
$w_1,\ldots,w_m$ such that the quadrature rule
\begin{align}\label{eq: gauss-radau scalar function}
    \int\limits_{0}^{1}  \mathrm{d}t \ g(t) 
    =
    \sum_{i=1}^{m-1} w_i \, g(t_i) + w_m \, g(1)
\end{align}
is exact for all polynomials $g$ of degree up to $2m-2$. Furthermore, the coefficients $w_m$ are defined as $w_m = m^{-2}$} \\

\noindent
\textbf{Proof.} The proof of Theorem~\ref{thm: gauss radau} can be found in \cite{brown2024device}. \\

In the context of QKD, we must adapt the Gauss-Radau quadratures for the relative entropy functional of Eq.~(\ref{eq: relative entropy}). The expansion of Eq. (\ref{eq: gauss-radau scalar function}) for the integral representation of the operator logarithm yields the following approximation.

\thm{\textbf{(Theorem 2.1 of \cite{brown2024device})}}{\label{thm: gauss radau quantum}
Let $\mathcal{H}$ be a finite-dimensional\footnote{We recall that in a DI scenario, the optimization should be agnostic to the dimension of the underlying systems (which could potentially be infinite-dimensional). Therefore, in order to adequately apply the equation appearing in Theorem \ref{thm: gauss radau quantum} to such cases, it is necessary to consider a generalization proved in \cite{brown2024device}, which makes use of quasi-relative entropies and arguments of functional analysis.} Hilbert space, and let $\rho,\sigma$ be positive semi-definite operators on $\mathcal{H}$. Assume there exists $\lambda > 0$ such that $\rho \leq \lambda \sigma$. Then for any $m \in \mathbb{N}$, there exist nodes $t_1,\ldots,t_m \in (0,1]$ and positive weights $w_1,\ldots,w_m$ such that
\begin{subequations}
\begin{align}
D(\rho \| \sigma)
&\leq 
-c_m
-\sum_{i=1}^{m-1} 
    \frac{w_i}{t_i \ln 2}\,
    \inf_{Z,\rho,\sigma}\, \xi(\rho,\sigma,Z) \label{eq: relative entropy tensor product} \\
&\text{s.t.}\quad 
\|Z\| \leq \frac{3}{2}\,
\max\left\{
    \frac{1}{t_i},
    \frac{\lambda}{1-t_i}
\right\}, \label{eq: relative entropy bound}
\end{align}
\end{subequations}
where the objective function is
\begin{align}
    \xi(\rho,\sigma,Z)
    \equiv
    \langle Z + Z^* + (1-t_i) Z^*Z \rangle_\rho
    + t_i \langle ZZ^* \rangle_\sigma,
\end{align}
and the expectations are taken with respect to the density matrices indicated in the subscripts of each moment term. The coefficient $c_m$ appear from the Gauss-Radau approximation and are given by\footnote{The constant $c_m$ includes all contributions coming from the Gauss-Radau approximation of the logarithm at the different nodes, including the endpoint $t_m = 1$. Note that in this theorem the operators $\rho$ and $\sigma$ are only assumed to be positive semi-definite, and are not required to be normalized. This is why the term in parentheses in the definition of $c_m$ appears multiplied by $\text{Tr}(\rho)$ where $\lambda$ is any constant such that $\rho \leq \lambda \sigma$. In a QKD setting, we always have $\text{Tr}(\rho)=1$, so $c_m$ reduces to a constant determined by the choice of quadrature and the bound $\lambda$. The generality in \cite{brown2024device} allowing unnormalized $\rho$ is convenient for the functional analysis treatment used to prove the validity of the method for DI scenarios.} 
\begin{align}
    c_m= \text{Tr}(\rho)\left(\sum_{i=1}^{m}\frac{w_i}{t_i \ln 2}- \frac{\lambda}{m^2 \ln 2}\right).
\end{align}
Moreover, as $m \to \infty$, the right-hand side saturates the inequality.  \\

\noindent
\textbf{Proof.} The proof of Theorem~\ref{thm: gauss radau quantum} can be found in~\cite{brown2024device}. \\

The use of Gauss-Radau quadratures \cite{davis2007methods} proposed in \cite{brown2024device} to lower bound the conditional von Neumann entropy in DI scenarios (which we detail in Appendix \ref{sec: basic optimizations}) has proven useful also for fully characterized protocols. It was shown in \cite{araujo2023quantum} that one can make use of the \textit{body expansion} method proposed in \cite{navascues2014characterization} in order to convert a noncomutative polynomial optimization problem (see objective function given in Eq. (\ref{eq: Brown objective function}) in Appendix \ref{sec: basic optimizations}) into a SDP by adding bounds on the dimension of the underlying physical systems, thus being suitable for device-dependent situations.

In contrast to the DI formulation of Eq.~\eqref{eq: Brown objective function}, this approach does not require the use of NPA hierarchy as the optimization is performed directly over the true set of states compatible with the data (Eq. \eqref{Eq: measurement equality constraints}), and its complexity scales essentially with the local Hilbert space dimensions of the authenticated parties rather than with their number of measurement settings. This allows the method proposed in \cite{araujo2023quantum} to efficiently treat high-dimensional protocols and to incorporate imperfections simply as additional linear constraints on $\sigma$ (or, equivalently, on the measurement data constraints).

Transforming Equations (\ref{eq: relative entropy tensor product}) and (\ref{eq: relative entropy bound}) into an SDP relies on the fact that, in the device-dependent setting, the measurement operators are assumed to be fully characterized. This means that Alice's key-generating POVM $\{M_{a_0 \vert 0}\}_{a=0\dots d_A-1}$ and Bob's measurement operators (which we collectively denote by $\{\Gamma_k\}_k$ - defined previously in Eq. (\ref{Eq: measurement equality constraints})) are fixed and act on a known finite-dimensional Hilbert space $\mathcal{H}_A \otimes \mathcal{H}_B$. The only unknown object is the joint state $\rho_{ABE}$, which is constrained by the observed statistics. 

By using the identity connecting the conditional entropy and the Umegaki divergence, Eq. (\ref{eq: relative entropy conditional entropy}), together with the Gauss-Radau expansion for the relative entropy, Eq. (\ref{eq: relative entropy tensor product}), Araújo et al. \cite{araujo2023quantum} rewrite the lower bound as an SDP which includes the eavesdropper's measurements, in contrast with the method of Winick et al., where the identity Eq. (\ref{eq: equality conditional entropy relative entropy}) was used to remove the malicious party from the objective function. The corresponding objective function with the Gauss-Radau expansion for the relative entropy is then
\begin{widetext}
\begin{align}
\alpha \geq  c_m + \min_{\substack{
  \rho \\
  \{Z_{a,i}\}_{a,i}
}}
\sum_{i=1}^{m-1}
\sum_{a=0}^{d_A-1}
\frac{w_i}{t_i \log 2}\,
\langle M_{a_0 \vert 0} \otimes \mathbb{1}_B \otimes
\big(Z_{a,i} + Z_{a,i}^\dagger + (1-t_i)\, Z_{a,i}^\dagger Z_{a,i}\big) +\, t_i\, \mathbb{1}_{AB} \otimes Z_{a,i} Z_{a,i}^\dagger
\rangle_{\rho},
\label{eq:Arajo nonlinear}
\end{align}
\end{widetext}
where the infimum is taken over all states $\rho_{ABE}$ compatible with the experimental constraints Eq. (\ref{Eq: measurement equality constraints}) and over arbitrary complex matrices $Z_{a,i}$ acting on Eve's Hilbert space. The nodes $t_i$ and weights $w_i$ are the Gauss--Radau parameters and $c_m = \sum_{i=1}^m w_i/(t_i \log 2)$.

The difficulty in \eqref{eq:Arajo nonlinear} is that the objective function is not linear in the optimization variables, because of the quadratic terms $Z_{a,i}^\dagger Z_{a,i}$ and $Z_{a,i} Z_{a,i}^\dagger$. In the DI case, this non-linearity was absorbed by treating everything as a non-commutative polynomial optimization over pure state (as there's no bound for the dimension in the DI case) and promoting it to a relaxed SDP via the NPA hierarchy (see Appendix \ref{sec: NPA}). Here, however, the dimension of
$\mathcal{H}_A \otimes \mathcal{H}_B$ is fixed, so one would like to avoid any hierarchy and end up with a single SDP for each choice of Gauss--Radau order $m$.

The idea of the numerical technique developed in ~\cite{araujo2023quantum} is to move all the dependence on the auxiliary operators $Z_{a,i}$ into a completely positive map and then use the \emph{body expansion} method introduced in \cite{navascues2014characterization} to define a variable whose constraints and objective function define an SDP. Given a state $\rho_{ABE}$ and an operator $Z$ on Eve's system, define the map
\begin{align}
\Xi :\  \mathcal{B}(\mathcal{H}_E) &\longrightarrow \mathcal{B}(\mathcal{H}_{AB}), \\
Z &\longmapsto \text{Tr}_E \big[ \rho_{ABE} \big(\mathbb{1}_{AB} \otimes Z^T\big) \big],
\label{eq:Xi map}
\end{align}
where $T$ denotes the transpose in a fixed basis of $\mathcal{H}_E$. Using this map, we introduce
new variables on Alice and Bob's space:
\begin{align}
\sigma &\equiv \Xi(\mathbb{1}_E),
\label{eq:sigma-def}\\
\zeta_{a,i} &\equiv \Xi(Z_{a,i}),
\label{eq:zeta-def}\\
\eta_{a,i} &\equiv \Xi(Z_{a,i}^\dagger Z_{a,i}),
\label{eq:eta-def}\\
\theta_{a,i} &\equiv \Xi(Z_{a,i} Z_{a,i}^\dagger).
\label{eq:theta-def}
\end{align}
By construction, $\sigma$ is a density operator on $\mathcal{H}_{AB}$, and the observed statistics
are now encoded as linear constraints as in \eqref{Eq: measurement equality constraints}. Moreover, if we rewrite the objective function in terms of
Equations~\eqref{eq:zeta-def}–\eqref{eq:theta-def}, then its dependence on the new variables, implicitly depending on $Z_{a,i}$, becomes linear (i.e., no products of unknown variables):
\begin{align}
\text{Tr}\Big[
\big(M_{a_0 \vert 0} \otimes \mathbb{1}_B\big) \cdot
\big(\zeta_{a,i} + \zeta_{a,i}^\dagger + (1-t_i)\, \eta_{a,i}\big)
+ t_i\, \theta_{a,i}
\Big].
\label{eq:linear-objective}
\end{align}
At this point, all the non-linearity has been hidden in the relations between
$\sigma, \zeta_{a,i}, \eta_{a,i}, \theta_{a,i}$, which must still be compatible with an underlying
state $\rho_{ABE}$ and operators $Z_{a,i}$.

Instead of explicitly enforcing that, $\eta_{a,i} = \Xi(Z_{a,i}^\dagger Z_{a,i})$ and $\theta_{a,i} = \Xi(Z_{a,i} Z_{a,i}^\dagger)$ are products of the same $Z_{a,i}$ that appears in $\zeta_{a,i}$, each of these objects become independent variables whose consistency is encoded with the positivity of certain block matrices. In order to do this, it is shown in \cite{araujo2023quantum} that the existence of some $\rho_{ABE}$ and $\{Z_{a,i}\}_{a,i}$ such that Eqs.~\eqref{eq:sigma-def}–\eqref{eq:theta-def} hold is equivalent to the existence of operators $\sigma, \zeta_{a,i}, \eta_{a,i}, \theta_{a,i}$ on $\mathcal{H}_{AB}$ satisfying the linear constraints \eqref{Eq: measurement equality constraints} together with
\begin{align}
\Gamma^{(1)}_{a,i} &\equiv 
\begin{pmatrix}
\sigma      & \zeta_{a,i} \\
\zeta_{a,i}^\dagger & \eta_{a,i}
\end{pmatrix}
\geq 0,
 \hspace{0.25cm}
\Gamma^{(2)}_{a,i} &\equiv 
\begin{pmatrix}
\sigma      & \zeta_{a,i}^\dagger \\
\zeta_{a,i} & \theta_{a,i}
\end{pmatrix}
\geq 0,
\label{eq:Gamma2}
\end{align}
for all $a$ and $i$. Each block matrix $\Gamma^{(g)}_{a,i}$ is the image under the completely positive map $\Xi$ of a simple positive operator on $\mathcal{H}_E$ built from $\mathbb{1}_E$ and $Z_{a,i}$. The positivity constraints \eqref{eq:Gamma2} can be seen as a ``single relaxation" (in the same spirit as moment matrices in the NPA hierarchy, explained in Appendix \ref{sec: NPA}, but in the case considered here this is a single relaxation, while NPA is based on a hierarchy of infinite relaxations). Considering the constraints given in Equations (\ref{eq:Gamma2}), the following SDP is the desired optimization for the key rate.
\begin{widetext}
\begin{subequations}
\begin{align}
\alpha \geq  &\min_{\substack{ \sigma \in \mathbf{S} \\ \{\zeta_{a,i},\eta_{a,i},\theta_{a,i}\}_{a,i}}}\quad \sum_{i=1}^{m-1} \sum_{a=0}^{d_A-1} \frac{w_i}{t_i \log 2}\,
\text{Tr}\Big[ \big(M_{a_0 \vert 0} \otimes \mathbb{1}_B\big)
\big(\zeta_{a,i} + \zeta_{a,i}^\dagger + (1-t_i)\, \eta_{a,i}\big)
+ t_i\, \theta_{a,i} \Big] + c_m ,
\label{eq:Araujo SDP}\\
&\hspace{1cm}\text{s.t.}\hspace{1.225cm} \Gamma^{(j)}_{a,i} \succeq 0,
\hfill \forall \ a, i, j.
\label{eq:Araujo SDP constraints-2}
\end{align}
\end{subequations}
\end{widetext}
For each fixed choice of nodes $\{t_i\}$ and weights $\{w_i\}$ in the Gauss--Radau approximation, solving the SDP~\eqref{eq:Araujo SDP}--\eqref{eq:Araujo SDP constraints-2} yields a valid lower bound on $H(A|E)$ in the fully device-dependent scenario.

Below we provide an example of optimization with the Gauss-Radau radau expansion with the inclusion of semidefinite constraints representing the confidence region explained in Sec. \ref{sec: confidence regions}.

\exam{\textbf{(Lower bounds for key rates with imperfect data \cite{araujo2023quantum})}} {Let $\boldsymbol{f}$ be a vector containing the measured frequencies in a certain QKD protocol, and $\boldsymbol{p}$ be the vector containing the ideally measured statistics defined according to Eq. (\ref{Eq: measurement equality constraints}). A lower bound for the key rate via Gauss-Radau quadratures compatible with a confidence region constraining $\boldsymbol{p}$ with $\boldsymbol{f}$ under a covariance matrix $\Sigma$ and an ellipsoid radius $\chi^{2}$ (characterizing the confidence interval as described in Eq. (\ref{eq: confidence region SDP})) is given by the following SDP.}

\begin{widetext}
\begin{subequations}
\begin{align}
&\min_{\sigma\in \mathbf{S}, \boldsymbol{p},\left\{\zeta_i^a, \eta_i^a, \theta_i^a\right\}_{a, i}} c_m+\sum_{i=1}^m \sum_{a=0}^{d_A-1} \frac{w_i}{t_i \log 2} \text{Tr}\left[(M_{a_0 \vert 0} \otimes \mathbb{1}_B)(\zeta_{a,i}+\zeta^{\dagger}_{a,i \dagger}+\left(1-t_i\right) \eta_i^a)+t_i \theta_i^a\right] \\
&\hspace{1.15cm}\text{s.t.} \hspace{0.7cm} \quad\left\langle\boldsymbol{p}-\boldsymbol{f}, \Sigma^{-1}(\boldsymbol{p}-\boldsymbol{f})\right\rangle \leq \chi^2 \\
&\hspace{2.75cm}\Gamma^{(j)}_{a,i} \succeq 0,
\hfill \forall \ a, i, j.
\end{align}
\end{subequations}
\end{widetext}

More recently, a numerical technique based on a different expansion for the relative entropy \cite{frenkel2023integral} has been shown to outperform the Gauss-Radau expansions of Eq. (\ref{eq:Araujo SDP}). This technique also appeared initially in the context of approaching DIQKD protocols \cite{kossmann2024reliable}, being afterwards adapted to fully characterized scenarios \cite{kossmann2024optimising}.

\subsection{Key rates from non-symmetric conic optimization}
\label{sec: Lorente}

More recently, a different way of exploring the minimization of the relative entropy was proposed in \cite{lorente2025quantum}, which is based on the idea of converting the convex problem of Eq.~\eqref{optimization alpha} into a \emph{conic} problem. The advantage of this consists on using the relative entropy to define the cone in which the optimization is performed, rather than being the objective function itself. With this, it is possible to solve the problem with fewer explicit constraints, often resulting in faster computations in comparison with methods that need to encode the relative entropy in the objective.

Conic programs are a generalization of SDP's, where the latter is recovered as the special case in which the cone is the set of positive semi-definite matrices. A cone is defined as follows \cite{watrous2020advanced}.

\dfn{\textbf{(Cone)}} {Let $\mathcal{K}$ be a subset of a finite-dimensional real inner product space. $\mathcal{K}$ is said to be a cone if, for every vector $k \in \mathcal{K}$ and scalar $\lambda \geq 0$, one has $\lambda k \in \mathcal{K}$. Furthermore, if $\mathcal{K}$ satisfies convexity, then it is said to be a \emph{convex cone}.} \\

This definition recovers the usual cone that the reader might picture when thinking about it - for example the sets depicted on Fig. \ref{fig:Cones} -, but more abstract sets also fit in this category, such as the set of non-negative continuous functions or the cone of PSD matrices. 

\begin{center}
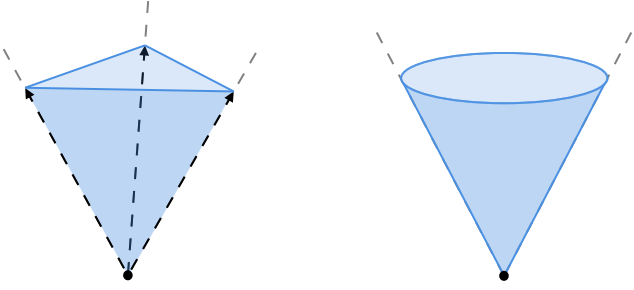
\begin{figure}

\tikzset{every picture/.style={line width=0.75pt}} 

\begin{tikzpicture}[x=0.75pt,y=0.75pt,yscale=-1,xscale=1,scale=0.575]

\draw [color={rgb, 255:red, 128; green, 128; blue, 128 }  ,draw opacity=1 ] [dash pattern={on 4.5pt off 4.5pt}]  (277,64.25) -- (164.71,259.47) ;
\draw [color={rgb, 255:red, 128; green, 128; blue, 128 }  ,draw opacity=1 ] [dash pattern={on 4.5pt off 4.5pt}]  (606.25,46.35) -- (494.74,259.94) ;
\draw [color={rgb, 255:red, 128; green, 128; blue, 128 }  ,draw opacity=1 ] [dash pattern={on 4.5pt off 4.5pt}]  (383,46.35) -- (494.59,259.94) ;
\draw [color={rgb, 255:red, 128; green, 128; blue, 128 }  ,draw opacity=1 ] [dash pattern={on 4.5pt off 4.5pt}]  (55.75,61) -- (164.71,259.47) ;
\draw [color={rgb, 255:red, 128; green, 128; blue, 128 }  ,draw opacity=1 ] [dash pattern={on 4.5pt off 4.5pt}]  (182.5,18.75) -- (164.71,259.47) ;
\draw  [draw opacity=0][fill={rgb, 255:red, 74; green, 144; blue, 226 }  ,fill opacity=0.2 ] (404.38,86.37) -- (585.09,86.37) -- (494.74,259.94) -- cycle ;
\draw  [dash pattern={on 4.5pt off 4.5pt}]  (179.47,60.61) -- (164.71,259.47) ;
\draw [shift={(179.7,57.62)}, rotate = 94.25] [fill={rgb, 255:red, 0; green, 0; blue, 0 }  ][line width=0.08]  [draw opacity=0] (8.93,-4.29) -- (0,0) -- (8.93,4.29) -- cycle    ;
\draw  [draw opacity=0][fill={rgb, 255:red, 74; green, 144; blue, 226 }  ,fill opacity=0.2 ] (164.33,257.92) -- (74.55,94.85) -- (179.71,57.62) -- cycle ;
\draw  [draw opacity=0][fill={rgb, 255:red, 74; green, 144; blue, 226 }  ,fill opacity=0.2 ] (164.1,261.1) -- (179.71,57.62) -- (257.6,98.1) -- cycle ;
\draw  [draw opacity=0][fill={rgb, 255:red, 74; green, 144; blue, 226 }  ,fill opacity=0.2 ] (165.17,260.24) -- (74.65,95.21) -- (256.66,98.23) -- cycle ;
\draw  [fill={rgb, 255:red, 0; green, 0; blue, 0 }  ,fill opacity=1 ] (161.47,260.19) .. controls (161.09,258.28) and (162.25,256.41) .. (164.04,256.02) .. controls (165.83,255.62) and (167.58,256.85) .. (167.96,258.76) .. controls (168.33,260.66) and (167.18,262.53) .. (165.38,262.93) .. controls (163.59,263.32) and (161.84,262.09) .. (161.47,260.19) -- cycle ;
\draw  [dash pattern={on 4.5pt off 4.5pt}]  (256.08,100.7) -- (164.71,259.47) ;
\draw [shift={(257.57,98.1)}, rotate = 119.92] [fill={rgb, 255:red, 0; green, 0; blue, 0 }  ][line width=0.08]  [draw opacity=0] (8.93,-4.29) -- (0,0) -- (8.93,4.29) -- cycle    ;
\draw  [dash pattern={on 4.5pt off 4.5pt}]  (76.09,97.84) -- (164.71,259.47) ;
\draw [shift={(74.65,95.21)}, rotate = 61.27] [fill={rgb, 255:red, 0; green, 0; blue, 0 }  ][line width=0.08]  [draw opacity=0] (8.93,-4.29) -- (0,0) -- (8.93,4.29) -- cycle    ;
\draw  [draw opacity=0][fill={rgb, 255:red, 74; green, 144; blue, 226 }  ,fill opacity=0.2 ] (404.38,86.37) -- (585.09,86.37) -- (494.74,259.94) -- cycle ;
\draw  [color={rgb, 255:red, 74; green, 144; blue, 226 }  ,draw opacity=0.75 ][fill={rgb, 255:red, 255; green, 255; blue, 255 }  ,fill opacity=1 ] (404.38,86.37) .. controls (404.38,74.18) and (444.95,64.3) .. (494.98,64.3) .. controls (545.02,64.3) and (585.58,74.18) .. (585.58,86.37) .. controls (585.58,98.56) and (545.02,108.44) .. (494.98,108.44) .. controls (444.95,108.44) and (404.38,98.56) .. (404.38,86.37) -- cycle ;
\draw  [color={rgb, 255:red, 74; green, 144; blue, 226 }  ,draw opacity=0.75 ][fill={rgb, 255:red, 74; green, 144; blue, 226 }  ,fill opacity=0.2 ] (404.38,86.37) .. controls (404.38,74.18) and (444.83,64.3) .. (494.74,64.3) .. controls (544.64,64.3) and (585.09,74.18) .. (585.09,86.37) .. controls (585.09,98.56) and (544.64,108.44) .. (494.74,108.44) .. controls (444.83,108.44) and (404.38,98.56) .. (404.38,86.37) -- cycle ;
\draw [color={rgb, 255:red, 74; green, 144; blue, 226 }  ,draw opacity=1 ]   (74.55,94.85) -- (257.57,98.1) ;
\draw [color={rgb, 255:red, 74; green, 144; blue, 226 }  ,draw opacity=1 ]   (74.55,94.85) -- (179.68,57.62) ;
\draw [color={rgb, 255:red, 74; green, 144; blue, 226 }  ,draw opacity=1 ]   (179.7,57.62) -- (257.57,98.1) ;
\draw [color={rgb, 255:red, 74; green, 144; blue, 226 }  ,draw opacity=1 ]   (406.25,90.75) -- (494.74,259.94) ;
\draw [color={rgb, 255:red, 74; green, 144; blue, 226 }  ,draw opacity=1 ]   (582.75,91.5) -- (494.74,259.94) ;
\draw  [fill={rgb, 255:red, 0; green, 0; blue, 0 }  ,fill opacity=1 ] (491.34,260.66) .. controls (490.97,258.75) and (492.12,256.88) .. (493.92,256.49) .. controls (495.71,256.09) and (497.46,257.32) .. (497.83,259.23) .. controls (498.2,261.14) and (497.05,263) .. (495.26,263.4) .. controls (493.47,263.79) and (491.71,262.57) .. (491.34,260.66) -- cycle ;

\end{tikzpicture}
    \caption{Geometric representation of two sorts of cones, represented by the regions spanned from the origin (the black dots) to the entire subset delimited by the gray dashed lines. The left cone is generated by a conic hull of three linearly independent vectors (dashed black arrows), while right cone can not be written as the conic hull of a finite amount of vectors is represented.}
    \label{fig:Cones}
\end{figure}
\end{center}

A general conic program is written as follows: Consider a finite dimensional real inner product space $V$ and a closed\footnote{A set in which every Cauchy sequence converges is said to be closed. In other words, a closed set contain its boundaries. In practical numerical implementations, this means that the optimal value could lie in the boundary of the feasible set so that it is generally important for the feasible set to be closed and bounded.}} cone $\mathcal{K} \subseteq V$. Let $W$ be another finite dimensional real inner product space, $\phi : V \rightarrow W$ a linear function, and vectors $a\in V$, $b \in W$. The form of a conic program is then \cite{watrous2020advanced}
\begin{subequations}
\begin{align}
    \min_{x} \quad & \langle a, x \rangle, \\
    \text{s.t.} \quad & \phi(x) = b, \\
                     & x \in \mathcal{K}.
\end{align}
\label{eq: conic standard form}
\end{subequations}

As the set of density matrices forms a convex cone, problems of the form of Eq.~\eqref{optimization alpha} can often be formulated over this sort of feasible set. For the purpose of QKD, as we are interested in optimizing the relative entropy in the form of \ref{thm: removing eve} (which depends on two entries determined by the maps $\mathcal{G}$ and $\mathcal{Z}$ acting over $\rho$), it is relevant to highlight that if $\mathcal{K}$ and $\mathcal{K}^{\prime}$ are convex cones, then the set defined by composition via cartesian product
\begin{align}
    \mathcal{K} \times \mathcal{K}^{\prime} \equiv \{(k,k^{\prime}); k\in \mathcal{K}, k^{\prime} \in \mathcal{K}^{\prime}  \}
\end{align}
is again a convex cone, as this composition preserves both convexity and the conic aspect of the new set \cite{watrous2020advanced}. These observations allow us to construct cones via cartesian products that characterize an adequate feasible set for the relative entropy. 

When the feasible cone is a composition of different sorts of cones, a \textit{non-symmetric cone} is characterized. It was shown in \cite{lorente2025quantum} that a non-symmetric cone can be defined for the task of optimizing the relative entropy, allowing the evaluation of asymptotic key rates. Instead of minimizing over density matrices as variables, in this space we minimize a real parameter $h$ satisfying $h \geq D(\rho\|\sigma)$. Following \cite{lorente2025quantum}, the \emph{relative entropy cone} is defined as the composition
\begin{align}
\begin{aligned}
        \mathcal{K}_{\mathrm{RE}}
    = \operatorname{cl}\big\{
        (h, \rho, \sigma) \in &\mathbb{R} \times \mathbb{H}^d \times \mathbb{H}^d  ; \\
        &\rho, \sigma \succ 0, h \geq D(\rho \| \sigma)
      \big\},
    \label{eq: relative entropy cone}
\end{aligned}
\end{align}
where $\mathbb{H}^d$ denotes the space of $d-$dimensional Hermitian matrices, and $\operatorname{cl}$ denotes the topological closure. Intuitively, $\mathcal{K}_{\mathrm{RE}}$ collects all triples $(h,\rho,\sigma)$ for which $h$ upper-bounds the relative entropy between $\rho$ and $\sigma$. As we will see next, this cone allows us to reformulate the key-rate problem as a conic optimization problem\footnote{JuMP and QICS are the most recent conic optimization modelling tools \cite{he2024qics,lubin2023jump} and allow a relatively easy implementation of the cones relevant for QKD.} in the sense of Eq.~\eqref{eq: conic standard form}. 

\thm{\textbf{(Non-symmetric conic optimization for QKD rates \cite{lorente2025quantum})}} {\label{thm: nonsymmetric conic opt key rate} Consider a device-dependent QKD protocol specified by a feasible set $\mathbf{S}$ of density operators $\rho$ on $\mathcal{H}_{AB}$, encoding the linear constraints \eqref{Eq: measurement equality constraints}; The CP map $\mathcal{G}$ that relates $\rho$ with the classical key register (Eq. \eqref{eq: map G}) and the CPTP pinching map $\mathcal{Z}$ (Eq. \eqref{eq: map Z}). Then the optimization problem represented in Eq. \eqref{optimization alpha} is equivalent to the conic program
\begin{subequations}
    \begin{align}
        \alpha\geq  \min_{\substack{h \in \mathbb{R} , \rho \in \mathbf{S} }} \quad & h \\
        &\hspace{-1.10cm}\text{s.t.}\hspace{0.55cm}
        \big(h,\mathcal{G}(\rho),\mathcal{Z}(\mathcal{G}(\rho))\big) \in \mathcal{K}_{\mathrm{RE}},
        \label{eq: conic RE formulation}
    \end{align}
\end{subequations}
where $\mathcal{K}_{\mathrm{RE}}$ is the relative entropy cone defined in Eq.~\eqref{eq: relative entropy cone}.} \\

\noindent
\textbf{Proof.} The proof of Theorem~\ref{thm: nonsymmetric conic opt key rate} can be found in~\cite{lorente2025quantum}. \\

In particular, at the optimum we have $h^{*} = D\big(\mathcal{G}(\rho^{*})\| \mathcal{Z}(\mathcal{G}(\rho^{*})))$. It was shown in \cite{lorente2025quantum} that after applying facial reduction to the feasible set $\mathbf{S}$ and to the images of $\mathcal{G}$ and $\mathcal{Z}$, there exists a reduced state space (encoded by an isometry), reduced POVM elements $\{F_k\}_k$, and reduced maps $\widehat{\mathcal{G}}$ and $\widehat{\mathcal{Z}}$ such that the problem can be further written as
\begin{subequations}
\begin{align}\label{eq: conic QKD formulation}
     \alpha \geq &\min_{h, \sigma \in \mathbf{S}^{\prime}} \hspace{0.35cm} h \\
    &\hspace{0.3cm}\text{s.t.}\hspace{0.5cm}
     (h,\sigma) \in \mathcal{K}^{\widehat{\mathcal{G}},\widehat{\mathcal{Z}}}_{\mathrm{QKD}},
\end{align}
\end{subequations}
where $\mathbf{S}^{\prime}$ is the reduced feasible set and the QKD cone, defined as 
\begin{align}
\begin{aligned}
\mathcal{K}^{\widehat{\mathcal{G}},\widehat{\mathcal{Z}}}_{\mathrm{QKD}}\equiv  \big\{&(h,\rho)\in \mathbb{R}\times\mathbb{H}^n  ;
        \rho \succeq 0, \\
        &\hspace{1.5cm}h \geq H(\widehat{\mathcal{Z}}(\rho)) - H(\widehat{\mathcal{G}}(\rho)) \big\},
        \label{eq: QKD cone}
\end{aligned}
\end{align}
is a closed convex cone depending explicitly on the reduced maps $\widehat{\mathcal{G}}$ and $\widehat{\mathcal{Z}}$. For any fixed protocol, solving the conic program
\eqref{eq: conic QKD formulation} yields a valid lower bound on $H(A|E)$, and hence on the
asymptotic Devetak--Winter key rate.

In Eq.~\eqref{eq: conic RE formulation} the relative entropy cone $\mathcal{K}_{\mathrm{RE}}$ appears directly as part of the feasible set, so that the objective is simply linear in the auxiliary scalar $h$. However, in order to solve such problems efficiently we have already seen (Sec. \ref{sec: Winick}) that it is essential to have a method of facial reduction to solve the problem. We recall that there must exist a point $(h,\rho)$ that satisfies the constraints and lies in the interior of the QKD cone to guarantee strict feasibility, and they can be guaranteed by using a method of facial reduction. The standard notion of facial reduction consists of identifying the smallest subspace of the cone that contains all feasible points, and then rewriting the problem on this lower-dimensional space so that the optimizer becomes an interior point of the new cone, as illustrated in Fig. \ref{fig: facial reduction}

\begin{center}
\begin{figure}
\begin{centering}

  
\tikzset {_5epg4vvsn/.code = {\pgfsetadditionalshadetransform{ \pgftransformshift{\pgfpoint{0 bp } { 0 bp }  }  \pgftransformrotate{0 }  \pgftransformscale{2 }  }}}
\pgfdeclarehorizontalshading{_vnev684nl}{150bp}{rgb(0bp)=(0.18,0.18,0.84);
rgb(37.5bp)=(0.18,0.18,0.84);
rgb(45.357142857142854bp)=(0.18,0.18,0.84);
rgb(50.714285714285715bp)=(0.18,0.18,0.84);
rgb(53.39285714285714bp)=(0.18,0.18,0.84);
rgb(57.03543526785714bp)=(0.18,0.18,0.84);
rgb(62.5bp)=(0.18,0.18,0.84);
rgb(100bp)=(0.18,0.18,0.84)}
\tikzset{_mjskqem0g/.code = {\pgfsetadditionalshadetransform{\pgftransformshift{\pgfpoint{0 bp } { 0 bp }  }  \pgftransformrotate{0 }  \pgftransformscale{2 } }}}
\pgfdeclarehorizontalshading{_t3os8rj16} {150bp} {color(0bp)=(transparent!80);
color(37.5bp)=(transparent!80);
color(45.357142857142854bp)=(transparent!70);
color(50.714285714285715bp)=(transparent!60);
color(53.39285714285714bp)=(transparent!51);
color(57.03543526785714bp)=(transparent!40);
color(62.5bp)=(transparent!0);
color(100bp)=(transparent!0) } 
\pgfdeclarefading{_vijcciuia}{\tikz \fill[shading=_t3os8rj16,_mjskqem0g] (0,0) rectangle (50bp,50bp); } 
\tikzset{every picture/.style={line width=0.75pt}} 

\begin{tikzpicture}[x=0.75pt,y=0.75pt,yscale=-1,xscale=1,scale=0.84]

\draw [color={rgb, 255:red, 74; green, 144; blue, 226 }  ,draw opacity=0.25 ] [dash pattern={on 4.5pt off 4.5pt}]  (13.16,44.81) -- (269.95,33.99) ;
\draw [color={rgb, 255:red, 74; green, 144; blue, 226 }  ,draw opacity=0.25 ] [dash pattern={on 4.5pt off 4.5pt}]  (3.28,130.81) -- (260.06,119.99) ;
\path  [shading=_vnev684nl,_5epg4vvsn,path fading= _vijcciuia ,fading transform={xshift=2}] (233.57,79.31) -- (183.74,150.39) -- (119.13,126.26) -- (129.02,40.27) -- (199.74,11.25) -- cycle ; 
 \draw   (233.57,79.31) -- (183.74,150.39) -- (119.13,126.26) -- (129.02,40.27) -- (199.74,11.25) -- cycle ; 

\draw  [fill={rgb, 255:red, 0; green, 0; blue, 0 }  ,fill opacity=1 ] (144.36,73.47) .. controls (144.5,72.22) and (145.45,71.18) .. (146.48,71.14) .. controls (147.52,71.1) and (148.24,72.08) .. (148.1,73.33) .. controls (147.96,74.58) and (147.01,75.62) .. (145.97,75.66) .. controls (144.94,75.7) and (144.21,74.72) .. (144.36,73.47) -- cycle ;
\draw [color={rgb, 255:red, 74; green, 144; blue, 226 }  ,draw opacity=0.51 ] [dash pattern={on 4.5pt off 4.5pt}]  (83.89,15.79) -- (340.67,4.96) ;
\draw [color={rgb, 255:red, 74; green, 144; blue, 226 }  ,draw opacity=0.5 ] [dash pattern={on 4.5pt off 4.5pt}]  (67.89,154.93) -- (324.67,144.11) ;
\draw [color={rgb, 255:red, 0; green, 0; blue, 0 }  ,draw opacity=0.76 ]   (146.23,73.4) -- (164.88,84.39) ;
\draw [shift={(157.8,80.21)}, rotate = 210.5] [fill={rgb, 255:red, 0; green, 0; blue, 0 }  ,fill opacity=0.76 ][line width=0.08]  [draw opacity=0] (5.36,-2.57) -- (0,0) -- (5.36,2.57) -- cycle    ;
\draw [color={rgb, 255:red, 0; green, 0; blue, 0 }  ,draw opacity=0.81 ][line width=0.75]    (164.88,84.39) -- (187.47,88.05) ;
\draw [shift={(178.74,86.63)}, rotate = 189.2] [fill={rgb, 255:red, 0; green, 0; blue, 0 }  ,fill opacity=0.81 ][line width=0.08]  [draw opacity=0] (5.36,-2.57) -- (0,0) -- (5.36,2.57) -- cycle    ;
\draw [color={rgb, 255:red, 0; green, 0; blue, 0 }  ,draw opacity=0.89 ][line width=0.75]    (187.47,88.05) -- (208.27,86.42) ;
\draw [shift={(200.46,87.03)}, rotate = 175.52] [fill={rgb, 255:red, 0; green, 0; blue, 0 }  ,fill opacity=0.89 ][line width=0.08]  [draw opacity=0] (5.36,-2.57) -- (0,0) -- (5.36,2.57) -- cycle    ;
\draw [line width=0.75]    (208.27,86.42) -- (233.73,79.24) ;
\draw [shift={(223.5,82.12)}, rotate = 164.27] [fill={rgb, 255:red, 0; green, 0; blue, 0 }  ][line width=0.08]  [draw opacity=0] (5.36,-2.57) -- (0,0) -- (5.36,2.57) -- cycle    ;
\draw [color={rgb, 255:red, 74; green, 144; blue, 226 }  ,draw opacity=1 ][line width=1.5]  [dash pattern={on 5.63pt off 4.5pt}]  (117.71,83.84) -- (374.49,73.02) ;
\draw  [fill={rgb, 255:red, 208; green, 2; blue, 27 }  ,fill opacity=1 ] (230.44,79.37) .. controls (230.69,77.18) and (232.36,75.34) .. (234.18,75.27) .. controls (235.99,75.2) and (237.26,76.92) .. (237.02,79.12) .. controls (236.77,81.31) and (235.1,83.15) .. (233.28,83.22) .. controls (231.46,83.29) and (230.19,81.57) .. (230.44,79.37) -- cycle ;
\draw [color={rgb, 255:red, 74; green, 144; blue, 226 }  ,draw opacity=1 ] [dash pattern={on 4.5pt off 4.5pt}]  (340.67,4.96) -- (374.49,73.02) ;
\draw [color={rgb, 255:red, 74; green, 144; blue, 226 }  ,draw opacity=1 ] [dash pattern={on 4.5pt off 4.5pt}]  (374.49,73.02) -- (324.67,144.11) ;
\draw [color={rgb, 255:red, 74; green, 144; blue, 226 }  ,draw opacity=0.52 ] [dash pattern={on 4.5pt off 4.5pt}]  (260.06,119.99) -- (324.67,144.11) ;
\draw [color={rgb, 255:red, 74; green, 144; blue, 226 }  ,draw opacity=0.51 ] [dash pattern={on 4.5pt off 4.5pt}]  (269.95,33.99) -- (260.06,119.99) ;
\draw [color={rgb, 255:red, 74; green, 144; blue, 226 }  ,draw opacity=0.52 ] [dash pattern={on 4.5pt off 4.5pt}]  (340.67,4.96) -- (269.95,33.99) ;
\draw [color={rgb, 255:red, 74; green, 144; blue, 226 }  ,draw opacity=1 ][line width=1.5]  [dash pattern={on 5.63pt off 4.5pt}]  (83.89,15.79) -- (117.71,83.84) ;
\draw [color={rgb, 255:red, 74; green, 144; blue, 226 }  ,draw opacity=1 ][line width=1.5]  [dash pattern={on 5.63pt off 4.5pt}]  (117.71,83.84) -- (67.89,154.93) ;
\draw [color={rgb, 255:red, 74; green, 144; blue, 226 }  ,draw opacity=0.52 ][line width=1.5]  [dash pattern={on 5.63pt off 4.5pt}]  (83.89,15.79) -- (13.16,44.81) ;
\draw [color={rgb, 255:red, 74; green, 144; blue, 226 }  ,draw opacity=0.51 ][line width=1.5]  [dash pattern={on 5.63pt off 4.5pt}]  (13.16,44.81) -- (3.28,130.81) ;
\draw [color={rgb, 255:red, 74; green, 144; blue, 226 }  ,draw opacity=0.52 ][line width=1.5]  [dash pattern={on 5.63pt off 4.5pt}]  (3.28,130.81) -- (67.89,154.93) ;

\draw (135.38,47.74) node [anchor=north west][inner sep=0.75pt]  [font=\large]  {$\sigma _{0}$};
\draw (232.46,56.32) node [anchor=north west][inner sep=0.75pt]  [font=\large]  {$\sigma^{*}$};
\draw (38.13,3) node [anchor=north west][inner sep=0.75pt]  [font=\Large,color={rgb, 255:red, 74; green, 144; blue, 226 }  ,opacity=1]  {$\mathbf{S}$};
\draw (211.47,109.33) node [anchor=north west][inner sep=0.75pt]  [font=\Large,color={rgb, 255:red, 0; green, 0; blue, 0 }  ,opacity=1]  {$\mathbf{S'}$};

\end{tikzpicture}
        \caption{Pictorial representation of the facial reduction method employed in Lorente et al. \cite{lorente2025quantum}. The optimization space performed over the feasible set $\mathbf{S}$ (Eq. (\ref{Eq: measurement equality constraints})) with (not necessarily full rank) density matrices is reduced to $\mathbf{S}^{\prime}$ after applying an isometry that allows an equivalent optimization to be performed in the reduced face $\mathbf{S^\prime}$ over full rank operators $\sigma$, and redundant constraints to be dropped.}
    \label{fig: facial reduction}
\end{centering}
\end{figure}
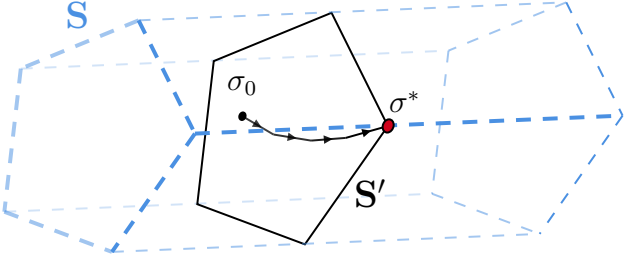
\end{center}

Facial reductions must always be part of primal-dual optimization techniques, because without the possibility of ensuring strict feasibility the optimality conditions may be meaningless in certain situations, as shown in \cite{drusvyatskiy2017many}. In \cite{lorente2025quantum}, facial reduction is implemented as follows: the feasible set $\mathbf{S}$ defined in Eq.~\eqref{Eq: measurement equality constraints} is reduced by finding an isometry $U$ whose range contains the support of all feasible states, so that any $\rho\in\mathbf{S}$ can be written as $\rho = U\sigma U^{\dagger}$ for some $\sigma\succeq 0$ on the reduced space (see Fig. \ref{fig: facial reduction}). The statistics are expressed in terms of $\sigma$ through reduced operators $F_k \equiv U^{\dagger}E_k U$ (and redundant constraints may be dropped). Similarly, the maps $\mathcal{G}$ and $\mathcal{Z}\circ\mathcal{G}$ are restricted to the supports of their respective domains, yielding reduced maps $\widehat{\mathcal{G}}$ and $\widehat{\mathcal{Z}}$. In order to employ these maps in the numerical optimization, one needs to make use of the identity\footnote{The use of this identity is necessary because $D( \mathcal{G}(\rho)\Vert \mathcal{Z}(\mathcal{G}(\rho))) \neq D( \hat{\mathcal{G}}(\rho) \Vert \hat{\mathcal{Z}}(\rho))$. On the other hand, the von Neumann entropy satisfies $H(\mathcal{G}(\rho)) = H (\hat{\mathcal{G}}(\rho))$ and $H(\mathcal{Z}(\mathcal{G}(\rho))) = H (\hat{\mathcal{Z}}(\rho))$.}
\begin{align}
    D(\mathcal{G}(\rho)\big\|\mathcal{Z}(\mathcal{G}(\rho)))
    =  H(\mathcal{Z}(\mathcal{G}(\rho))) - H(\mathcal{G}(\rho)).
\end{align}
This motivates the definition of the QKD cone $\mathcal{K}^{\widehat{\mathcal{G}},\widehat{\mathcal{Z}}}_{\mathrm{QKD}}$ in Eq.~\eqref{eq: QKD cone}, where the constraint $h \geq -H(\widehat{\mathcal{G}}(\sigma)) + H(\widehat{\mathcal{Z}}(\sigma))$ replaces the relative-entropy bound $h \geq D(\rho\|\sigma)$ used in $\mathcal{K}_{\mathrm{RE}}$.

In contrast to the reliability of solutions guaranteed by duality discussed for the method of Winick et al. \cite{winick2018reliable}, the conic optimization method guarantees avoids overestimating the key rate with the use of a so-called logarithmically homogeneous self-concordant barrier (LHSCB)\footnote{Without facial reductions, these functions ight not be well-defined \cite{araujo2023comment}.} function \cite{nesterov2000squared}. LHSCB functions act as a sort of ``potential" which tends to infinity when approaching boundary points and the curve of the relative entropy itself. In this way, this method provides a certified lower bound for the asymptotic key rate. The LHSCB function used by Lorente \textit{et al.} \cite{lorente2025quantum} is\footnote{In principle, many different LHSCB functions could be chosen. This choice doesn't affect the optimal solution (which is defined by the objective function and the constraints of the conic program), but instead only the performance of the optimization.}
\begin{align}
\begin{aligned}
f(h,\sigma) = &- \log\big(h + H(\widehat{\mathcal{G}}(\sigma))  - H(\widehat{\mathcal{Z}}(\sigma))\big) - \\ &- \log\det(\sigma).
\end{aligned}
\end{align}
The definition above ensures that this interior-point algorithm follows a trajectory that remains strictly inside the reduced cone, as illustrated in Fig.~\ref{fig: interior sets}, being therefore a conceptually different of the perturbed map $\mathcal{G}_\epsilon$ used in \cite{winick2018reliable}. \\

\exam{\textbf{(Lower bounds for key rates via conic optimization under partial state characterization \cite{pereira2025optimal})}} {Consider a PM scenario in which Alice's unknown signals $\{\rho_j\}_j$ are written, without loss of generality, as $\rho_j = \ket{\psi_j} \bra{\psi_j}$, where each state $\ket{\psi_j}$ can be expressed in terms of a set of known reference states $\{\ket{\phi_j}\}_j$ and coefficients $\{\epsilon_j\}_j$, without loss of generality, as
\begin{align}\label{eq: pure partial state characterization}
    \ket{\psi_j} = \sqrt{1-\epsilon_j} \ket{\phi_j} + \sqrt{\epsilon_j} \ket{\phi^{\perp}_j},
\end{align}
where $\ket{\phi^{\perp}_j}$ is an unknown state satisfying $\langle \phi_j^{\perp}\vert \phi_j \rangle = 0$. As shown in Appendix A of \cite{pereira2025optimal}, any set of states satisfying Eq. (\ref{eq: partial state characterization mixed state}) can be derived by applying a CPTP map on the set $\{\ket{\psi_j}\}_j$ described by Eq. (\ref{eq: pure partial state characterization}) (which also satisfy the partial state characterization constraint, so that the equation above does not constitutes a relaxation of the original problem), implying that a security proof for the set of pure states above extends to the entire set of density matrices. Therefore, the key rate can be lower bounded by the non-symmetric conic optimization under partial constraints with pure states as follows.
}
\begin{widetext}
\begin{subequations}
\begin{align}
&\min_{h,\rho_{AB},\left\{\left|\phi_j^{\perp}\right\rangle\right\}_j} h
\\
&\begin{aligned}
\hspace{0.9cm}
\text{s.t.}\hspace{0.66cm}
\frac{\langle i|\rho_A|j\rangle}{\sqrt{p_i p_j}}
&=
\sqrt{(1-\epsilon_i)(1-\epsilon_j)}
\left\langle\phi_j \middle| \phi_i\right\rangle
+
\sqrt{(1-\epsilon_i)\epsilon_j}
\left\langle\phi_j^{\perp} \middle| \phi_i\right\rangle +
\\
&\quad
+
\sqrt{\epsilon_i(1-\epsilon_j)}
\left\langle\phi_j \middle| \phi_i^{\perp}\right\rangle
+
\sqrt{\epsilon_i\epsilon_j}
\left\langle\phi_j^{\perp} \middle| \phi_i^{\perp}\right\rangle,
\end{aligned} \hspace{3.5cm}\forall i,j,
\\
&\hspace{2.0cm}\operatorname{Tr}\left[
\left(
|j\rangle\!\left\langle j\right|_A \otimes \Gamma_k
\right)\rho_{AB}
\right]=p_jY_{k|j}, \hspace{7.75cm} \forall j,k,
\\
&\hspace{2cm}\left\langle\phi_j^{\perp} \middle| \phi_j^{\perp}\right\rangle=1,
\quad
\left\langle\phi_j^{\perp} \middle| \phi_j\right\rangle=0, \hspace{8.525cm} \forall j,\\
&\hspace{2cm}\left(h,\rho_{AB}\right)\in
\mathcal{K}_{\mathrm{QKD}}^{\widehat{\mathcal{G}},\widehat{\mathcal{Z}}}.
\end{align}
\end{subequations}

The partial constraints in the optimization above can be further simplified in terms of the positive-semidefinite condition of a Gram matrix with entries defined in terms of the overlaps of the states $\{\ket{\phi_j^{(\perp)}}\}_j$.

\subsection{Open-source codes}
\label{sec: open sourcer codes}

In the table below we exhibit the methods discussed in the previous section, and provide the links for their open source codes, together with the language in which they are written and the modelling tool used for the optimization.
\begin{center}
\begin{table}[H]
\centering
\caption{Open source codes for evaluating asymptotic keys in fully characterized QKD protocols.}
\label{tab:open_source_codes}
\renewcommand{\arraystretch}{1.6} 
\setlength{\tabcolsep}{21.5pt} 


\begin{tabular}{c c c c c} 
\toprule
\makecell{\textbf{Methods} \\ \textbf{(Sec.~\ref{sec: algorithms})}} & 
\makecell{\textbf{Underlying} \\ \textbf{algorithm}} & 
\textbf{Package} & 
\textbf{Repository} & 
\makecell{\textbf{Language} \\ \textbf{(modeling tool)}} \\
\midrule 

\makecell{Winick \textit{et al.} \\ \cite{winick2018reliable}}  
& \makecell{Frank-Wolfe \\ \cite{frank1956algorithm}} & \cite{burniston_2024_14262569} & \href{https://github.com/Optical-Quantum-Communication-Theory/openQKDsecurity/tree/main}{Github} & \makecell{MATLAB (CVX) \\ Octave (YALMIP \cite{lofberg2004yalmip})} \\[12pt]




\makecell{Araújo \textit{et al.} \\ \cite{araujo2023quantum}} 
& \makecell{Gauss-Radau \\ \cite{davis2007methods}} & \cite{araujo2023quantum} & \href{https://github.com/araujoms/qkd}{Github} & \makecell{MATLAB (CVX) \\ Octave (YALMIP \cite{lofberg2004yalmip})} \\[12pt]

\makecell{Lorente \textit{et al.} \\ \cite{lorente2025quantum}} 
& \makecell{Skajaa-Ye \\ \cite{skajaa2015homogeneous,papp2017homogeneous}} & \cite{lorente2025quantum} & \href{https://github.com/araujoms/ConicQKD.jl}{Github} & \makecell{Julia \\ (JuMP \cite{lubin2023jump})} \\ [12pt]
\bottomrule
\end{tabular}
\end{table}
\end{center}
\end{widetext}

\section{Analytical methods for finite-length keys}
\label{sec: finite keys}

The methods for numerical security proofs considered so far allow one to derive reliable lower bounds to the asymptotic key rates of QKD protocols under various imperfections. In real-world implementations, a protocol is also subject to uncertainties arising from finite statistics as well as the possibility of Eve applying arbitrary strategies that may deviate from the iid. assumption of collective attacks. The main approaches to derive key rates accounting for finite-size effects and coherent attacks consist in powerful theorems that reduce the secret key analysis to the estimation of single round quantities.

The techniques discussed here do not replace the numerical methods for key-rate estimation presented in the previous section. Rather, they constitute an initial step in the security proof of a QKD protocol, bridging the computation of the key rate for a finite number of rounds with the asymptotic quantities that can be estimated using the numerical methods described in Section~\ref{sec: algorithms}.  In this way, one can combine different techniques to develop numerical security proofs for a wide range of realistic protocols that account for finite-size effects and various imperfections.

In this section, we discuss three techniques for the security analysis of QKD protocols in the finite regime. We start with the \textit{post-selection} technique, which is based on \textit{de Finetti theorems}, and allows one to use symmetries of a QKD protocol to infer how secure a protocol is against coherent attacks based on the security of the protocol against collective attacks (i.e., when the protocol implementation is described by iid states).

\subsection{Postselection technique}
\label{sec: postselection}

De Finetti theorems  use symmetries of (classical and quantum) systems to infer how close a small fraction of a big system is to an iid distribution. It therefore provides an interesting connection between the iid property and permutation invariance, and, as we shall see, this has useful applications in the context of QKD\footnote{De Finetti theorems are also useful within other subjects of Quantum Information \cite{brandao2013quantum,belzig2024studying,gross2021schur,costa2025finetti}.} \cite{renner2008security}.

We start by presenting the classical and the corresponding quantum generalizations of the de Finetti theorems to provide intuition for how they can be applied to the security proof of a QKD protocol.

A classical system of $m$ random variables $\{X_i\}_{i=1\dots,m}$ is \textit{exchangeable} if the global probability distribution is preserved under an arbitrary permutation operation $\pi$ that relabels the variables $\{X_i\}_{i=1,\dots,m}$,
\begin{align}
    p_{X_1, \dots, X_m}(x_1, \dots, x_m) = p_{X_{1}, \dots, X_{m}}(x_{\pi(1)}, \dots, x_{\pi(m)}).
\end{align}

The physical interpretation of this is that permutation symmetry leads to invariance of the observed statistics if the systems are permuted. 

The original de Finetti theorem \cite{de1937prevision} shows that infinite exchangeable sequences ($m\rightarrow \infty$) have global probability distributions that can be written as a convex mixture of iid distributions. 

The reasoning of the original de Finetti theorem can be extended to finite sequences ($m< \infty$), as shown in \cite{diaconis1980finite}. In this case, we have a bound on the distance between the iid mixture and the exchangeable probability, a generalization that recovers the original de Finetti result when taking the limit $m\rightarrow \infty$. 

\thm{\textbf{(Classical de Finetti theorem for finite sequences \cite{diaconis1980finite})} \label{thm: classical finite de finetti} {
Let $ \{X_i\}_{i=1}^m$ be an exchangeable sequence of random variables, each taking values in a finite alphabet $\mathcal{X}$.
Let $p_{X_1, \ldots, X_k}$ be the marginal distribution of the first $k \leq m$ variables.
Then, there exists a probability measure\footnote{A measurable space $(\Omega,\mathcal{F})$ consists of a set $\Omega$ and a $\sigma$-algebra $\mathcal{F}$ of subsets of $\Omega$.
A measure $\mu$ on $(\Omega,\mathcal{F})$ is a function $\mu:\mathcal{F}\to[0,\infty]$ such that $\mu(\emptyset)=0$ and $\mu$ is countably additive, i.e. $\mu(\cup_{i} A_i)=\sum_i \mu(A_i)$ for any countable family of disjoint sets $A_i\in\mathcal{F}$.
A probability measure additionally satisfies $\mu(\Omega)=1$.}
$\mu$ such that
\begin{align}\label{eq: classical finite de finetti}
    \left\| p_{X_1, \ldots, X_k} - \int \mu(d\theta) \, \prod_{i=1}^{k} 
    p_\theta(x_i) \right\|_1 \leq \frac{2 k |\mathcal{X}|}{m}.
\end{align}
}} \\

\noindent
\textbf{Proof.} The proof of Theorem \ref{thm: classical finite de finetti} can be found in \cite{diaconis1980finite}. \\

The finite version of the classical de Finetti's theorem also shows that if a big sample of the total amount of random variables is chosen ($k\approx m$), then the distance between the exchangeable probability and the iid mixture can increase, so that it is necessary to consider a large number of variables $m$ or equivalently a very small number of samples $k$.

\begin{center}
\begin{figure}
\begin{centering}
\includegraphics[]{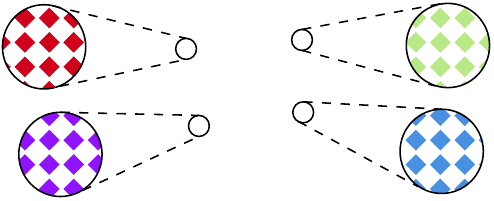}
    \caption{Pictorial intuition of de Finetti's theorem: A global system which is symmetric under permutation (represented by the square filled with particles of different colors in a non-iid distribution) whose small fractions can be locally seen as being iid.}
    \label{fig: de Finetti}
\end{centering}
\end{figure}
\end{center}

An extension of de Finetti's theorem can be applied to quantum systems by assuming that the random variables are now density matrices. This generalization was worked out in \cite{renner2008security}. The notion of exchangeability here is given by the action of projectors $P_\pi$ representing the permutation operation $\pi$ between different parties $A_{i=1,\dots,m}$, so that a $m-$partite quantum state is symmetric \cite{marconi2025symmetric} if the equality 
\begin{align}\label{eq: symmetric state}
    \rho_{A_1 \ldots A_m} = P_\pi \, \rho_{A_1 \ldots A_m} \, P_\pi^\dagger 
\end{align}
holds for all permutations $\pi$. Lifting Theorem \ref{thm: classical finite de finetti} to its quantum version involves considering a global system of $m + k$ parties (with $m\ll k$), whose sample of $m$ parties admits an iid decomposition \cite{renner2007symmetry,renner2008security}.

The quantum version of de Finetti’s theorem requires introducing some notation. For a Hilbert space $\mathcal{H}$, we denote by $\mathcal{D}(\mathcal{H})$ the set of density operators on $\mathcal{H}$, and by $\mathcal{D}_1(\mathcal{H})$ the subset of pure states.  
For $m$ subsystems, the fully symmetric subspace $\operatorname{Sym}(\mathcal{H}^{\otimes m})$ is the span of all density matrices satisfying Eq. (\ref{eq: symmetric state}).  
Let $\rho_{A_1\ldots A_{m+k}}$ be a symmetric global state supported on $\operatorname{Sym}(\mathcal{H}^{\otimes (m+k)})$.  
Denoting by $\rho_m$ its marginal on the first $m$ systems, i.e. 
\begin{align}
    \rho_m \equiv \rho_{A_1\ldots A_m} = \operatorname{Tr}_{A_{m+1}\ldots A_{m+k}}(\rho_{A_1\ldots A_{m+k}}),
\end{align}
The quantum de Finetti representation expresses this reduced state approximately as a convex mixture of iid states.  
For each pure state $|\theta\rangle \in \mathcal{D}_1(\mathcal{H})$, one defines an operator $\bar{\rho}^{|\theta\rangle}_m$ acting on the subspace of $\mathcal{H}^{\otimes m}$ which is symmetric with respect to $|\theta\rangle^{\otimes (m-r)}$.  
The iid mixture is then an average with respect to a probability measure $\nu$ over $\mathcal{D}_1(\mathcal{H})$.

\thm{\textbf{(Quantum de Finetti theorem, Theorem 4.3.2 of \cite{renner2008security})}} {\label{thm: quantum de finetti}
Let $\rho_{m+k}$ be a state supported on the symmetric subspace 
$\operatorname{Sym}(\mathcal{H}^{\otimes (m+k)})$, and let $0 \le r \le m$.  
Then there exists a probability measure $\mu$ on the set of pure states 
$\mathcal{D}_1(\mathcal{H})$ and, for each $|\theta\rangle \in \mathcal{D}_1(\mathcal{H})$, 
an operator $\bar{\rho}^{|\theta\rangle}_m$ acting on the subspace 
$\operatorname{Sym}(\mathcal{H}^{\otimes m},\,|\theta\rangle^{\otimes (m-r)})$ such that
\begin{align}
\left\Vert 
    \rho_{m}
    - \int_{\mathcal{D}_1(\mathcal{H})} \mu(d\theta) 
        \bar{\rho}^{|\theta\rangle}_m   
\right\Vert_{\text{Tr}}
    \leq 2\, e^{\left[ 
        \frac{1}{2}\,\mathrm{dim}(\mathcal{H}) \ln k -\frac{k(r+1)}{2(m+k)}
    \right]}.
\end{align}
}

\noindent\textbf{Proof.} The proof of Theorem \ref{thm: quantum de finetti} can be found in \cite{renner2008security}. \\

To build some intuition about how the quantum de Finetti representation behaves, it is useful to look at a few concrete symmetric states and examine how their reduced marginals relate to mixtures of iid states.  
The examples below are adapted from \cite{renner2007symmetry} and illustrate both the situations in which the representation is exact and the situations where symmetry alone is too weak to guarantee a good iid approximation.

\exam{\textbf{(Example 1 of \cite{renner2007symmetry})}}{
Consider the $m+k$--partite state 
\begin{align}
\rho_{m+k} = \frac{1}{2}\bigl(|0\rangle\langle 0|^{\otimes m+k} + |1\rangle\langle 1|^{\otimes m+k}\bigr),
\end{align}
corresponding to the uniform mixture of the two iid product states $|0\rangle^{\otimes m+k}$ and $|1\rangle^{\otimes m+k}$.  
Because of its symmetry, every $m$--partite marginal $\rho_{m}$ has exactly the same structure:
\begin{align}
\rho_{m} = \frac{1}{2}\bigl(|0\rangle\langle 0|^{\otimes m} + |1\rangle\langle 1|^{\otimes m}\bigr).
\end{align}
In this case, the de Finetti representation is exact rather than approximate: the state is already a convex mixture of two perfectly iid components.  
At the same time, the example shows that one typically cannot hope to approximate a symmetric state by a \emph{single} product state; the best one can do in general is a convex combination of such product states.}

\exam{\textbf{(Example 2 of \cite{renner2007symmetry}) \label{exam: example 2 de finetti}}}{
Let the global system consist of $m+k$ qubits, and define $\rho_{m+k}$ as the uniform mixture over all computational-basis vectors
$\ket{b_1 \dots b_{m+k}}$  
whose bit string $(b_1,\dots,b_{m+k}) \in \{0,1\}^{m+k}$ contains an even number of $1$'s. This can be described as a density matrix as 
\begin{align}
    \rho_{m+k}
    = \frac{1}{2^{\,m+k-1}}
      \sum_{\substack{b_1,\dots,b_{m+k} \\ \text{even \# of } 1\text{'s}}}
      \ket{b_1 \dots b_{m+k}}\bra{b_1 \dots b_{m+k}} .
\end{align}
This state is symmetric under permutations of all $m+k$ subsystems. However, every $m$-partite marginal has a perfectly iid structure: Tracing out any $k$ subsystems yields
\begin{align}
    \rho_m
    = \left( \frac{\ket{0}\bra{0} + \ket{1}\bra{1}}{2} \right)^{\otimes m}.
\end{align}
Thus, all reduced states $\rho_m$ are exactly iid. Nevertheless, the global state $\rho_{m+k}$ itself is not iid and cannot be written as $\sigma^{\otimes (m+k)}$ for some $\sigma$.} \\

This example illustrates that, even when all local marginals look perfectly iid and the global state is symmetric, the full multipartite state may still be far from any iid structure.

The quantum de Finetti theorem allows one to exploit the symmetry of QKD protocols in order to prove security against general attacks. In particular, global permutation symmetry of the $n$ distributed quantum systems implies that the corresponding attack strategy can be approximated by a convex mixture of iid structures due to the quantum de Finetti theorem. With this, it is possible to reduce proofs against coherent attacks to proofs against collective attacks (which are typically much easier to analyze). The quantitative version of this reduction is given in \cite{christandl2009postselection}, where the scaling of $\varepsilon$-security under postselection is derived for permutation-invariant protocols.

\thm{\textbf{(de Finetti postselection method for QKD, Theorem 1 of \cite{christandl2009postselection})}}{\label{thm: de finetti QKD}
Assume that a QKD protocol is invariant under any permutation of the input subsystems.
If the protocol is $\varepsilon$-secure against collective attacks and produces a key of size $\ell$, then the same protocol is $\varepsilon'$-secure against coherent attacks provided the final key is shortened to size $\ell'$, where
\begin{align}
\varepsilon' = (n+1)^{d^2 - 1} \varepsilon 
\end{align}
and
\begin{align}
\ell' = \ell - 2(d^2 - 1)\log(n+1) .
\end{align}
Here $n$ denotes the number of distributed subsystems and $d$ the local Hilbert space dimension of each subsystem shared between Alice and Bob.
} \\

\noindent
\textbf{Proof.} The proof of Theorem \ref{thm: de finetti QKD} can be found in \cite{christandl2009postselection}. 

\subsection{Entropic Uncertainty Relations}
\label{sec: EUR}

Uncertainty relations are constraints on the knowledge that can be learned about distinct measurable quantities in some (potentially multipartite) system \cite{coles2017entropic}. Classical quantities are not restricted fundamentally by uncertainty relations, but information about incompatible quantum observables is inherently bounded by Heisenberg's relations \cite{heisenberg1927anschaulichen}, due to the necessity of disturbing such systems in order to extract information about them.

From the perspective of information theory, it is more natural to quantify uncertainty in terms of entropies rather than variances, context in which entropic uncertainty relations (EURs) appear. The simplest EUR constraints the information of observables in a single physical system. 

\thm{\textbf{(EUR for a single system \cite{maassen1988generalized,krishna2002entropic})}} {\label{thm: eur 1}Let $x_{1,2}$ label non-degenerate observables acting on the same $d$-dimensional Hilbert space $\mathcal{H}_A$, and $\{M_{a_{x}\vert x}\}$ be their associated POVM elements. The simplest EUR that can be written is given by
\begin{align}
    H(A_{x_1}) + H(A_{x_2}) \geq - \log c,
    \label{eq: MU EUR}
\end{align}
where the constant $c$ is defined as 
\begin{align}
    c \equiv \max_{a_{x_{1,2}}}
    \left\Vert \sqrt{M_{a_{x_1} \vert x_1}} \sqrt{M_{a_{x_2}\vert x_2}} \right\Vert_\infty^2,
    \label{eq: c POVM def}
\end{align}
and quantifies the incompatibility of the two POVMs - they commute if and only if $c=1$, while $c=d^{-1}$ corresponds to maximal incompatibility. Here $\|\cdot\|_\infty$ denotes the operator norm (largest singular value), so that, if the POVM's are projective and rank one ($M_{a_{x_i}\vert x_i}= \ket{a_{x_i}}\bra{a_{x_i}}$), then Eq. (\ref{eq: c POVM def}) reduces to the maximum overlap\footnote{In terms of observables, the constant $c$ can be similarly written as the maximization of the overlap of the eigenbases $\{\vert x_1\rangle\}$ and $\{\vert x_2\rangle\}$.} $\max_{a_{x_1, x_2}} \vert \langle a_{x_1} \vert a_{x_2} \rangle\vert^{2}$.} \\

\noindent
\textbf{Proof.} The proof of Theorem \ref{thm: eur 1} can be found in \cite{maassen1988generalized,krishna2002entropic}. \\

Multipartite entropic uncertainty relations have proven themselves to be very useful in the context of QKD. If Alice encodes information in one of two conjugate bases, then any adversary that interferes in the system must induce a trade-off between learning about one basis and the other. If Bob can predict well the outcome of one of Alice's measurements, then the uncertainty principle forces an eavesdropper to have large uncertainty about the conjugate one. In modern security proofs, these EUR's can be related to conditional entropies of the authenticated parties given the eavesdropper's information, which then enter in the calculation of key rates through the Devetak-Winter relation (Eq. \ref{eq: devetak-winter}).

The mathematical apparatus of EUR's for QKD is based on extending Eq. \eqref{eq: MU EUR} to the case where the eavesdropper has access to a \textit{quantum memory} correlated with the measured system. 

For practical QKD applications, we need to reformulate Eq. (\ref{eq: MU EUR}) for a tripartite pure state in order to include the purification of the eavesdropper.

\thm{\textbf{(Tripartite EUR \cite{renes2009conjectured,coles2011information,berta2010uncertainty,coles2017entropic})}} {\label{thm: eur 3}Consider $\ket{\psi}_{ABE} \in \mathcal{H}_{A}\otimes \mathcal{H}_{B} \otimes \mathcal{H}_{E}$ as the purification of the bipartite state $\rho = \text{Tr}_E (\ket{\psi} \bra{\psi}_{ABE})$. In this setting, the conditional entropies $H(A\vert B)$ and $H(A\vert E)$ satisfy the inequality
\begin{align}
    H(A_{x_1} \vert B)_{\Lambda_{x_1}(\rho)} + H(A_{x_2} \vert E)_{\Lambda_{x_2}(\rho)}
    \geq - \log c,
    \label{eq: Coles EUR}
\end{align}
where $\Lambda_{x_{1,2}}$ are the measure-and-prepare channels introduced in Eq. (\ref{eq: MP map}).} \\

\noindent
\textbf{Proof.} For a proof of Theorem \ref{thm: eur 3} the reader can check \cite{coles2017entropic}. \\

So far, the discussion of EURs was phrased in terms of conditional von Neumann entropies, which are appropriate for iid scenarios. For finite-size QKD, one needs single-round entropic quantities with a direct operational meaning in terms of guessing probabilities and data compression with quantum side information. These are provided by the smooth min- and max-entropies \cite{konig2009operational,tomamichel2015quantum} (see also Appendix \ref{sec: entropic quantities}). The generalization of EUR for these entropic quantities was established in \cite{tomamichel2011uncertainty}.

\thm{\textbf{(EUR for smooth entropies \cite{tomamichel2011uncertainty,murta2023lecture})}}
{\label{thm: eur 4} Let $\rho_{ABE}$ be a tripartite quantum state, and let
$\{M_{a_{x_{1,2}\vert x_{1,2}}}\}$ be two POVMs on $A$ associated to observables $x_1$ and $x_{2}$, described respectively by the measurement-and-preparation channels
$\Lambda_{x_{1,2}}$. Then, for any $\epsilon \geq 0$,
\begin{align}
    H_{\min}^\epsilon\left(A_{x_1} \vert E\right)_{\Lambda_{x_1}(\rho)}
    + H_{\max }^\epsilon\left(A_{x_2} \vert B\right)_{\Lambda_{x_2}(\rho)}
    \geq -\log c,
    \label{eq: smooth EUR}
\end{align}
where the constant $c$ is defined  as in Eq. (\ref{eq: c POVM def}).} \\

\noindent
\textbf{Proof.} A proof of Theorem \ref{thm: eur 4} can be found in \cite{tomamichel2011uncertainty}. \\

Equation~\eqref{eq: smooth EUR} can be viewed as a single shot version of the relation given in Eq. \eqref{eq: Coles EUR}. Indeed, by applying the quantum AEP (\ref{thm: AEP}) \cite{tomamichel2009fully,tomamichel2015quantum} we recover the von Neumann entropy version in the iid limit:
\begin{align}
    \lim_{\substack{n \to \infty \\ \epsilon \to 0}} \frac{1}{n}\,
    H_{\min/\max}^\epsilon(A_1^n\vert B_1^n)_{\rho^{\otimes n}}
    = H(A\vert B)_\rho.
\end{align}

For QKD, Eq.~\eqref{eq: smooth EUR} provides a direct tool to prove the
security of BB84-type protocols against coherent attacks in the- finite-key
regime. Roughly speaking, $H_{\min}^\epsilon(A_Z\vert E)$ quantifies Eve's
(conditional) guessing probability of Alice's $Z$-basis outcomes (in the same spirit of Eq. \eqref{eq: hmin via pguess}), while
$H_{\max}^\epsilon(A_X\vert B)$ can be bounded using the observed statistics in
the $X$-basis. The uncertainty relation then translates the observed error rate
into a bound on Eve's knowledge (as shown in Eq. (\ref{eq: EUR example})).

As an example, we consider the security proof of the BB84 key rate, which can be derived directly from \ref{thm: eur 3}, where we consider the entanglement-based picture.

\exam{\textbf{(Security of BB84 against coherent attacks using EUR \cite{tomamichel2012tight,murta2023lecture})}} 
{Alice and Bob share a
state $\rho_{ABE}$, where $E$ is held by the eavesdropper. Alice measures her system either in the $Z$-basis (for key generation) or in the $X$-basis (for parameter estimation). Bob performs the corresponding measurements on his system. Applied to the $X$ and $Z$ measurements on $A$, the tripartite EUR Eq. \eqref{eq: Coles EUR} tells us that in BB84 we have $c = 1/2$, so $\log c = -1$. Moreover, because Bob's $X$-basis measurement is classical post-processing of $A_X$ (plus noise), the data-processing inequality yields the chain
\begin{align}
    H(A_X \vert B) \leq H(A_X \vert B_X) \leq  h(Q_X)
\end{align}
where $Q_X$ is the expected quantum bit error rate in the $X$-basis, corresponding to the hypothetical scenario in which the key-generation rounds were measured in the X
X-basis. This quantity can be inferred from the observed quantum bit error rate in the $X$-basis. Hence Eq. \eqref{eq: Coles EUR} gives
\begin{align}\label{eq: EUR example}
    H(A_Z \vert E) \geq 1 - h(Q_X).
\end{align}
On the other hand, using the Devetak-Winter relation (Eq. \eqref{eq: devetak-winter}), by lower bounding the term $H(A_Z \vert B) \leq h(Q_Z)$ we obtain
\begin{align}
    r_\infty \geq 1 - h(Q_X) - h(Q_Z).
\end{align}
For depolarizing noise, we have that $Q_X = Q_Z = Q$, which reduced this equation to the expression $r_\infty \geq 1 - 2 h(Q)$.} \\

This example of security proof based on EUR is restricted to protocols in which Alice performs only two measurements (here $X$ and $Z$). The post-selection technique (Sec. \ref{sec: postselection}), on the other hand, can be applied to more general protocols, such as the six-state protocol in which Alice and Bob measure in three bases $X$, $Y$, and $Z$ \cite{renner2008security}. For the BB84, in the asymptotic limit, both techniques yield the same key rate and demonstrate that collective attacks are optimal (see the examples mentioned in Sec. \ref{sec: connecting}). In the finite-key regime, however, the EUR-based analysis, involving the direct application of Theorem~\ref{thm: eur 4}, typically leads to tighter results \cite{staffieri2026finite}, as it has smaller overhead terms and therefore better rates.

\subsection{Entropy Accumulation Theorems}
\label{sec: EAT}

Entropy accumulation theorems (EATs) are tools that allow one to lower bound the smooth min-entropy $H^{\epsilon}_\text{min}$ of a sequence of outcomes of quantum systems produced in sequential processes, even in situations where the devices of the protocol are not characterized.

In contrast to the AEP, discussed in Theorem~\ref{thm: AEP}, EAT does not require the rounds to be iid, and therefore are suitable to analyze cryptographic protocols under general coherent attacks (Sec.~\ref{sec: eavesdropping strategies}). In this section, we discuss the first formulation of the EAT \cite{dupuis2020entropy}, together with its generalizations, which include the so-called generalized entropy accumulation theorem (GEAT) \cite{metger2024generalised} and a version that allows the use of marginal probability constraints - the so-called marginal-constrained entropy accumulation theorem (MEAT) \cite{arqand2025marginal}. The latter is currently the most adequate route to combine finite size considerations with the numerical techniques discussed in Sec. \ref{sec: algorithms} for prepare-and-measure protocols.

The general idea of EAT is to use the sequential ordering of QKD protocols to show that, in each of the rounds, some entropy is gained, contributing with the increase of the smooth entropy. With this reasoning, EAT theorems can provide a lower bound for $H^{\epsilon}_\text{min}$ based on the sum of single-round contributions and with a penalty $\propto \sqrt{n}$, under some reasonable assumptions that we will detail in the next paragraphs.

The first version of EAT was introduced in \cite{dupuis2020entropy}, which was then used to prove the security of DIQKD in~ \cite{arnon2019simple}. Its generalization~\cite{metger2024generalised} was recently applied for standard QKD in \cite{metger2023security}. Its setup works as follows. Consider a process that runs for $n$ rounds, in which every round $i$ has an associated register $R_i$, which contains the systems of Alice and Bob, $A_i$ and $B_i$. The main requirement of the EAT is that, given the information of the string $B_1^{i}$, the element $A_i$ can depend on $B_1^{i}$ and on the eavesdropper influence $E$, but not on $B_{i+1}$. In this context, this condition is usually denoted in the language of quantum Markov chains (see Sec.2.2 of \cite{dupuis2020entropy} for a more
precise definition), which in this case takes the form
\begin{align}\label{eq: EAT markov}
    A_1^{i} \leftrightarrow B_1^i E \leftrightarrow B_{i+1},
\end{align}
where $\leftrightarrow$ denotes a \emph{quantum Markov chain} relation: conditioned on the middle system $B_1^{i}E$, the systems $A_1^{i}$ and $B_{i+1}$ have no additional correlations. In terms of channels, the theorem takes some initial state $\rho_{R_0 E}$, where $E$ denotes the adversary's side information and $R_0$ is an internal memory register, and in each round $i=1,\dots,n$ a CPTP map
\begin{align}\label{eq: markov maps}
    \mathcal{M}_i : \mathcal{D}(\mathcal{H}_{R_{i-1}}) \longrightarrow
    \mathcal{D}(\mathcal{H}_{R_i} \otimes \mathcal{H}_{A_i} \otimes \mathcal{H}_{B_i} \otimes \mathcal{H}_{X_i})
\end{align}
is applied to $R_{i-1}$. The map $\mathcal{M}_i$ produces three output systems:
\begin{itemize}
    \item $A_i$: the system whose entropy we ultimately want to bound;
    \item $B_i$: registers available to the honest (authenticated) parties;
    \item $X_i$: a classical register that stores the data used for tests
    (e.g. in parameter estimation).
\end{itemize}
The memory $R_i$ is then passed to the next round. After $n$ applications, we
obtain a global state of the form
\begin{align}\label{eq: eat map}
    \rho_{A_1^n B_1^n X_1^n E}
    =
    (\mathcal{M}_n \circ \dots \circ \mathcal{M}_1 \otimes \mathcal{I}_E)
    (\rho_{R_0 E}).
\end{align}
The iid regime would correspond to the case when $R$ and $\mathcal{M}$ are trivial (i.e. mutual independence of the pair $A_i B_i$ and $\mathcal{M}_i$ being always the same for every round $i$). In QKD, the map $\mathcal{M}_i$ represents the generation of the $i$th bit of the raw key, and can depend on Eve's strategy (being potentially unknown). The string of classical registered values $X_{i=1,\dots,n}$ which are globally used to infer the statistics of the protocol.

To connect the global min-entropy of $A_1^n$ to single-round quantities, the EAT uses the notion of a \emph{min-tradeoff function}. Let $\mathcal{X}$ denote the alphabet of the classical registers $X_i$, and let $\mathbb{P}(\mathcal{X})$ be the set of probability distributions on $\mathcal{X}$. For each map $\mathcal{M}_i$, we can consider the marginal distribution $q_i \in \mathbb{P}(\mathcal{X})$ of $X_i$ that is induced by a given input state to that round.

\dfn{\textbf{(Affine min-tradeoff function, Definition 4.1 of \cite{dupuis2020entropy})}}
{Let $\mathcal{M}_1,\dots,\mathcal{M}_n$ be the maps defined in
Eq.~\eqref{eq: markov maps}. An affine function
$f : \mathbb{P}(\mathcal{X}) \to \mathbb{R}$ is called a \emph{min-tradeoff
function} for $\mathcal{M}_1,\dots,\mathcal{M}_n$ if, for every
$i \in \{1,\dots,n\}$ and for every state $\omega_{R_{i-1} R}$ on $R_{i-1} R$
(with $R$ isomorphic to $R_{i-1}$), the state
\begin{align}
    \nu_{A_i B_i X_i R} \equiv (\mathcal{M}_i \otimes \mathcal{I}_R)(\omega_{R_{i-1} R})
\end{align}
satisfies
\begin{align}
    H(A_i \vert B_i R)_{\nu}\geq f(q),
\end{align}
where $q \in \mathbb{P}(\mathcal{X})$ is the distribution of $X_i$ in the state
$\nu$.}
One typically constructs such a min-tradeoff function from a single-round
analysis of the protocol (i.e. it is necessary to find a "good" min-tradeoff function), and the observed statistics $X_1^n$ are then used to restrict the set of admissible states $\omega_{R_{i-1} R}$.

\thm{\textbf{(Entropy accumulation, Theorem 4.4 of \cite{dupuis2020entropy})}
\label{thm: EAT}}
{Let $\mathcal{M}_1,\ldots,\mathcal{M}_n$ and
$\rho_{A_1^n B_1^n X_1^n E}$ be as previously defined and assume that the Markov condition\eqref{eq: EAT markov} holds for each round. Let $f$ be an affine min-tradeoff function for $\mathcal{M}_1,\dots,\mathcal{M}_n$ and let $\varepsilon \in (0,1)$. Consider any event
$\Omega \subseteq \mathcal{X}^n$ (representing, e.g., successful parameter
estimation) such that
\begin{align}
    f(\text{freq}(X_1^n)) \geq h
    \quad\forall\quad X_1^n \in \Omega,
    \label{eq: EAT freq condition}
\end{align}
for some real number $h \in \mathbb{R}$. Here
$\text{freq}(X_1^n)$ denotes the empirical distribution of the string $X_1^n$.
Then the following bounds hold on the state conditioned on $\Omega$:
\begin{align}
    H_{\min}^{\epsilon}(A_1^n \vert B_1^n E)_{\rho_{\vert \Omega}}
    &> n h - c_\text{EAT} \sqrt{n},
    \label{eq: EAT min}\\
    H_{\max}^{\epsilon}\left(A_1^n \vert B_1^n E\right)_{\rho_{\vert \Omega}}
    &< n h + c_\text{EAT} \sqrt{n},
    \label{eq: EAT max}
\end{align}
where
\begin{align}
    c_\text{EAT} \equiv 2\big[\log\big(1+2 d_A\big)+ \big\lceil \|\nabla f\|_{\infty} \big\rceil\big]
      \sqrt{1 - 2 \log\big(\epsilon \rho[\Omega]\big)}.
    \label{eq: EAT constant}
\end{align}
In the expression above, $d_A$ is the maximum dimension of the register $A_i$, $\nabla f$ denotes the gradient of $f$, and $\rho[\Omega]$ is the probability of the event $\Omega$ in the state $\rho_{X_1^n}$.} \\

\noindent
\textbf{Proof.} The proof of EAT can be found in \cite{dupuis2020entropy}. \\

Theorem~\ref{thm: EAT} shows that, under very general assumptions, the smooth min-entropy of the full sequence $A_1^n$ accumulates through every rund, growing linearly with $n$ and with a penalty second-order correction of order $\sqrt{n}$. The single-round behaviour enters through the min-tradeoff function $f$, which is typically determined by the same constraints that appear in the asymptotic key-rate analysis.

The constant $c_\text{EAT}$ in Eq.~\eqref{eq: EAT constant} depends on the norm of the gradient of the min-tradeoff function $f$. In many QKD protocols, parameter estimation is performed by sampling each round with a small probability, which can make $\|\nabla f\|_\infty$ scale linearly with $n$ and thus render the second-order term overly pessimistic. This issue was addressed in \cite{dupuis2019entropy}, where an improved version of the theorem is shown, where second order term are less penalizing. 

More general assumptions regarding how the side information can be altered were explored in \cite{metger2024generalised}. The Theorem \ref{thm: EAT} demands a relatively rigid structure on the eavesdropper information, as newly acquired side information generated in round $i$ cannot modify the side information produced in round $i-1$. In many cryptographic settings, however, it is natural to consider more general models, where the adversary’s side information is updated adaptively at each step, which is particularly relevant for security analyses of PM scenarios. GEAT \cite{metger2024generalised} allows an adaptative model for Eve's side channel by considering sequential process in which, at each step $i$, the channel $\mathcal{M}_i$ maps the register $R_{i-1}$ and the side information $E_{i-1}$ into the next round systems\footnote{In the GEAT framework, it is common to omit $B_i$ by incorporating the public
transcript into the evolving adversarial register $E_i$.
} $E_i$, $A_i$ and $X_i$
\begin{align}\label{eq: maps GEAT}
    \mathcal{M}_i : \mathcal{D}(\mathcal{H}_{R_{i-1}}\otimes \mathcal{H}_{E_{i-1}}) \longrightarrow
    \mathcal{D}(\mathcal{H}_{A_i} \otimes \mathcal{H}_{R_i} \otimes \mathcal{H}_{E_i})
\end{align}
acts directly on the adversary’s side information $E_{i-1}$, producing an
output system $A_i$, a new side-information register $E_i$, and classical data $X_i$, in contrast with Eq. (\ref{eq: eat map}), where the side channel $E$ was independent of the round $i$). Instead of the Markov-chain condition used in the original EAT, the GEAT assumes a \emph{no-signalling} condition for the evolution of the adversary's side
information. In each round $i$ a channel
$\mathcal{M}_i$ acts on the internal memory $R_{i-1}$ and the adversary's current register $E_{i-1}$ and outputs $A_i$, an updated memory $R_i$, and an
updated side-information register $E_i$. The no-signalling requirement states that the marginal update of $E_i$ cannot depend on the internal memory
$R_{i-1}$: there exists a channel $\mathcal{R}_i$ acting only on $E_{i-1}$ such
that \cite{metger2024generalised}:
\begin{align}\label{eq: no signalling GEAT}
   \text{Tr}_{A_i R_i} \circ \mathcal{M}_i = \mathcal{R}_i \circ \text{Tr}_{R_{i-1}} .
\end{align}
Equivalently, after discarding $A_i$ and the next-round memory $R_i$, Eve's updated register $E_i$ can be generated from $E_{i-1}$ alone, so that no information can be ``stored'' in $R_{i-1}$ and leaked to Eve only at a later time.

Under this more general side-information model, the GEAT shows that the smooth min-entropy of $A_1^n$ conditioned on the final side information $E_n$ still accumulates linearly, up to $\mathcal{O}(\sqrt{n})$ finite-size corrections.

\thm{\textbf{(Generalised entropy accumulation (GEAT) \cite{metger2024generalised})} \label{thm: GEAT}} {Let $\mathcal{M}_i$ be CPTP maps as defined in Eq. (\ref{eq: maps GEAT}) describing a sequential process and satisfying the no-signalling condition of Eq. (\ref{eq: no signalling GEAT}), and let $f$ be an appropriate min-tradeoff function. Then, for any event $\Omega$ defined in terms of the classical registers $X_1^n$ that implies $f(\operatorname{freq}(X_1^n)) \geq h$, the smooth min-entropy of $A_1^n$ conditioned on the final side information $E_n$ and on $\Omega$ admits a lower bound of the form
\begin{align}
    H_{\min}^{\epsilon}\left(A_1^n \vert E_n\right)_{\rho_{\vert \Omega}}
    \geq
    n h - c_{\text{GEAT}} \sqrt{n},
\end{align}
for an explicit constant $c_{\text{GEAT}}$ that depends only on the local
dimensions, the tradeoff function, and the smoothing parameter.}

Because the side information can now be updated in every round, GEAT can be applied to a broader range of cryptographic protocols and it is particularly useful for PM protocols (as shown in \cite{metger2023security}). \\

\noindent
\textbf{Proof.} The proof of GEAT can be found in \cite{metger2024generalised}. \\

\begin{figure}[H]
    \centering
    \begin{minipage}{1\linewidth}
        \centering

\tikzset{every picture/.style={line width=0.75pt}} 

\begin{tikzpicture}[x=0.75pt,y=0.75pt,yscale=-1,xscale=1,scale=0.8]

\draw    (6,88) -- (66.13,28.33) ;
\draw    (6,88) -- (66.33,148.67) ;
\draw    (66.13,28.33) -- (406,28.47) ;
\draw  [fill={rgb, 255:red, 74; green, 144; blue, 226 }  ,fill opacity=0.5 ] (116.55,132) .. controls (116.55,129.79) and (118.34,128) .. (120.55,128) -- (152,128) .. controls (154.21,128) and (156,129.79) .. (156,132) -- (156,164) .. controls (156,166.21) and (154.21,168) .. (152,168) -- (120.55,168) .. controls (118.34,168) and (116.55,166.21) .. (116.55,164) -- cycle ;
\draw    (66.33,148.67) -- (102.67,148.67) ;
\draw [shift={(105.67,148.67)}, rotate = 180] [fill={rgb, 255:red, 0; green, 0; blue, 0 }  ][line width=0.08]  [draw opacity=0] (5.36,-2.57) -- (0,0) -- (5.36,2.57) -- cycle    ;
\draw    (166.67,148.08) -- (193,148.01) ;
\draw [shift={(196,148)}, rotate = 179.84] [fill={rgb, 255:red, 0; green, 0; blue, 0 }  ][line width=0.08]  [draw opacity=0] (5.36,-2.57) -- (0,0) -- (5.36,2.57) -- cycle    ;
\draw  [fill={rgb, 255:red, 74; green, 144; blue, 226 }  ,fill opacity=0.5 ] (206.3,132) .. controls (206.3,129.79) and (208.09,128) .. (210.3,128) -- (241.75,128) .. controls (243.96,128) and (245.75,129.79) .. (245.75,132) -- (245.75,164) .. controls (245.75,166.21) and (243.96,168) .. (241.75,168) -- (210.3,168) .. controls (208.09,168) and (206.3,166.21) .. (206.3,164) -- cycle ;
\draw  [fill={rgb, 255:red, 74; green, 144; blue, 226 }  ,fill opacity=0.5 ] (366.22,132.33) .. controls (366.22,130.12) and (368.01,128.33) .. (370.22,128.33) -- (401.67,128.33) .. controls (403.88,128.33) and (405.67,130.12) .. (405.67,132.33) -- (405.67,164.33) .. controls (405.67,166.54) and (403.88,168.33) .. (401.67,168.33) -- (370.22,168.33) .. controls (368.01,168.33) and (366.22,166.54) .. (366.22,164.33) -- cycle ;
\draw    (256.67,148.08) -- (283,148.01) ;
\draw [shift={(286,148)}, rotate = 179.84] [fill={rgb, 255:red, 0; green, 0; blue, 0 }  ][line width=0.08]  [draw opacity=0] (5.36,-2.57) -- (0,0) -- (5.36,2.57) -- cycle    ;
\draw    (327,148.08) -- (353.33,148.01) ;
\draw [shift={(356.33,148)}, rotate = 179.84] [fill={rgb, 255:red, 0; green, 0; blue, 0 }  ][line width=0.08]  [draw opacity=0] (5.36,-2.57) -- (0,0) -- (5.36,2.57) -- cycle    ;
\draw    (135.2,173.7) -- (127.05,195.44) ;
\draw [shift={(126,198.25)}, rotate = 290.54] [fill={rgb, 255:red, 0; green, 0; blue, 0 }  ][line width=0.08]  [draw opacity=0] (5.36,-2.57) -- (0,0) -- (5.36,2.57) -- cycle    ;
\draw    (144.74,195.26) -- (135.2,173.7) ;
\draw [shift={(145.95,198)}, rotate = 246.14] [fill={rgb, 255:red, 0; green, 0; blue, 0 }  ][line width=0.08]  [draw opacity=0] (5.36,-2.57) -- (0,0) -- (5.36,2.57) -- cycle    ;
\draw    (225.45,173.45) -- (217.3,195.19) ;
\draw [shift={(216.25,198)}, rotate = 290.54] [fill={rgb, 255:red, 0; green, 0; blue, 0 }  ][line width=0.08]  [draw opacity=0] (5.36,-2.57) -- (0,0) -- (5.36,2.57) -- cycle    ;
\draw    (234.99,195.01) -- (225.45,173.45) ;
\draw [shift={(236.2,197.75)}, rotate = 246.14] [fill={rgb, 255:red, 0; green, 0; blue, 0 }  ][line width=0.08]  [draw opacity=0] (5.36,-2.57) -- (0,0) -- (5.36,2.57) -- cycle    ;
\draw    (385.45,173.2) -- (377.3,194.94) ;
\draw [shift={(376.25,197.75)}, rotate = 290.54] [fill={rgb, 255:red, 0; green, 0; blue, 0 }  ][line width=0.08]  [draw opacity=0] (5.36,-2.57) -- (0,0) -- (5.36,2.57) -- cycle    ;
\draw    (394.99,194.76) -- (385.45,173.2) ;
\draw [shift={(396.2,197.5)}, rotate = 246.14] [fill={rgb, 255:red, 0; green, 0; blue, 0 }  ][line width=0.08]  [draw opacity=0] (5.36,-2.57) -- (0,0) -- (5.36,2.57) -- cycle    ;
\draw   (106.5,219.5) .. controls (106.5,224.17) and (108.83,226.5) .. (113.5,226.5) -- (126.5,226.5) .. controls (133.17,226.5) and (136.5,228.83) .. (136.5,233.5) .. controls (136.5,228.83) and (139.83,226.5) .. (146.5,226.5)(143.5,226.5) -- (159.5,226.5) .. controls (164.17,226.5) and (166.5,224.17) .. (166.5,219.5) ;
\draw   (198.83,220.83) .. controls (198.83,225.5) and (201.16,227.83) .. (205.83,227.83) -- (218.83,227.83) .. controls (225.5,227.83) and (228.83,230.16) .. (228.83,234.83) .. controls (228.83,230.16) and (232.16,227.83) .. (238.83,227.83)(235.83,227.83) -- (251.83,227.83) .. controls (256.5,227.83) and (258.83,225.5) .. (258.83,220.83) ;
\draw   (356.83,220.5) .. controls (356.83,225.17) and (359.16,227.5) .. (363.83,227.5) -- (376.83,227.5) .. controls (383.5,227.5) and (386.83,229.83) .. (386.83,234.5) .. controls (386.83,229.83) and (390.16,227.5) .. (396.83,227.5)(393.83,227.5) -- (409.83,227.5) .. controls (414.5,227.5) and (416.83,225.17) .. (416.83,220.5) ;

\draw (292.75,141.75) node [anchor=north west][inner sep=0.75pt]    {$\dotsc $};
\draw (122,140) node [anchor=north west][inner sep=0.75pt]    {$\mathcal{M}_{1}$};
\draw (212,140) node [anchor=north west][inner sep=0.75pt]    {$\mathcal{M}_{2}$};
\draw (371,140) node [anchor=north west][inner sep=0.75pt]    {$\mathcal{M}_{n}$};
\draw (231,8.5) node [anchor=north west][inner sep=0.75pt]    {$E$};
\draw (75,129.5) node [anchor=north west][inner sep=0.75pt]    {$R_{0}$};
\draw (169,129.5) node [anchor=north west][inner sep=0.75pt]    {$R_{1}$};
\draw (257,129.5) node [anchor=north west][inner sep=0.75pt]    {$R_{2}$};
\draw (322.5,129.5) node [anchor=north west][inner sep=0.75pt]    {$R_{n-1}$};
\draw (112.5,199.5) node [anchor=north west][inner sep=0.75pt]    {$A_{1}$};
\draw (141,200) node [anchor=north west][inner sep=0.75pt]    {$B_{1}$};
\draw (202.5,199.5) node [anchor=north west][inner sep=0.75pt]    {$A_{2}$};
\draw (231.5,199.5) node [anchor=north west][inner sep=0.75pt]    {$B_{2}$};
\draw (362.5,199.5) node [anchor=north west][inner sep=0.75pt]    {$A_{n}$};
\draw (391,200) node [anchor=north west][inner sep=0.75pt]    {$B_{n}$};
\draw (125.33,240) node [anchor=north west][inner sep=0.75pt]    {$X_{1}$};
\draw (217,240) node [anchor=north west][inner sep=0.75pt]    {$X_{2}$};
\draw (375,240) node [anchor=north west][inner sep=0.75pt]    {$X_{n}$};
\end{tikzpicture}
        \caption*{(a) Representation of the original EAT, where Eve can't adapt previously acquired information in future rounds.}
        \vspace{0.5cm}
        \centering

\tikzset{every picture/.style={line width=0.75pt}} 

\begin{tikzpicture}[x=0.75pt,y=0.75pt,yscale=-1,xscale=1,scale=0.75]

\draw    (100,94) -- (160.13,34.33) ;
\draw    (100,94) -- (160.33,154.67) ;
\draw  [fill={rgb, 255:red, 74; green, 144; blue, 226 }  ,fill opacity=0.5 ] (210.55,18.4) .. controls (210.55,16.19) and (212.34,14.4) .. (214.55,14.4) -- (246,14.4) .. controls (248.21,14.4) and (250,16.19) .. (250,18.4) -- (250,170) .. controls (250,172.21) and (248.21,174) .. (246,174) -- (214.55,174) .. controls (212.34,174) and (210.55,172.21) .. (210.55,170) -- cycle ;
\draw    (160.33,154.67) -- (196.67,154.67) ;
\draw [shift={(199.67,154.67)}, rotate = 180] [fill={rgb, 255:red, 0; green, 0; blue, 0 }  ][line width=0.08]  [draw opacity=0] (5.36,-2.57) -- (0,0) -- (5.36,2.57) -- cycle    ;
\draw    (260.67,154.08) -- (287,154.01) ;
\draw [shift={(290,154)}, rotate = 179.84] [fill={rgb, 255:red, 0; green, 0; blue, 0 }  ][line width=0.08]  [draw opacity=0] (5.36,-2.57) -- (0,0) -- (5.36,2.57) -- cycle    ;
\draw  [fill={rgb, 255:red, 74; green, 144; blue, 226 }  ,fill opacity=0.5 ] (300.3,18) .. controls (300.3,15.79) and (302.09,14) .. (304.3,14) -- (335.75,14) .. controls (337.96,14) and (339.75,15.79) .. (339.75,18) -- (339.75,170) .. controls (339.75,172.21) and (337.96,174) .. (335.75,174) -- (304.3,174) .. controls (302.09,174) and (300.3,172.21) .. (300.3,170) -- cycle ;
\draw  [fill={rgb, 255:red, 74; green, 144; blue, 226 }  ,fill opacity=0.5 ] (460.22,18) .. controls (460.22,15.79) and (462.01,14) .. (464.22,14) -- (495.67,14) .. controls (497.88,14) and (499.67,15.79) .. (499.67,18) -- (499.67,170.33) .. controls (499.67,172.54) and (497.88,174.33) .. (495.67,174.33) -- (464.22,174.33) .. controls (462.01,174.33) and (460.22,172.54) .. (460.22,170.33) -- cycle ;
\draw    (350.67,154.08) -- (377,154.01) ;
\draw [shift={(380,154)}, rotate = 179.84] [fill={rgb, 255:red, 0; green, 0; blue, 0 }  ][line width=0.08]  [draw opacity=0] (5.36,-2.57) -- (0,0) -- (5.36,2.57) -- cycle    ;
\draw    (421,154.08) -- (447.33,154.01) ;
\draw [shift={(450.33,154)}, rotate = 179.84] [fill={rgb, 255:red, 0; green, 0; blue, 0 }  ][line width=0.08]  [draw opacity=0] (5.36,-2.57) -- (0,0) -- (5.36,2.57) -- cycle    ;
\draw    (159.67,34.67) -- (196,34.67) ;
\draw [shift={(199,34.67)}, rotate = 180] [fill={rgb, 255:red, 0; green, 0; blue, 0 }  ][line width=0.08]  [draw opacity=0] (5.36,-2.57) -- (0,0) -- (5.36,2.57) -- cycle    ;
\draw    (260,34.08) -- (286.33,34.01) ;
\draw [shift={(289.33,34)}, rotate = 179.84] [fill={rgb, 255:red, 0; green, 0; blue, 0 }  ][line width=0.08]  [draw opacity=0] (5.36,-2.57) -- (0,0) -- (5.36,2.57) -- cycle    ;
\draw    (350,34.08) -- (376.33,34.01) ;
\draw [shift={(379.33,34)}, rotate = 179.84] [fill={rgb, 255:red, 0; green, 0; blue, 0 }  ][line width=0.08]  [draw opacity=0] (5.36,-2.57) -- (0,0) -- (5.36,2.57) -- cycle    ;
\draw    (420.33,34.08) -- (446.67,34.01) ;
\draw [shift={(449.67,34)}, rotate = 179.84] [fill={rgb, 255:red, 0; green, 0; blue, 0 }  ][line width=0.08]  [draw opacity=0] (5.36,-2.57) -- (0,0) -- (5.36,2.57) -- cycle    ;
\draw    (510.53,34.28) -- (536.87,34.21) ;
\draw [shift={(539.87,34.2)}, rotate = 179.84] [fill={rgb, 255:red, 0; green, 0; blue, 0 }  ][line width=0.08]  [draw opacity=0] (5.36,-2.57) -- (0,0) -- (5.36,2.57) -- cycle    ;
\draw    (509.53,154.08) -- (535.87,154.01) ;
\draw [shift={(538.87,154)}, rotate = 179.84] [fill={rgb, 255:red, 0; green, 0; blue, 0 }  ][line width=0.08]  [draw opacity=0] (5.36,-2.57) -- (0,0) -- (5.36,2.57) -- cycle    ;
\draw    (230.5,180.5) -- (230.5,197.5) ;
\draw [shift={(230.5,200.5)}, rotate = 270] [fill={rgb, 255:red, 0; green, 0; blue, 0 }  ][line width=0.08]  [draw opacity=0] (5.36,-2.57) -- (0,0) -- (5.36,2.57) -- cycle    ;
\draw    (320.5,180.72) -- (320.5,197.72) ;
\draw [shift={(320.5,200.72)}, rotate = 270] [fill={rgb, 255:red, 0; green, 0; blue, 0 }  ][line width=0.08]  [draw opacity=0] (5.36,-2.57) -- (0,0) -- (5.36,2.57) -- cycle    ;
\draw    (480.75,180.47) -- (480.75,197.47) ;
\draw [shift={(480.75,200.47)}, rotate = 270] [fill={rgb, 255:red, 0; green, 0; blue, 0 }  ][line width=0.08]  [draw opacity=0] (5.36,-2.57) -- (0,0) -- (5.36,2.57) -- cycle    ;

\draw (388.75,151.15) node [anchor=north west][inner sep=0.75pt]    {$\dotsc $};
\draw (214.5,85.9) node [anchor=north west][inner sep=0.75pt]    {$\mathcal{M}_{1}$};
\draw (305,86.4) node [anchor=north west][inner sep=0.75pt]    {$\mathcal{M}_{2}$};
\draw (463.5,86.9) node [anchor=north west][inner sep=0.75pt]    {$\mathcal{M}_{n}$};
\draw (164.5,132.9) node [anchor=north west][inner sep=0.75pt]    {$R_{0}$};
\draw (261.5,133.9) node [anchor=north west][inner sep=0.75pt]    {$R_{1}$};
\draw (350,134.4) node [anchor=north west][inner sep=0.75pt]    {$R_{2}$};
\draw (414.5,132.9) node [anchor=north west][inner sep=0.75pt]    {$R_{n-1}$};
\draw (224,206.15) node [anchor=north west][inner sep=0.75pt]    {$A_{1}$};
\draw (388.08,31.15) node [anchor=north west][inner sep=0.75pt]    {$\dotsc $};
\draw (163.83,12.9) node [anchor=north west][inner sep=0.75pt]    {$E_{0}$};
\draw (260.83,13.9) node [anchor=north west][inner sep=0.75pt]    {$E_{1}$};
\draw (349.33,14.4) node [anchor=north west][inner sep=0.75pt]    {$E_{2}$};
\draw (413.83,12.9) node [anchor=north west][inner sep=0.75pt]    {$E_{n-1}$};
\draw (512.03,13.1) node [anchor=north west][inner sep=0.75pt]    {$E_{n}$};
\draw (510.03,132.9) node [anchor=north west][inner sep=0.75pt]    {$R_{n}$};
\draw (314,206.37) node [anchor=north west][inner sep=0.75pt]    {$A_{2}$};
\draw (474.25,206.12) node [anchor=north west][inner sep=0.75pt]    {$A_{n}$};

\end{tikzpicture}
        \caption*{(b) Representation of the GEAT, where the Markovian channels allow Eve to modify her information in each round.}
    \end{minipage}
    \caption{Visual representations of the Markovian maps in EAT (Theorem \ref{thm: EAT}) and GEAT (Theorem \ref{thm: GEAT}). Figure inspired on \cite{metger2024generalised}.}
    \label{fig:EAT_GEAT}
\end{figure} 

Both EAT and GEAT treat the input state to each round's channel in a similar way: besides the sequential condition (Markov in EAT and non-signalling in GEAT) and the observed classical data $X_i$, one typically imposes no further assumptions on the per-round input states beyond those encoded in the model. In PM protocols, however, additional information is available because the source is trusted and its emitted signal state is characterized, e.g. the reduced state on a given subsystem (or an average-state constraint) is known (as detailed in Sec. \ref{sec: Distribution of states}). It is then natural to ask whether such marginal information can be incorporated already at the level of entropy accumulation. This leads to the marginal-constrained EAT (MEAT).

The MEAT is a generalization of the entropy accumulation theorem that allows one to incorporate \emph{marginal state constraints}~\cite{arqand2025marginal}. For each round $i$, a set $\mathcal{S}^{(i)}$ of allowed reduced states on a designated input subsystem (e.g. the emitted signal system $S_i$) is specified, and assumes that the actual input state to the channel
$\mathcal{M}_i$ has its marginal on $S_i$ contained in $\mathcal{S}^{(i)}$. These sets may encode, for example, that the emitted reduced state is a fixed, characterized source state, as in Eq. (\ref{eq: expanded state sourcerep}), while still allowing correlations with the adversary's side information. Under suitable convexity assumptions on the sets $\mathcal{S}^{(i)}$, it is possible to obtain a bound in which the min-tradeoff condition needs to hold only for input states consistent with the marginal constraints similar to the one previously found for EAT and GEAT.

\thm{\textbf{(Marginal-constrained entropy accumulation (MEAT),
Theorems~4.2a and~4.2b of~\cite{arqand2025marginal})}}
{\label{thm: MEAT} Consider a sequential process described by channels $\mathcal{M}_1,\dots,\mathcal{M}_n$. In each round $i$, the channel $\mathcal{M}_i$ acts on an internal memory $R_{i-1}$, the adversary's side information $E_{i-1}$, and a designated input\footnote{Here $A_{i-1}$ denotes the designated \emph{input} subsystem whose marginal is constrained, following the notation
of~\cite{arqand2025marginal}.} subsystem $A_{i-1}$ (on which a marginal constraint is imposed), and outputs a raw-key register $S_i$, classical registers $C_i,X_i$, updated memory $R_i$, and updated side information $E_i$.}

Suppose that for each $i$ the input marginal on $A_{i-1}$ belongs to a prescribed convex set $\mathcal{S}^{(i-1)}$, and that the process satisfies a GEAT-type non-signalling condition preventing the internal memory from directly signalling into the updated adversarial register $E_i$. Let $f$ be a min-tradeoff function whose validity is required only for single-round input states consistent with the marginal constraints. Then, for any event $\Omega$ on $X_1^n$ such that $f(\mathrm{freq}(X_1^n)) \ge h$ for all $X_1^n\in\Omega$, the smooth min-entropy of $S_1^n$ conditioned on the final side information and public transcript satisfies
\begin{align}
    H_{\min}^{\epsilon}\left(S_1^n \vert E_n C_1^n\right)_{\rho_{\vert \Omega}}
    \geq n h - c_{\text{MEAT}} \sqrt{n},
\end{align}
for a constant $c_{\text{MEAT}}$ depending on the marginal
constraints, the tradeoff function, and $\epsilon$.
 \\

\noindent
\textbf{Proof.} The proof of MEAT can be found in \cite{arqand2025marginal}. \\

The important aspect of MEAT is that it allows us to combine the generality of GEAT (allowing an adaptative side-channel, with round-by-round information) with additional information about trusted subsystems, in particular trusted sources in PM protocols, making it particularly relevant for finite-key analyses of this sort of protocol (as recently demonstrated in \cite{navarro2025finite}). Further results simplify the use of the min-tradeoff functions and allow the use of more general entropic quantities \cite{arqand2024generalized}.

\section{Connecting analytical and numerical techniques}\label{sec: connecting}

The methods presented in Sections \ref{sec: algorithms} and \ref{sec: finite keys} can be combined to provide a versatile framework for the analysis of realistic QKD protocols. In particular, the techniques discussed in Section~\ref{sec: finite keys} allow one to address the most general class of eavesdropping strategies -- coherent attacks -- in protocols with a finite number of rounds. On the other hand, the numerical methods explored in Section~\ref{sec: algorithms}, provide powerful tools to incorporate realistic device imperfections into the key-rate optimization.

Combining these two approaches can provide a powerful framework to recover strong security guarantees while reflecting practical implementations. However, this integration is not straightforward: finite-size security techniques, such as the postselection technique (Sec.~\ref{sec: postselection}) and entropy accumulation theorems (Sec.~\ref{sec: EAT}), often rely on technical steps that must be carefully adapted to ensure compatibility with numerical optimization procedures. Nevertheless, significant progress has been made in recent years toward combining these approaches for the analysis of specific protocols. In Table~\ref{tab: connecting analytical and numerical}, we list representative works that successfully combine these methods, enabling security proofs for a wide range of QKD protocols and yielding key-rate estimates that better match the practical implementations.

We note that, as illustrated in  Table~\ref{tab: connecting analytical and numerical}, the number of representative works differs significantly across the different finite-size techniques. This reflects the fact that different approaches accommodate different classes of protocols and parameter-estimation schemes.

Entropic uncertainty relation methods (Sec.~\ref{sec: EUR}), for instance, are suited for protocols with only two measurement settings per party. Consequently, EUR techniques have been successfully applied to BB84-type protocols and related variants, where the relevant parameters reduce essentially to error rates in two conjugate bases. In these cases, fully analytical security proofs are often possible, and therefore EUR-based approaches rarely interface with numerical optimization methods as represented in Table~\ref{tab: connecting analytical and numerical}. By contrast, protocols involving a broader set of inputs or a richer parameter-estimation stage—such as those estimating multiple correlators, Bell inequalities, or high-dimensional statistics—do not naturally fit into the EUR framework, and extending these techniques to such scenarios remains challenging.

The postselection technique (Sec.~\ref{sec: postselection}), while powerful and conceptually elegant, , relies on strict symmetry assumptions. In practice, ensuring the required symmetry can be difficult, since QKD protocols must allow aborts if certain conditions are not met. Conditioning on the event that the protocol does not abort can introduce correlations that violate the symmetry assumptions (see discussion in~\cite{tomamichel2017largely}). Moreover post-selection technique typically leads to finite-size corrections that scale unfavorably with the dimension of the underlying Hilbert space or the size of the parameter set~\cite{staffieri2026finite}.

From a more general perspective, entropy accumulation techniques appear better suited for combination with numerical security analyses, as illustrated in Table~\ref{tab: connecting analytical and numerical}. EAT's (Sec.~\ref{sec: EAT}) provide a flexible framework in which security can be related to the accumulation of entropy contributions from individual rounds, expressed through tradeoff functions that depend on experimentally accessible statistics. These tradeoff functions can often be evaluated or bounded using numerical optimization, making techniques based on EAT particularly compatible with semidefinite optimization methods to evaluate key-rates, and with parameter-estimation procedures involving multiple observables. For these reasons, recent progress in combining analytical finite-size security tools with numerical key-rate optimization has largely relied on entropy accumulation–based approaches, which offer a more scalable and adaptable framework for the analysis of realistic QKD implementations.

\begin{widetext}
\begin{center}
\begin{table}[H]
\centering

\captionsetup{
    width=0.9225\textwidth,
    justification=justified,
    singlelinecheck=false
}

\caption{%
Examples of numerical security proofs of a few specific QKD protocols which make use of the algorithms explained in Section \ref{sec: algorithms} and analytical techniques exposed in Section \ref{sec: finite keys}.
}%
\label{tab: connecting analytical and numerical}

\renewcommand{\arraystretch}{5}
\setlength{\tabcolsep}{5.5pt}

\begin{tabular}{c || c c c c} 
\toprule
\backslashbox{\makecell{\textbf{Algorithms} \\  \textbf{(Sec. \ref{sec: algorithms})}}}{\makecell{\textbf{Analytical} \\ \textbf{tools} \\ \textbf{(Sec. \ref{sec: finite keys})}}} & \makecell{\textbf{Asymptotic}} & \makecell{\textbf{Postselection} \\ (Sec. \ref{sec: postselection}) }  &
\makecell{\textbf{EUR}  \\ (Sec. \ref{sec: EUR}) } & 
\makecell{\textbf{EAT} \\ (Sec. \ref{sec: EAT}) }  \\
\midrule \midrule  

\makecell{\textbf{Frank-Wolfe} \cite{winick2018reliable} \\ (Sec. \ref{sec: Winick})} & \makecell{BB84 \cite{winick2018reliable} \\ DMCV \cite{lin2019asymptotic}} &  BB84 \cite{george2021numerical}  & - &  \makecell{BB84 \cite{george2022finite,kamin2025finite,chung2025generalized} \\ B92 \cite{metger2023security} \\ Decoy \cite{kamin2025finite} \\ HD \cite{george2022finite} \\ Six-state \cite{george2022finite}} \\ [12pt]

\makecell{\textbf{Gauss-Radau} \cite{brown2024device,araujo2023quantum} \\ (Sec. \ref{sec: Araujo}, Appendix \ref{sec: appendix gauss radau DI})} & \makecell{BB84 \cite{araujo2023quantum} \\ DIQKD \cite{brown2024device,rivera2025device} \\
HD \cite{araujo2023quantum}  \\  1SDI \cite{masini2024one}} & - & - & \makecell{DMCV \cite{primaatmaja2024discrete}}\\[12pt]

\makecell{\textbf{Conic} \cite{lorente2025quantum} \\ (Sec. \ref{sec: Lorente})} & \makecell{BB84 \cite{lorente2025quantum} \\ DMCV \cite{lorente2025quantum} \\ HD \cite{lorente2025quantum}} & - & - & \makecell{BB84 \cite{navarro2025finite}  \\ DMCV \cite{navarro2025finite,Pascual_Garc_a_2025} \\ HD \cite{navarro2025finite}}  \\ [12pt]

\makecell{\textbf{Min-entropy} \cite{masanes2011secure} \\ (Appendix \ref{sec: appendix Hmin})} & \makecell{BB84 \cite{bratzik2011min} \\ DIQKD \cite{masanes2011secure} \\ HD \cite{kanitschar2025composable} \\
Six-state \cite{bratzik2011min} } & \makecell{HD \cite{kanitschar2024practical}} & - & - \\ [12pt]

\makecell{\textbf{Other (or no)} \\ \textbf{numerical technique}} & \makecell{BB84 \cite{christandl2004generic,hu2022robust} \\
B92 \cite{christandl2004generic} \\ Decoy \cite{cai2009finite} \\ DIQKD \cite{kossmann2024reliable} \\ DMCV \cite{matsuura2021finite} \\
Six-state \cite{christandl2004generic} \\  1SDI \cite{branciard2012one}} & \makecell{BB84 \cite{scarani2008quantum,sheridan2010finite} \\ Decoy \cite{christandl2009postselection,nahar2024postselection,kamin2025improved} \\ Six-state \cite{scarani2008quantum} } & \makecell{BB84 \cite{tomamichel2012tight,tomamichel2017largely} \\ Decoy \cite{lim2014concise,rusca2018finite,tupkary2024phase,wiesemann2024consolidated} \\ 1SDI \cite{wang2013finite}}  & \makecell{ Decoy \cite{tupkary2026rigorous} \\ DIQKD \cite{arnon2018practical} } \\[12pt]

\bottomrule
\end{tabular}
\end{table}
\end{center}
\end{widetext}

\section{Acknowledgements}

We thank Florian Kanitschar and Mateus Araújo for helpful comments and clarifications on parts of the text, and Gereon Koßmann for pointing out the existence of references \cite{frenkel2023integral,kossmann2024reliable,kossmann2024optimising}. We also thank Marcus Huber, Monika Mothsara, Peter Brown and Ramona Wolf for useful discussions. This project was funded by the Austrian Research Promotion Agency (FFG) through the Project NSPT-QKD FO999915265.

\newpage
\bibliography{bibliography}

@article{bennett2014quantum,
  title={Quantum cryptography: Public key distribution and coin tossing},
  author={Bennett, Charles H and Brassard, Gilles},
  journal={Theoretical computer science},
  volume={560},
  pages={7--11},
  year={2014},
  publisher={Elsevier},
  url={https://www.sciencedirect.com/science/article/pii/S0304397514004241}
}

@article{Guo2025,
    author = {Mingxuan Guo  and Peng Huang  and Le Huang  and Xiaojuan Liao  and Xueqin Jiang  and Tao Wang  and Guihua Zeng },
    title = {Discrete-Modulated Coherent-State Quantum Key Distribution with Basis-Encoding},
    journal = {Research},
    volume = {8},
    number = {},
    pages = {0691},
    year = {2025},
    doi = {10.34133/research.0691},
    URL = {https://spj.science.org/doi/abs/10.34133/research.0691},
    eprint = {https://spj.science.org/doi/pdf/10.34133/research.0691}
}

@article{Tang2023,
    author = {Yan-Lin Tang and Chun Zhou and Dong-Dong Li and Zhi-Lin Xie and Mu-Lan Xu and Jian Sun and Ze-Xu Zhang and Lian-Jun Jiang and Li-Wei Wang and Guo-Qing Liu and Kun Wu and Yan Ma and Bo-Ran Zheng and Mu-Sheng Jiang and Yang Wang and Yu-Kang Zhao and Qing-Li Ma and Dexiang Zhang and Mei-Sheng Zhao and Wan-Su Bao and Shi-Biao Tang},
    journal = {Opt. Express},
    number = {16},
    pages = {26335--26343},
    publisher = {Optica Publishing Group},
    title = {Time-bin phase-encoding quantum key distribution using Sagnac-based optics and compatible electronics},
    volume = {31},
    month = {Jul},
    year = {2023},
    url = {https://opg.optica.org/oe/abstract.cfm?URI=oe-31-16-26335},
    doi = {10.1364/OE.496723},
    }

@article{Cocchi2025,
    author = {Sebastiano Cocchi and Domenico Ribezzo and Giulia Guarda and Pietro Centorrino and Tommaso Occhipinti and Alessandro Zavatta and Davide Bacco},
    journal = {Optica Quantum},
    number = {4},
    pages = {346--350},
    publisher = {Optica Publishing Group},
    title = {Time-bin encoding quantum key distribution in free-space horizontal links during nighttime and daytime},
    volume = {3},
    month = {Aug},
    year = {2025},
    url = {https://opg.optica.org/opticaq/abstract.cfm?URI=opticaq-3-4-346},
    doi = {10.1364/OPTICAQ.553977},
}

@article{sayat2024,
  author={Sayat, Mikhael T. and Shajilal, Biveen and Kish, Sebastian P. and Assad, Syed M. and Symul, Thomas and Lam, Ping Koy and Rattenbury, Nicholas J. and Cater, John E.},
  journal={IEEE Transactions on Communications}, 
  title={Satellite-to-Ground Continuous Variable Quantum Key Distribution: The Gaussian and Discrete Modulated Protocols in Low Earth Orbit}, 
  year={2024},
  volume={72},
  number={6},
  pages={3244-3255},
  doi={10.1109/TCOMM.2024.3359295}}

@article{Pan2022,
    author = {Yan Pan and Heng Wang and Yun Shao and Yaodi Pi and Yang Li and Bin Liu and Wei Huang and Bingjie Xu},
    journal = {Opt. Lett.},
    number = {13},
    pages = {3307--3310},
    publisher = {Optica Publishing Group},
    title = {Experimental demonstration of high-rate discrete-modulated continuous-variable quantum key distribution system},
    volume = {47},
    month = {Jul},
    year = {2022},
    url = {https://opg.optica.org/ol/abstract.cfm?URI=ol-47-13-3307},
    doi = {10.1364/OL.456978},
}

@article{grunenfelder2018polarization,
    author = {Grünenfelder, Fadri and Boaron, Alberto and Rusca, Davide and Martin, Anthony and Zbinden, Hugo},
    title = {Simple and high-speed polarization-based QKD},
    journal = {Applied Physics Letters},
    volume = {112},
    number = {5},
    pages = {051108},
    year = {2018},
    month = {01},
    issn = {0003-6951},
    doi = {10.1063/1.5016931},
    url = {https://doi.org/10.1063/1.5016931}
}

@article{pathak2023phase,
  title={Phase encoded quantum key distribution up to 380 km in standard telecom grade fiber enabled by baseline error optimization},
  author={Pathak, Nishant Kumar and Chaudhary, Sumit and Sangeeta and Kanseri, Bhaskar},
  journal={Scientific Reports},
  volume={13},
  number={1},
  pages={15868},
  year={2023},
  publisher={Nature Publishing Group UK London},
  url={https://www.nature.com/articles/s41598-023-42445-y}
}

@article{inoue25difphase,
author = {Kyo Inoue and Toshimori Honjo},
journal = {J. Opt. Soc. Am. B},
number = {9},
pages = {2116--2120},
publisher = {Optica Publishing Group},
title = {Differential-phase-shift quantum key distribution with intense decoy pulses},
volume = {42},
month = {Sep},
year = {2025},
url = {https://opg.optica.org/josab/abstract.cfm?URI=josab-42-9-2116},
doi = {10.1364/JOSAB.564939},
}

@article{boaron2018timebin,
    author = {Boaron, Alberto and Korzh, Boris and Houlmann, Raphael and Boso, Gianluca and Rusca, Davide and Gray, Stuart and Li, Ming-Jun and Nolan, Daniel and Martin, Anthony and Zbinden, Hugo},
    title = {Simple 2.5 GHz time-bin quantum key distribution},
    journal = {Applied Physics Letters},
    volume = {112},
    number = {17},
    pages = {171108},
    year = {2018},
    month = {04},
    issn = {0003-6951},
    doi = {10.1063/1.5027030},
    url = {https://doi.org/10.1063/1.5027030}
}

@article{laudenbach2018gaussianMod,
author = {Laudenbach, Fabian and Pacher, Christoph and Fung, Chi-Hang Fred and Poppe, Andreas and Peev, Momtchil and Schrenk, Bernhard and Hentschel, Michael and Walther, Philip and Hübel, Hannes},
title = {Continuous-Variable Quantum Key Distribution with Gaussian Modulation—The Theory of Practical Implementations},
journal = {Advanced Quantum Technologies},
volume = {1},
number = {1},
pages = {1800011},
doi = {https://doi.org/10.1002/qute.201800011},
url={https://advanced.onlinelibrary.wiley.com/doi/abs/10.1002/qute.201800011},
year = {2018}
}

@article{wang2005review,
  title={A review on the decoy-state method for practical quantum key distribution},
  author={Wang, Xiang-Bin},
  journal={arXiv preprint quant-ph/0509084},
  year={2005},
  url={https://arxiv.org/abs/quant-ph/0509084}
}

@article{upadhyaya2021dimension,
  title={Dimension reduction in quantum key distribution for continuous-and discrete-variable protocols},
  author={Upadhyaya, Twesh and Van Himbeeck, Thomas and Lin, Jie and L{\"u}tkenhaus, Norbert},
  journal={PRX Quantum},
  volume={2},
  number={2},
  pages={020325},
  year={2021},
  publisher={APS},
  url={https://journals.aps.org/prxquantum/abstract/10.1103/PRXQuantum.2.020325}
}

@article{wang2005beating,
  title={Beating the photon-number-splitting attack in practical quantum cryptography},
  author={Wang, Xiang-Bin},
  journal={Physical Review Letters},
  volume={94},
  number={23},
  pages={230503},
  year={2005},
  publisher={APS},
  url={https://journals.aps.org/prl/abstract/10.1103/PhysRevLett.94.230503}
}

@article{mueller2025performance,
  title={Performance of cascade and ldpc codes for information reconciliation on industrial quantum key distribution systems},
  author={Mueller, Ronny and De Lazzari, Claudia and Chirici, Fernando and Vagniluca, Ilaria and Oxenl{\o}we, Leif Katsuo and Forchhammer, S{\o}ren and Zavatta, Alessandro and Bacco, Davide},
  journal={IET Quantum Communication},
  volume={6},
  number={1},
  pages={e70003},
  year={2025},
  publisher={Wiley Online Library},
  url={https://arxiv.org/abs/2408.15758}
}

@article{martinez2012blind,
  title={Blind reconciliation},
  author={Martinez-Mateo, Jesus and Elkouss, David and Martin, Vicente},
  journal={arXiv preprint arXiv:1205.5729},
  year={2012},
  url={https://dl.acm.org/doi/abs/10.5555/2481580.2481585}
}

@inproceedings{brassard1993secret,
  title={Secret-key reconciliation by public discussion},
  author={Brassard, Gilles and Salvail, Louis},
  booktitle={Workshop on the Theory and Application of Cryptographic Techniques},
  pages={410--423},
  year={1993},
  organization={Springer},
  url={https://link.springer.com/chapter/10.1007/3-540-48285-7_35#:~:text=Cite%20this%20paper,3-540-48285-7_35}
}

@article{xu2014measurement,
  title={Measurement-device-independent quantum cryptography},
  author={Xu, Feihu and Curty, Marcos and Qi, Bing and Lo, Hoi-Kwong},
  journal={IEEE Journal of Selected Topics in Quantum Electronics},
  volume={21},
  number={3},
  pages={148--158},
  year={2014},
  publisher={IEEE},
  url={https://ieeexplore.ieee.org/document/6985598/}
}

@article{grosshans2003gaussianMod,
  title={Quantum key distribution using gaussian-modulated coherent states},
  author={Grosshans, Fr{\'e}d{\'e}ric and Van Assche, Gilles and Wenger, J{\'e}r{\^o}me and Brouri, Rosa and Cerf, Nicolas J and Grangier, Philippe},
  journal={Nature},
  volume={421},
  number={6920},
  pages={238--241},
  year={2003},
  publisher={Nature Publishing Group UK London},
  url={https://www.nature.com/articles/nature01289}
}

@article{tan2024entropy,
  title={Entropy bounds for device-independent quantum key distribution with local Bell test},
  author={Tan, Ernest Y-Z and Wolf, Ramona},
  journal={Physical Review Letters},
  volume={133},
  number={12},
  pages={120803},
  year={2024},
  publisher={APS},
  url={https://journals.aps.org/prl/abstract/10.1103/PhysRevLett.133.120803}
}

@article{ghorai2019discreteMod,
  title = {Asymptotic Security of Continuous-Variable Quantum Key Distribution with a Discrete Modulation},
  author = {Ghorai, Shouvik and Grangier, Philippe and Diamanti, Eleni and Leverrier, Anthony},
  journal = {Phys. Rev. X},
  volume = {9},
  issue = {2},
  pages = {021059},
  numpages = {11},
  year = {2019},
  month = {Jun},
  publisher = {American Physical Society},
  doi = {10.1103/PhysRevX.9.021059},
  url = {https://link.aps.org/doi/10.1103/PhysRevX.9.021059}
}

@article{berta2010uncertainty,
  title={The uncertainty principle in the presence of quantum memory},
  author={Berta, Mario and Christandl, Matthias and Colbeck, Roger and Renes, Joseph M and Renner, Renato},
  journal={Nature Physics},
  volume={6},
  number={9},
  pages={659--662},
  year={2010},
  publisher={Nature Publishing Group UK London},
  url={https://www.nature.com/articles/nphys1734}
}

@article{scarani2009security,
  title={The security of practical quantum key distribution},
  author={Scarani, Valerio and Bechmann-Pasquinucci, Helle and Cerf, Nicolas J and Du{\v{s}}ek, Miloslav and L{\"u}tkenhaus, Norbert and Peev, Momtchil},
  journal={Reviews of modern physics},
  volume={81},
  number={3},
  pages={1301--1350},
  year={2009},
  publisher={APS},
  url={https://journals.aps.org/rmp/abstract/10.1103/RevModPhys.81.1301}
}

@article{winick2018reliable,
  title={Reliable numerical key rates for quantum key distribution},
  author={Winick, Adam and L{\"u}tkenhaus, Norbert and Coles, Patrick J},
  journal={Quantum},
  volume={2},
  pages={77},
  year={2018},
  publisher={Verein zur F{\"o}rderung des Open Access Publizierens in den Quantenwissenschaften},
  url={https://quantum-journal.org/papers/q-2018-07-26-77/}
}

@article{devetak2005distillation,
  title={Distillation of secret key and entanglement from quantum states},
  author={Devetak, Igor and Winter, Andreas},
  journal={Proceedings of the Royal Society A: Mathematical, Physical and engineering sciences},
  volume={461},
  number={2053},
  pages={207--235},
  year={2005},
  publisher={The Royal Society},
  url={https://royalsocietypublishing.org/doi/10.1098/rspa.2004.1372}
}

@article{skajaa2015homogeneous,
  title={A homogeneous interior-point algorithm for nonsymmetric convex conic optimization},
  author={Skajaa, Anders and Ye, Yinyu},
  journal={Mathematical Programming},
  volume={150},
  pages={391--422},
  year={2015},
  publisher={Springer},
  url={https://link.springer.com/article/10.1007/s10107-014-0773-1}
}

@article{tavakoli2024semidefinite,
  title={Semidefinite programming relaxations for quantum correlations},
  author={Tavakoli, Armin and Pozas-Kerstjens, Alejandro and Brown, Peter and Ara{\'u}jo, Mateus},
  journal={Reviews of Modern Physics},
  volume={96},
  number={4},
  pages={045006},
  year={2024},
  publisher={APS},
  url={https://journals.aps.org/rmp/abstract/10.1103/RevModPhys.96.045006}
}

@book{skrzypczyk2023semidefinite,
  title={Semidefinite programming in quantum information science},
  author={Skrzypczyk, Paul and Cavalcanti, Daniel},
  year={2023},
  publisher={IOP Publishing},
  url={https://iopscience.iop.org/book/mono/978-0-7503-3343-6}
}

@article{watrous2020advanced,
  title={Advanced topics in quantum information theory},
  author={Watrous, John},
  journal={Lecture notes},
  year={2020},
}

@article{brunner2014bell,
  title={Bell nonlocality},
  author={Brunner, Nicolas and Cavalcanti, Daniel and Pironio, Stefano and Scarani, Valerio and Wehner, Stephanie},
  journal={Reviews of modern physics},
  volume={86},
  number={2},
  pages={419--478},
  year={2014},
  publisher={APS},
  url={https://journals.aps.org/rmp/abstract/10.1103/RevModPhys.86.419}
}

@article{navascues2007bounding,
  title={Bounding the set of quantum correlations},
  author={Navascu{\'e}s, Miguel and Pironio, Stefano and Ac{\'\i}n, Antonio},
  journal={Physical Review Letters},
  volume={98},
  number={1},
  pages={010401},
  year={2007},
  publisher={APS},
  url={https://journals.aps.org/prl/abstract/10.1103/PhysRevLett.98.010401}
}

@article{pironio2010convergent,
  title={Convergent relaxations of polynomial optimization problems with noncommuting variables},
  author={Pironio, Stefano and Navascu{\'e}s, Miguel and Acin, Antonio},
  journal={SIAM Journal on Optimization},
  volume={20},
  number={5},
  pages={2157--2180},
  year={2010},
  publisher={SIAM},
  url={https://epubs.siam.org/doi/10.1137/090760155}
}

@article{papp2017homogeneous,
  title={On" A Homogeneous Interior-Point Algorithm for Non-Symmetric Convex Conic Optimization"},
  author={Papp, D{\'a}vid and Y{\i}ld{\i}z, Sercan},
  journal={arXiv preprint arXiv:1712.00492},
  year={2017},
  url={https://arxiv.org/abs/1712.00492}
}

@article{lasserre2001global,
  title={Global optimization with polynomials and the problem of moments},
  author={Lasserre, Jean B},
  journal={SIAM Journal on optimization},
  volume={11},
  number={3},
  pages={796--817},
  year={2001},
  publisher={SIAM},
  url={https://epubs.siam.org/doi/10.1137/S1052623400366802}
}

@article{coles2012unification,
  title={Unification of different views of decoherence and discord},
  author={Coles, Patrick J},
  journal={Physical Review A—Atomic, Molecular, and Optical Physics},
  volume={85},
  number={4},
  pages={042103},
  year={2012},
  publisher={APS},
  url={https://journals.aps.org/pra/abstract/10.1103/PhysRevA.85.042103}
}

@article{frank1956algorithm,
  title={An algorithm for quadratic programming},
  author={Frank, Marguerite and Wolfe, Philip and others},
  journal={Naval research logistics quarterly},
  volume={3},
  number={1-2},
  pages={95--110},
  year={1956},
  publisher={Wiley Subscription Services, Inc., A Wiley Company New York},
  url={https://onlinelibrary.wiley.com/doi/10.1002/nav.3800030109}
}

@article{branciard2012one,
  title={One-sided device-independent quantum key distribution: Security, feasibility, and the connection with steering},
  author={Branciard, Cyril and Cavalcanti, Eric G and Walborn, Stephen P and Scarani, Valerio and Wiseman, Howard M},
  journal={Physical Review A—Atomic, Molecular, and Optical Physics},
  volume={85},
  number={1},
  pages={010301},
  year={2012},
  publisher={APS},
  url={https://arxiv.org/abs/1109.1435}
}

@inproceedings{scarani2008security,
  title={Security bounds for quantum cryptography with finite resources},
  author={Scarani, Valerio and Renner, Renato},
  booktitle={Workshop on Quantum Computation, Communication, and Cryptography},
  pages={83--95},
  year={2008},
  organization={Springer},
  url={https://arxiv.org/abs/0806.0120}
}

@article{huber2013weak,
  title={Weak randomness in device-independent quantum key distribution and the advantage of using high-dimensional entanglement},
  author={Huber, Marcus and Paw{\l}owski, Marcin},
  journal={Physical Review A—Atomic, Molecular, and Optical Physics},
  volume={88},
  number={3},
  pages={032309},
  year={2013},
  publisher={APS},
  url={https://arxiv.org/abs/1301.2455}
}

@article{Diamanti2016,
  author    = {Elena Diamanti and Hoi-Kwong Lo and Bing Qi and Zhiliang Yuan},
  title     = {Practical challenges in quantum key distribution},
  journal   = {npj Quantum Information},
  volume    = {2},
  pages     = {16025},
  year      = {2016},
  doi       = {10.1038/npjqi.2016.25},
  url       = {https://doi.org/10.1038/npjqi.2016.25}
}

@article{murta2020quantum,
  title={Quantum conference key agreement: A review},
  author={Murta, Gl{\'a}ucia and Grasselli, Federico and Kampermann, Hermann and Bru{\ss}, Dagmar},
  journal={Advanced Quantum Technologies},
  volume={3},
  number={11},
  pages={2000025},
  year={2020},
  publisher={Wiley Online Library},
  url={https://advanced.onlinelibrary.wiley.com/doi/full/10.1002/qute.202000025}
}

@article{brown2024device,
  title={Device-independent lower bounds on the conditional von Neumann entropy},
  author={Brown, Peter and Fawzi, Hamza and Fawzi, Omar},
  journal={Quantum},
  volume={8},
  pages={1445},
  year={2024},
  publisher={Verein zur F{\"o}rderung des Open Access Publizierens in den Quantenwissenschaften},
  url={https://quantum-journal.org/papers/q-2024-08-27-1445/}
}

@article{lorente2025quantum,
  title={Quantum key distribution rates from non-symmetric conic optimization},
  author={Lorente, Andr{\'e}s Gonz{\'a}lez and Parellada, Pablo V and Castillo-Celeita, Miguel and Ara{\'u}jo, Mateus},
  journal={Quantum},
  volume={9},
  pages={1657},
  year={2025},
  publisher={Verein zur F{\"o}rderung des Open Access Publizierens in den Quantenwissenschaften},
  url={https://quantum-journal.org/papers/q-2025-03-10-1657/#}
}

@article{masanes2011secure,
  title={Secure device-independent quantum key distribution with causally independent measurement devices},
  author={Masanes, Lluis and Pironio, Stefano and Ac{\'\i}n, Antonio},
  journal={Nature communications},
  volume={2},
  number={1},
  pages={238},
  year={2011},
  publisher={Nature Publishing Group UK London},
  url={https://www.nature.com/articles/ncomms1244}
}

@article{acin2007device,
  title={Device-independent security of quantum cryptography against collective attacks},
  author={Ac{\'\i}n, Antonio and Brunner, Nicolas and Gisin, Nicolas and Massar, Serge and Pironio, Stefano and Scarani, Valerio},
  journal={Physical Review Letters},
  volume={98},
  number={23},
  pages={230501},
  year={2007},
  publisher={APS},
  url={https://journals.aps.org/prl/abstract/10.1103/PhysRevLett.98.230501}
}

@article{acin2006bell,
  title={From Bell’s theorem to secure quantum key distribution},
  author={Acin, Antonio and Gisin, Nicolas and Masanes, Lluis},
  journal={Physical Review Letters},
  volume={97},
  number={12},
  pages={120405},
  year={2006},
  publisher={APS},
  url={https://journals.aps.org/prl/abstract/10.1103/PhysRevLett.97.120405}
}

@article{ekert1991quantum,
  title={Quantum cryptography based on Bell’s theorem},
  author={Ekert, Artur K},
  journal={Physical Review Letters},
  volume={67},
  number={6},
  pages={661},
  year={1991},
  publisher={APS},
  url={https://journals.aps.org/prl/abstract/10.1103/PhysRevLett.67.661}
}

@article{kanitschar2025composable,
  title={Composable finite-size security of high-dimensional quantum-key-distribution protocols},
  author={Kanitschar, Florian and Huber, Marcus},
  journal={Physical Review Applied},
  volume={24},
  number={5},
  pages={054028},
  year={2025},
  publisher={APS},
  url={https://journals.aps.org/prapplied/abstract/10.1103/v51y-vkfr}
}

@article{cavalcanti2016quantum,
  title={Quantum steering: a review with focus on semidefinite programming},
  author={Cavalcanti, Daniel and Skrzypczyk, Paul},
  journal={Reports on Progress in Physics},
  volume={80},
  number={2},
  pages={024001},
  year={2016},
  publisher={IOP Publishing},
  url={https://iopscience.iop.org/article/10.1088/1361-6633/80/2/024001}
}

@article{uola2020quantum,
  title={Quantum steering},
  author={Uola, Roope and Costa, Ana CS and Nguyen, H Chau and G{\"u}hne, Otfried},
  journal={Reviews of Modern Physics},
  volume={92},
  number={1},
  pages={015001},
  year={2020},
  publisher={APS},
  url={https://journals.aps.org/rmp/abstract/10.1103/RevModPhys.92.015001}
}

@article{tomamichel2017largely,
  title={A largely self-contained and complete security proof for quantum key distribution},
  author={Tomamichel, Marco and Leverrier, Anthony},
  journal={Quantum},
  volume={1},
  pages={14},
  year={2017},
  publisher={Verein zur F{\"o}rderung des Open Access Publizierens in den Quantenwissenschaften},
  url={https://quantum-journal.org/papers/q-2017-07-14-14/}
}

@book{vidick2023introduction,
  title={Introduction to quantum cryptography},
  author={Vidick, Thomas and Wehner, Stephanie},
  year={2023},
  publisher={Cambridge University Press}
}

@article{bennett1992quantum,
  title={Quantum cryptography without Bell’s theorem},
  author={Bennett, Charles H and Brassard, Gilles and Mermin, N David},
  journal={Physical Review Letters},
  volume={68},
  number={5},
  pages={557},
  year={1992},
  publisher={APS},
  url={https://journals.aps.org/prl/abstract/10.1103/PhysRevLett.68.557}
}

@article{marconi2025symmetric,
  title={Symmetric quantum states: a review of recent progress},
  author={Marconi, Carlo and M{\"u}ller-Rigat, Guillem and Romero-Pallej{\`a}, Jordi and Tura Brugu{\'e}s, Jordi and Sanpera, Anna},
  journal={Reports on Progress in Physics},
  year={2025},
  url={https://iopscience.iop.org/article/10.1088/1361-6633/ae440a}
}

@article{slepian2003noiseless,
  title={Noiseless coding of correlated information sources},
  author={Slepian, David and Wolf, Jack},
  journal={IEEE Transactions on information Theory},
  volume={19},
  number={4},
  pages={471--480},
  year={2003},
  publisher={IEEE},
  url={https://ieeexplore.ieee.org/document/1055037}
}

@inproceedings{pacher2015information,
  title={An information reconciliation protocol for secret-key agreement with small leakage},
  author={Pacher, Christoph and Grabenweger, Philipp and Martinez-Mateo, Jesus and Martin, Vicente},
  booktitle={2015 IEEE international symposium on information theory (ISIT)},
  pages={730--734},
  year={2015},
  organization={IEEE},
  url={https://publications.ait.ac.at/en/publications/an-information-reconciliation-protocol-for-secret-key-agreement-w-9/}
}

@article{marcomini2025loss,
  title={Loss-tolerant quantum key distribution with detection efficiency mismatch},
  author={Marcomini, Alessandro and Mizutani, Akihiro and Gr{\"u}nenfelder, Fadri and Curty, Marcos and Tamaki, Kiyoshi},
  journal={Quantum Science and Technology},
  volume={10},
  number={3},
  pages={035002},
  year={2025},
  publisher={IOP Publishing},
  url={https://iopscience.iop.org/article/10.1088/2058-9565/adc8cc}
}

@article{mironowicz2024semi,
  title={Semi-definite programming and quantum information},
  author={Mironowicz, Piotr},
  journal={Journal of Physics A: Mathematical and Theoretical},
  volume={57},
  number={16},
  pages={163002},
  year={2024},
  publisher={IOP Publishing},
  url={https://iopscience.iop.org/article/10.1088/1751-8121/ad2b85}
}

@article{lo2005decoy,
  title={Decoy state quantum key distribution},
  author={Lo, Hoi-Kwong and Ma, Xiongfeng and Chen, Kai},
  journal={Physical Review Letters},
  volume={94},
  number={23},
  pages={230504},
  year={2005},
  publisher={APS},
  url={https://journals.aps.org/prl/abstract/10.1103/PhysRevLett.94.230504}
}

@article{liao2018satellite,
  title={Satellite-relayed intercontinental quantum network},
  author={Liao, Sheng-Kai and Cai, Wen-Qi and Handsteiner, Johannes and Liu, Bo and Yin, Juan and Zhang, Liang and Rauch, Dominik and Fink, Matthias and Ren, Ji-Gang and Liu, Wei-Yue and others},
  journal={Physical Review Letters},
  volume={120},
  number={3},
  pages={030501},
  year={2018},
  publisher={APS},
  url={https://journals.aps.org/prl/abstract/10.1103/PhysRevLett.120.030501}
}

@article{yin2017satellite,
  title={Satellite-based entanglement distribution over 1200 kilometers},
  author={Yin, Juan and Cao, Yuan and Li, Yu-Huai and Liao, Sheng-Kai and Zhang, Liang and Ren, Ji-Gang and Cai, Wen-Qi and Liu, Wei-Yue and Li, Bo and Dai, Hui and others},
  journal={Science},
  volume={356},
  number={6343},
  pages={1140--1144},
  year={2017},
  publisher={American Association for the Advancement of Science},
  url={https://www.science.org/doi/10.1126/science.aan3211}
}

@article{nadlinger2022experimental,
  title={Experimental quantum key distribution certified by Bell's theorem},
  author={Nadlinger, David P and Drmota, Peter and Nichol, Bethan C and Araneda, Gabriel and Main, Dougal and Srinivas, Raghavendra and Lucas, David M and Ballance, Christopher J and Ivanov, Kirill and Tan, EY-Z and others},
  journal={Nature},
  volume={607},
  number={7920},
  pages={682--686},
  year={2022},
  publisher={Nature Publishing Group UK London},
  url={https://www.nature.com/articles/s41586-022-04941-5}
}

@article{navarro2025finite,
  title={Finite-size quantum key distribution rates from R$\backslash$'enyi entropies using conic optimization},
  author={Navarro, Mariana and Lorente, Andr{\'e}s Gonz{\'a}lez and Parellada, Pablo V and Pascual-Garc{\'\i}a, Carlos and Ara{\'u}jo, Mateus},
  journal={arXiv preprint arXiv:2511.10584},
  year={2025},
  url={https://arxiv.org/abs/2511.10584}
}

@article{usenko2025continuous,
  title={Continuous-variable quantum communication},
  author={Usenko, Vladyslav C and Ac{\'\i}n, Antonio and All{\'e}aume, Romain and Andersen, Ulrik L and Diamanti, Eleni and Gehring, Tobias and Hajomer, Adnan AE and Kanitschar, Florian and Pacher, Christoph and Pirandola, Stefano and others},
  journal={arXiv preprint usenkoarXiv:2501.12801},
  year={2025},
  url={https://arxiv.org/abs/2501.12801}
}

@article{diamanti2015distributing,
  title={Distributing secret keys with quantum continuous variables: principle, security and implementations},
  author={Diamanti, Eleni and Leverrier, Anthony},
  journal={Entropy},
  volume={17},
  number={9},
  pages={6072--6092},
  year={2015},
  publisher={MDPI},
  url={https://www.mdpi.com/1099-4300/17/9/6072}
}

@article{lim2014concise,
  title={Concise security bounds for practical decoy-state quantum key distribution},
  author={Lim, Charles Ci Wen and Curty, Marcos and Walenta, Nino and Xu, Feihu and Zbinden, Hugo},
  journal={Physical Review A},
  volume={89},
  number={2},
  pages={022307},
  year={2014},
  publisher={APS},
  url={https://journals.aps.org/pra/abstract/10.1103/PhysRevA.89.022307}
}

@article{rusca2018finite,
  title={Finite-key analysis for the 1-decoy state QKD protocol},
  author={Rusca, Davide and Boaron, Alberto and Gr{\"u}nenfelder, Fadri and Martin, Anthony and Zbinden, Hugo},
  journal={Applied Physics Letters},
  volume={112},
  number={17},
  year={2018},
  publisher={AIP Publishing},
  url={https://pubs.aip.org/aip/apl/article/112/17/171104/312107/Finite-key-analysis-for-the-1-decoy-state-QKD}
}

@article{tupkary2024phase,
  title={Phase error rate estimation in QKD with imperfect detectors},
  author={Tupkary, Devashish and Nahar, Shlok and Sinha, Pulkit and L{\"u}tkenhaus, Norbert},
  journal={arXiv preprint arXiv:2408.17349},
  year={2024},
  url={https://arxiv.org/abs/2408.17349}
}

@article{wiesemann2024consolidated,
  title={A consolidated and accessible security proof for finite-size decoy-state quantum key distribution},
  author={Wiesemann, Jerome and Krause, Jan and Tupkary, Devashish and L{\"u}tkenhaus, Norbert and Rusca, Davide and Walenta, Nino},
  journal={arXiv preprint arXiv:2405.16578},
  year={2024},
  url={https://arxiv.org/abs/2405.16578}
}

@article{kamin2025improved,
  title={Improved finite-size effects in QKD protocols with applications to decoy-state QKD},
  author={Kamin, Lars and Tupkary, Devashish and L{\"u}tkenhaus, Norbert},
  journal={arXiv preprint arXiv:2502.05382},
  year={2025},
  url={https://arxiv.org/abs/2502.05382}
}

@article{nahar2024postselection,
  title={Postselection technique for optical Quantum Key Distribution with improved de Finetti reductions},
  author={Nahar, Shlok and Tupkary, Devashish and Zhao, Yuming and L{\"u}tkenhaus, Norbert and Tan, Ernest Y-Z},
  journal={PRX Quantum},
  volume={5},
  number={4},
  pages={040315},
  year={2024},
  publisher={APS},
  url={https://journals.aps.org/prxquantum/abstract/10.1103/PRXQuantum.5.040315}
}

@article{anka2025introductoryreviewtheorycontinuousvariable,
      title={An introductory review of the theory of continuous-variable quantum key distribution: Fundamentals, protocols, and security}, 
      author={Maron F Anka and John A. Mora Rodríguez and Douglas F. Pinto and Lucas Q. Galvão and Micael A. Dias and Alexandre B. Tacla},
      year={2025},
      journal={arXiv preprint},
      eprint={2512.01758},
      archivePrefix={arXiv},
      primaryClass={quant-ph},
      url={https://arxiv.org/abs/2512.01758}, 
}

@article{gottesman2003proof,
  title={Proof of security of quantum key distribution with two-way classical communications},
  author={Gottesman, Daniel and Lo, Hoi-Kwong},
  journal={IEEE Transactions on Information Theory},
  volume={49},
  number={2},
  pages={457--475},
  year={2003},
  publisher={IEEE},
  url={https://arxiv.org/abs/quant-ph/0105121}
}

@article{tupkary2023using,
  title={Using cascade in quantum key distribution},
  author={Tupkary, Devashish and L{\"u}tkenhaus, Norbert},
  journal={Physical Review Applied},
  volume={20},
  number={6},
  pages={064040},
  year={2023},
  publisher={APS},
  url={https://journals.aps.org/prapplied/abstract/10.1103/PhysRevApplied.20.064040}
}

@article{krishna2002entropic,
  title={An entropic uncertainty principle for quantum measurements},
  author={Krishna, M and Parthasarathy, KR},
  journal={Sankhy{\=a}: The Indian Journal of Statistics, Series A},
  pages={842--851},
  year={2002},
  publisher={JSTOR},
  url={https://www.jstor.org/stable/25051432?seq=1}
}

@article{zhang2025towards,
  title={Towards global quantum key distribution},
  author={Zhang, Haoran and Zhu, Haotao and He, Ruihua and Zhang, Yan and Ding, Chao and Hanzo, Lajos and Gao, Weibo},
  journal={Nature Reviews Electrical Engineering},
  pages={1--13},
  year={2025},
  publisher={Nature Publishing Group UK London},
  url={https://www.nature.com/articles/s44287-025-00238-7}
}

@article{renes2009conjectured,
  title={Conjectured strong complementary information tradeoff},
  author={Renes, Joseph M and Boileau, Jean-Christian},
  journal={Physical Review Letters},
  volume={103},
  number={2},
  pages={020402},
  year={2009},
  publisher={APS},
  url={https://journals.aps.org/prl/abstract/10.1103/PhysRevLett.103.020402}
}

@article{george2021numerical,
  title={Numerical calculations of the finite key rate for general quantum key distribution protocols},
  author={George, Ian and Lin, Jie and L{\"u}tkenhaus, Norbert},
  journal={Physical Review Research},
  volume={3},
  number={1},
  pages={013274},
  year={2021},
  publisher={APS},
  url={https://journals.aps.org/prresearch/abstract/10.1103/PhysRevResearch.3.013274}
}

@article{mannalath2025sharp,
  title={Sharp finite statistics for quantum key distribution},
  author={Mannalath, Vaisakh and Zapatero, V{\'\i}ctor and Curty, Marcos},
  journal={Physical Review Letters},
  volume={135},
  number={2},
  pages={020803},
  year={2025},
  publisher={APS},
  url={https://journals.aps.org/prl/abstract/10.1103/l735-x48g}
}

@article{tomamichel2012tight,
  title={Tight finite-key analysis for quantum cryptography},
  author={Tomamichel, Marco and Lim, Charles Ci Wen and Gisin, Nicolas and Renner, Renato},
  journal={Nature communications},
  volume={3},
  number={1},
  pages={634},
  year={2012},
  publisher={Nature Publishing Group UK London},
  url={https://www.nature.com/articles/ncomms1631}
}

@article{sheridan2010finite,
  title={Finite-key security against coherent attacks in quantum key distribution},
  author={Sheridan, Lana and Le, Thinh Phuc and Scarani, Valerio},
  journal={New Journal of Physics},
  volume={12},
  number={12},
  pages={123019},
  year={2010},
  publisher={IOP Publishing},
  url={https://arxiv.org/abs/1008.2596}
}

@article{coles2011information,
  title={Information-theoretic treatment of tripartite systems and quantum channels},
  author={Coles, Patrick J and Yu, Li and Gheorghiu, Vlad and Griffiths, Robert B},
  journal={Physical Review A—Atomic, Molecular, and Optical Physics},
  volume={83},
  number={6},
  pages={062338},
  year={2011},
  publisher={APS},
  url={https://journals.aps.org/pra/abstract/10.1103/PhysRevA.83.062338}
}

@article{belzig2024studying,
  title={Studying stabilizer de Finetti theorems and possible applications in quantum information processing},
  author={Belzig, Paula},
  journal={arXiv preprint arXiv:2403.10592},
  year={2024},
  url={https://arxiv.org/abs/2403.10592}
}

@article{gross2021schur,
  title={Schur--Weyl duality for the Clifford group with applications: Property testing, a robust Hudson theorem, and de Finetti representations},
  author={Gross, David and Nezami, Sepehr and Walter, Michael},
  journal={Communications in Mathematical Physics},
  volume={385},
  number={3},
  pages={1325--1393},
  year={2021},
  publisher={Springer},
  url={https://link.springer.com/article/10.1007/s00220-021-04118-7}
}

@article{costa2025finetti,
  title={A de Finetti theorem for quantum causal structures},
  author={Costa, Fabio and Barrett, Jonathan and Shrapnel, Sally},
  journal={Quantum},
  volume={9},
  pages={1628},
  year={2025},
  publisher={Verein zur F{\"o}rderung des Open Access Publizierens in den Quantenwissenschaften},
  url={https://quantum-journal.org/papers/q-2025-02-11-1628/}
}

@article{pereira2025optimal,
  title={Optimal key rates for quantum key distribution with partial source characterization},
  author={Pereira, Margarida and Curr{\'a}s-Lorenzo, Guillermo and Ara{\'u}jo, Mateus},
  journal={arXiv preprint arXiv:2510.13085},
  year={2025},
  url={https://arxiv.org/abs/2510.13085}
}

@inproceedings{gottesman2004security,
  title={Security of quantum key distribution with imperfect devices},
  author={Gottesman, Daniel and Lo, H-K and Lutkenhaus, Norbert and Preskill, John},
  booktitle={International Symposium onInformation Theory, 2004. ISIT 2004. Proceedings.},
  pages={136},
  year={2004},
  organization={IEEE},
  url={https://arxiv.org/abs/quant-ph/0212066}
}

@article{staffieri2026finite,
  title={Finite-size security of QKD: comparison of three proof techniques},
  author={Staffieri, Gabriele and Scala, Giovanni and Lupo, Cosmo},
  journal={arXiv preprint arXiv:2601.03829},
  year={2026},
  url={https://arxiv.org/abs/2601.03829}
}

@article{scarani2004quantum,
  title={Quantum Cryptography Protocols Robust against Photon Number Splitting Attacks for Weak Laser Pulse Implementations},
  author={Scarani, Valerio and Acin, Antonio and Ribordy, Gr{\'e}goire and Gisin, Nicolas},
  journal={Physical Review Letters},
  volume={92},
  number={5},
  pages={057901},
  year={2004},
  publisher={APS},
  url={https://journals.aps.org/prl/abstract/10.1103/PhysRevLett.92.057901}
}

@book{scarani2019bell,
  title={Bell nonlocality},
  author={Scarani, Valerio},
  year={2019},
  publisher={Oxford University Press}
}

@article{tamaki2013loss,
  title={Loss-tolerant quantum cryptography with imperfect sources},
  author={Tamaki, Kiyoshi and Curty, Marcos and Kato, Go and Lo, Hoi-Kwong and Azuma, Koji},
  journal={arXiv preprint arXiv:1312.3514},
  year={2013},
  url={https://arxiv.org/abs/1312.3514}
}

@article{pereira2019quantum,
  title={Quantum key distribution with flawed and leaky sources},
  author={Pereira, Margarida and Curty, Marcos and Tamaki, Kiyoshi},
  journal={npj Quantum Information},
  volume={5},
  number={1},
  pages={62},
  year={2019},
  publisher={Nature Publishing Group UK London},
  url={https://www.nature.com/articles/s41534-019-0180-9}
}

@article{pereira2020quantum,
  title={Quantum key distribution with correlated sources},
  author={Pereira, Margarida and Kato, Go and Mizutani, Akihiro and Curty, Marcos and Tamaki, Kiyoshi},
  journal={Science Advances},
  volume={6},
  number={37},
  pages={eaaz4487},
  year={2020},
  publisher={American Association for the Advancement of Science},
  url={https://arxiv.org/abs/1908.08261}
}

@article{juvencio2025digital,
  title={Digital Signal Processing from Classical Coherent Systems to Continuous-Variable QKD: A Review of Cross-Domain Techniques, Applications, and Challenges},
  author={Juv{\^e}ncio Gomes de Sousa, Davi and da Silva Morais Alves, Caroline and Loureiro da Silva, Val{\'e}ria and Alves Ferreira Neto, Nelson},
  journal={arXiv e-prints},
  pages={arXiv--2509},
  year={2025},
  url={https://arxiv.org/abs/2509.20141}
}

@inproceedings{de1937prevision,
  title={La pr{\'e}vision: ses lois logiques, ses sources subjectives},
  author={De Finetti, Bruno},
  booktitle={Annales de l'institut Henri Poincar{\'e}},
  volume={7},
  pages={1--68},
  year={1937},
  url={https://eudml.org/doc/79004}
  }

@article{murta2023lecture,
  title={Lecture notes: Security proofs of Quantum Key Distribution},
  author={Murta, Gl{\'a}ucia},
  year={2023},
  url={https://scholar.google.com/citations?view_op=view_citation&hl=en&user=Jm0mDVgAAAAJ&sortby=pubdate&citation_for_view=Jm0mDVgAAAAJ:IWHjjKOFINEC},
  journal={unpublished lecture notes},
}

@article{arqand2025marginal,
  title={Marginal-constrained entropy accumulation theorem},
  author={Arqand, Amir and Tan, Ernest Y-Z},
  journal={arXiv preprint arXiv:2502.02563},
  year={2025},
  url={https://arxiv.org/abs/2502.02563}
}

@article{maassen1988generalized,
  title={Generalized entropic uncertainty relations},
  author={Maassen, Hans and Uffink, Jos BM},
  journal={Physical Review Letters},
  volume={60},
  number={12},
  pages={1103},
  year={1988},
  publisher={APS},
  url={https://journals.aps.org/prl/abstract/10.1103/PhysRevLett.60.1103}
}

@inproceedings{brandao2013quantum,
  title={Quantum de Finetti theorems under local measurements with applications},
  author={Brandao, Fernando GSL and Harrow, Aram W},
  booktitle={Proceedings of the forty-fifth annual ACM symposium on Theory of computing},
  pages={861--870},
  year={2013},
  url={https://link.springer.com/article/10.1007/s00220-017-2880-3}
}

@article{diaconis1980finite,
  title={Finite exchangeable sequences},
  author={Diaconis, Persi and Freedman, David},
  journal={The Annals of Probability},
  pages={745--764},
  year={1980},
  publisher={JSTOR},
  url={https://projecteuclid.org/journals/annals-of-probability/volume-8/issue-4/Finite-Exchangeable-Sequences/10.1214/aop/1176994663.full}
}

@article{renner2008security,
  title={Security of quantum key distribution},
  author={Renner, Renato},
  journal={International Journal of Quantum Information},
  volume={6},
  number={01},
  pages={1--127},
  year={2008},
  publisher={World Scientific},
  url={https://arxiv.org/abs/quant-ph/0512258}
}

@article{renner2007symmetry,
  title={Symmetry of large physical systems implies independence of subsystems},
  author={Renner, Renato},
  journal={Nature Physics},
  volume={3},
  number={9},
  pages={645--649},
  year={2007},
  publisher={Nature Publishing Group UK London},
  url={}
}

@article{christandl2009postselection,
  title={Postselection technique for quantum channels with applications to quantum cryptography},
  author={Christandl, Matthias and K{\"o}nig, Robert and Renner, Renato},
  journal={Physical Review Letters},
  volume={102},
  number={2},
  pages={020504},
  year={2009},
  publisher={APS},
  url={https://journals.aps.org/prl/abstract/10.1103/PhysRevLett.102.020504}
}

@article{zhang2022device,
  title={A device-independent quantum key distribution system for distant users},
  author={Zhang, Wei and van Leent, Tim and Redeker, Kai and Garthoff, Robert and Schwonnek, Ren{\'e} and Fertig, Florian and Eppelt, Sebastian and Rosenfeld, Wenjamin and Scarani, Valerio and Lim, Charles C-W and others},
  journal={Nature},
  volume={607},
  number={7920},
  pages={687--691},
  year={2022},
  publisher={Nature Publishing Group UK London},
  url={https://www.nature.com/articles/s41586-022-04891-y}
}

@article{liu2022toward,
  title={Toward a photonic demonstration of device-independent quantum key distribution},
  author={Liu, Wen-Zhao and Zhang, Yu-Zhe and Zhen, Yi-Zheng and Li, Ming-Han and Liu, Yang and Fan, Jingyun and Xu, Feihu and Zhang, Qiang and Pan, Jian-Wei},
  journal={Physical Review Letters},
  volume={129},
  number={5},
  pages={050502},
  year={2022},
  publisher={APS},
  url={https://journals.aps.org/prl/abstract/10.1103/PhysRevLett.129.050502}
}

@article{ma2005practical,
  title={Practical decoy state for quantum key distribution},
  author={Ma, Xiongfeng and Qi, Bing and Zhao, Yi and Lo, Hoi-Kwong},
  journal={Physical Review A—Atomic, Molecular, and Optical Physics},
  volume={72},
  number={1},
  pages={012326},
  year={2005},
  publisher={APS},
  url={https://journals.aps.org/pra/abstract/10.1103/PhysRevA.72.012326}
}

@article{dupuis2020entropy,
  title={Entropy accumulation},
  author={Dupuis, Frederic and Fawzi, Omar and Renner, Renato},
  journal={Communications in Mathematical Physics},
  volume={379},
  number={3},
  pages={867--913},
  year={2020},
  publisher={Springer},
  url={https://link.springer.com/article/10.1007/s00220-020-03839-5}
}

@article{dupuis2019entropy,
  title={Entropy accumulation with improved second-order term},
  author={Dupuis, Fr{\'e}d{\'e}ric and Fawzi, Omar},
  journal={IEEE Transactions on information theory},
  volume={65},
  number={11},
  pages={7596--7612},
  year={2019},
  publisher={IEEE},
  url={https://ieeexplore.ieee.org/document/8765829}
}

@article{metger2024generalised,
  title={Generalised entropy accumulation},
  author={Metger, Tony and Fawzi, Omar and Sutter, David and Renner, Renato},
  journal={Communications in Mathematical Physics},
  volume={405},
  number={11},
  pages={261},
  year={2024},
  publisher={Springer},
  url={https://ieeexplore.ieee.org/document/9996821}
}

@article{arnon2018practical,
  title={Practical device-independent quantum cryptography via entropy accumulation},
  author={Arnon-Friedman, Rotem and Dupuis, Fr{\'e}d{\'e}ric and Fawzi, Omar and Renner, Renato and Vidick, Thomas},
  journal={Nature communications},
  volume={9},
  number={1},
  pages={459},
  year={2018},
  publisher={Nature Publishing Group UK London},
  url={https://www.nature.com/articles/s41467-017-02307-4}
}

@article{metger2023security,
  title={Security of quantum key distribution from generalised entropy accumulation},
  author={Metger, Tony and Renner, Renato},
  journal={Nature Communications},
  volume={14},
  number={1},
  pages={5272},
  year={2023},
  publisher={Nature Publishing Group UK London},
  url={https://www.nature.com/articles/s41467-023-40920-8}
}

@article{george2022finite,
  title={Finite-key analysis of quantum key distribution with characterized devices using entropy accumulation},
  author={George, Ian and Lin, Jie and van Himbeeck, Thomas and Fang, Kun and L{\"u}tkenhaus, Norbert},
  journal={arXiv preprint arXiv:2203.06554},
  year={2022},
  url={https://arxiv.org/abs/2203.06554}
}

@article{zapatero2023advances,
  title={Advances in device-independent quantum key distribution},
  author={Zapatero, V{\'\i}ctor and van Leent, Tim and Arnon-Friedman, Rotem and Liu, Wen-Zhao and Zhang, Qiang and Weinfurter, Harald and Curty, Marcos},
  journal={npj quantum information},
  volume={9},
  number={1},
  pages={10},
  year={2023},
  publisher={Nature Publishing Group UK London},
  url={https://www.nature.com/articles/s41534-023-00684-x}
}

@article{schwonnek2021device,
  title={Device-independent quantum key distribution with random key basis},
  author={Schwonnek, Ren{\'e} and Goh, Koon Tong and Primaatmaja, Ignatius W and Tan, Ernest Y-Z and Wolf, Ramona and Scarani, Valerio and Lim, Charles C-W},
  journal={Nature communications},
  volume={12},
  number={1},
  pages={2880},
  year={2021},
  publisher={Nature Publishing Group UK London},
  url={https://www.nature.com/articles/s41467-021-23147-3?fromPaywallRec=false}
}

@book{cover1999elements,
  title={Elements of information theory},
  author={Cover, Thomas M},
  year={1999},
  publisher={John Wiley \& Sons}
}

@article{curras2025numerical,
  title={Numerical security analysis for quantum key distribution with partial state characterization},
  author={Curr{\'a}s-Lorenzo, Guillermo and Navarrete, {\'A}lvaro and N{\'u}{\~n}ez-Bon, Javier and Pereira, Margarida and Curty, Marcos},
  journal={arXiv preprint arXiv:2503.07223},
  year={2025},
  url={https://arxiv.org/abs/2309.06686}
}

@article{trushechkin2022security,
  title={Security of quantum key distribution with detection-efficiency mismatch in the multiphoton case},
  author={Trushechkin, Anton},
  journal={Quantum},
  volume={6},
  pages={771},
  year={2022},
  publisher={Verein zur F{\"o}rderung des Open Access Publizierens in den Quantenwissenschaften},
  url={https://quantum-journal.org/papers/q-2022-07-22-771/}
}

@article{wolf2021quantum,
  title={Quantum key distribution},
  author={Wolf, Ramona},
  journal={Lecture notes in physics},
  volume={988},
  year={2021},
  publisher={Springer}
}

@misc{veeren2024semi,
    title = {Semi-Defined Programming and Applications in Quantum Information},
    year = {2024},
    organization = {Youtube},
    author = {Isadora Veeren},
    url = {https://www.youtube.com/playlist?list=PLqTLz9G2bGs2jH-dutOoFGGIxMRvL0jw2},
    note={lectures ministred at the International Institute of Physics (Natal, Brazil)}
}

@article{gonzales2021device,
  title={Device-independent quantum key distribution based on Bell inequalities with more than two inputs and two outputs},
  author={Gonzales-Ureta, Junior R and Predojevi{\'c}, Ana and Cabello, Ad{\'a}n},
  journal={Physical Review A},
  volume={103},
  number={5},
  pages={052436},
  year={2021},
  publisher={APS},
  url={https://journals.aps.org/pra/abstract/10.1103/PhysRevA.103.052436}
}

@article{gisin2002quantum,
  title={Quantum cryptography},
  author={Gisin, Nicolas and Ribordy, Gr{\'e}goire and Tittel, Wolfgang and Zbinden, Hugo},
  journal={Reviews of modern physics},
  volume={74},
  number={1},
  pages={145},
  year={2002},
  publisher={APS},
  url={https://journals.aps.org/rmp/abstract/10.1103/RevModPhys.74.145}
}

@article{pirandola2020advances,
  title={Advances in quantum cryptography},
  author={Pirandola, Stefano and Andersen, Ulrik L and Banchi, Leonardo and Berta, Mario and Bunandar, Darius and Colbeck, Roger and Englund, Dirk and Gehring, Tobias and Lupo, Cosmo and Ottaviani, Carlo and others},
  journal={Advances in optics and photonics},
  volume={12},
  number={4},
  pages={1012--1236},
  year={2020},
  publisher={Optical Society of America},
  url={https://opg.optica.org/aop/abstract.cfm?uri=aop-12-4-1012}
}

@article{lo2014secure,
  title={Secure quantum key distribution},
  author={Lo, Hoi-Kwong and Curty, Marcos and Tamaki, Kiyoshi},
  journal={Nature Photonics},
  volume={8},
  number={8},
  pages={595--604},
  year={2014},
  publisher={Nature Publishing Group UK London},
  url={https://www.nature.com/articles/nphoton.2014.149}
}

@article{primaatmaja2023security,
  title={Security of device-independent quantum key distribution protocols: a review},
  author={Primaatmaja, Ignatius W and Goh, Koon Tong and Tan, Ernest Y-Z and Khoo, John T-F and Ghorai, Shouvik and Lim, Charles C-W},
  journal={Quantum},
  volume={7},
  pages={932},
  year={2023},
  publisher={Verein zur F{\"o}rderung des Open Access Publizierens in den Quantenwissenschaften}, 
  url={https://quantum-journal.org/papers/q-2023-03-02-932/}
}

@article{grasselli2021quantum,
  title={Quantum cryptography},
  author={Grasselli, Federico},
  journal={Quantum science and technology. Cham: Springer},
  year={2021},
  publisher={Springer},
  url={https://link.springer.com/book/10.1007/978-3-030-64360-7}
}

@article{murta2019towards,
  title={Towards a realization of device-independent quantum key distribution},
  author={Murta, Gl{\'a}ucia and van Dam, Suzanne B and Ribeiro, J{\'e}r{\'e}my and Hanson, Ronald and Wehner, Stephanie},
  journal={Quantum Science and Technology},
  volume={4},
  number={3},
  pages={035011},
  year={2019},
  publisher={IOP Publishing},
  url={https://iopscience.iop.org/article/10.1088/2058-9565/ab2819}
}

@article{werner1989quantum,
  title={Quantum states with Einstein-Podolsky-Rosen correlations admitting a hidden-variable model},
  author={Werner, Reinhard F},
  journal={Physical Review A},
  volume={40},
  number={8},
  pages={4277},
  year={1989},
  publisher={APS},
  url={https://journals.aps.org/pra/abstract/10.1103/PhysRevA.40.4277}
}

@article{van2017photodetector,
  title={Photodetector figures of merit in terms of POVMs},
  author={van Enk, Steven J},
  journal={Journal of Physics Communications},
  volume={1},
  number={4},
  pages={045001},
  year={2017},
  publisher={IOP Publishing},
  url={https://arxiv.org/abs/1705.09640}
}

@article{huttner1995quantum,
  title={Quantum cryptography with coherent states},
  author={Huttner, Bruno and Imoto, Nobuyuki and Gisin, Nicolas and Mor, Tsafrir},
  journal={Physical Review A},
  volume={51},
  number={3},
  pages={1863},
  year={1995},
  publisher={APS},
  url={https://journals.aps.org/pra/abstract/10.1103/PhysRevA.51.1863}
}

@article{slutsky1998security,
  title={Security of quantum cryptography against individual attacks},
  author={Slutsky, Boris A and Rao, Ramesh and Sun, Pang-Chen and Fainman, Yeshaiahu},
  journal={Physical Review A},
  volume={57},
  number={4},
  pages={2383},
  year={1998},
  publisher={APS},
  url={https://journals.aps.org/pra/abstract/10.1103/PhysRevA.57.2383}
}

@article{bennett1988privacy,
  title={Privacy amplification by public discussion},
  author={Bennett, Charles H and Brassard, Gilles and Robert, Jean-Marc},
  journal={SIAM journal on Computing},
  volume={17},
  number={2},
  pages={210--229},
  year={1988},
  publisher={SIAM},
  url={https://epubs.siam.org/doi/10.1137/0217014}
}

@article{araujo2023quantum,
  title={Quantum key distribution rates from semidefinite programming},
  author={Ara{\'u}jo, Mateus and Huber, Marcus and Navascu{\'e}s, Miguel and Pivoluska, Matej and Tavakoli, Armin},
  journal={Quantum},
  volume={7},
  pages={1019},
  year={2023},
  publisher={Verein zur F{\"o}rderung des Open Access Publizierens in den Quantenwissenschaften},
  url={https://quantum-journal.org/papers/q-2023-05-24-1019/}
}

@article{cabello2023logical,
  title={Logical possibilities for physics after MIP*= RE},
  author={Cabello, Ad{\'a}n and Quintino, Marco T{\'u}lio and Kleinmann, Matthias},
  journal={arXiv preprint arXiv:2307.02920},
  year={2023},
  url={https://arxiv.org/abs/2307.02920}
}

@article{wegman1981new,
  title={New hash functions and their use in authentication and set equality},
  author={Wegman, Mark N and Carter, J Lawrence},
  journal={Journal of computer and system sciences},
  volume={22},
  number={3},
  pages={265--279},
  year={1981},
  publisher={Elsevier},
  url={https://www.sciencedirect.com/science/article/pii/0022000081900337}
}

@incollection{nesterov2000squared,
  title={Squared functional systems and optimization problems},
  author={Nesterov, Yurii},
  booktitle={High performance optimization},
  pages={405--440},
  year={2000},
  publisher={Springer},
  url={https://link.springer.com/chapter/10.1007/978-1-4757-3216-0_17}
}

@misc{burniston_2024_14262569,
  author       = {Burniston, John and
                  Wang, Wenyuan and
                  Kamin, Lars and
                  Lin, Jie and
                  Coles, Patrick and
                  Metodiev, Eric and
                  George, Ian and
                  Li, Nicky Kai Hong and
                  Fang, Kun and
                  Chemtov, Max and
                  Zhang, Yanbao and
                  Böhm, Christopher and
                  Winick, Adam and
                  van Himbeeck, Thomas and
                  Johnstun, Scott and
                  Nahar, Shlok and
                  Tupkary, Devashish and
                  Pan, Shihong and
                  Wang, Zhiyao and
                  Corrigan, Aodhan and
                  Kanitschar, Florian and
                  Gracie, Laura and
                  Gu, Shouzhen and
                  Mathur, Natansh and
                  Upadhyaya, Twesh and
                  Lutkenhaus, Norbert},
  title        = {Open QKD Security: Version 2.0.2},
  month        = dec,
  year         = 2024,
  publisher    = {Zenodo},
  version      = {v2.0.2},
  doi          = {10.5281/zenodo.14262569},
  url          = {https://doi.org/10.5281/zenodo.14262569},
  swhid        = {swh:1:dir:ce63165f716a15a425fbadc208e27934cc66be10
                   ;origin=https://doi.org/10.5281/zenodo.14262568;vi
                   sit=swh:1:snp:ea6ece4519d009abf5ae6b7c084f97ba9d3f
                   14c2;anchor=swh:1:rel:54e7860c1d1e613f2e7a075653d1
                   2fa44f1226fa;path=/
                  },
}

@article{kamin2025finite,
  title={Finite-Size Analysis of Prepare-and-Measure and Decoy-State Quantum Key Distribution via Entropy Accumulation},
  author={Kamin, Lars and Arqand, Amir and George, Ian and L{\"u}tkenhaus, Norbert and Tan, Ernest Y-Z},
  journal={PRX Quantum},
  volume={6},
  number={2},
  pages={020342},
  year={2025},
  publisher={APS},
  url={https://journals.aps.org/prxquantum/abstract/10.1103/PRXQuantum.6.020342}
}

@article{masini2024one,
  title={One-sided di-qkd secure against coherent attacks over long distances},
  author={Masini, Michele and Sarkar, Shubhayan},
  journal={arXiv preprint arXiv:2403.11850},
  year={2024},
  url={https://arxiv.org/abs/2403.11850}
}

@article{zhou2022numerical,
  title={Numerical method for finite-size security analysis of quantum key distribution},
  author={Zhou, Hongyi and Sasaki, Toshihiko and Koashi, Masato},
  journal={Physical Review Research},
  volume={4},
  number={3},
  pages={033126},
  year={2022},
  publisher={APS},
  url={https://journals.aps.org/prresearch/abstract/10.1103/PhysRevResearch.4.033126}
}

@article{nahar2023imperfect,
  title={Imperfect phase randomization and generalized decoy-state quantum key distribution},
  author={Nahar, Shlok and Upadhyaya, Twesh and L{\"u}tkenhaus, Norbert},
  journal={Physical Review Applied},
  volume={20},
  number={6},
  pages={064031},
  year={2023},
  publisher={APS},
  url={https://journals.aps.org/prapplied/abstract/10.1103/PhysRevApplied.20.064031}
}

@article{mizutani2023finite,
  title={Finite-key security analysis of differential-phase-shift quantum key distribution},
  author={Mizutani, Akihiro and Takeuchi, Yuki and Tamaki, Kiyoshi},
  journal={Physical Review Research},
  volume={5},
  number={2},
  pages={023132},
  year={2023},
  publisher={APS},
  url={https://journals.aps.org/prresearch/abstract/10.1103/PhysRevResearch.5.023132}
}

@article{sandfuchs2025security,
  title={Security of differential phase shift QKD from relativistic principles},
  author={Sandfuchs, Martin and Haberland, Marcus and Vilasini, Venkatesh and Wolf, Ramona},
  journal={Quantum},
  volume={9},
  pages={1611},
  year={2025},
  publisher={Verein zur F{\"o}rderung des Open Access Publizierens in den Quantenwissenschaften},
  url={https://quantum-journal.org/papers/q-2025-01-27-1611/}
}

@article{mizutani2019quantum,
    author={Mizutani, Akihiro
            and Sasaki, Toshihiko
            and Takeuchi, Yuki
            and Tamaki, Kiyoshi
            and Koashi, Masato},
    title={Quantum key distribution with simply characterized light sources},
    journal={npj Quantum Information},
    year={2019},
    month={Oct},
    day={11},
    volume={5},
    number={1},
    pages={87},
    issn={2056-6387},
    doi={10.1038/s41534-019-0194-3},
    url={https://doi.org/10.1038/s41534-019-0194-3}
}

@article{koashi2003secure,
  title={Secure quantum key distribution with an uncharacterized source},
  author={Koashi, Masato and Preskill, John},
  journal={Physical Review Letters},
  volume={90},
  number={5},
  pages={057902},
  year={2003},
  publisher={APS},
  url={https://journals.aps.org/prl/abstract/10.1103/PhysRevLett.90.057902}
}

@article{rivera2025device,
  title={Device-independent quantum key distribution beyond qubits},
  author={Rivera-Dean, Javier and Steffinlongo, Anna and Parker-S{\'a}nchez, Neil and Ac{\'\i}n, Antonio and Oudot, Enky},
  journal={New Journal of Physics},
  volume={27},
  number={5},
  pages={054512},
  year={2025},
  publisher={IOP Publishing},
  url={https://arxiv.org/abs/2402.00161}
}

@article{mizutani2020quantum,
  title={Quantum key distribution with any two independent and identically distributed states},
  author={Mizutani, Akihiro},
  journal={Physical Review A},
  volume={102},
  number={2},
  pages={022613},
  year={2020},
  publisher={APS},
  url={https://journals.aps.org/pra/abstract/10.1103/PhysRevA.102.022613}
}

@book{holevo2019quantum,
  title={Quantum systems, channels, information: a mathematical introduction},
  author={Holevo, Alexander S},
  year={2019},
  publisher={Walter de Gruyter GmbH \& Co KG}
}

@article{chaturvedi2024extending,
  title={Extending loophole-free nonlocal correlations to arbitrarily large distances},
  author={Chaturvedi, Anubhav and Viola, Giuseppe and Paw{\l}owski, Marcin},
  journal={npj Quantum Information},
  volume={10},
  number={1},
  pages={7},
  year={2024},
  publisher={Nature Publishing Group UK London},
  url={https://www.nature.com/articles/s41534-023-00799-1}
}

@article{le2025device,
  title={Device-independent quantum key distribution based on routed Bell tests},
  author={Le Roy-Deloison, Tristan and Lobo, Edwin Peter and Pauwels, Jef and Pironio, Stefano},
  journal={PRX Quantum},
  volume={6},
  number={2},
  pages={020311},
  year={2025},
  publisher={APS},
  url={https://arxiv.org/abs/2404.01202}
}

@article{kossmann2024optimising,
  title={Optimising the relative entropy under semidefinite constraints},
  author={Ko{\ss}mann, Gereon and Schwonnek, Ren{\'e}},
  journal={arXiv preprint arXiv:2404.17016},
  year={2024},
  url={https://www.nature.com/articles/s41534-026-01184-4}
}

@article{kossmann2024reliable,
  title={Reliable Entropy Estimation from Observed Statistics for Device-Independent Quantum Cryptography},
  author={Ko{\ss}mann, Gereon and Schwonnek, Ren{\'e}},
  journal={arXiv preprint arXiv:2411.04858},
  year={2024},
  url={https://arxiv.org/abs/2411.04858}
}

@article{frenkel2023integral,
  title={Integral formula for quantum relative entropy implies data processing inequality},
  author={Frenkel, P{\'e}ter E},
  journal={Quantum},
  volume={7},
  pages={1102},
  year={2023},
  publisher={Verein zur F{\"o}rderung des Open Access Publizierens in den Quantenwissenschaften},
  url={https://quantum-journal.org/papers/q-2023-09-07-1102/}
}

@article{he2024qics,
  title={QICS: Quantum information conic solver},
  author={He, Kerry and Saunderson, James and Fawzi, Hamza},
  journal={arXiv preprint arXiv:2410.17803},
  year={2024},
  url={https://arxiv.org/abs/2410.17803}
}

@article{kossmann2025routed,
  title={Routed Bell tests with arbitrarily many local parties},
  author={Ko{\ss}mann, Gereon and Berta, Mario and Schwonnek, Ren{\'e}},
  journal={arXiv preprint arXiv:2510.08405},
  year={2025},
  url={https://arxiv.org/abs/2510.08405}
}

@article{bruss1998optimal,
  title={Optimal eavesdropping in quantum cryptography with six states},
  author={Bru{\ss}, Dagmar},
  journal={Physical Review Letters},
  volume={81},
  number={14},
  pages={3018},
  year={1998},
  publisher={APS}
}

@article{chung2025generalized,
  title={Generalized numerical framework for improved finite-sized key rates with R{\'e}nyi entropy},
  author={Chung, Rebecca RB and Ng, Nelly HY and Cai, Yu},
  journal={Physical Review A},
  volume={112},
  number={1},
  pages={012612},
  year={2025},
  publisher={APS},
  url={https://arxiv.org/abs/2502.02319}
}

@book{davis2007methods,
  title={Methods of numerical integration},
  author={Davis, Philip J and Rabinowitz, Philip},
  year={2007},
  publisher={Courier Corporation},
  url={}
}

@article{buhrman2009non,
  title={Non-locality and communication complexity},
  author={Buhrman, Harry and Cleve, Richard and Massar, Serge and De Wolf, Ronald},
  journal={arXiv preprint arXiv:0907.3584},
  year={2009},
  url={https://journals.aps.org/rmp/abstract/10.1103/RevModPhys.82.665}
}

@article{matsuura2021finite,
  title={Finite-size security of continuous-variable quantum key distribution with digital signal processing},
  author={Matsuura, Takaya and Maeda, Kento and Sasaki, Toshihiko and Koashi, Masato},
  journal={Nature communications},
  volume={12},
  number={1},
  pages={252},
  year={2021},
  publisher={Nature Publishing Group UK London},
  url={https://www.nature.com/articles/s41467-020-19916-1}
}

@article{tupkary2025qkd,
  title={QKD security proofs for decoy-state BB84: protocol variations, proof techniques, gaps and limitations},
  author={Tupkary, Devashish and Tan, Ernest Y-Z and Nahar, Shlok and Kamin, Lars and L{\"u}tkenhaus, Norbert},
  journal={arXiv preprint arXiv:2502.10340},
  year={2025},
  url={https://arxiv.org/abs/2502.10340}
}

@article{hu2022robust,
  title={Robust interior point method for quantum key distribution rate computation},
  author={Hu, Hao and Im, Jiyoung and Lin, Jie and L{\"u}tkenhaus, Norbert and Wolkowicz, Henry},
  journal={Quantum},
  volume={6},
  pages={792},
  year={2022},
  publisher={Verein zur F{\"o}rderung des Open Access Publizierens in den Quantenwissenschaften},
  url={https://quantum-journal.org/papers/q-2022-09-08-792/}
}

@article{lubin2023jump,
  title={JuMP 1.0: Recent improvements to a modeling language for mathematical optimization},
  author={Lubin, Miles and Dowson, Oscar and Garcia, Joaquim Dias and Huchette, Joey and Legat, Beno{\^\i}t and Vielma, Juan Pablo},
  journal={Mathematical Programming Computation},
  volume={15},
  number={3},
  pages={581--589},
  year={2023},
  publisher={Springer},
  url={https://link.springer.com/article/10.1007/s12532-023-00239-3}
}

@inproceedings{lofberg2004yalmip,
  title={YALMIP: A toolbox for modeling and optimization in MATLAB},
  author={Lofberg, Johan},
  booktitle={2004 IEEE international conference on robotics and automation (IEEE Cat. No. 04CH37508)},
  pages={284--289},
  year={2004},
  organization={IEEE},
  url={https://ieeexplore.ieee.org/document/1393890}
}

@article{ghoreishi2025future,
  title={The future of secure communications: device independence in quantum key distribution},
  author={Ghoreishi, Seyed Arash and Scala, Giovanni and Renner, Renato and Tacca, Let{\u{A}}cia Lira and Bouda, Jan and Walborn, Stephen Patrick and others},
  journal={arXiv preprint arXiv:2504.06350},
  year={2025},
  url={https://arxiv.org/abs/2504.06350}
}

@article{portmann2022security,
  title={Security in quantum cryptography},
  author={Portmann, Christopher and Renner, Renato},
  journal={Reviews of Modern Physics},
  volume={94},
  number={2},
  pages={025008},
  year={2022},
  publisher={APS},
  url={https://journals.aps.org/rmp/abstract/10.1103/RevModPhys.94.025008}
}

@article{portmann2014cryptographic,
  title={Cryptographic security of quantum key distribution},
  author={Portmann, Christopher and Renner, Renato},
  journal={arXiv preprint arXiv:1409.3525},
  year={2014},
  url={https://arxiv.org/abs/1409.3525}
}

@inproceedings{carter1977universal,
  title={Universal classes of hash functions},
  author={Carter, J Lawrence and Wegman, Mark N},
  booktitle={Proceedings of the ninth annual ACM symposium on Theory of computing},
  pages={106--112},
  year={1977},
  url={https://www.sciencedirect.com/science/article/pii/0022000079900448}
}

@mastersthesis{li2020application,
  title={Application of the flag-state squashing model to numerical quantum key distribution security analysis},
  author={Li, Nicky Kai Hong},
  year={2020},
  publisher={University of Waterloo},
  school={University of Waterloo},
  url={https://uwspace.uwaterloo.ca/items/0848a639-2d47-4a40-854b-94e103334561}
}

@article{arslan2025device,
  title={Device-Independent Quantum Key Distribution: Protocols, Quantum Games, and Security},
  author={Arslan, Syed M and Al-Kuwari, Saif and Rahim, MT and Kuniyal, Hashir},
  journal={arXiv preprint arXiv:2505.14243},
  year={2025},
  url={https://arxiv.org/abs/2505.14243}
}

@article{navascues2014characterization,
  title={Characterization of quantum correlations with local dimension constraints and its device-independent applications},
  author={Navascu{\'e}s, Miguel and de la Torre, Gonzalo and V{\'e}rtesi, Tam{\'a}s},
  journal={Physical Review X},
  volume={4},
  number={1},
  pages={011011},
  year={2014},
  publisher={APS},
  url={https://journals.aps.org/prx/abstract/10.1103/PhysRevX.4.011011}
}

@article{scholz2008tsirelson,
  title={Tsirelson's problem},
  author={Scholz, Volkher B and Werner, Reinhard F},
  journal={arXiv preprint arXiv:0812.4305},
  year={2008},
  url={https://arxiv.org/abs/0812.4305}
}

@article{konig2009operational,
  title={The operational meaning of min-and max-entropy},
  author={Konig, Robert and Renner, Renato and Schaffner, Christian},
  journal={IEEE Transactions on Information theory},
  volume={55},
  number={9},
  pages={4337--4347},
  year={2009},
  publisher={IEEE},
  url={https://ieeexplore.ieee.org/document/5208530}
}

@article{coles2017entropic,
  title={Entropic uncertainty relations and their applications},
  author={Coles, Patrick J and Berta, Mario and Tomamichel, Marco and Wehner, Stephanie},
  journal={Reviews of Modern Physics},
  volume={89},
  number={1},
  pages={015002},
  year={2017},
  publisher={APS},
  url={https://journals.aps.org/rmp/abstract/10.1103/RevModPhys.89.015002}
}

@article{arnon2019simple,
  title={Simple and tight device-independent security proofs},
  author={Arnon-Friedman, Rotem and Renner, Renato and Vidick, Thomas},
  journal={SIAM Journal on Computing},
  volume={48},
  number={1},
  pages={181--225},
  year={2019},
  publisher={SIAM},
  url={https://epubs.siam.org/doi/10.1137/18M1174726}
}

@article{xu2020secure,
  title={Secure quantum key distribution with realistic devices},
  author={Xu, Feihu and Ma, Xiongfeng and Zhang, Qiang and Lo, Hoi-Kwong and Pan, Jian-Wei},
  journal={Reviews of modern physics},
  volume={92},
  number={2},
  pages={025002},
  year={2020},
  publisher={APS},
  url={https://journals.aps.org/rmp/abstract/10.1103/RevModPhys.92.025002}
}

@book{boyd2004convex,
  title={Convex optimization},
  author={Boyd, Stephen P and Vandenberghe, Lieven},
  year={2004},
  publisher={Cambridge university press},
  url={https://stanford.edu/~boyd/cvxbook}
}

@article{mukherjee2021role,
  title={Role of Steering Inequality In Quantum Key Distribution Protocol},
  author={Mukherjee, Kaushiki and Patro, Tapaswini and Ganguly, Nirman},
  journal={arXiv preprint arXiv:2106.12759},
  year={2021},
  url={https://dankogeorgiev.com/ojs/index.php/quanta/article/view/74}
}

@article{wiseman2007steering,
  title={Steering, entanglement, nonlocality, and the Einstein-Podolsky-Rosen paradox},
  author={Wiseman, Howard M and Jones, Steve James and Doherty, Andrew C},
  journal={Physical Review Letters},
  volume={98},
  number={14},
  pages={140402},
  year={2007},
  publisher={APS},
  url={https://journals.aps.org/prl/abstract/10.1103/PhysRevLett.98.140402}
}

@article{fine1982hidden,
  title={Hidden variables, joint probability, and the Bell inequalities},
  author={Fine, Arthur},
  journal={Physical Review Letters},
  volume={48},
  number={5},
  pages={291},
  year={1982},
  publisher={APS},
  url={https://journals.aps.org/prl/abstract/10.1103/PhysRevLett.48.291}
}

@article{curty2004entanglement,
  title={Entanglement as a precondition for secure quantum key distribution},
  author={Curty, Marcos and Lewenstein, Maciej and L{\"u}tkenhaus, Norbert},
  journal={Physical Review Letters},
  volume={92},
  number={21},
  pages={217903},
  year={2004},
  publisher={APS},
  url={https://arxiv.org/abs/quant-ph/0307151}
}

@article{navascues2015bounding,
  title={Bounding the set of finite dimensional quantum correlations},
  author={Navascu{\'e}s, Miguel and V{\'e}rtesi, Tam{\'a}s},
  journal={Physical Review Letters},
  volume={115},
  number={2},
  pages={020501},
  year={2015},
  publisher={APS},
  url={https://journals.aps.org/prl/abstract/10.1103/PhysRevLett.115.020501}
}

@article{lobo2023certifying,
  title={Certifying long-range quantum correlations through routed Bell tests},
  author={Lobo, Edwin Peter and Pauwels, Jef and Pironio, Stefano},
  journal={arXiv preprint arXiv:2310.07484},
  year={2023},
  url={https://quantum-journal.org/papers/q-2024-05-02-1332}
}

@article{wang2013finite,
  title={Finite-key analysis for one-sided device-independent quantum key distribution},
  author={Wang, Yang and Bao, Wan-su and Li, Hong-wei and Zhou, Chun and Li, Yuan},
  journal={Physical Review A—Atomic, Molecular, and Optical Physics},
  volume={88},
  number={5},
  pages={052322},
  year={2013},
  publisher={APS},
  url={https://journals.aps.org/pra/abstract/10.1103/PhysRevA.88.052322}
}

@article{du2024advantage,
  title={Advantage distillation for quantum key distribution},
  author={Du, Zhenyu and Liu, Guoding and Zhang, Xingjian and Ma, Xiongfeng},
  journal={Quantum Science and Technology},
  volume={10},
  number={1},
  pages={015050},
  year={2024},
  publisher={IOP Publishing},
  url={https://iopscience.iop.org/article/10.1088/2058-9565/ad9d75}
}

@article{tupkary2026authentication,
      title={Authentication in Security Proofs for Quantum Key Distribution}, 
      author={Devashish Tupkary and Shlok Nahar and Ernest Y. -Z. Tan},
      year={2026},
      eprint={2601.17960},
      archivePrefix={arXiv},
      primaryClass={quant-ph},
      journal={arxiv},
      url={https://arxiv.org/abs/2601.17960}
}

@article{xin2020one,
  title={One-sided device-independent quantum key distribution for two independent parties},
  author={Xin, Jun and Lu, Xiao-Ming and Li, Xingmin and Li, Guolong},
  journal={Optics Express},
  volume={28},
  number={8},
  pages={11439--11450},
  year={2020},
  publisher={Optical Society of America},
  url={https://opg.optica.org/oe/fulltext.cfm?uri=oe-28-8-11439}
}

@inproceedings{tomamichel2013one,
  title={One-sided device-independent QKD and position-based cryptography from monogamy games},
  author={Tomamichel, Marco and Fehr, Serge and Kaniewski, J{\k{e}}drzej and Wehner, Stephanie},
  booktitle={Annual International Conference on the Theory and Applications of Cryptographic Techniques},
  pages={609--625},
  year={2013},
  organization={Springer},
  url={https://link.springer.com/chapter/10.1007/978-3-642-38348-9_36}
}

@article{hoeffding1963probability,
  title={Probability inequalities for sums of bounded random variables},
  author={Hoeffding, Wassily},
  journal={Journal of the American statistical association},
  volume={58},
  number={301},
  pages={13--30},
  year={1963},
  publisher={Taylor \& Francis},
  url={https://www.jstor.org/stable/2282952?seq=1}
}

@article{jain2014trojan,
  title={Trojan-horse attacks threaten the security of practical quantum cryptography},
  author={Jain, Nitin and Anisimova, Elena and Khan, Imran and Makarov, Vadim and Marquardt, Christoph and Leuchs, Gerd},
  journal={New Journal of Physics},
  volume={16},
  number={12},
  pages={123030},
  year={2014},
  publisher={IOP Publishing},
  url={https://iopscience.iop.org/article/10.1088/1367-2630/16/12/123030/meta}
}

@article{jain2014risk,
  title={Risk analysis of Trojan-horse attacks on practical quantum key distribution systems},
  author={Jain, Nitin and Stiller, Birgit and Khan, Imran and Makarov, Vadim and Marquardt, Christoph and Leuchs, Gerd},
  journal={IEEE Journal of Selected Topics in Quantum Electronics},
  volume={21},
  number={3},
  pages={168--177},
  year={2014},
  publisher={IEEE},
url={https://ieeexplore.ieee.org/document/6948230/}
}

@article{bugge2014laser,
  title={Laser damage helps the eavesdropper in quantum cryptography},
  author={Bugge, Audun Nystad and Sauge, Sebastien and Ghazali, Aina Mardhiyah M and Skaar, Johannes and Lydersen, Lars and Makarov, Vadim},
  journal={Physical Review Letters},
  volume={112},
  number={7},
  pages={070503},
  year={2014},
  publisher={APS},
  url={https://journals.aps.org/prl/abstract/10.1103/PhysRevLett.112.070503}
}

@article{drusvyatskiy2017many,
  title={The many faces of degeneracy in conic optimization},
  author={Drusvyatskiy, Dmitriy and Wolkowicz, Henry and others},
  journal={Foundations and Trends{\textregistered} in Optimization},
  volume={3},
  number={2},
  pages={77--170},
  year={2017},
  publisher={Now Publishers, Inc.},
  url={https://arxiv.org/abs/1706.03705}
}

@article{araujo2023comment,
  title={Comment on “Geometry of the quantum set on no-signaling faces”},
  author={Ara{\'u}jo, Mateus},
  journal={Physical Review A},
  volume={107},
  number={3},
  pages={036201},
  year={2023},
  publisher={APS},
  url={https://arxiv.org/abs/2302.03529}
}

@article{tupkary2024security,
  title={Security proof for variable-length quantum key distribution},
  author={Tupkary, Devashish and Tan, Ernest Y-Z and L{\"u}tkenhaus, Norbert},
  journal={Physical Review Research},
  volume={6},
  number={2},
  pages={023002},
  year={2024},
  publisher={APS},
  url={https://journals.aps.org/prresearch/abstract/10.1103/PhysRevResearch.6.023002}
}

@inproceedings{ben2005universal,
  title={The universal composable security of quantum key distribution},
  author={Ben-Or, Michael and Horodecki, Micha{\l} and Leung, Debbie W and Mayers, Dominic and Oppenheim, Jonathan},
  booktitle={Theory of Cryptography Conference},
  pages={386--406},
  year={2005},
  organization={Springer},
  url={https://arxiv.org/abs/quant-ph/0409078}
}

@article{hayashi2012concise,
  title={Concise and tight security analysis of the Bennett--Brassard 1984 protocol with finite key lengths},
  author={Hayashi, Masahito and Tsurumaru, Toyohiro},
  journal={New Journal of Physics},
  volume={14},
  number={9},
  pages={093014},
  year={2012},
  publisher={IOP Publishing},
  url={https://iopscience.iop.org/article/10.1088/1367-2630/14/9/093014}
}

@article{pironio2009device,
  title={Device-independent quantum key distribution secure against collective attacks},
  author={Pironio, Stefano and Ac{\'\i}n, Antonio and Brunner, Nicolas and Gisin, Nicolas and Massar, Serge and Scarani, Valerio},
  journal={New Journal of Physics},
  volume={11},
  number={4},
  pages={045021},
  year={2009},
  publisher={IOP Publishing},
  url={https://iopscience.iop.org/article/10.1088/1367-2630/11/4/045021}
}

@inproceedings{sasaki2015key,
  title={Key rate of the B92 quantum key distribution protocol with finite qubits},
  author={Sasaki, Hiroaki and Matsumoto, Ryutaroh and Uyematsu, Tomohiko},
  booktitle={2015 IEEE International Symposium on Information Theory (ISIT)},
  pages={696--699},
  year={2015},
  organization={IEEE},
  url={https://ieeexplore.ieee.org/document/7282544}
}

@article{sano2010secure,
  title={Secure key rate of the BB84 protocol using finite sample bits},
  author={Sano, Yousuke and Matsumoto, Ryutaroh and Uyematsu, Tomohiko},
  journal={Journal of Physics A: Mathematical and Theoretical},
  volume={43},
  number={49},
  pages={495302},
  year={2010},
  publisher={IOP Publishing},
  url={https://iopscience.iop.org/article/10.1088/1751-8113/43/49/495302}
}

@article{bunandar2020numerical,
  title={Numerical finite-key analysis of quantum key distribution},
  author={Bunandar, Darius and Govia, Luke CG and Krovi, Hari and Englund, Dirk},
  journal={npj Quantum Information},
  volume={6},
  number={1},
  pages={104},
  year={2020},
  publisher={Nature Publishing Group UK London},
  url={https://www.nature.com/articles/s41534-020-00322-w}
}

@article{clauser1969proposed,
  title={Proposed experiment to test local hidden-variable theories},
  author={Clauser, John F and Horne, Michael A and Shimony, Abner and Holt, Richard A},
  journal={Physical Review Letters},
  volume={23},
  number={15},
  pages={880},
  year={1969},
  publisher={APS},
  url={https://journals.aps.org/prl/abstract/10.1103/PhysRevLett.23.880}
}

@article{gerhardt2011experimentally,
  title={Experimentally faking the violation of Bell’s inequalities},
  author={Gerhardt, Ilja and Liu, Qin and Lamas-Linares, Ant{\'\i}a and Skaar, Johannes and Scarani, Valerio and Makarov, Vadim and Kurtsiefer, Christian},
  journal={Physical Review Letters},
  volume={107},
  number={17},
  pages={170404},
  year={2011},
  publisher={APS},
  url={https://journals.aps.org/prl/abstract/10.1103/PhysRevLett.107.170404}
}

@article{gehring2015implementation,
  title={Implementation of continuous-variable quantum key distribution with composable and one-sided-device-independent security against coherent attacks},
  author={Gehring, Tobias and H{\"a}ndchen, Vitus and Duhme, J{\"o}rg and Furrer, Fabian and Franz, Torsten and Pacher, Christoph and Werner, Reinhard F and Schnabel, Roman},
  journal={Nature communications},
  volume={6},
  number={1},
  pages={8795},
  year={2015},
  publisher={Nature Publishing Group UK London},
  url={https://www.nature.com/articles/ncomms9795}
}

@article{sun2024enhancing,
  title={Enhancing Quantum Key Distribution with Entanglement Distillation and Classical Advantage Distillation},
  author={Sun, Shin and Goodenough, Kenneth and Bhatti, Daniel and Elkouss, David},
  journal={arXiv preprint arXiv:2410.19334},
  year={2024},
  url={https://arxiv.org/abs/2410.19334}
}

@article{abushgra2022variations,
  title={Variations of QKD protocols based on conventional system measurements: A literature review},
  author={Abushgra, Abdulbast A},
  journal={Cryptography},
  volume={6},
  number={1},
  pages={12},
  year={2022},
  publisher={MDPI},
  url={https://www.mdpi.com/2410-387X/6/1/12}
}

@article{tupkary2026rigorous,
      title={A rigorous and complete security proof of decoy-state BB84 quantum key distribution}, 
      author={Devashish Tupkary and Shlok Nahar and Amir Arqand and Ernest Y. -Z. Tan and Norbert Lütkenhaus},
      year={2026},
      journal={arxiv},
      eprint={2601.18035},
      archivePrefix={arXiv},
      primaryClass={quant-ph},
      url={https://arxiv.org/abs/2601.18035}
}

@article{scarani2008quantum,
  title={Quantum Cryptography with Finite Resources: Unconditional Security Bound for Discrete-Variable Protocols with One-Way Postprocessing},
  author={Scarani, Valerio and Renner, Renato},
  journal={Physical Review Letters},
  volume={100},
  number={20},
  pages={200501},
  year={2008},
  publisher={APS},
  url={https://journals.aps.org/prl/abstract/10.1103/PhysRevLett.100.200501}
}

@article{rosset2018resource,
  title={Resource theory of quantum memories and their faithful verification with minimal assumptions},
  author={Rosset, Denis and Buscemi, Francesco and Liang, Yeong-Cherng},
  journal={Physical Review X},
  volume={8},
  number={2},
  pages={021033},
  year={2018},
  publisher={APS},
  url={https://journals.aps.org/prx/abstract/10.1103/PhysRevX.8.021033}
}

@article{Pascual_Garc_a_2025,
   title={Improved finite-size key rates for discrete-modulated continuous-variable quantum key distribution under coherent attacks},
   volume={111},
   ISSN={2469-9934},
   url={http://dx.doi.org/10.1103/PhysRevA.111.022610},
   number={2},
   journal={Physical Review A},
   publisher={American Physical Society (APS)},
   author={Pascual-García, Carlos and Bäuml, Stefan and Araújo, Mateus and Liss, Rotem and Acín, Antonio},
   year={2025},
   url={https://journals.aps.org/pra/abstract/10.1103/PhysRevA.111.022610}
}

@article{sena2025tutorial,
  title={Um tutorial sobre Distribui{\c{c}}{\~a}o Qu{\^a}ntica de Chaves: dos fundamentos {\`a}s tecnologias modernas},
  author={Sena, Vitor L and de Melo, Fernando and Dias, Micael A and Tacla, Alexandre B and Chaves, Rafael},
  journal={Revista Brasileira de Ensino de F{\'\i}sica},
  volume={47},
  pages={e20250373},
  year={2025},
  publisher={SciELO Brasil},
  url={https://www.scielo.br/j/rbef/a/WDyd8NmMJPVYNqRSfhT643b/?lang=pt}
}

@article{vsupic2020self,
  title={Self-testing of quantum systems: a review},
  author={{\v{S}}upi{\'c}, Ivan and Bowles, Joseph},
  journal={Quantum},
  volume={4},
  pages={337},
  year={2020},
  publisher={Verein zur F{\"o}rderung des Open Access Publizierens in den Quantenwissenschaften},
  url={https://quantum-journal.org/papers/q-2020-09-30-337/}
}

@article{tsirelson1993some,
  title={Some results and problems on quantum Bell-type inequalities},
  author={Tsirelson, Boris S},
  journal={Hadronic Journal Supplement},
  volume={8},
  number={4},
  pages={329--345},
  year={1993},
  url={https://www.semanticscholar.org/paper/Some-results-and-problems-on-quan-tum-Bell-type-Tsirelson/214ee464d660139732a07239db8cf67d16cce038}
}

@article{bratzik2011min,
  title={Min-entropy and quantum key distribution: Nonzero key rates for “small” numbers of signals},
  author={Bratzik, Sylvia and Mertz, Markus and Kampermann, Hermann and Bru{\ss}, Dagmar},
  journal={Physical Review A—Atomic, Molecular, and Optical Physics},
  volume={83},
  number={2},
  pages={022330},
  year={2011},
  publisher={APS},
  URL={https://arxiv.org/abs/1011.1190}
}

@article{primaatmaja2024discrete,
  title={Discrete-modulated continuous-variable quantum key distribution secure against general attacks},
  author={Primaatmaja, Ignatius William and Kon, Wen Yu and Lim, Charles},
  journal={arXiv preprint arXiv:2409.02630},
  year={2024},
  url={https://arxiv.org/abs/2409.02630}
}

@article{christandl2004generic,
  title={A generic security proof for quantum key distribution},
  author={Christandl, Matthias and Renner, Renato and Ekert, Artur},
  journal={arXiv preprint quant-ph/0402131},
  year={2004},
  url={https://arxiv.org/abs/quant-ph/0402131}
}

@article{cai2009finite,
  title={Finite-key analysis for practical implementations of quantum key distribution},
  author={Cai, Raymond YQ and Scarani, Valerio},
  journal={New Journal of Physics},
  volume={11},
  number={4},
  pages={045024},
  year={2009},
  publisher={IOP Publishing},
  url={https://iopscience.iop.org/article/10.1088/1367-2630/11/4/045024}
}

@article{wang2025advances,
  title={Advances in continuous variable measurement-device-independent quantum key distribution},
  author={Wang, Pu and Tian, Yan and Li, Yongmin},
  journal={Science China Information Sciences},
  volume={68},
  number={8},
  pages={180501},
  year={2025},
  publisher={Springer},
  url={https://arxiv.org/abs/2502.16448}
}

@article{chou2023satellite,
  title={Satellite-based Quantum Network: Security and Challenges over Atmospheric Channel},
  author={Chou, Hong-fu and Ha, Vu Nguyen and Al-Hraishawi, Hayder and Garces-Socarras, Luis Manuel and Gonzalez-Rios, Jorge Luis and Merlano-Duncan, Juan Carlos and Chatzinotas, Symeon},
  journal={arXiv preprint arXiv:2308.00011},
  year={2023},
  url={https://arxiv.org/abs/2308.00011}
}

@article{kanitschar2024practical,
  title={A practical framework for analyzing high-dimensional QKD setups},
  author={Kanitschar, Florian and Huber, Marcus},
  journal={arXiv preprint arXiv:2406.08544},
  year={2024},
  url={https://arxiv.org/abs/2406.08544}
}

@article{tomamichel2011uncertainty,
  title={Uncertainty relation for smooth entropies},
  author={Tomamichel, Marco and Renner, Renato},
  journal={Physical Review Letters},
  volume={106},
  number={11},
  pages={110506},
  year={2011},
  publisher={APS},
  url={https://journals.aps.org/prl/abstract/10.1103/PhysRevLett.106.110506}
}

@article{arqand2024generalized,
  title={Generalized R$\backslash$'enyi entropy accumulation theorem and generalized quantum probability estimation},
  author={Arqand, Amir and Hahn, Thomas A and Tan, Ernest Y-Z},
  journal={arXiv preprint arXiv:2405.05912},
  year={2024},
  url={https://arxiv.org/abs/2405.05912}
}

@article{heisenberg1927anschaulichen,
  title={{\"U}ber den anschaulichen Inhalt der quantentheoretischen Kinematik und Mechanik},
  author={Heisenberg, Werner},
  journal={Zeitschrift f{\"u}r Physik},
  volume={43},
  number={3},
  pages={172--198},
  year={1927},
  publisher={Springer},
  url={https://link.springer.com/article/10.1007/BF01397280}
}

@article{lydersen2010hacking,
  title={Hacking commercial quantum cryptography systems by tailored bright illumination},
  author={Lydersen, Lars and Wiechers, Carlos and Wittmann, Christoffer and Elser, Dominique and Skaar, Johannes and Makarov, Vadim},
  journal={Nature photonics},
  volume={4},
  number={10},
  pages={686--689},
  year={2010},
  publisher={Nature Publishing Group UK London},
url={https://www.nature.com/articles/nphoton.2010.214}
}

@article{tomamichel2009fully,
  title={A fully quantum asymptotic equipartition property},
  author={Tomamichel, Marco and Colbeck, Roger and Renner, Renato},
  journal={IEEE Transactions on information theory},
  volume={55},
  number={12},
  pages={5840--5847},
  year={2009},
  publisher={IEEE},
url={https://arxiv.org/abs/0811.1221}
}

@inproceedings{renner2005simple,
  title={Simple and tight bounds for information reconciliation and privacy amplification},
  author={Renner, Renato and Wolf, Stefan},
  booktitle={International conference on the theory and application of cryptology and information security},
  pages={199--216},
  year={2005},
  organization={Springer},
  url={https://link.springer.com/chapter/10.1007/11593447_11}
}

@book{nielsen2010quantum,
  title={Quantum computation and quantum information},
  author={Nielsen, Michael A and Chuang, Isaac L},
  year={2010},
  publisher={Cambridge university press}
}

@book{tomamichel2015quantum,
  title={Quantum information processing with finite resources: mathematical foundations},
  author={Tomamichel, Marco},
  volume={5},
  year={2015},
  publisher={Springer},
  url={https://arxiv.org/abs/1504.00233}
}

@article{ma2016quantum,
  title={Quantum random number generation},
  author={Ma, Xiongfeng and Yuan, Xiao and Cao, Zhu and Qi, Bing and Zhang, Zhen},
  journal={npj Quantum Information},
  volume={2},
  number={1},
  pages={1--9},
  year={2016},
  publisher={Nature Publishing Group},
  url={https://www.nature.com/articles/npjqi201621}
}

@article{einstein1935can,
  title={Can quantum-mechanical description of physical reality be considered complete?},
  author={Einstein, Albert and Podolsky, Boris and Rosen, Nathan},
  journal={Physical review},
  volume={47},
  number={10},
  pages={777},
  year={1935},
  publisher={APS},
  url={https://journals.aps.org/pr/abstract/10.1103/PhysRev.47.777}
}

@article{bell1964einstein,
  title={On the Einstein Podolsky Rosen paradox},
  author={Bell, John S},
  journal={Physics Physique Fizika},
  volume={1},
  number={3},
  pages={195},
  year={1964},
  publisher={APS},
  url={https://journals.aps.org/ppf/abstract/10.1103/PhysicsPhysiqueFizika.1.195}
}

@article{li2014space,
  title={Space-bound optical source for satellite-ground decoy-state quantum key distribution},
  author={Li, Yang and Liao, Sheng-Kai and Chen, Xie-Le and Chen, Wei and Cheng, Kun and Cao, Yuan and Yong, Hai-Lin and Wang, Tao and Yang, Hua-Qiang and Liu, Wei-Yue and others},
  journal={Optics express},
  volume={22},
  number={22},
  pages={27281--27289},
  year={2014},
  publisher={Optical Society of America},
  url={https://opg.optica.org/oe/fulltext.cfm?uri=oe-22-22-27281}
}

@article{liao2017satellite,
  title={Satellite-to-ground quantum key distribution},
  author={Liao, Sheng-Kai and Cai, Wen-Qi and Liu, Wei-Yue and Zhang, Liang and Li, Yang and Ren, Ji-Gang and Yin, Juan and Shen, Qi and Cao, Yuan and Li, Zheng-Ping and others},
  journal={Nature},
  volume={549},
  number={7670},
  pages={43--47},
  year={2017},
  publisher={Nature Publishing Group UK London},
  url={https://www.nature.com/articles/nature23655}
}

@article{hu2021decoy,
  title={Decoy-state quantum key distribution over a long-distance high-loss air-water channel},
  author={Hu, Cheng-Qiu and Yan, Zeng-Quan and Gao, Jun and Li, Zhan-Ming and Zhou, Heng and Dou, Jian-Peng and Jin, Xian-Min},
  journal={Physical Review Applied},
  volume={15},
  number={2},
  pages={024060},
  year={2021},
  publisher={APS},
  url={https://journals.aps.org/prapplied/abstract/10.1103/PhysRevApplied.15.024060}
}

@article{meyer2011implement,
  title={How to implement decoy-state quantum key distribution for a satellite uplink with 50-dB channel loss},
  author={Meyer-Scott, Evan and Yan, Zhizhong and MacDonald, Allison and Bourgoin, Jean-Philippe and H{\"u}bel, Hannes and Jennewein, Thomas},
  journal={Physical Review A—Atomic, Molecular, and Optical Physics},
  volume={84},
  number={6},
  pages={062326},
  year={2011},
  publisher={APS},
  url={https://journals.aps.org/pra/abstract/10.1103/PhysRevA.84.062326}
}

@article{lin2019asymptotic,
  title={Asymptotic security analysis of discrete-modulated continuous-variable quantum key distribution},
  author={Lin, Jie and Upadhyaya, Twesh and L{\"u}tkenhaus, Norbert},
  journal={Physical Review X},
  volume={9},
  number={4},
  pages={041064},
  year={2019},
  publisher={APS},
  url={https://journals.aps.org/prx/abstract/10.1103/PhysRevX.9.041064}
}

@article{navascues2008convergent,
  title={A convergent hierarchy of semidefinite programs characterizing the set of quantum correlations},
  author={Navascu{\'e}s, Miguel and Pironio, Stefano and Ac{\'\i}n, Antonio},
  journal={New Journal of Physics},
  volume={10},
  number={7},
  pages={073013},
  year={2008},
  publisher={IOP Publishing},
  url={https://iopscience.iop.org/article/10.1088/1367-2630/10/7/073013}
}

@article{ferradini2025defining,
  title={Defining Security in Quantum Key Distribution},
  author={Ferradini, Carla and Sandfuchs, Martin and Wolf, Ramona and Renner, Renato},
  journal={arXiv preprint arXiv:2509.13405},
  year={2025},
  url={https://arxiv.org/abs/2509.13405}
}

@TechReport{Wiesemann2024,
  author           = {Wiesemann, Jerome and Krause, Jan and Rusca, Davide and Walenta, Nino},
  title            = {A consolidated and accessible security proof for finite-size decoy-state quantum key distribution},
  year             = {2024},
  month            = may,
  note             = {arXiv:2405.16578 [quant-ph] type: article},
  archiveprefix    = {arxiv},
  copyright        = {arXiv.org perpetual, non-exclusive license},
  creationdate     = {2025-05-13T11:33:06},
  doi              = {10.48550/arxiv.2405.16578},
  eprint           = {2405.16578},
  modificationdate = {2025-05-13T11:33:06},
  primaryclass     = {quant-ph},
  publisher        = {arXiv},
  school           = {arXiv},
  institution      = {arXiv}
}

@article{Mizutani2025,
  author           = {Mizutani, Akihiro and Sasaki, Toshihiko and Kato, Go},
  title            = {Protocol-level description and self-contained security proof of decoy-state BB84 QKD protocol},
  year             = {2025},
  month            = apr,
  archiveprefix    = {arXiv},
  copyright        = {arXiv.org perpetual, non-exclusive license},
  creationdate     = {2025-05-05T17:48:23},
  journal={arXiv preprint},
  eprint           = {2504.20417},
  modificationdate = {2025-05-05T17:48:26},
  primaryclass     = {quant-ph},
  publisher        = {arXiv},
}

@article{Sun2015,
  title = {Effect of source tampering in the security of quantum cryptography},
  author = {Sun, Shi-Hai and Xu, Feihu and Jiang, Mu-Sheng and Ma, Xiang-Chun and Lo, Hoi-Kwong and Liang, Lin-Mei},
  journal = {Phys. Rev. A},
  volume = {92},
  issue = {2},
  pages = {022304},
  numpages = {8},
  year = {2015},
  month = {Aug},
  publisher = {American Physical Society},
  doi = {10.1103/PhysRevA.92.022304},
  url = {https://link.aps.org/doi/10.1103/PhysRevA.92.022304}
}

@article{Vernan,
    author = {G. Vernam} ,
    title = {Cipher printing telegraph systems for secret wire and radio telegraphic communi-
cations},
    journal = {T. Am. Inst. Elec. Eng., 55:109},
    year = {1926}
}

\newpage 

\begin{widetext}

\begin{center}
\textbf{APPENDIX}
\addparttoc{Appendix}
\end{center}

\appendix
\section{Entropies}
\label{sec: entropic quantities}

\renewcommand{\thethm}{A\arabic{thm}} 
\setcounter{thm}{0}
\renewcommand{\theequation}{A.\arabic{equation}} 
\setcounter{equation}{0}

\begin{table}[H]
\centering
\renewcommand{\arraystretch}{1.25} 
\setlength{\tabcolsep}{10pt} 
\begin{tabular}{@{} c l @{}} 
\toprule
\textbf{Symbol} & \textbf{Definition} \\
\midrule 
$H(\bullet)$ & Eq. \eqref{eq: shannon entropy},\eqref{eq: vN entropy} - Shannon/von Neumann entropy of (sub)system $\bullet$ \\[8pt]
$h(\bullet)$ & Eq. \eqref{eq: binary entropy} - Binary entropy of (sub)system $\bullet$ \\[8pt]
$H(\bullet \vert \circ)$ & Eq. \eqref{eq: conditional vN entropy} - Conditional entropy of $\bullet$ given $\circ$\\[8pt]
$D(\bullet \Vert \circ)$ & Eq. (\ref{eq: relative entropy}) - Relative entropy of $\bullet$ given $\circ$ \\[8pt]
$H_\text{min}(\bullet \vert \circ)$ & Eq. (\ref{eq: hmin via pguess}) - Min-entropy of $\bullet$ given $\circ$ \\[8pt]
$H_\text{max}(\bullet \vert \circ)$ & Eq. (\ref{eq: def max entropy}) - Max-entropy of $\bullet$ given $\circ$ \\[8pt]
$H^{\epsilon}_\text{min}(\bullet \vert \circ)$ & Eq. (\ref{eq:smooth-min-entropy}) - Smooth min-entropy of $\bullet$ given $\circ$\\[8pt]
$H^{\epsilon}_\text{max}(\bullet \vert \circ)$ & Eq. (\ref{eq:smooth-min-entropy}) - Smooth max-entropy of $\bullet$ given $\circ$\\[8pt]
\bottomrule
\end{tabular}
\caption{For detailed definition and properties, the reader may check ~\cite{watrous2020advanced,tomamichel2015quantum}.}\label{tab:entropic_quantities}
\end{table}

\dfn{\textbf{(Classical-quantum states)}} 
{
In the quantum formalism, classical information is modelled using an orthonormal
basis. A random variable $X$ that takes values $x$ with probabilities $p(x)$ can
be represented by the state
\begin{align}
    \rho_X=\sum_x p(x)\,\vert x\rangle\langle x\vert,
\end{align}
where $\{\vert x\rangle\}_x$ is an orthonormal basis satisfying
$\langle x \vert x^{\prime}\rangle=\delta_{x,x^{\prime}}$. More generally, one can describe hybrid systems in which a classical variable is
correlated with a quantum system. In particular, when analysing the security of a
QKD protocol we are often interested in \emph{classical--quantum} (cq) states of
the form
\begin{align}
    \rho_{A E}
    =\sum_x p(x)\,\vert x\rangle\langle x\vert_A \otimes \rho_{E \vert x},
\end{align}
where system $A$ encodes the classical variable $X$ in the basis $\{\vert
x\rangle\}$ and each $\rho_{E\vert x}$ is a quantum state of system $E$ that may depend on the value $x$.}

\dfn{\textbf{(Shannon entropy)} \label{def: shannon entropy}} 
{The Shannon entropy quantifies the uncertainty about a random variable. If $X$ is a discrete and classical random variable that assumes the value $x$ with probability $p(x)$ then the entropy of $X$ is defined as
\begin{align}\label{eq: shannon entropy}
    H(X)=-\sum_x p(x)\,\log p(x).
\end{align}} 

\dfn{\textbf{(Binary entropy)} \label{def: binary shannon entropy}} 
{The binary entropy corresponds to Shannon entropy for base 2 logarithm,
\begin{align}\label{eq: binary entropy}
    h(X) = - \sum_x p(x)\log_2 p(x),
\end{align}
typically employed when the alphabet of the system under analysis has two characters (e.g. 0 and 1).} 

\dfn{\textbf{(Conditional Shannon entropy)} \label{def: conditional shannon entropy}} 
{The \emph{conditional} Shannon entropy quantifies the remaining uncertainty about $X$ when the value of another variable $Y$ is known:
\begin{align}
    H(X \vert Y) = -\sum_{x, y} p(x, y) \log p(x \vert y) = H(X, Y)-H(Y).
\end{align}} \\

\dfn{\textbf{(von Neumann entropy)}} 
{The von Neumann entropy is the quantum analogue of Shannon entropy. For a quantum system $X$ in state $\rho_X$, its entropy is given by
\begin{align}\label{eq: vN entropy}
    H(X)_\rho=-\text{Tr}\big(\rho_X \log \rho_X\big).
\end{align}
If $\rho_X$ has spectral decomposition
$\rho_X=\sum_i \lambda_i\vert \lambda_i\rangle\langle \lambda_i\vert$, then
\begin{align}\label{eq: vN entropy decomposition}
    H(X)_\rho = -\sum_i \lambda_i \log \lambda_i.
\end{align}
In particular, if $\rho_X$ is diagonal in the basis
$\{\vert x\rangle\}$ with eigenvalues $p(x)$, then Eq. (\ref{eq: vN entropy decomposition}) reduces to the classical expression of Eq. (\ref{eq: shannon entropy})} \\

\dfn{\textbf{(Conditional von Neumann entropy)}} 
{Let $\rho_{AE}$ be a bipartite quantum state. The entropy of system $A$
conditioned on $E$ is defined as
\begin{align}\label{eq: conditional vN entropy}
    H(A \vert E)_\rho = H(AE)_\rho - H(E)_\rho,
\end{align}
where $H(E)_\rho=-\text{Tr}\big(\rho_E \log \rho_E\big)$ is the von Neumann
entropy of the reduced state $\rho_E = \text{Tr}_A(\rho_{AE})$, and similarly
for $H(AE)_\rho$. When $A$ and $E$ are both classical, this reduces to the
conditional Shannon entropy.} \\

\prop{} 
{
The conditional von Neumann entropy satisfies the following properties
\cite{tomamichel2015quantum}:
\begin{itemize}
    \item[\textit{1)}] \textbf{Positivity for separable states (Lemma 5.11 of \cite{tomamichel2015quantum})}:  
    If $\rho_{AB}$ is separable then
    \begin{align}
        H(A \vert B)_\rho \geq 0.
    \end{align}

    \item[\textit{2)}] \textbf{Data processing (Corollary 5.5 of \cite{tomamichel2015quantum})}:  
    Let $\tau_{A B^{\prime}}=(\mathbb{1}_A \otimes \mathcal{E}_B)\big(\rho_{A B}\big)$,
    where $\mathcal{E}_B$ is a CPTP map from $B$ to $B^{\prime}$. Then
    \begin{align}
        H(A \vert B)_\rho \leq H\left(A \vert B^{\prime}\right)_\tau.
    \end{align}

    \item[\textit{3)}] \textbf{Additivity (Corollary 5.9 of \cite{tomamichel2015quantum})}:  
    For a product state $\rho_{A B} \otimes \tau_{A^{\prime} B^{\prime}}$ it holds that
    \begin{align}
        H\left(A A^{\prime} \vert B B^{\prime}\right)_{\rho \otimes \tau}
        =H(A \vert B)_\rho+H\left(A^{\prime} \vert B^{\prime}\right)_\tau.
    \end{align}

    \item[\textit{4)}] \textbf{Conditioning on classical information (Proposition 5.4 of \cite{tomamichel2015quantum})}:  
    Let $\rho_{A B X}$ be a cq-state of the form
    $\rho_{A B X}=\sum_x p(x)\vert x\rangle\langle x\vert_X \otimes \rho_{A B \vert x}$.
    Then
    \begin{align}
        H(A \vert B X)_\rho
        =\sum_x p(x)\, H(A \vert B)_{\rho_{\vert x}},
    \end{align}
    where $\rho_{\vert x}$ is shorthand for $\rho_{A B\vert x}$.

    \item[\textit{5)}] \textbf{Removing classical information (Lemma 5.15 of \cite{tomamichel2015quantum})}:  
    For any state $\rho_{A B X}$ classical on $X$,
    \begin{align}
        H(A \vert X B)_\rho \geq H(A \vert B)_\rho -\log |X|,
    \end{align}
    where $|X|$ is the dimension of the system $X$.
\end{itemize}
}

\dfn{\textbf{(Min-entropy)}}
{\label{def:min-entropy}
Let $\rho_{AE} \in \mathcal{D}(\mathcal{H}_A \otimes \mathcal{H}_E)$ be a (possibly sub-normalized) bipartite state. The conditional min-entropy of $A$ given $E$ is defined as
\begin{align}\label{eq:min-entropy-def}
    H_{\min}(A\vert E)
    \equiv \sup\Bigl\{\lambda\in\mathbb{R}\,:\,\exists\,\rho_E\in\mathcal{H}_B\ \text{s.t.}\ \rho_{AE}\le 2^{-\lambda}\,\mathbb{1}_A\otimes \rho_E\Bigr\}.
\end{align}} \\

\noindent
Another entropy often appearing in security analysis is the max-entropy, which is defined as follows.

\dfn{\textbf{(Max-entropy)}} 
{For a bipartite state $\rho_{AE} \in \mathcal{D}(\mathcal{H}_A \otimes \mathcal{H}_E)$, the conditional max-entropy of $A$ given $E$ is defined as
\begin{align}\label{eq: def max entropy}
    H_{\max }(A \vert E)
    \equiv \sup _{\rho_E \in \mathcal{H}_E} \log F \left(\rho_{A E}, \mathbb{1}_A \otimes \sigma_E\right),
\end{align}
where $F$ denotes the (generalized) fidelity.} \\

We now introduce generalizations of the entropies defined so far, as one-parameter families of quantities which recover the given definitions in certain limits. We start with \emph{smooth} conditional min- and max-entropies, which are relevant for the finite-size analysis of QKD protocols.

The intuition behind the smooth entropies is that they are obtained by optimizing the min- and max-entropies over a small neighborhood of the reference state, effectively accounting for statistical fluctuations and finite-size effects in quantum systems. This provides a more robust measure of uncertainty that is meaningful in  non-asymptotic scenarios. The neighborhood is defined using the purified distance and the optimization is taken over (sub-normalised) operators that are $\epsilon$-close to the state of interest.

\dfn{\textbf{(Smooth entropies \cite{tomamichel2017largely})}} 
{Given a state $\rho_{AE}$, an $\epsilon$-ball around $\rho_{AE}$ as
\begin{align}
    \mathcal{B}^\epsilon(\rho_{AE}) \equiv  \bigl\{
        \tilde{\rho}_{AE} \in \mathcal{L}(AE) :
        \tilde{\rho}_{AE} \geq 0,\;
        \operatorname{Tr}(\tilde{\rho}_{AE}) \leq 1,\;
        D_P(\rho_{AE},\tilde{\rho}_{AE}) \leq \epsilon
    \bigr\},
\end{align}
where $D_P$ denotes the purified distance. The smooth conditional min- and max-entropies of $A$ given $E$ are then
defined as
\begin{align}
    H_{\min}^{\epsilon}(A \vert E)_\rho
    &\equiv \max_{\tilde{\rho}_{AE} \in \mathcal{B}^\epsilon(\rho_{AE})}
       H_{\min}(A \vert E)_{\tilde{\rho}}, \label{eq:smooth-min-entropy} \\
    H_{\max}^{\epsilon}(A \vert E)_\rho
    &\equiv \min_{\tilde{\rho}_{AE} \in \mathcal{B}^\epsilon(\rho_{AE})}
       H_{\max}(A \vert E)_{\tilde{\rho}}. \label{eq:smooth-max-entropy}
\end{align}}

\section{Secret key derivation}\label{app: secretkey}

Here we provide more details regarding the results presented in Sec. \ref{sec: QKD protocols}, where quantum and classical steps are performed over a string of size $n$ to distill a secure (secret and correct) key of $\ell$ bits. The main result allowing one to associate the key $\ell$ to smooth min-entropy is the quantum leftover hashing lemma, which allows us to connect $n$ and $\ell$ in terms of this entropic quantity.

\thm{\textbf{\textbf{(Quantum leftover hashing lemma with smooth min-entropy, Proposition 9 of \cite{tomamichel2017largely})}}} {Let $\rho_{A_1^n E}$ be a cq-state, where the classical register $A_1^n$ stores an $n$-bit string, and let $\rho_{K_A F E}$ be the state after applying a 2-universal hash function from $\{0,1\}^n$ to $\{0,1\}^{\ell}$, which maps $A_1^n$ into $K_A$. Then the following inequality holds
\begin{align}\label{eq: smooth quantum leftover}
    \left\|\rho_{K_A F E} - 2^{-\ell}\mathbb{1}_{K_A} \otimes \rho_{F E}\right\|_{\mathrm{Tr}} \leq  2^{-1-\frac{1}{2}(H_{\min }^\epsilon(\overline{A_1^n} \vert E \tilde{A}_{1}^{n} \tilde{B}_1^{n})_\rho-\ell)}+2 \epsilon
\end{align}
}

\noindent
\textbf{Proof.} The proof of this theorem can be found in \cite{tomamichel2017largely}. \\

The results given by Equation (\ref{eq: smooth quantum leftover}) and the definition of $\varepsilon_\text{s}-$secrecy (Eq. (\ref{eq: secrecy def})) allow us to write the key length as a function of the smooth min-entropy generated in the $n$ rounds of the protocol minus penalty terms:
\begin{align}\label{eq: key general formula 2}
    \ell=H_{\min }^\epsilon(\overline{A_1^n} \vert E\tilde{A}_{1}^{n} \tilde{B}_1^{n})_\rho-2 \log \left(\frac{1}{2 \varepsilon_\text{PA}}\right).
\end{align}

In order to estimate the real parameter $\varepsilon_{\text{IR}}$ associated to the information reconciliation step, it is possible to use inequalities satisfied by the smooth min-entropy to remove the dependence on the information publicly announced by the authenticated parties ($\tilde{A}_{1}^{n} \tilde{B}_1^{n}$) in the process of error correction. More precisely, the inequality allowing the removal of such information is
\begin{align}
H_\text{min}^\epsilon(\overline{A_1^n} \vert E\tilde{A}_{1}^{n} \tilde{B}_1^{n})_{\rho} \geq H_{\min }^\epsilon\left(\overline{A_1^n} \vert E\right)_{\rho}-\vert \text{leak}_{\text{IR}} \vert,
\end{align}
where $\text{leak}_{\mathrm{IR}}$ is the minimum amount of leaked elements of the string necessary to perform one-way information reconciliation, derived in \cite{bennett1988privacy,renner2005simple}.

The use of two-universal hashing functions in the information reconciliation subroutine allows us to bound the probability of that the strings that Alice and Bob hold by the end of this phase phase are different. Consider the event in which the error correction subroutine does not abort, given that the syndromes of Alice and Bob are distinct. In this case, we can write the equality 
\begin{align}
    p(f_{\mathrm{IR}}(\overline{B_1^n}) &=f_{\mathrm{IR}}(\overline{A_1^n}) \vert \overline{A_1^n} \neq \overline{B_1^n} )=\varepsilon_{\mathrm{IR}},
\end{align}
so that
\begin{align}
p\left(K_A \neq K_B\right) & =p(K_A \neq K_B \wedge f_{\mathrm{IR}}(\overline{B_1^n})=f_{\mathrm{IR}}(\overline{A_1^n})) \\
& \leq p(\overline{A_1^n} \neq \overline{B_1^n} \wedge f_{\mathrm{IR}}(\overline{B_1^n})=f_{\mathrm{IR}}(\overline{A_1^n}) ) \\
& \leq \varepsilon_{\mathrm{IR}}.
\end{align}
As we saw before (Sec. \ref{sec: security}), the definition of $\varepsilon_\text{c}$-correctness of a protocol coincides with the parameter of $\varepsilon_\text{IR}$ given above.

\section{Basic optimization methods}
\label{sec: basic optimizations}

\renewcommand{\thethm}{B\arabic{thm}} 
\setcounter{thm}{0}
\renewcommand{\theequation}{B.\arabic{equation}} 
\setcounter{equation}{0}

In this Appendix, we briefly review some basic aspects of optimization theory \cite{tavakoli2024semidefinite,skrzypczyk2023semidefinite,veeren2024semi,mironowicz2024semi}, which contain essential ingredients for the more advanced methods exposed in Sec. \ref{sec: algorithms}. We start explaining \textit{linear programs}, tasks of optimizing convex-linear multivariable functions under linear constraints of the function itself and its variables. We then generalize this optimization class to \textit{semi-definite programs}, which are optimization problems performed over convex-linear functions of matrices (being naturally an important optimization class for Quantum Information and related subjects).

\subsection{Linear Programs (LP's)}

Linear programs (LPs) consist of optimization problems where the objective function is a multi-variable linear object $f:\mathbb{R}^k\rightarrow \mathbb{R}$ subject to linear constraints consisting of equalities and inequalities of $f$ and its variables \cite{boyd2004convex}. A general LP can always be formulated as 
\begin{subequations}
\begin{align}\label{eq: LP-1}
    \min_{x_{1},\dots,x_k} \hspace{0.25cm} &f(x_1, \dots, x_k) \\
    \text{s.t} \hspace{0.55cm}  &g_i(x_1, \dots, x_k) = b_i & i\in \{1,\dots, m\}; \label{eq: equality 1 LP} \\
    &h_j (x_1, \dots, x_k) \geq c_j & j\in \{1,\dots, n\}. \label{eq: inequality 1 LP}
\end{align}
\end{subequations}
Due to linearity, Equations (\ref{eq: LP-1}) can be always rewritten as 
\begin{subequations}
\begin{align}
    \min_{\vec{x}} \hspace{0.25cm} & \Vec{a}\cdot \Vec{x} \label{eq: LP 2} \\
    \text{s.t} \hspace{0.35cm}  &B \Vec{x} = \Vec{b}, \label{eq: equality 2 LP}\\
    & C \Vec{x} \geq \Vec{c}. \label{eq: inequality 2 LP}
\end{align}
\end{subequations}

The set of all variables satisfying constraints represented in Eq. (\ref{eq: equality 2 LP}) and (\ref{eq: inequality 2 LP}) composes the \textit{feasibility set} of a linear program, denoted by $\mathcal{F}$. As a consequence of the linearity of the constraints, the feasible set is convex, and it is non-empty whenever constraints are non-contradictory. Another way to denote an LP is to write it in terms of feasible points:
\begin{subequations}
\begin{align}
    \min_{\vec{x}} \hspace{0.25cm} & \Vec{a}\cdot \Vec{x} \\
    \text{s.t} \hspace{0.3cm} &\Vec{x}\in \mathcal{F}.
\end{align}
\end{subequations}

If a particular $\Vec{x}^{*}$ realizes the optimization $\Vec{a}\cdot \Vec{x}^{*} =  \max_{\Vec{x}} \hspace{0.25cm} \Vec{a}\cdot \Vec{x}$, then $\Vec{x}^{*}$ is said to be the optimal variable, while $\alpha^{*} \equiv \Vec{a}\cdot \Vec{x}^{*}$ is the optimal value. The set of all optimal values, defined as $\mathcal{O} \equiv \{\Vec{x}^{*}; \Vec{a}\cdot \Vec{x}^{*} =  \min_{\Vec{x}} \hspace{0.25cm} \Vec{a}\cdot \Vec{x}\}$ is referred to the optimal set. $\mathcal{O}$ inherits convexity of $\mathcal{F}$, and has either cardinality zero (when there's no optimal variable, e.g. when the program is infeasible), one (when the optimal variable exists and is unique) or infinity (if there are two distinct variables that realize the optimization).

Optimization problems such as LPs have a property named \textit{duality}. This
characteristic allows us to express the same optimization problem in two
different versions, named \textit{primal} and \textit{dual}. A primal
minimization problem is related to a dual maximization problem, and vice
versa. The idea to relate these two formulations is that constraints of the
primal program become variables of the dual, and the primal objective becomes
a constraint in the dual. Depending on the number of constraints and variables, one of these formulations might be more adequate for numerical implementations.

To construct the dual program of a primal LP, we consider the optimization written in as in Equations (\ref{eq: LP 2}) - (\ref{eq: inequality 2 LP}). one can define a dual variables $\vec{y}$ for the equalities and $\vec{z} \ge 0$ for the inequalities (the non-negative restriction here is taken as convention without loss of generality), and the global function containing primal and dual variables (the Lagrangian) is given by
\begin{subequations}
\begin{align}
  \mathcal{L}(\vec{x},\vec{y},\vec{z})
  &\equiv \vec{a}\cdot\vec{x}
   + \vec{y}\cdot(\vec{b} - B\vec{x})
   + \vec{z}\cdot(\vec{c} - C\vec{x})  \\
  &= \big(\vec{a} - B^\top\vec{y} - C^\top\vec{z}\big)\cdot\vec{x}
     + \vec{b}\cdot\vec{y} + \vec{c}\cdot\vec{z}.
  \label{eq:Lagrangian-LP}
\end{align}
\end{subequations}
For any primal feasible $\vec{x}$ (satisfying the constraints of Equations (\ref{eq: equality 2 LP}) - (\ref{eq: inequality 2 LP})) and any $\vec{z}\ge 0$, the last term $\vec{z}\cdot(\vec{c} - C\vec{x})$ is non-positive, so
$\mathcal{L}(\vec{x},\vec{y},\vec{z}) \le \vec{a}\cdot\vec{x}$. In other
words, every dual feasible pair $(\vec{y},\vec{z})$ provides a \emph{lower} bound on the primal objective. By choosing the dual variables so that the dependence on $\vec{x}$ cancels, one arrives at the dual optimization problem
\begin{subequations}
\begin{align}
  \beta^*
  = \max_{\vec{y},\vec{z}} \hspace{0.25cm} 
    & \vec{b}\cdot\vec{y} + \vec{c}\cdot\vec{z} \\
    \text{s.t.} \hspace{0.3cm}& B^\top \vec{y} + C^\top \vec{z} = \vec{a}, \\
                 & \vec{z} \ge 0,
  \label{eq:dual-LP}
\end{align}
\end{subequations}
which is itself an LP. The construction of duality allows us to connect the primal and dual values via \emph{weak} and \emph{strong} duality. Weak duality states that for any primal feasible $\vec{x}$ and dual feasible $(\vec{y},\vec{z})$, the following inequality holds
\begin{align}
  \vec{b}\cdot\vec{y} + \vec{c}\cdot\vec{z}
  \leq \vec{a}\cdot\vec{x}.
\end{align}
A particular consequence of this is that the optimal values of the primal and dual satisfy $\beta^* \leq \alpha^*$. In particular, when the primal, or equivalently the dual, is feasible and bounded, then \emph{strong duality} holds: the optimal values coincide, $\alpha^\star = \beta^\star$. In this case, finding a pair of primal and dual feasible solutions with equal objective values certifies optimality for both.

\subsection{Semi-definite Programs (SDPs)}

A natural generalization of linear programs should be considered when the variables are matrices instead of variables in $\mathbb{R}^n$ \cite{cavalcanti2016quantum}. More precisely, our variables are now assumed to be hermitian operators $X=X^\dagger$, while the objective function translates into the trace of an operator $A$ acting over $X$. The constraints are lifted to equalities and inequalities of the respective maps $\xi$ and $\zeta$, which preserve hermiticity of the variable.
\begin{subequations}
\begin{align}
    \Vec{x} \ &\rightarrow \ X = X^\dagger, \\
    \Vec{a}\cdot \Vec{x} \ &\rightarrow \ \text{Tr} (A X), \\
    B \Vec{x} = \Vec{b} \ &\rightarrow \ \xi_i (X) = B_i \label{eq: sdp constraint equality 1} & \forall \  i; \\ 
    C \Vec{x}\geq \Vec{c} \ &\rightarrow \ \zeta_j (X) \succeq C_j & \forall \  j. \label{eq: sdp constraint psd 1}
\end{align} 
\end{subequations}
In the constraints above, $B$ and $C$ denote matrices for every inequality indices $i,j$. In particular, the inequality constraint should be understood as the positive semi-definite condition $\zeta_j(X) - C_j\succeq 0$ instead of an element-wise inequality between elements of $\zeta$ and $C$. With this, a general SDP can be written as
\begin{subequations}
\begin{align}
    \min_{X} \hspace{0.25cm} & \text{Tr} (A X) \\
    \text{s.t} \hspace{0.35cm} &\xi_i (X) = B_i & \forall \  i; \label{eq: sdp constraint equality 2} \\
    \hspace{0.35cm} &\zeta_j (X) \succeq C_j & \forall \  j. \label{eq: sdp constraint psd 2}
\end{align}
\end{subequations}

Similar to LPs, SDPs admit a dual formulation. To construct it, we introduce Hermitian Lagrange-multiplier operators $Y_i$ for the equality constraints and
$Z_j \succeq 0$ for the inequality constraints. The associated Lagrangian is
\begin{align}
    \mathcal{L}(X,\{Y_i\},\{Z_j\})
    &\equiv \text{Tr}(A X)
      + \sum_{i=1}^m \text{Tr}\bigl[ Y_i\,(B_i - \xi_i(X)) \bigr]
      + \sum_{j=1}^n \text{Tr}\bigl[ Z_j\,(C_j - \zeta_j(X)) \bigr]. \label{eq: SDP lagrangian 1} \\
    &= \text{Tr}\Biggl[
        X\Bigl(
          A - \sum_{i=1}^m \xi_i^\dagger(Y_i)
            - \sum_{j=1}^n \zeta_j^\dagger(Z_j)
        \Bigr)
      \Biggr]
      + \sum_{i=1}^m \text{Tr}(Y_i B_i)
      + \sum_{j=1}^n \text{Tr}(Z_j C_j). \label{eq: SDP lagrangian 2}
\end{align}
where in the second line we used  the adjoint maps $\xi_i^\dagger$ and $\zeta_j^\dagger$, defined by
$\text{Tr}\bigl( Y_i\,\xi_i(X) \bigr) = \text{Tr}\bigl( \xi_i^\dagger(Y_i)\,X \bigr)$
(similarly for $\zeta_j$). For any primal variable $X$ satisfying Equations (\ref{eq: sdp constraint equality 2})–(\ref{eq: sdp constraint psd 2}) and any choice of $Z_j \succeq 0$, the term last term of Eq. (\ref{eq: SDP lagrangian 1}) is non-positive, so $\mathcal{L}(X,\{Y_i\},\{Z_j\}) \leq \text{Tr}(A X)$. In other words, every dual provides a lower (upper) bound on the primal minimization (maximization) objective.

By restricting the Lagrangian to $X=0$, the first term of Eq. (\ref{eq: SDP lagrangian 2}) vanishes and we're left with the dual problem 
\begin{subequations}
\begin{align}
    \beta^{*} =
    \max_{\{Y_i\},\{Z_j \succeq 0\}} \quad
        & \sum_{i=1}^m \text{Tr}(Y_i B_i)
        + \sum_{j=1}^n \text{Tr}(Z_j C_j),
        \label{eq:SDP-dual-obj}\\
    &\hspace{-1.25cm}\text{s.t.}\quad
         A - \sum_{i=1}^m \xi_i^\dagger(Y_i)
            - \sum_{j=1}^n \zeta_j^\dagger(Z_j) = 0.
        \label{eq:SDP-dual-constraints}
\end{align}
\end{subequations}
The relationship between the primal value $\alpha^{*}$ and the dual value
$\beta^{*}$ is somewhat analogous to the LP case. \emph{Weak duality} holds for any primal feasible $X$ and dual feasible $\{Y_i,Z_j\}$, so that $\beta^* \leq \alpha^*$. Strong duality depends on the primal SDP being bounded and admitting a \emph{strictly feasible} point $X \succ 0$. Under these conditions, the optimal primal and dual values coincide, $\alpha^{*} = \beta^{*}$.

\subsection{Relaxations of polynomial optimization problems}
\label{sec: NPA}

In the previous section, we considered optimization classes in which the function to be optimized had a linear dependence on the variables (finite-dimensional vectors or matrices). We may consider a broader class of problems in which the function is a general polynomial of the variables \cite{tavakoli2024semidefinite}. In this case, finding the solution for such functions is an NP-hard problem \cite{nesterov2000squared}.

A common way to address this sort of optimization is based on the concept of a \textit{relaxation}. It means that instead of exactly solving the original hard problem, we can consider less constrained optimizations (the relaxations) - which in some limit converge to the solution of the original polynomial problem - that provide an aproximate solution for the desired problem (see Fig. \ref{fig:NPA}). This type of strategy works for sufficiently small scenarios (i.e., few parties, measurements, and outcomes), where the relaxations may achieve results very close to the optimal value of the original problem. In more complex situations, relaxations of polynomial optimization problems become computationally expensive very quickly.

The class of polynomial optimization problems can be divided into commutative and noncommutative problems. The relaxations for the commutative case - when polynomials are functions of scalars - are known as the Lasserre hierarchy \cite{lasserre2001global}, while relaxations of non-commutative polynomial optimizations - those with matrix variables - were developed by Navascués, Pironio, and Acín (NPA) \cite{navascues2007bounding,pironio2010convergent,navascues2008convergent}. As the latter is more related to the subject of DIQKD, in this section, we'll provide a brief introduction to this hierarchy.

A general non-commutative optimization problem has its objective function defined in terms of a hermitian polynomial $f=f(X_1,\dots,X_N)$, where $\{ X_n\}_{n=1,\dots,N}$ is a collection of operators in a Hilbert space $\mathcal{H}$ with a certain number of inputs and outputs (the number of outputs doesn't coincides with the dimension of the operator).

\begin{subequations}
\begin{align}
\min_{\rho, \{ X_n\}} \hspace{0.25cm} & \left\langle f\left\{ X_n\}\right)\right\rangle_{\rho} \label{eq: objective function NCPO}  \\
\text{s.t.} \hspace{0.35cm} & \left\langle h_i(\{ X_n\})\right\rangle_\rho \geq 0 & \forall \  i; \\
& g_j\left(\{ X_n\}\right) \succeq 0 &  \forall \  j; \label{eq: NCPO operator PSD constraint} \\
& \operatorname{Tr}(\rho)=1, \\
& \rho \succeq 0.  \label{eq: PSD NCPO}
\end{align}
\end{subequations}

Here $f$, $h$ and $g_j$ are Hermitian polynomials in the non-commuting
variables $X_1,\dots,X_N$ acting on some Hilbert space $\mathcal{H}$, and
$\langle\cdot\rangle_\rho \equiv \text{Tr}(\rho\,\cdot)$ denotes the expectation
with respect to the state $\rho$. 

In device-independent QKD settings, the polynomial is typically characterized by the unknown state shared and measurements used by both authenticated parties and the eavesdropped. In these situations, the operators $\{X\}$ are usually identified with the measurement operators of the parties, $\{M_{a_x \vert x}\}$ and $\{N_{b_y \vert y}\}$, which we may assume to be projectors acting on some Hilbert space of \textit{unspecified dimension} - the only known information about the boxes of Alice and Bob are the amount of inputs and outputs. A generic quantum correlation is then written as in Eq. (\ref{eq: probabilities EB}), but for the numerical implementations necessary for DIQKD scenarios it is useful to substitute the tensor-product structure between Alice and Bob measurement operators by the commutation condition 
\begin{align}
    [M_{a_x\vert x}, N_{b_y\vert y}] = 0 \quad \forall \ a,b,x,y.
\end{align}
In this way, joint probabilities are moments of products of operators living in the same global Hilbert space $\mathcal{H}_{AB}$ instead of the usually considered composed spaces $\mathcal{H}_A\otimes \mathcal{H}_B$:
\begin{align}
    p(ab\vert xy) = \text{Tr}(M_{a_x \vert x}\cdot N_{b_y \vert y} \cdot \rho) \equiv \langle M_{a_x \vert x} N_{b_y \vert y} \rangle_\rho.
\end{align}

The strategy to tackle problems of the form of Equations (\ref{eq: objective function NCPO}) - (\ref{eq: PSD NCPO}) is to optimize over not over the variables $M_{a_x \vert x}$ and $N_{b_y \vert y}$ themselves, but instead over matrices constructed with their moments. This allows us to work without fixed Hilbert spaces and bounded dimensions, which is the general idea of the NPA hierarchy \cite{navascues2015bounding}. One first collects all linearly independent measurement operators into a list
\begin{align}
    L = \{ \mathbb{1}, \mathbf{M}, \mathbf{N} \},
\end{align}
where $\mathbf{M} \equiv \{M_{0_0\vert 0}, \dots, M_{\text{max} (a_x)\vert \text{max}(x)} \}$ and $\mathbf{N} \equiv \{N_{0_0\vert 0}, \dots, N_{\text{max} (b_y)\vert \text{max}(y)} \}$. For a fixed integer $k\geq 1$, let $S_k$ denote the set of all products of length at most $k$ of the operators appearing in $L$ (including the identity). Elements of $S_k$ are often called sequences of operators. To every level $k$ we associate a \(|S_k|\times|S_k|\) \emph{moment matrix} $\Gamma^{(k)}$ whose rows and columns are indexed by sequences $u,v\in S_k$, and
whose entries are defined by
\begin{align}
    \Gamma^{(k)}(u,v)
    \equiv \langle  u^\dagger v \rangle.
\end{align}
By construction, the moment matrix satisfies some properties:
\begin{itemize}
    \item[1.] $\Gamma^{(k)}(\mathbb{1},\mathbb{1}) = 1$;

    \item[2.] For any family of complex coefficients
    $\{\alpha_u\}_{u\in S_k}$ we have
    \begin{align}
        \sum_{u,v\in S_k} \alpha_u^*\, \Gamma^{(k)}(u,v)\, \alpha_v
        &= \left\langle 
             \Big(\sum_{u\in S_k} \alpha_u u\Big)^\dagger
             \Big(\sum_{v\in S_k} \alpha_v v\Big)
          \right\rangle
        \geq 0,
    \end{align}
    which is equivalent to the matrix constraint
    \begin{align}
        \Gamma^{(k)} \succeq 0;
    \end{align}
    \item[3.] Coefficients associated with probabilities: products of individual measurement operators of each party and products between measurement operators and identity reproduce joint and marginal probabilities
    \begin{align}
    \Gamma^{(k)}\left(M_{a_x\vert x}, N_{b_y\vert y}\right)
    &= p(ab\vert xy), \label{eq: mom matrix entries 1} \\
    \Gamma^{(k)}\left(M_{a_x\vert x}, \mathbb{1}\right) &= p_A(a\vert x),  \label{eq: mom matrix entries 2} \\
    \Gamma^{(k)}\left(\mathbb{1}, N_{b_y\vert y}\right)
    &= p_B(b\vert y).  \label{eq: mom matrix entries 3}
\end{align}
\end{itemize}

All remaining entries of $\Gamma^{(k)}$, which can't be interpreted as probabilities listed in Equations (\ref{eq: mom matrix entries 1}) - (\ref{eq: mom matrix entries 3}), are treated as the free variables of the relaxed optimization and are subject to the linear constraints listed above and the semidefinite condition $\Gamma^{(k)} \succeq 0$.

Any Hermitian polynomial in the measurement operators, such as the objective function $f$ or the constraint polynomials $h_i$ and $g_j$, can be written as a linear combination of products of sequences in $S_k$. More precisely, for a level $k$ such that $\deg(f),\deg(h_i) \leq 2k$ \cite{tavakoli2024semidefinite}, we can expand these functions as
\begin{align}
    f &= \sum_{u,v\in S_k} f_{uv}\, u^\dagger v, \\
    h_i &= \sum_{u,v\in S_k} h^{(i)}_{uv}\, u^\dagger v &  \forall \ i . \label{eq:poly-expansion}
\end{align}
Using the definition of the moment matrix, the corresponding expectation values are linear functionals of $\Gamma^{(k)}$:
\begin{align}
    \langle f \rangle
    &= \sum_{u,v\in S_k} f_{uv}\, \Gamma^{(k)}(u,v), \\
    \langle h_i \rangle
    &= \sum_{u,v\in S_k} h^{(i)}_{uv}\, \Gamma^{(k)}(u,v) & \forall \ i .
\end{align}
Hence, the minimization of $\langle f \rangle$ and constraints of the form $\langle h_i \rangle \geq 0$ can be written as an objective function and linear inequalities in terms of the moment matrix of level $k$. A similar procedure can be done for the case of operator PSD constraints (Eq. (\ref{eq: NCPO operator PSD constraint})), leading to another set of moment matrices (named "localized moment matrices"), and denoted by $\Gamma_{g_j}^{(k)}$ \cite{tavakoli2024semidefinite}. A level-$k$ relaxation of the original non-commutative polynomial optimization problem can then be written as the relaxed SDP
\begin{subequations}
\begin{align}\label{eq: relaxed NCPO}
\min_{\Gamma^{(k)}} \quad
        & \sum_{u,v\in S_k} f_{uv}\, \Gamma^{(k)}(u,v) \\
    \text{s.t.}\quad
        & \sum_{u,v\in S_k} h^{(i)}_{uv}\, \Gamma^{(k)}(u,v) \geq 0
          & \forall i; \\
        & \Gamma_{g_j}^{(k)} \succeq 0 & \forall j; \\
        & \Gamma^{(k)} \succeq 0.
\end{align}
\end{subequations}
together with any additional linear equalities that encode algebraic relations between the measurement operators (projector identities, commutation relations, etc.). For each fixed $k$, this is a finite-dimensional SDP whose optimal value $\beta_k$ provides a relaxed bound on the original non-commutative problem; as $k$ increases, the relaxations become tighter and converge to the optimum within the set of commuting correlations \cite{tsirelson1993some,scholz2008tsirelson} under suitable conditions \cite{navascues2007bounding,pironio2010convergent}.

\begin{center}
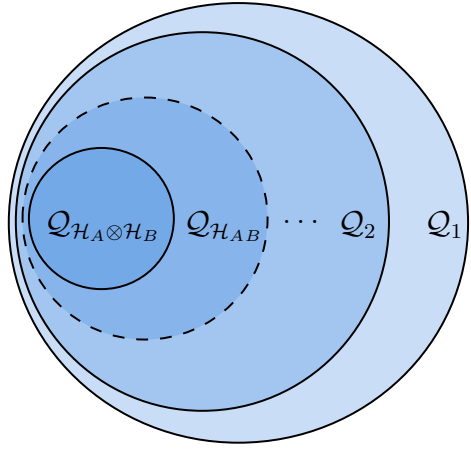
\begin{figure}
\begin{centering}

\tikzset{every picture/.style={line width=0.75pt}} 

\begin{tikzpicture}[x=0.75pt,y=0.75pt,yscale=-1,xscale=1,scale=0.7]

\draw  [fill={rgb, 255:red, 74; green, 144; blue, 226 }  ,fill opacity=0.3 ] (99.87,239.1) .. controls (99.87,151.56) and (173.95,80.6) .. (265.33,80.6) .. controls (356.72,80.6) and (430.8,151.56) .. (430.8,239.1) .. controls (430.8,326.64) and (356.72,397.6) .. (265.33,397.6) .. controls (173.95,397.6) and (99.87,326.64) .. (99.87,239.1) -- cycle ;
\draw  [color={rgb, 255:red, 0; green, 0; blue, 0 }  ,draw opacity=1 ][fill={rgb, 255:red, 74; green, 144; blue, 226 }  ,fill opacity=0.3 ] (104.94,238.1) .. controls (104.94,162.71) and (165.12,101.6) .. (239.37,101.6) .. controls (313.61,101.6) and (373.8,162.71) .. (373.8,238.1) .. controls (373.8,313.49) and (313.61,374.6) .. (239.37,374.6) .. controls (165.12,374.6) and (104.94,313.49) .. (104.94,238.1) -- cycle ;
\draw  [fill={rgb, 255:red, 74; green, 144; blue, 226 }  ,fill opacity=0.31 ][dash pattern={on 4.5pt off 4.5pt}][line width=0.75]  (109.8,235.97) .. controls (109.8,187.78) and (149.32,148.71) .. (198.07,148.71) .. controls (246.81,148.71) and (286.33,187.78) .. (286.33,235.97) .. controls (286.33,284.16) and (246.81,323.23) .. (198.07,323.23) .. controls (149.32,323.23) and (109.8,284.16) .. (109.8,235.97) -- cycle ;
\draw  [fill={rgb, 255:red, 74; green, 144; blue, 226 }  ,fill opacity=0.3 ][line width=0.75]  (114.36,235.97) .. controls (114.36,207.86) and (137.74,185.07) .. (166.58,185.07) .. controls (195.42,185.07) and (218.8,207.86) .. (218.8,235.97) .. controls (218.8,264.08) and (195.42,286.87) .. (166.58,286.87) .. controls (137.74,286.87) and (114.36,264.08) .. (114.36,235.97) -- cycle ;

\draw (123.7,228.45) node [anchor=north west][inner sep=0.75pt]  [font=\large]  {$\mathcal{Q}_{\mathcal{H}_{A} \otimes \mathcal{H}_{B}}$};
\draw (224.01,228.45) node [anchor=north west][inner sep=0.75pt]  [font=\large]  {$\mathcal{Q}_{\mathcal{H}_{A}{}_{B}}$};
\draw (397.5,228.45) node [anchor=north west][inner sep=0.75pt]  [font=\large]  {$\mathcal{Q}_{1}$};
\draw (335,228.45) node [anchor=north west][inner sep=0.75pt]  [font=\large]  {$\mathcal{Q}_{2}$};
\draw (295,235) node [anchor=north west][inner sep=0.75pt]  [font=\large]  {$\dotsc $};

\end{tikzpicture}
    \caption{Hierarchy of quantum correlation sets. $\mathcal{Q}_{\mathcal{H}_A\otimes \mathcal{H}_B}$ denotes the achievable correlations with operators living in the bipartite tensor product of Alice's and Bob's Hilbert spaces. $\mathcal{Q}_{\mathcal{H}_{AB}}$ denotes the set of probabilities that can be obtained when compatibility relations are imposed between operations of these two parties. $\mathcal{Q}_1 \supset \mathcal{Q}_2 \supset \dots$ represents feasible sets of each corresponding level of NPA hierarchy converging to $\mathcal{Q}_{\mathcal{H}_{AB}}$.}
    \label{fig:NPA}
\end{centering}
\end{figure}
\end{center}

\section{Device-independent key rate algorithms}
\label{sec: DI methods}

The main challenge to bound numerically the key rate in a device-independent setting is the fact that no assumption is made regarding the distributed state and the measurements performed in each laboratory (Sec. \ref{sec: device trutability}). This typically characterizes problems in which the objective function is a polynomial of the variables that must be optimized over, as in Eq. (\ref{eq: objective function NCPO}). Therefore, in DIQKD it is common to use the apparatus of NPA hierarchy in order to find lower bounds for asymptotic key rates given a protocol of interest. In this section we provide details of two methods to calculate such keys, based on the method proposed by Masanes et al. \cite{masanes2011secure} and Brown et al. \cite{brown2024device}.

\subsection{Lower bounds on the key rate via optimization of $H_\text{min}$}
\label{sec: appendix Hmin}

The first practical numerical method to compute asymptotic key rates in DI settings was proposed by Masanes et al. in \cite{masanes2011secure}. In this approach, in order to provide a lower bound for $r_\infty$, we use the fact that the conditional von Neumann entropy (that can't be directly computed) is always lower bounded by the \textit{min-entropy} (defined in Eq. (\ref{def:min-entropy}) of Appendix \ref{sec: entropic quantities}):
\begin{align}\label{eq: min entropy lower bound}
    H(A \vert E) \geq H_\text{min}(A\vert E).
\end{align}
The min-entropy can be written in terms of Eve's guessing probability over Alice's outcomes. This figure of merit that quantifies how much information of one of the authenticated parties the eavesdropper could probabilistically guess, which is defined as follows.

\dfn{\textbf{(Guessing probability)}} {\label{dfn: pguess}
The guessing probability, $p_{\text {guess}}(A \vert E)$, is the optimal probability that the eavesdropper, with system system $E$, by measuring a POVM's $Z_e$ can correctly guess the value of the variable $A$: 
\begin{align} \label{eq: pguess definition}
p_\text{guess}(A \vert E)_\rho=\sup _{\{Z_{e}\}} \sum_a p(a) \text{Tr}\left(Z_{e=a} \rho_{E \vert a}\right),
\end{align}
where the supremum is over all possible measurements, described by the set of POVMs $\left\{Z_{e}\right\}_a$ on the system $E$.} \\

The connection between the min-entropy (Definition \ref{def:min-entropy}) and the guessing probability was established in \cite{konig2009operational}, and is stated as follows.

\thm{(Theorem 1 of \cite{konig2009operational})} {\label{thm: Hmin pguess}Let $\rho_{A E}=\sum_a p(a)\vert a\rangle\langle a \vert\otimes \rho_{E\vert a}$ be classical on $\mathcal{H}_A$ and $p_{\text{guess}}(A \vert E)_\rho$ the maximal guessing probability $A$ from $E$ with a POVM $\left\{Z_e\right\}_{e=a}$ on $\mathcal{H}_E$. Then the conditional min-entropy $H_\text{min}(A\vert E)$ and the guessing probability satisfy
\begin{align} \label{eq: hmin via pguess}
    H_{\min}(A \vert E)_\rho=-\log p_{\text{guess}}(A \vert E)_\rho.
\end{align}} 

\noindent
\textbf{Proof.} For a proof of Theorem \ref{thm: Hmin pguess} the reader can check \cite{konig2009operational}. \\

The combination of Equations (\ref{eq: min entropy lower bound}) and (\ref{eq: hmin via pguess}) gives us a route to lower bound the conditional von Neumann entropy, which then can be input in the Winter-Devetak relation, synthesized by the maximization problem defined in Eq. (\ref{eq: pguess definition}). The main problem to perform such optimization is the fact that instead of a simple semi-definite program (in which the objective function depends linearly in matrix variables constrained by PSD relations) we have a rather more complex problem in which our objective function depends on a product of unknown matrices (shared states and measurements of Alice and Eve). This essentially constitutes a non-commutative polynomial optimization (NCPO) problem, which can't be exactly solved in a single optimization step, as is the case for SDPs. The standard way to approach polynomial optimization problems such as the one given by Eq. (\ref{eq: pguess definition}) is through the use of the NPA hierarchy \cite{navascues2007bounding} (whose formulation is explained in more detail in Appendix \ref{sec: NPA}). 

The idea of this method is that the NP-hard task of finding the global optimum of the polynomial function of matrices can be converted into a hierarchy os convergent SDPs. We can approximate the solution to the NCPO by the solution of a single SDP performed over a larger, \textit{relaxed}, feasible set. The \textit{level} of the hierarchy (an integer number $k\geq 1$, see Eq. (\ref{eq: relaxed NCPO})) parametrizes how relaxed is the set $\mathcal{Q}_k$, so that the set of commuting correlations is recovered in the limit 
\begin{align}
    \lim_{k\rightarrow \infty} \mathcal{Q}_k = \mathcal{Q}_{\mathcal{H}_{AB}}.
\end{align}
Higher levels of the hierarchy then provide better approximations, but are also more computationally costly as the dimension of the moment matrix increases. In some situations, low levels of the hierarchy are able to produce very good approximations of bounds for the NCPO optimal value. The complete optimization problem to be solved for the computation of $p_\text{guess}$ is written as follows. \\ 

\opt{ \textbf{(Polynomial optimization for $p_\text{guess}$ with tensor product structure \cite{masanes2011secure}) \label{thm: pguess tensor product optimization} } {Let $\rho \in \mathcal{H} = \mathcal{H}_A \otimes \mathcal{H}_B \otimes \mathcal{H}_E$ and denote by $\{M_{a_x\vert x}\}$, $\{N_{b_y\vert y}\}$} and $\{Z_{e}\}$ the respective POVM's in each party. Given lower bounds $\{w_j\}$ of a Bell polynomial $\mathcal{B}(\rho_{ABE},\{M_{a_x\vert x}\},\{N_{b_y\vert y}\})$, The guessing probability $p_\text{guess} (A\vert E)$ can be estimated with the non-commutative polynomial optimization problem characterized by}

\begin{subequations}
\begin{align}
\max_{\substack{\rho, \mathcal{H} \\ \{Z_a,M_{a_x\vert x}\}}} & \sum_a \langle M_{a_x \vert x} \otimes \mathbb{1} \otimes Z_{a}\rangle_{\rho}; \label{Eq:pguess_Masanes} \\
\text{s.t.} & \sum_a Z_a= \sum_a M_{a_x \vert x}=\sum_b N_{b_y \vert y}= \mathbb{1} & \forall \ x, y; \label{Eq:completeness_Masanes} \\
& Z_e \geq 0, \quad M_{a_x \vert x} \geq 0, \quad N_{b_y \vert y} \geq 0 & \forall \ a, b, x, y; \label{Eq:PSD_Masanes} \\
&\sum_{a b x y} c_{a b x y} \langle M_{a_x \vert x} \otimes N_{b_y \vert y} \otimes \mathbb{1}\rangle_\rho \geq w_j & \forall j \label{Eq:Bell_Masanes}.
\end{align}
\end{subequations}

The objective function in Eq. (\ref{Eq:pguess_Masanes}) corresponds to Eq. (\ref{eq: pguess definition}); The constraints of Eq. (\ref{Eq:completeness_Masanes}) and Eq. (\ref{Eq:PSD_Masanes}) correspond to completeness and positive semi-definiteness of the measurement operators of each party, while the constraint of Eq. (\ref{Eq:Bell_Masanes}) represents the violation of the Bell inequality by the statistics obtained in the test rounds of the protocol (discussed previously in Eq. (\ref{eq: bell inequality})).

An additional difficulty arising from the device-independence assumption of the protocol relies on algorithmically defining a tensor product structure between measurement operators acting in different parts of the experiment. As stated in Equations (\ref{Eq:pguess_Masanes}) - (\ref{Eq:Bell_Masanes}), the computation of the guessing probability is not itself a computationally tractable problem, as the NPA hierarchy relies solely on expectation values of products of matrices (being agnostic regarding the tensor composition of states and measurements). One must then replace the tensor product structure of Theorem \ref{thm: pguess tensor product optimization} by sets of measurements that commute when acting on different parties. 

\opt{\textbf{(Polynomial optimization for $p_\text{guess}$ with commuting operators)} \label{thm: pguess commuting problem}} {Let $\rho_{ABE}$ be a density operator acting a \textit{global} Hilbert space $\mathcal{H}_{ABE}$, and $\{M_{a_x\vert x}\}$, $\{N_{b_y\vert y}\}$} and $\{Z_{e}\}$ denote the respective POVM's of Alice, Bob and Eve. The guessing probability $p_\text{guess} (A\vert E)$ given lower bounds $\{w_j\}$ of a Bell polynomial $\mathcal{B}(\rho_{ABE},\{M_{a_x\vert x}\},\{N_{b_y\vert y}\})$ can be estimated with the non-commutative polynomial optimization characterized by

\begin{subequations}
\begin{align}
 \max_{\substack{\rho, \mathcal{H} \\ \{Z_a,M_{a_x\vert x}\}}} & \sum_a \langle M_{a_x \vert x} \cdot \mathbb{1} \cdot Z_{a}\rangle_\rho \label{Eq:pguess_Masanes_relaxed} \\
\text{s.t.} \hspace{0.45cm} & \sum_e Z_e= \sum_a M_{a_x \vert x}=\sum_b N_{b_y \vert y}= \mathbb{1} & \forall \ x, y ; \label{Eq:completeness_Masanes_relaxed} \\
& Z_a \geq 0, \quad M_{a_x \vert x} \geq 0, \quad N_{b_y \vert y} \geq 0 & \forall a, b, x, y ;\label{Eq:PSD_Masanes_relaxed} \\
& {\left[M_{a_x \vert x}, N_{b_y \vert y}\right]=\left[M_{a_x \vert x}, Z_{c}\right]=\left[N_{b_y \vert y}, Z_{c}\right]=0} & \forall \ a, b, x, y, c; \\
&\sum_{a b x y} c_{a b x y} \langle M_{a_x \vert x} \cdot N_{b_y \vert y} \cdot \mathbb{1}\rangle_\rho \geq w_i & \forall \ i \label{Eq:Bell_Masanes_relaxed}. 
\end{align}
\label{Eq:NCPO_Hmin_relaxed}
\end{subequations}

As every tensor product strategy trivially satisfies such commutation relations (but commuting sets of observables don't necessarily satisfy a tensor product decomposition), the replacement of tensor product by commutators is also a relaxation of the original problem. In other words, the set of correlations achievable under tensor product composition of subsystems, $\mathcal{Q}_{\mathcal{H}_A \otimes \mathcal{H}_B}$, is a strict subset of the commuting set of correlations $\mathcal{Q}_{AB}$ \cite{navascues2008convergent,navascues2007bounding,cabello2023logical}. As a consequence, the NPA hierarchy supplies an approximation for the commuting strategy guessing probability. An intuition for the approximations attained in this hierarchy of SDPs is provided pictorially in Fig. \ref{fig:NPA}.

As no assumption is made about the underlying dimension of the Hilbert space (potentially being infinite-dimensional), one may also assume the global state to be pure and the measurement operators to be projective \cite{holevo2019quantum}. It was also demonstrated in \cite{masanes2011secure} that the use of a single Bell inequality as a witness of nonlocality allows one to trace out Eve from the above optimization, making it considerably less computationally costly.

Furthermore, the method of using the min-entropy as a lower bound for the key rate is not only useful in device-independent situations: Under the assumption of fixed Hilbert spaces and known measurements, the $p_\text{guess}$ optimization can also often be used in fully characterized scenarios \cite{kanitschar2024practical} and 1SDI protocols \cite{branciard2012one,cavalcanti2016quantum}.

\subsection{Lower bounding device-independent key rates via Gauss-Radau expansions}
\label{sec: appendix gauss radau DI}

An alternative approach to evaluate DIQKD key rates was proposed by Brown et al. in \cite{brown2024device}, allowing improved values in comparison with the min-entropy method detailed in the previous section. The idea here is to lower-bound the von Neumann entropy with the Gauss-Radau expansion of the relative entropy (explained in \ref{sec: Araujo} in the context of fully characterized protocols), which can also be evaluated through the NPA hierarchy for DI scenarios.

The idea here is that, within a certain level of the NPA hierarchy, more terms in the Gauss-Radau expansion allow increasingly better rates. Therefore, the number of \textit{nodes} employed in the expansion appears as another parameter that (together with the hierarchy level) can increase the accuracy of the key rate lower bound, and in some simple situations can lead to results very close to the analytical prediction.

While this method is still characterized as a non-commutative polynomial optimization problem - therefore requiring the NPA hierarchy and being limited to scenarios with few parties, number of inputs or outputs, it is advantageous in comparison with the min-entropy approach as one can use the number $m$ of Gauss-Radau radau nodes (a parameter independent of the hierarchy level) to regulate the tightness of the approximation within a certain relaxation of the quantum set.

In the same spirit that the polynomial optimization provided in Theorem \ref{thm: pguess tensor product optimization} with tensor product structure was converted into a computationally solvable optimization with commuting constraints in a global Hilbert space (\ref{thm: pguess commuting problem}), below we state the polynomial optimization problem for the minimization stated in Eq. (\ref{eq: relative entropy tensor product}) using this same strategy:

\begin{subequations}
\begin{align}
c_m+ &\max_{\substack{\rho, \mathcal{H} \\ \{Z^{(\dagger)}_{a,i},M_{a_x\vert x}\}}} \sum_{i=1}^{m-1} \frac{w_i}{t_i \ln 2} \sum_a\langle M_{a_x \vert x^*}(Z_{a, i}+Z_{a, i}^\dagger+\left(1-t_i\right) Z_{a, i}^{\dagger} Z_{a, i})+t_i Z_{a, i} Z_{a, i}^{\dagger}\rangle_\rho, \label{eq: Brown objective function} \\
&\hspace{0.685cm}\text{s.t.}\hspace{0.69cm}  \sum_a M_{a_x \vert x}=\sum_b N_{b_y \vert y}= \mathbb{1} & \forall x, y; \\
&\hspace{1.9cm} M_{a_x \vert x} \geq 0, \quad N_{b_y \vert y} \geq 0 & \forall a, b, x, y; \\
&\hspace{1.9cm}Z_{a, i}^\dagger Z_{a, i} \leq \alpha_i, \quad Z_{a, i} Z_{a, i}^\dagger \leq \alpha_i & \forall a, i=1, \ldots, m-1; \\
&\hspace{1.9cm}{\left[M_{a_x \vert x}, N_{b_y \vert y}\right]=\left[M_{a_x \vert x}, Z_{b, i}^{(\dagger)}\right]=\left[N_{b_y \vert y}, Z_{a, i}^{(\dagger) }\right]=0} & \forall a, b, x, y, i; \\
 &\hspace{1.9cm}\sum_{a b x y} c_{a b x y} \langle M_{a_x \vert x} \cdot N_{b_y \vert y} \cdot \mathbb{1}\rangle_\rho \geq w_i & \forall i ;
\end{align}
\end{subequations}
where $c_m$ are constants related to the Gauss-Radau expansion and are given by $c_m=\sum_{i=1}^m \frac{w_i}{t_i \log 2}$.

In a more recent approach, another  expansion for the relative entropy \cite{frenkel2023integral} has been shown to allow the optimization of this functional under semi-definite constraints \cite{kossmann2024optimising}, in a similar logic to Eq. (\ref{eq: Brown objective function}). It was shown that it can outperforming the Gauss-Radau expansion technique \cite{kossmann2024reliable}.

\end{widetext}
\end{document}